\documentclass{article}
\usepackage{graphicx} % Required for inserting images

\usepackage{mathtools}

\usepackage{color,xcolor,ucs}
\usepackage{mathtools}   \usepackage{tikz} 
\usepackage{ amssymb }
\usepackage{extarrows} 
\usepackage{pgf,tikz}
\usepackage{float}
\usetikzlibrary{positioning}
\usetikzlibrary{shapes.geometric}
\usetikzlibrary{shapes.misc}
\usetikzlibrary{arrows}
\usepackage{caption}
\usepackage{mathrsfs}
\usetikzlibrary{arrows,shapes,automata,backgrounds,petri,positioning}
\usetikzlibrary{decorations.pathmorphing}
\usetikzlibrary{decorations.shapes}
\usetikzlibrary{decorations.text}
\usetikzlibrary{decorations.fractals}
\usetikzlibrary{decorations.footprints}
\usetikzlibrary{shadows}
\usetikzlibrary{calc}
\usetikzlibrary{spy}
\usepackage{amsmath}
\usepackage{array}
\usepackage{ amssymb }
\usepackage{braket}
\usepackage{qcircuit}
\usepackage{soul}
\usepackage{braket} 
\usepackage{relsize}

\usepackage{amsmath}
\usepackage{ amssymb }
\usepackage{braket}
\usepackage{qcircuit}
\usepackage{soul}
\usepackage{braket}

\title{{\LARGE Error correction, authentication, and false acceptance, probabilities for communication over noisy quantum channels: converse upper bounds on the bit transmission rate}}
\author{Pete Rigas \footnote{Newport Beach, CA 92625, pbr43@cornell.edu, ORCiD: 0009-0003-1053-9720}}
\date{}

\begin{document}

\noindent

\bigskip

\maketitle

\noindent \textbf{Abstract} We obtain strict upper bounds on the bit transmission rate for communication of Classical bit codewords over Quantum channels. Albeit previous arguments in arXiv: 1804.01797 which have demonstrated that lower bounds can be shown to hold for the bit transmission rate without the presence of significant noise over the channel shared by Alice and Bob for the purposes of encoding, decoding, transmission and authentication, the author suggests that upper bounding the bit transmission rate could be of use towards classifying paradoxical aspects of communication protocols, as well as constructing error correcting codes which are resilient to noise. The upper bound that is obtained in this work for the bit transmission rate, as a converse result, is dependent upon the natural logarithm of the size of each player's alphabet, as well as smaller alphabets, which can be leveraged for simultaneously realizing Quantum advantage for maximizing error correction and minimizing false acceptance. Crucially, the upper bound to the bit transmission rate is dependent upon a \textit{pruning} procedure, which seeks to determine whether letters from player's alphabets can be removed for maintaining prospective Quantum advantage, in order for Alice and Bob to implement error correction protocols with high probability, despite the fact that there is more noise over the channel between Alice and Bob in comparison to that between Bob and Eve. \footnote{\textbf{MSC Class}: 81P02; 81Q02}

\bigskip


\section{Introduction}

\subsection{Overview}

Quantum Information theory, as an extension of Classical Information theory, has attracted significant attention, with regards to dimensionality reduction, [10], which can be of significant use for inference and classification tasks related to Machine Learning, quantum embeddings, [19], exact, and approximate, optimality, in several game-theoretic settings [38,44,46], in addition to several closely related topics. In game-theoretic settings in which players, typically Alice and Bob for the 2-player case, can make use of entanglement, several proposed sources of Quantum advantage have been identified, and rigorously characterized. By and large, such proposed sources of Quantum advantage are reliant upon constrained optimization problems, formulated through semidefinite programs, which seek to determine primal feasible solutions from a partial ordering of the positive semidefinite cone. In other areas of Quantum Information theory, entropy, as a quantity which can impact the degree of optimality of quantum strategies, can be used to characterize communication of Classical information over a Quantum channel.

For such communication protocols, one must carefully consider the manner in which the Mutual Information, and Conditional Shannon, entropies, impact how a third paritipant, Eve, can tamper with information transmitted between Alice and Bob. The information-theoretic condition that is placed upon the Mutual Information entropy, and the conditional Shannon entropy, from some perspectives can be viewed as a variant of constrained optimization procedures that have previously been characterized extensively by the author in previous work for 2-player, and multiplayer, games [44,46]. As an optimization problem that is well-posed under the choice of some probability distribution that is separate from those introduced for the computation of the Mutual Information and conditional Shannon entropies, several parallels emerge with the formulation of primal feasible solutions for semidefinite programs initially in the two-player game-theoretic setting [37,44]. Despite the fact that the formulation of an optimization over the Mutual Information and conditional Shannon entropies is not dependent upon the partial ordering of the positive semidefinite cone as the semidefinite program is, each constrained optimization procedure raises implications about physical properties of inforation that are being measured. Albeit the fact that several efforts along such lines for performing tomography on Quantum systems have been employed, it continues to remain of interest to further build upon fundamental properties of Information-theoretic measures, beyond error bounds, upper bounds on systems of Frobenius norm inequalities, and approximality, as previously investigated by the author [44,46]. Such efforts not only focus upon potential sources of Quantum advantage for agents seeking to maximize his or her respective utility, but also instrinsic aspects of Quantum information that are not present in Classical information. 

To determine how the probabilities of false decoding, and false acceptance, of messages that players can accept over a Quantum channel of communication depend on the number of codewords, and entanglement, of the message, we seek to determine how Mutual Information, and conditional Shannon, entropies behave under the choice of different probability distributions that are optimized over. Regardless, as an optimization problem, one nevertheless encounters limitations of classical information, particularly through classical-quantum gaps which determine potential sources of advantage, and potential classes of strategies, for players that are seeking to maximize the winning probably for some game. 

For communication between two, or even more, players of a noisy quantum channel, one must take into account the \textit{transmission rate} between players, in addition to the potential action of a third nefarious actor, Eve, who is attempting to evesdrop on the communication, and hence the content of the message, that is being shared between the two players. Such a transmission rate over the potentially noisy quantum channel was shown to have a strict upper bound in terms of a minimum of the Mutual Information, and conditional Shannon, entropies, as provided in [38], which reflects upon previously introduced constrained optimization problems, also taken over probability distributions, for the capacity of quantum channels [38]. As one potential direction of future research interest, the author of [38] raised the possibility of more extensively understanding whether a converse result would hold for the transmission rate over the noisy quantum channel, namely whether the upper bound for the transmission rate between the two players could in fact, under different circumstances, constitute a lower bound for the transmission rate over the quantum channel. To further characterize a converse result on the lower bound to the transmission rate, we elaborate upon the intuition provided in [38] for conditions that one may hope to place on the converse result of the transmission rate. In a simplified example provided in [38], the author discusses how the lower bound for the transmission rate, from which Alice and Bob can hope to simultaneously achieve error correction and authentication. With Alice having an alphabet consisting of two letters only, when communicating with Bob and Eve an expression for transition probabilities between Alice's alphabet is expressed in terms of the elements of the alphabets of Bob and Eve, respectively. Paradoxically, properties of the quantum channel between Alice and Bob are more noisy than the channel between Bob and the evesdropper Eve; moreover, to characterize the range of possible upper bounds for the transmission rate over the noisy quantum channel between Alice and Bob, the players can still continue to achieve higher probabilities of being able to not only authenticate some message received from the other player over the channel, but also execute error correction scheme. Error correction schemes ultimately determine whether quantum algorithms can provide polynomial, or, far less realistically, exponential, speedup over classical counterparts [45], with potential adaptations to a wide variety of other fields [1,2,3,4,6,7,16,19,20,21,22,24,25,26,27,29,30,31,33,39,47,48,49,50,51]. Albeit the fact that some prospects for quantum advantage have emerged in several industrial applications, [9,34,35,36,40,41,42], determining whether such advantages continue to remain in the presence of noise is paramount. In two-player, and multiplayer, game-theoretic settings, several perspectives, [5,8,11,12,13,14,15,17,23,28,32,37], have been previously analyzed in the author from a work of this year, [46]. Such efforts not only seek elaborate upon the set of all possible transformations that can be performed on tensor product operators, but also upon a combinatorially larger space of possible optimal strategies.

Under other assumptions that one can place on quantum channel communication, irrespective of how noisy the channel between two players is, proposals of quantum advantage are still expected to hold. Specifically, while the presence of noise in a quantum channel can impact the probability of a player who is sending, and receiving, a message over the channel, additional communication protocols depending upon the transmission rate, which inherently relies on various forms of entropy, can be further described. The description of possible upper bounds on the transmission rate, as a function of the mutual information and conditional Shannon entropies, further elaborates upon unexpected sources of quantum advantage in the presence of noise. While noise can also negatively impact the ability of a player to detect whether he or she receives an encoding that has been tampered with, at the same time it permits for Alice and Bob to maximize the rates at which error detection and authentication can simultaneously be obtained, over the channel from Bob to Eve. Altogether, the presence of noise may not end up imposing as much of a deterrence on communication over quantum channel as one would expect.

In the next section, we recapitulate the primary game-theoretic notions previously manipulated by the author in [46]. Introducing, and reiterating, the importance of such notions is significant for illustrating how the referee can help players, Alice and Bob, be able to communicate realiably over a quantum channel for maximizing error correction and authentication. Despite the fact that secure communication between Alice and Bob is not initially described with a referee being present, as in multiplayer game-theoretic settings, [46], and extensions of various two-player settings, [37,44], the referee can determine whether the ability of players to error correct, and authenticate, messages that are shared over the quantum channel is degraded. Irrespective of the presence of noise over the quantum channel, and hence the states that are distributed for communication across the channel, Alice and Bob can participate in enhanced communication protocols. In the absence of classical information, such protocols can depend upon entanglement, error bounds and representation-theoretic intertwiners. Furthermore, in the presence of noise, the maximum probability of a player winning the game, denoted throughout the literature with the optimal value $\omega$, can be related to the probability of a player correctly authenticating a message sent over a quantum channel which occurs with probability $0$ or $1$. With regard to the ability of players to perform error correction across the channel, the referee can also be a useful asset for players for the following reasons: (1) determining the sequence of transformations that the adversarial player, Eve, can apply to messages communicated between Alice and Bob over the channel; (2) minimizing the probability of false acceptance of a message that has been adversarially tampered; (3) distinguishing between changes to the Mutual Information, and conditional Shannon, entropies before, and after, the eavesdropper Eve intervenes.

The bit transmission rate, $r$, over the quantum channel being dependent upon the Mutual Information and conditional Shannon, entropies simultaneously informs the possible upper bounds that are expected to hold in the Converse result (namely, the inequality that one obtains for the bit transmission rate from the minimziation of the Mutual Information and conditional Shannon entropies). Depending upon which contribution plays the least role in a minimization between the Mutual Information, or conditional Shannon, entropy, the bit transmission rate also depends on the capacity of the quantum channel. The capacity of classical, and quantum, channels alike not only determines the complexity of the information that Alice and Bob can share, but also the expected number of ways in which the adversary Eve can corrupt shared information encoded in the message. In the presence of the referree, having an additional party examine messages that are distributed across the channel for secure communication can be accounted for with similar classes of scoring functions which are used for determining whether player's individual responses constitute a winning, and hence optimal, strategy. Deviations from optimal strategies, in several game-theoretic settings alike, are captured with a multiplicative factor, $1-\epsilon$, for $\epsilon$ taken to be sufficiently small depending upon the number of players in the game. With several considerations of Quantum information, and transmission of Quantum information, expressed over the past few pages, we introduce the following objects surrounding communication over quantum channels.

\subsection{This paper's contributions}

\subsubsection{The broadcast channel model}

\noindent We describe the broadcast channel model before providing a description of the contributions of this paper. Such a model is dependent upon the encoding of classical bits that are distributed through a discrete, memoryless system that is characterized through the probability transition matrix,

{\small \begin{align*}
    \textbf{P} {\tiny \big[ y , z \big| x \big] } 
, \end{align*} }

\noindent where $y,z$ denote the inputs that are received by Bob and Eve, respectively, $x$ denotes the initial bit of information that is transmitted over the channel by Alice and $ \textbf{P} {\tiny\big[ \cdot \big]}$ denotes the probability measure over bits of information manipulated by Alice, Bob and Eve. Through her choice of a suitable encoding, Alice is not only capable of distributing information across the broadcast channel securely to Bob if the noise across her shared channel with Bob is less noisy than the channel shared between Bob and Eve, but also of preventing Eve form passing off messages that she has tampered with to Bob from their shared channel. To quantitatively characterize how the physical noise asymmetry is related to expressions for the bit transmission rate and its converse discussed in the next section we demonstrate how the noise gap between Alice and Bob's channel, and Bob and Eve's channel, impacts the error correction, authentication and false acceptance probabilities. When Classical information is distributed across a Quantum channel Eve encounters difficulties associated with: performing measurements; the fact that her superior noise level, in comparison to those of Alice and Bob across their shared channel, is inevitably degraded as she attempts to decode information from Alice's encoding; the manner in which the transmission mechanism of information across the channel enforces physical constraints on Eve's ability to eavesdrop, hence impacting the probabilities that Bob can error correct information received from Alice or Eve, in addition to minimizing his false acceptance probability as much as possible. 

\subsubsection{Description}

Examining the circumstances under which Classical information can be transmitted, and authenticated, over Quantum channels is of great interest to explore. Under the context of Classical bit information being transmitted over a Quantum channel, previous works have identified prospects of Quantum advantage, particularly from the perspective of Alice and Bob being able to securely transmit information within an eavesdropper, Eve, tampering with the bit codewords underlying their communication. Unexpectedly, other sources of proposed Quantum advantage have been identified, [38], particularly in the presence of noise, which primarily revolve around the existence of resilient communication protocols, under the assumption,

{\small \begin{align*}
N_{A \longleftrightarrow B}  >  N_{B \longleftrightarrow E}     \text{, }
\end{align*}
 } 

\noindent for the noise thresholds of the channels,

{\small \begin{align*}
N_{A \longleftrightarrow B} \equiv    \text{Noise over the Quantum channel between Alice and Bob}        \text{,} \\ \\  N_{B \longleftrightarrow E} \equiv       \text{Noise over the Quantum channel between Bob and Eve} \text{. }
\end{align*} }

\noindent For transmission over a Quantum channel, denote $\textbf{P} {\tiny \big[ \cdot \big]}$ as the probability measure over bit codewords that Alice, Bob, and Eve, receive. Bit transmission rates, $r$, over the Quantum channel, which were shown to have lower bounds of the form, [38],

{\small \begin{align*}
r < \underset{P_{\textbf{X}}}{\mathrm{sup}}  \big\{ \mathrm{min} \big\{  I {\tiny \big(} \textbf{X} , \textbf{Y} {\tiny \big)}  , \underset{z}{\mathrm{min}}  \big\{ H_Q {\tiny \big(} \textbf{Y} \big| \textbf{Z} = z {\tiny \big)}   -  H_P {\tiny \big(}  \textbf{Y} \big| \textbf{X} {\tiny \big)}  \big\}  \big\}    \big\} \text{, }
\end{align*} }

\noindent from analyses of the decoding error, and false acceptance, probabilities, given the Mutual Information entropy, $I {\tiny \big( \cdot \big)}$, and conditional Shannon entropy, $H {\tiny \big( \cdot \big| \cdot \big)}$. Such probabilities, qualitatively speaking, are not only employed to quantify the complexity of bit codewords that Alice and Bob can transmit over the Quantum channel, but also the capacity of Quantum channels, per channel use. For converse results on the bit transmission rate, which would be of the form,

{\small \begin{align*}
   r >  \underset{P_{\textbf{X}}}{\mathrm{sup}}   \big\{ \mathrm{min} \big\{  I {\tiny \big(} \textbf{X} , \textbf{Y} {\tiny\big)}  , \underset{z}{\mathrm{min}}  \big\{ H_Q {\tiny  \big( }  \textbf{Y} \big| \textbf{Z} = z {\tiny \big)}    -  H_P {\tiny \big( } \textbf{Y} \big| \textbf{X} {\tiny \big) }   \big\}    \big\}   \big\}     \text{. }
\end{align*} } 

\noindent From an extremely simple example provided in [38], namely with Alice's alphabet consisting of two letters, Bob's alphabet consisting of three letters, and Eve's alphabet consisting of the same two letters in Alice's alphabet, Quantum advantage, particularly through the stochastic domination,

{\small \begin{align*}
   p_{\text{Error correction}, A \longleftrightarrow B} >    p_{\text{Error correction}, B\longleftrightarrow E}  \text{,} \\    p_{\text{Authentication}, A \longleftrightarrow B} >    p_{\text{Authentication}, B\longleftrightarrow E} \text{, }
\end{align*}    
} 

\noindent was identified, in spite of the fact that,

{\small \begin{align*}
N_{A \longleftrightarrow B}  >  N_{B \longleftrightarrow E}   \text{, }
\end{align*}
} 

\noindent for the probabilities,

{\small \begin{align*}
    p_{\text{Error correction}, A \longleftrightarrow B} \equiv p_{\mathrm{EC}, A \longleftrightarrow B} \text{,} \\ p_{\text{Error correction}, B \longleftrightarrow E} \equiv p_{\mathrm{EC}, B\longleftrightarrow E} \text{,} \\ p_{\text{Authentication}, A \longleftrightarrow B}  \equiv p_{\mathcal{A}, A \longleftrightarrow B}  \text{,} \\  p_{\text{Authentication}, B \longleftrightarrow E}  \equiv p_{\mathcal{A}, B \longleftrightarrow E}  \text{, }
\end{align*} }

\noindent which, over infinitely many bit codewords, take the form,

{\small \begin{align*}
   p_{\text{Error correction}, A \longleftrightarrow B}   \equiv         \underset{n \longrightarrow + \infty}{\mathrm{lim}} \bigg\{  \underset{\mathrm{ec} \in \mathrm{EC}}{\bigcup} \textbf{P} \big[ \text{Alice and Bob implement } n\text{-bit error correcting codes ec over} \\ \text{ the channel between them, } A \longleftrightarrow B      \big]    \bigg\}        \text{,} \\  \\   p_{\text{Error correction}, B\longleftrightarrow E}  \equiv \underset{n \longrightarrow + \infty}{\mathrm{lim}}   \bigg\{ \underset{\mathrm{ec} \in \mathrm{EC}}{\bigcup}        \textbf{P} \big[    \text{Bob and Eve implement } n\text{-bit error correcting codes ec over} \\  \text{the channel between them, } B \longleftrightarrow E             \big]    \bigg\}    \text{,} \end{align*}

   \begin{align*}  p_{\text{Authentication}, A \longleftrightarrow B}   \equiv         \underset{n \longrightarrow + \infty}{\mathrm{lim}} \bigg\{ \underset{A \in \mathcal{A}}{\bigcup} \textbf{P} \big[       \text{Alice and Bob implement } n\text{-bit authentication protocols, A, over} \\ \text{the channel between them, } A \longleftrightarrow B        \big]  \bigg\}    \text{,} \\ \\ p_{\text{Authentication}, B\longleftrightarrow E}  \equiv  \underset{n \longrightarrow + \infty}{\mathrm{lim}} \bigg\{   \underset{A \in \mathcal{A}}{\bigcup}          \textbf{P} \big[   \text{Bob and Eve implement } n\text{-bit authentication protocols, A, over} \\ \text{the channel between them, } B \longleftrightarrow E    \big]  \bigg\}   \text{. }
\end{align*}     }

\noindent In forthcoming arguments, $\mathcal{A}$ is used to define the authenticated spaces of codewords, whether with finite or infinitely many bits, of Alice, Bob and Eve. Over $A \longleftrightarrow B$, and $B \longleftrightarrow E$, we denote such spaces of codewords with,

{\small \begin{align*}
 \mathcal{A}^{rn}_{A \longleftrightarrow B}   \text{,  } \\\mathcal{A}^{rn}_{B \longleftrightarrow E}  \text{. }
 \end{align*} } 

\noindent Determining the circumstances under which resilient communication protocols can be enacted by Alice, and Bob, in the presence of Eve is significant for Information processing tasks. In providing arguments for obtaining one possible class of upper bounds on $r$, through the converse result which provides upper bounds on the bit transmission rate, Alice and Bob can realize proposed sources of Quantum advantage, albeit Eve's attempt to corrupt codewords transmitted between them.

In the case of alphabets for each player consisting of a few letters, in the presence of a larger noise threshold over Alice and Bob's channel, in comparison to that over Bob and Eve's channel, the overlap function,

{\small \begin{align*}
   \mathscr{O} {\tiny \big(} \textbf{X}, \textbf{Y}, \textbf{Z} {\tiny \big) }  \equiv  \underset{ x \in \textbf{X}, y \in \textbf{Y}, z \in  \textbf{Z}}{\bigcup} \mathscr{O} {\tiny \big( x, y, z \big) }  \text{, }
\end{align*} }

\noindent given the alphabets $\textbf{X}$, $\textbf{Y}$, and $\textbf{Z}$, of Alice, Bob and Eve, respectively, determines whether Eve uses any symbols from Alice's, or Bob's, alphabets. Namely, if,

{\small \begin{align*}
   \mathscr{O} {\tiny \big( } \textbf{X}, \textbf{Y}, \textbf{Z} {\tiny \big) }  \equiv \emptyset    \text{, }
\end{align*} } 

\noindent necessarily,

{\small \begin{align*}
  \bigg\{ {\tiny \mathscr{O} \big(}  \textbf{X}, \textbf{Y}, \textbf{Z} {\tiny \big) \equiv \emptyset}   \bigg\}  \Longleftrightarrow  \bigg\{ \underset{z \in \textbf{Z}}{\bigcap} \bigg\{            {\tiny  \big\{   z \big\}  }  \cap \bigg\{ \underset{y \in \textbf{Y}}{\bigcap} {\tiny \big\{   y \big\}  \cap \big\{}  \textbf{X} {\tiny \big\} }  \bigg\}  \bigg\}     \equiv \emptyset  \bigg\} \Longleftrightarrow  \bigg\{ \underset{z \in \textbf{Z}}{\bigcap} \bigg\{            {\tiny \big\{   z \big\}  }  \cap \bigg\{ \underset{x \in \textbf{X}}{\bigcap} {\tiny \big\{}   \textbf{Y} {\tiny \big\}  \cap \big\{  x \big\} }  \bigg\}  \bigg\}   \equiv \emptyset \bigg\}  \text{. }
\end{align*}
}

\noindent Despite the fact that the condition placed above on the overlap function, $\mathscr{O}$, is useful for anticipating whether pronounced Quantum advantage exists for Alice and Bob for the converse result,

{\small \begin{align*}
   r > \underset{P_{\textbf{X}}}{\mathrm{sup}}  \big\{ \mathrm{min} \big\{  I {\tiny \big(} \textbf{X} , \textbf{Y} {\tiny \big)}  , \underset{z}{\mathrm{min}}  \big\{ H_Q {\tiny \big(} \textbf{Y} \big| \textbf{Z} = z {\tiny \big)}   -  H_P {\tiny \big(}  \textbf{Y} \big| \textbf{X} {\tiny \big)}  \big\}  \big\}    \big\}    \text{, }
\end{align*}
 }

\noindent on the bit transmission rate, determining the cardinality of the alphabets for Alice, Bob, and Eve, namely the strictly positive threshold $c^{*}$, so that,

{\small \begin{align*}
   \mathscr{O} {\tiny \big( } \textbf{X}, \textbf{Y}, \textbf{Z} {\tiny \big) \equiv c^{*} }   \text{, }
\end{align*}
 }

\noindent can be used to characterize prospective Quantum advantage, specifically as to whether the probability of Alice and Bob performing error correction, and authentication, is extremely likely, from the set of conditions,

{\small \[
\left\{\!\begin{array}{ll@{}>{{}}l} 
  \big\{  \mathscr{O} {\tiny \big( } \textbf{X}, \textbf{Y}, \textbf{Z} {\tiny \big) \equiv c^{*} }  \big\}  \Longleftrightarrow   \big\{ N_{A\longrightarrow B} > N_{B\longrightarrow E} \big\}  \text{, }  \\ \\ \big\{   \mathscr{O} {\tiny \big( }  \textbf{X}, \textbf{Y}, \textbf{Z} {\tiny \big) > c^{*} }  \big\}  \Longleftrightarrow   \big\{ N_{A\longrightarrow B} <  N_{B\longrightarrow E}  \big\}   \text{. }
\end{array}\right.
\] }

\noindent For the converse result on the bit transmission rate, we determine whether any thresholds $c^{*}$ exist so that, given upper bounds on $r$, and some assumption on the total number of letters which Alice and Bob can use, so that the probability of false acceptance, and authentication, each occur with a very high likelihood.

Given this possible range of behavior relative to the noise thresholds, $N_{A\longrightarrow B}$ and $N_{B\longrightarrow E} $ of each Quantum channel, with respect to each noise threshold, the previous arguments have established, with high probability (whp), that Alice and Bob can map bit codewords into the authenticated space of their channel, $A \longleftrightarrow B$, $\mathcal{A}^{rn}_{A \longleftrightarrow B}$. In comparison to a previous result provided for mapping into the authenticated space that is provided in [38], one will notice that the bit transmission rate satisfies,

{\small \begin{align*}
 r < h \big( q \big) - h \big( p \big)    \text{, }
\end{align*} }

\noindent whereas the bit transmission rate, in \textbf{Theorem} \textit{3} provided in \textit{1.4}, satisfies,

{\small \begin{align*}
 r >  h \big( q \big) - h \big( p \big)    \text{. }
\end{align*} } 

\noindent This fundamentally divergent assumption on the bit transmission rate underlies the paradoxical, and unexpected, aspects of Quantum information which can allow Alice and Bob to not sacrifice error correction, and authentication, probabilities over $A \longleftrightarrow B$, albeit the fact that $N_{A \longleftrightarrow B} > N_{B \longleftrightarrow E}$. Moreover, the underlying aspects of Quantum information which guarantee that, as proposed sources of Quantum advantage, Alice and Bob can still execute error correction, and authentication, protocols whp, is closely related to the following series of observations, and questions:

\begin{itemize}
    \item[$\bullet$] \textit{Nonlocality and contextuality}. Information processing tasks making use of entanglement are known to satisfy nonlocality properties, which ultimately stems from the fact that the universe itself is locally not real. To this end, what other strict upper bounds on the bit transmission rate across Alice and Bob's shared Quantum channel, continue permitting for them to execute error correction, and authentication, protocols whp?
    \item[$\bullet$] \textit{Eve's attempts to send corrupted messages as authenticated ones to Alice's and Bob's shared Quantum channel}. The eavesdropper, Eve, attempts to corrupt bit codewords that Alice or Bob shares with the other player. How might Eve, to increase her probability of simultaneously corrupting a message into Alice's and Bob's authenticated space, determine which letters of Alice's, or Bob's, alphabets, have already been used for transmission?
    \item[$\bullet$] \textit{Equal noise thresholds across the Quantum channels shared between Alice and Bob, and between Bob and Eve}. What proposed sources of Quantum advantage would one expect to have when $N_{A \longleftrightarrow B} \equiv N_{B\longleftrightarrow E}$, in comparison to $N_{A \longleftrightarrow B} > N_{B\longleftrightarrow E}$?
\end{itemize}

\subsection{Information-theoretic objects}

For the following objects corresponding to communication and transmission of classical bits over a symmetric, binary, Quantum channel, introduce $0 \leq p < q \leq \frac{1}{2}$. For communication over a quantum channel, identical players Alice and Bob, can continue to share entangled information with each other as they have in previous game-theoretic objects. However, the aspects of entanglement appearing in the information that the players share can be interfered with from an additional player, Eve. As the eavesdropper to communication over the quantum channel, Eve seeks to maximize her utility function for successfully intercepting communication between Alice and Bob. She can perform such a task by decreasing the probability of Alice and Bob to perform error correction,

{\small \begin{align*}
P_{\mathrm{EC}} \equiv         \underset{\mathrm{ec} \in \mathrm{EC}}{\mathrm{sup}}    p_{\mathrm{ec}} \equiv          \mathrm{sup} \big\{ \text{success probability of a player using an error correcting code} \big\}               \text{, }
\end{align*}
 } 

\noindent is represented as a supremeum over all possible encodings $\mathrm{ec}$ over the set of all error correcting codes $\mathrm{EC}$,

{\small \begin{align*}
  \mathrm{EC} \equiv \underset{\mathrm{ec}}{\bigcup}  \big\{ \text{error correcting codes ec over a Quantum channel} \big\} \text{, }
\end{align*}
 } 

\noindent while increasing the probability of false acceptance,

{\small \begin{align*}
P_{\mathrm{FA}} \equiv          \underset{\mathrm{fa} \in \mathrm{FA}}{\mathrm{inf}}    p_{\mathrm{fa}} \equiv           \mathrm{inf} \big\{ \text{failure probability of a player accepting a message that should have not been} \\ \text{accepted} \big\}              \text{, }
\end{align*}
 } 

\noindent for the set of all instances of false acceptance over a Quantum channel,

{\small \begin{align*}
  \mathrm{FA} \equiv \underset{\mathrm{fa}}{\bigcup}  \big\{ \text{instances of false acceptance, fa, over a Quantum channel} \big\} \text{, }
\end{align*}
 }

\noindent is represented as a supremum over all possible false acceptance encodings that Alice, or Bob, can use for accepting a message that has been tampered with by Eve. First, denote $\textbf{X}$, and $\textbf{Y}$, as the alphabets for Alice, and Bob, respectively. To discuss upper, and lower, bounds on the bit transmission rate, denote, additionally,

{\small \begin{align*}
  X, Y \equiv \text{Two realizations of random variables from Alice's,} \\ \text{and Bob's, alphabets}  \text{, } \end{align*}

  \begin{align*} \textbf{X} \equiv \text{Collection of random variables } X  \text{ from Alice's alpha-} \\ \text{bet}       \text{, }  \end{align*}

  \begin{align*} \textbf{Y} \equiv \text{Collection of random variables } Y \text{ from Bob's alpha-} \\ \text{bet} \text{, }  \end{align*}

  \begin{align*} P_X \equiv \text{Probability measure on } X \text{, }  \end{align*}

  \begin{align*}
  P_{\textbf{X}} \equiv \underset{X \in \textbf{X}}{\bigcup}   P_X \equiv \text{Probability measures on } \textbf{X}  \text{, } \end{align*}
  
  \begin{align*}  P_Y \equiv \text{Probability measure on } Y \text{, }  \end{align*}

  \begin{align*} P_{\textbf{Y}} \equiv \underset{Y \in \textbf{Y}}{\bigcup}   P_Y \equiv \text{Probability measures on } \textbf{Y}  \text{, }  \end{align*}

  \begin{align*}  P_{X,Y} \equiv \text{Joint probability measure on } X, Y \text{, } \end{align*}

  \begin{align*}    P_{\textbf{X},\textbf{Y}} \equiv \text{Joint probability measures on } \textbf{X}, \textbf{Y} \equiv \underset{X \in \textbf{X}, Y \in \textbf{Y}}{\bigcup}  P_{X,Y} \text{. }           
\end{align*} }

\noindent The bit transmission rate, $r$, which determines the physical limit of how much information can be shared in an entangled information over a quantum channel, was shown to satisfy, [38],

{\small \begin{align*}
r < \underset{P_{\textbf{X}}}{\mathrm{sup}}  \big\{ \mathrm{min} \big\{  I {\tiny \big(} \textbf{X} , \textbf{Y} {\tiny \big)}  , \underset{z}{\mathrm{min}}  \big\{ H_Q {\tiny \big(} \textbf{Y} \big| \textbf{Z} = z {\tiny \big)}   -  H_P {\tiny \big(}  \textbf{Y} \big| \textbf{X} {\tiny \big)}  \big\}  \big\}    \big\} \text{, }
\end{align*} }

\noindent for the Mutual Information entropy, $I {\tiny \big( \cdot \big)}$, and conditional Shannon entropy, $H {\tiny \big( \cdot \big| \cdot \big)}$, and probability measures $P$, and $P_X$. The converse result for the bit transmission rate would entail,

{\small \begin{align*}
   r >  \underset{P_{\textbf{X}}}{\mathrm{sup}}  \big\{ \mathrm{min} \big\{  I {\tiny \big(} \textbf{X} , \textbf{Y} {\tiny \big)}  , \underset{z}{\mathrm{min}}  \big\{ H_Q {\tiny \big(} \textbf{Y} \big| \textbf{Z} = z {\tiny \big)}   -  H_P {\tiny \big(}  \textbf{Y} \big| \textbf{X} {\tiny \big)}  \big\}  \big\}    \big\}   \text{. }
\end{align*}
 }

\noindent To characterize potential upper bounds given the converse result on the bit transmission rate, one must consider upper bounds of the following form, for sufficiently small,

{\small \begin{align*}
\underset{P_{\textbf{X}}}{\mathrm{sup}}  \big\{ \mathrm{min} \big\{  I {\tiny \big(} \textbf{X} , \textbf{Y} {\tiny \big)}  , \underset{z}{\mathrm{min}}  \big\{ H_Q {\tiny \big(} \textbf{Y} \big| \textbf{Z} = z {\tiny \big)}   -  H_P {\tiny \big(}  \textbf{Y} \big| \textbf{X} {\tiny \big)}  \big\}  \big\}    \big\}
\end{align*}
\[ <  \left\{\!\begin{array}{ll@{}>{{}}l} 
     \mathrm{log} \text{ }  \mathrm{log} \bigg[   {  \mathrm{log} \big| \textbf{Y}^{*} \big|   } \big/ { \big| \textbf{X}^{*} \big|  }          \bigg]   +   \mathrm{log}  \bigg[ {   \mathrm{log}\big|  \textbf{Z}  \big|  } \big/ {  \big|  \textbf{Y}^{*}  \big|   } \bigg]  \Longleftrightarrow \big| \textbf{X} \big| > \big| \textbf{Y}^{*} \big| ,  \big| \textbf{Y}^{*} \big| > \big| \textbf{Z} \big|    ,  \\    \\ \mathrm{log}  \text{ }  \mathrm{log} \bigg[   {  \mathrm{log} \big| \textbf{X} \big|  } \big/ { \big| \textbf{Y}^{*} \big| }          \bigg]   +   \mathrm{log}  \bigg[  { \mathrm{log} \big|  \textbf{Y}^{*} \big|   } \big/ {  \big| \textbf{X} \big|   } \bigg]   \Longleftrightarrow \big| \textbf{X} \big| <  \big| \textbf{Y}^{*} \big| ,  \big|  \textbf{Y}^{*} \big| < \big| \textbf{Z} \big|        ,  \\ \\ \mathrm{log} \text{ }   \mathrm{log} \bigg[    {  \mathrm{log} \big| \textbf{Y}^{*} \big|  } \big/ { \big| \textbf{X}^{*} \big| }          \bigg]   +   \mathrm{log}  \bigg[  {  \mathrm{log}\big|  \textbf{Y}^{*} \big|   } \big/ { \big| \textbf{X} \big|   } \bigg]   \Longleftrightarrow \big| \textbf{X} \big| >   \big| \textbf{Y}^{*} \big| ,  \big|  \textbf{Y}^{*} \big| < \big| \textbf{Z} \big|        ,   \\   \\ \mathrm{log} \text{ }   \mathrm{log} \bigg[   {  \mathrm{log} \big| \textbf{Y}^{*} \big|  } \big/ {\big| \textbf{X}^{*} \big| }          \bigg]   +   \mathrm{log}  \bigg[  {  \mathrm{log}\big|  \textbf{Z}  \big|   } \big/ { \big| \textbf{Y}^{*} \big|   } \bigg]   \Longleftrightarrow \big| \textbf{X} \big| <    \big| \textbf{Y}^{*} \big| ,  \big|  \textbf{Y}^{*} \big| >   \big| \textbf{Z} \big|        .                          
\end{array}\right. \equiv  r  \\ \\ \tag{*} 
\]  }

\noindent For convenience, denote,

{\small 
\[   \left\{\!\begin{array}{ll@{}>{{}}l} 
  r_1 \equiv    \mathrm{log} \text{ }   \mathrm{log} \bigg[   {  \mathrm{log} \big| \textbf{Y}^{*} \big|   } \big/ { \big| \textbf{X}^{*} \big|  }          \bigg]   +   \mathrm{log}  \bigg[ {   \mathrm{log}\big|  \textbf{Z}  \big|  } \big/ {  \big|  \textbf{Y}^{*}  \big|   } \bigg]  \Longleftrightarrow \big| \textbf{X} \big| > \big| \textbf{Y}^{*} \big| ,  \big| \textbf{Y}^{*} \big| > \big| \textbf{Z} \big|    ,  \\    \\ r_2 \equiv \mathrm{log} \text{ }   \mathrm{log} \bigg[   {  \mathrm{log} \big| \textbf{X} \big|  } \big/ { \big| \textbf{Y}^{*} \big| }          \bigg]   +   \mathrm{log}  \bigg[  { \mathrm{log} \big|  \textbf{Y}^{*} \big|   } \big/ {  \big| \textbf{X} \big|   } \bigg]   \Longleftrightarrow \big| \textbf{X} \big| <  \big| \textbf{Y}^{*} \big| ,  \big|  \textbf{Y}^{*} \big| < \big| \textbf{Z} \big|        ,  \\ \\ r_3 \equiv \mathrm{log} \mathrm{log} \bigg[    {  \mathrm{log} \big| \textbf{Y}^{*} \big|  } \big/ { \big| \textbf{X}^{*} \big| }          \bigg]   +   \mathrm{log}  \bigg[  {  \mathrm{log}\big|  \textbf{Y}^{*} \big|   } \big/ { \big| \textbf{X} \big|   } \bigg]   \Longleftrightarrow \big| \textbf{X} \big| >   \big| \textbf{Y}^{*} \big| ,  \big|  \textbf{Y}^{*} \big| < \big| \textbf{Z} \big|        ,   \\   \\ r_4 \equiv  \mathrm{log}  \text{ }  \mathrm{log} \bigg[   {  \mathrm{log} \big| \textbf{Y}^{*} \big|  } \big/ {\big| \textbf{X}^{*} \big| }          \bigg]   +   \mathrm{log}  \bigg[  {  \mathrm{log}\big|  \textbf{Z}  \big|   } \big/ { \big| \textbf{Y}^{*} \big|   } \bigg]   \Longleftrightarrow \big| \textbf{X} \big| <    \big| \textbf{Y}^{*} \big| ,  \big|  \textbf{Y}^{*} \big| >   \big| \textbf{Z} \big|        .                          
\end{array}\right.  \\ 
\]  }

\noindent To ensure that the above closed form representation for the converse bit transmission rate is well defined (and, necessarily, a strictly positive quantity), we characterize the following possible scenarios concerning the asymptotic growth of the doubly logarithmic and logarithmic terms appearing in the above expression for $r$:

\begin{itemize}

    \item[$\bullet$] \underline{(1)}: \textit{Divergence of $r_1$ to $\pm \infty$}. For the bit transmission rate $r_1$ if:

\begin{itemize}

\item[$\bullet$] \textit{(1A):} Divergence of ${\small
{  \mathrm{log} \big| \textbf{Y}^{*} \big|   } \big/ { \big| \textbf{X}^{*} \big|  }    }$ (resp. ${\small
{  \mathrm{log} \big| \textbf{Z} \big|   } \big/ { \big| \textbf{Y}^{*} \big|  }    }$), as $\big| \textbf{X}^* \big| \longrightarrow 0^+$ (resp. $\big| \textbf{Y}^* \big| \longrightarrow 0^+$),

{\small \begin{align*}
  \mathrm{log}  \text{ }  \mathrm{log} \bigg[   {  \mathrm{log} \big| \textbf{Y}^{*} \big|   } \big/ { \big| \textbf{X}^{*} \big|  }          \bigg]  \longrightarrow - \infty  \bigg( \text{resp. }   \mathrm{log} \bigg[   {  \mathrm{log} \big| \textbf{Z} \big|   } \big/ { \big| \textbf{Y}^{*} \big|  }          \bigg]  \longrightarrow - \infty   \bigg) , 
\end{align*} }

\item[$\bullet$] \textit{(1B):} Divergence of ${\small
{  \mathrm{log} \big| \textbf{Y}^{*} \big|   } \big/ { \big| \textbf{X}^{*} \big|  }    }$ (resp. ${\small
{  \mathrm{log} \big| \textbf{Z} \big|   } \big/ { \big| \textbf{Y}^{*} \big|  }    }$), as $\big| \textbf{Y}^* \big| \longrightarrow +\infty$ (resp. $\big| \textbf{Z} \big| \longrightarrow +\infty$),

{\small \begin{align*}
  \mathrm{log} \text{ }   \mathrm{log} \bigg[   {  \mathrm{log} \big| \textbf{Y}^{*} \big|   } \big/ { \big| \textbf{X}^{*} \big|  }          \bigg]  \longrightarrow + \infty  \bigg( \text{resp. }   \mathrm{log} \bigg[   {  \mathrm{log} \big| \textbf{Z} \big|   } \big/ { \big| \textbf{Y}^{*} \big|  }          \bigg]  \longrightarrow +  \infty   \bigg)  ,
\end{align*}  } 

\end{itemize}

\noindent then ${\small r_1 \longrightarrow \pm  \infty      }$, corresponding to (\textit{1A}) and (\textit{1B}), respectively.

      \item[$\bullet$] \underline{(2)}: \textit{Divergence of $r_2$ to $\pm \infty$}. For the bit transmission rate $r_2$ if:

\begin{itemize}

\item[$\bullet$]

\textit{(2A):} Divergence of ${\small
{  \mathrm{log} \big| \textbf{X} \big|   } \big/ { \big| \textbf{Y}^{*} \big|  }    }$ (resp. ${\small
{  \mathrm{log} \big| \textbf{Y}^* \big|   } \big/ { \big| \textbf{X}  \big|  }    }$), as $\big| \textbf{Y}^* \big| \longrightarrow 0^+$ (resp. $\big| \textbf{X} \big| \longrightarrow 0^+$), 

{\small \begin{align*}
      \mathrm{log} \text{ }  \mathrm{log} \bigg[   {  \mathrm{log} \big| \textbf{X} \big|  } \big/ { \big| \textbf{Y}^{*} \big| }          \bigg]  \longrightarrow   - \infty \bigg( \text{resp. } \mathrm{log}  \bigg[  { \mathrm{log} \big|  \textbf{Y}^{*} \big|   } \big/ {  \big| \textbf{X} \big|   } \bigg]  \longrightarrow   - \infty    \bigg)  ,
\end{align*} } 

\textit{(2B):} Divergence of ${\small
{  \mathrm{log} \big| \textbf{X} \big|   } \big/ { \big| \textbf{Y}^{*} \big|  }    }$ (resp. ${\small
{  \mathrm{log} \big| \textbf{Y}^* \big|   } \big/ { \big| \textbf{X}  \big|  }    }$), as $\big| \textbf{Y}^* \big| \longrightarrow 0^+$ (resp. $\big| \textbf{X} \big| \longrightarrow 0^+$),

{\small \begin{align*}
        \mathrm{log} \text{ }  \mathrm{log} \bigg[   {  \mathrm{log} \big| \textbf{X} \big|  } \big/ { \big| \textbf{Y}^{*} \big| }          \bigg]  \longrightarrow   + \infty \bigg( \text{resp. } \mathrm{log}  \bigg[  { \mathrm{log} \big|  \textbf{Y}^{*} \big|   } \big/ {  \big| \textbf{X} \big|   } \bigg]  \longrightarrow   +  \infty    \bigg)       ,
\end{align*}
 } 

\noindent

\end{itemize}

\noindent then ${\small 
 r_2 \longrightarrow  \pm \infty } $, corresponding to (\textit{2A}) and (\textit{2B}), respectively.

        \item[$\bullet$] \underline{(3)}: \textit{Divergence of $r_3$ to $\pm \infty$}. For the bit transmission rate $r_3$ if:

\begin{itemize}
    \item[$\bullet$] \textit{(3A)}: Divergence of ${\small
{  \mathrm{log} \big| \textbf{Y}^* \big|   } \big/ { \big| \textbf{X}^{*} \big|  }    }$ (resp. ${\small
{  \mathrm{log} \big| \textbf{Y}^* \big|   } \big/ { \big| \textbf{X}  \big|  }    }$), as $\big| \textbf{X}^* \big| \longrightarrow 0^+$ (resp. $\big| \textbf{X} \big| \longrightarrow 0^+$),

{\small \begin{align*}
\mathrm{log}  \text{ }  \mathrm{log} \bigg[    {  \mathrm{log} \big| \textbf{Y}^{*} \big|  } \big/ { \big| \textbf{X}^{*} \big| }          \bigg]  \longrightarrow - \infty \bigg( \text{resp. }  \mathrm{log}  \bigg[  {  \mathrm{log}\big|  \textbf{Y}^{*} \big|   } \big/ { \big| \textbf{X} \big|   } \bigg]  \longrightarrow - \infty  \bigg)    ,
\end{align*} }

     \item[$\bullet$] \textit{(3B)}: Divergence of ${\small
{  \mathrm{log} \big| \textbf{X} \big|   } \big/ { \big| \textbf{Y}^{*} \big|  }    }$ (resp. ${\small
{  \mathrm{log} \big| \textbf{Y}^* \big|   } \big/ { \big| \textbf{X}  \big|  }    }$), as $\big| \textbf{X}^* \big| \longrightarrow 0^+$ (resp. $\big| \textbf{X} \big| \longrightarrow 0^+$),

{\small \begin{align*}
\mathrm{log} \text{ }   \mathrm{log} \bigg[    {  \mathrm{log} \big| \textbf{Y}^{*} \big|  } \big/ { \big| \textbf{X}^{*} \big| }          \bigg]  \longrightarrow + \infty \bigg( \text{resp. }  \mathrm{log}  \bigg[  {  \mathrm{log}\big|  \textbf{Y}^{*} \big|   } \big/ { \big| \textbf{X} \big|   } \bigg]  \longrightarrow + \infty  \bigg)    ,
\end{align*} }

\end{itemize}

\noindent then ${\small r_3   \longrightarrow  \pm \infty}$, corresponding to (\textit{3A}) and (\textit{3B}), respectively.

        \item[$\bullet$] \underline{(4)}: \textit{Divergence of $r_4$ to $\pm \infty$}. For the bit transmission rate $r_4$ if:

\begin{itemize}
    \item[$\bullet$] \textit{(4A)}: Divergence of ${\small
{  \mathrm{log} \big| \textbf{X} \big|   } \big/ { \big| \textbf{Y}^{*} \big|  }    }$ (resp. ${\small
{  \mathrm{log} \big| \textbf{Y}^* \big|   } \big/ { \big| \textbf{X}  \big|  }    }$), as $\big| \textbf{Y}^* \big| \longrightarrow 0^+$ (resp. $\big| \textbf{X} \big| \longrightarrow 0^+$),

{\small \begin{align*}
  \mathrm{log} \text{ }   \mathrm{log} \bigg[   {  \mathrm{log} \big| \textbf{Y}^{*} \big|  } \big/ {\big| \textbf{X}^{*} \big| }          \bigg]   \longrightarrow - \infty  \bigg(  \text{resp. } \mathrm{log}  \bigg[  {  \mathrm{log}\big|  \textbf{Z}  \big|   } \big/ { \big| \textbf{Y}^{*} \big|   } \bigg]  \longrightarrow - \infty   \bigg)   ,
\end{align*}
 }

     \item[$\bullet$] \textit{(4B)}: Divergence of ${\small
{  \mathrm{log} \big| \textbf{X} \big|   } \big/ { \big| \textbf{Y}^{*} \big|  }    }$ (resp. ${\small
{  \mathrm{log} \big| \textbf{Y}^* \big|   } \big/ { \big| \textbf{X}  \big|  }    }$), as $\big| \textbf{Y}^* \big| \longrightarrow 0^+$ (resp. $\big| \textbf{X} \big| \longrightarrow 0^+$),

{\small \begin{align*}
  \mathrm{log}  \text{ }  \mathrm{log} \bigg[   {  \mathrm{log} \big| \textbf{Y}^{*} \big|  } \big/ {\big| \textbf{X}^{*} \big| }          \bigg]   \longrightarrow +  \infty  \bigg(  \text{resp. } \mathrm{log}  \bigg[  {  \mathrm{log}\big|  \textbf{Z}  \big|   } \big/ { \big| \textbf{Y}^{*} \big|   } \bigg]  \longrightarrow + \infty   \bigg)   ,
\end{align*} }

\end{itemize}

\noindent then ${\small 
   r_4   \longrightarrow  \pm \infty }$, corresponding to (\textit{4A}) and (\textit{4B}), respectively.

\end{itemize}

\noindent Moreover, the following list of degeneracies between $\big| \textbf{X} \big|$ and $\big| \textbf{X}^* \big|$, $\big| \textbf{Y} \big|$ and $\big| \textbf{Y}^* \big|$ and $\big| \textbf{Z} \big|$ and $\big| \textbf{Z}^* \big|$, 

{\small \[
 \bigg\{ \mathrm{log} \text{ }  \mathrm{log} \bigg[   {  \mathrm{log} \big| \textbf{Y}^{*} \big|   } \big/ { \big| \textbf{X}^{*} \big|  }          \bigg]   +   \mathrm{log}  \bigg[ {   \mathrm{log}\big|  \textbf{Z}  \big|  } \big/ {  \big|  \textbf{Y}^{*}  \big|   } \bigg]   \longrightarrow  - \infty  \bigg\} \Longleftrightarrow \left\{\!\begin{array}{ll@{}>{{}}l} 
 {  \mathrm{log} \big| \textbf{Y}^{*} \big|   } \big/ { \big| \textbf{X}^{*} \big|  }     \equiv    \mathrm{log} \big\{ \big| \textbf{Y}^* \big| - \mathrm{exp} \big| \textbf{X}^* \big| \big\}    \approx - \infty    ,  \\    \\    {   \mathrm{log}\big|  \textbf{Z}  \big|  } \big/ {  \big|  \textbf{Y}^{*}  \big|   } \equiv    \mathrm{log} \big\{ \big| \textbf{Z} \big| - \mathrm{exp} \big| \textbf{Y}^* \big| \big\}   \approx - \infty         ,  \end{array}\right. \]

 \[ \bigg\{    \mathrm{log} \text{ }  \mathrm{log} \bigg[   {  \mathrm{log} \big| \textbf{X} \big|  } \big/ { \big| \textbf{Y}^{*} \big| }          \bigg]   +   \mathrm{log}  \bigg[  { \mathrm{log} \big|  \textbf{Y}^{*} \big|   } \big/ {  \big| \textbf{X} \big|   } \bigg]  \longrightarrow  - \infty    \bigg\} \Longleftrightarrow \left\{\!\begin{array}{ll@{}>{{}}l}  {  \mathrm{log} \big| \textbf{X} \big|   } \big/ { \big| \textbf{Y}^{*} \big|  }    \equiv    \mathrm{log} \big\{ \big| \textbf{X} \big| - \mathrm{exp} \big| \textbf{Y}^* \big| \big\}      \approx - \infty    ,     \\  \\  {  \mathrm{log} \big| \textbf{Y}^* \big|   } \big/ { \big| \textbf{X} \big|  }  \equiv     \mathrm{log} \big\{ \big| \textbf{Y}^* \big| - \mathrm{exp} \big| \textbf{X} \big| \big\}        \approx - \infty             ,    \end{array}\right. \]

 \[ \bigg\{ \mathrm{log}  \text{ } \mathrm{log} \bigg[    {  \mathrm{log} \big| \textbf{Y}^{*} \big|  } \big/ { \big| \textbf{X}^{*} \big| }          \bigg]   +   \mathrm{log}  \bigg[  {  \mathrm{log}\big|  \textbf{Y}^{*} \big|   } \big/ { \big| \textbf{X} \big|   } \bigg]    \longrightarrow  - \infty     \bigg\} \Longleftrightarrow 
\left\{\!\begin{array}{ll@{}>{{}}l}     {  \mathrm{log} \big| \textbf{Y}^* \big|   } \big/ { \big| \textbf{X}^* \big|  }  \equiv     \mathrm{log} \big\{ \big| \textbf{Y}^* \big| - \mathrm{exp} \big| \textbf{X}^* \big| \big\}      \approx - \infty     , \\ \\ {  \mathrm{log} \big| \textbf{Y}^* \big|   } \big/ { \big| \textbf{X} \big|  }  \equiv    \mathrm{log} \big\{ \big| \textbf{Y}^* \big| - \mathrm{exp} \big| \textbf{X} \big| \big\}     \approx - \infty      , \end{array}\right. \]

\[ \bigg\{   \mathrm{log} \text{ }  \mathrm{log} \bigg[   {  \mathrm{log} \big| \textbf{Y}^{*} \big|  } \big/ {\big| \textbf{X}^{*} \big| }          \bigg]   +   \mathrm{log}  \bigg[  {  \mathrm{log}\big|  \textbf{Z}  \big|   } \big/ { \big| \textbf{Y}^{*} \big|   } \bigg]   \longrightarrow  - \infty     \bigg\} \Longleftrightarrow 
\left\{\!\begin{array}{ll@{}>{{}}l}  {  \mathrm{log} \big| \textbf{Y}^* \big|   } \big/ { \big| \textbf{X}^*  \big|  }  \equiv    \mathrm{log} \big\{ \big| \textbf{Y}^* \big| - \mathrm{exp} \big| \textbf{X}^* \big| \big\}     \approx - \infty     , \\ \\ {  \mathrm{log} \big| \textbf{Z} \big|   } \big/ { \big| \textbf{Y}^* \big|  }  \equiv   \mathrm{log} \big\{ \big| \textbf{Z} \big| - \mathrm{exp} \big| \textbf{Y}^* \big| \big\}     \approx - \infty      .                 
\end{array}\right.  
\] }

\noindent The above quantities taken with respect to the natural logarithm controlling the asymptotic behavior of the bit transmission rate function,

{\small 
\[  r \equiv  \left\{\!\begin{array}{ll@{}>{{}}l} 
  r_1  \Longleftrightarrow \big\{ \big| \textbf{X} \big| > \big| \textbf{Y}^{*} \big| ,  \big| \textbf{Y}^{*} \big| > \big| \textbf{Z} \big|     \big\}   ,  \\    \\ r_2  \Longleftrightarrow \big\{ \big| \textbf{X} \big| <  \big| \textbf{Y}^{*} \big| ,  \big|  \textbf{Y}^{*} \big| < \big| \textbf{Z} \big|    \big\}        ,  \\ \\ r_3    \Longleftrightarrow  \big\{\big| \textbf{X} \big| >   \big| \textbf{Y}^{*} \big| ,  \big|  \textbf{Y}^{*} \big| < \big| \textbf{Z} \big|        \big\}    ,   \\   \\ r_4   \Longleftrightarrow  \big\{ \big| \textbf{X} \big| <    \big| \textbf{Y}^{*} \big| ,  \big|  \textbf{Y}^{*} \big| >   \big| \textbf{Z} \big|     \big\}       .                          
\end{array}\right.  \\ 
\]  }  

\noindent are related to the first and second derivatives,

{\small 
\[  r^{\prime} \equiv  \left\{\!\begin{array}{ll@{}>{{}}l} 
 \big\{  r^{\prime}_1 \equiv   {\partial} r_1   \big/  {\partial \big\{ \textbf{X}^*  , \textbf{Y}^* , \textbf{Z} \big\} } \big\} \Longleftrightarrow \big\{ \big| \textbf{X} \big| > \big| \textbf{Y}^{*} \big| ,  \big| \textbf{Y}^{*} \big| > \big| \textbf{Z} \big|   \big\} ,  \\    \\ \big\{ r^{\prime}_2  \equiv {\partial} r_2  \big/  {\partial \big\{ \textbf{X} , \textbf{Y}^*  \big\} } \big\} \Longleftrightarrow \big\{  \big| \textbf{X} \big| <  \big| \textbf{Y}^{*} \big| ,  \big|  \textbf{Y}^{*} \big| < \big| \textbf{Z} \big|    \big\}     ,  \\ \\ \big\{ r^{\prime}_3  \equiv     {\partial} r_3   \big/  {\partial \big\{ \textbf{X} , \textbf{X}^*   , \textbf{Y}^* \big\} } \big\}          \big\}   \Longleftrightarrow \big\{  \big| \textbf{X} \big| >   \big| \textbf{Y}^{*} \big| ,  \big|  \textbf{Y}^{*} \big| < \big| \textbf{Z} \big|   \big\}      ,   \\   \\ \big\{ r^{\prime}_4   \equiv  {\partial} r_4  \big/  {\partial \big\{ \textbf{X}^*   , \textbf{Y}^* , \textbf{Z} \big\} }  \big\}  \Longleftrightarrow  \big\{  \big| \textbf{X} \big| <    \big| \textbf{Y}^{*} \big| ,  \big|  \textbf{Y}^{*} \big| >   \big| \textbf{Z} \big|     \big\}    .                          
\end{array}\right.  \\ 
\]  }

\bigskip

{\small 
\[  r^{\prime\prime} \equiv  \left\{\!\begin{array}{ll@{}>{{}}l} 
 \big\{  r^{\prime\prime}_1 \equiv  {\partial^2} r_1   \big/  {\partial \big\{ \textbf{X}^*  , \textbf{Y}^* , \textbf{Z} \big\}^2 } \big\}  \Longleftrightarrow \big\{ \big| \textbf{X} \big| > \big| \textbf{Y}^{*} \big| ,  \big| \textbf{Y}^{*} \big| > \big| \textbf{Z} \big|  \big\} , \\  \\ 
 \big\{ r^{\prime\prime}_2 \equiv {\partial^2} r_2  \big/  {\partial \big\{ \textbf{X} , \textbf{Y}^*  \big\}^2 } \big\}  \Longleftrightarrow \big\{ \big| \textbf{X} \big| < \big| \textbf{Y}^{*} \big| ,  \big| \textbf{Y}^{*} \big| < \big| \textbf{Z} \big|  \big\}  , \\ \\ 
 \big\{ r^{\prime\prime}_3 \equiv {\partial^2} r_3   \big/  {\partial \big\{ \textbf{X} , \textbf{X}^*   , \textbf{Y}^* \big\}^2 } \big\} \Longleftrightarrow \big\{ \big| \textbf{X} \big| > \big| \textbf{Y}^{*} \big| ,  \big| \textbf{Y}^{*} \big| < \big| \textbf{Z} \big|   \big\}, \\ \\ 
 \big\{ r^{\prime\prime}_4 \equiv  {\partial^2} r_4  \big/  {\partial \big\{ \textbf{X}^*   , \textbf{Y}^* , \textbf{Z} \big\}^2 }  \big\} \Longleftrightarrow  \big\{ \big| \textbf{X} \big| < \big| \textbf{Y}^{*} \big| ,  \big| \textbf{Y}^{*} \big| > \big| \textbf{Z} \big|  \big\} .
\end{array}\right.  \\ 
\]  }

\noindent respectively, where,

{\small 
\[   \left\{\!\begin{array}{ll@{}>{{}}l} 
  r_1 \equiv    \mathrm{log} \text{ }  \mathrm{log} \bigg[   {  \mathrm{log} \big| {Y}^{*} \big|   } \big/ { \big| {X}^{*} \big|  }          \bigg]   +   \mathrm{log}  \bigg[ {   \mathrm{log}\big|  {Z}  \big|  } \big/ {  \big|  {Y}^{*}  \big|   } \bigg]    ,      
\end{array}\right.  \\ 
\]  }  

{\small 
\[   \left\{\!\begin{array}{ll@{}>{{}}l} 
   r_2 \equiv \mathrm{log} \text{ }   \mathrm{log} \bigg[   {  \mathrm{log} \big| {X} \big|  } \big/ { \big| {Y}^{*} \big| }          \bigg]   +   \mathrm{log}  \bigg[  { \mathrm{log} \big|  {Y}^{*} \big|   } \big/ {  \big| {X} \big|   } \bigg]      ,  \\ \\ r_3 \equiv \mathrm{log}  \text{ } \mathrm{log} \bigg[    {  \mathrm{log} \big| {Y}^{*} \big|  } \big/ { \big| {X}^{*} \big| }          \bigg]   +   \mathrm{log}  \bigg[  {  \mathrm{log}\big|  {Y}^{*} \big|   } \big/ { \big| {X} \big|   } \bigg]        ,   \\   \\ r_4 \equiv  \mathrm{log} \text{ }  \mathrm{log} \bigg[   {  \mathrm{log} \big| {Y}^{*} \big|  } \big/ {\big| {X}^{*} \big| }          \bigg]   +   \mathrm{log}  \bigg[  {  \mathrm{log}\big|  {Z}  \big|   } \big/ { \big| {Y}^{*} \big|   } \bigg]        .                          
\end{array}\right.  \\ 
\]  }

\noindent under the change of variables,

{\small 
\[   \left\{\!\begin{array}{ll@{}>{{}}l} 
   \textbf{X}  \longleftrightarrow   X  , \\   \textbf{X}^*  \longleftrightarrow   X^*    , \\   \textbf{Y}  \longleftrightarrow   Y   , \\   \textbf{Y}^*  \longleftrightarrow   Y^*  , \\   \textbf{Z}  \longleftrightarrow   Z   , \\   \textbf{Z}^*  \longleftrightarrow   Z^*    ,                         
\end{array}\right.  \\ 
\]  }

\noindent are related to the discontinuities of the first and second derivatives. Analtyically, the first derivative of the converse bit transmission rate being equal to $0$ implies,

{\small 
\[   \left\{\!\begin{array}{ll@{}>{{}}l} 
  \big|  Y^* \big|   =    +  \sqrt{  \big| X^* \big|^2    \cdot           \frac{   \bigg[   \frac{  \mathrm{log}\big|  {Z}  \big|  \cdot \big| Z \big| }{   \big|  {Y}^{*}  \big|^2  \cdot  Z^{\prime}     }    \bigg]^{-1}   +   \bigg[      \frac{1}{ \big| Y^* \big| \cdot \big\{ Y^* \big\}^{\prime} }  \bigg]^{-1}   }{     \mathrm{log} \bigg[     \frac{\mathrm{log} \big| {Y}^{*} \big|  }{ {\big| {X}^{*} \big| }    }         \bigg]        \cdot     \bigg\{            \frac{1}{\big| Y^{*} \big| }  \cdot  \big\{ Y^* \big\}^{\prime}  \cdot \big| X^* \big|      -    \mathrm{log}   \big| Y^{*} \big|  \cdot \big\{ X^* \big\}^{\prime }                  \bigg\}                                   }  }  \Longleftrightarrow \big\{ \big| {X} \big| > \big| {Y}^{*} \big| ,  \big| {Y}^{*} \big| > \big| {Z} \big|   \big\}   , \\    \\      \big| X \big| = +           \sqrt{  \big| Y^* \big|^2    \cdot           \frac{   \bigg[   \frac{  \mathrm{log}\big|  {Y^*}  \big|  \cdot \big| Y^*  \big| }{   \big|  {Y}^{*}  \big|^2  \cdot  Z^{\prime}     }    \bigg]^{-1}     +   \bigg[      \frac{1}{ \big| X \big| \cdot \big\{ X  \big\}^{\prime} }  \bigg]^{-1}   }{     \mathrm{log} \bigg[     \frac{\mathrm{log} \big| {X} \big|  }{ {\big| {Y}^{*} \big| }    }         \bigg]        \cdot     \bigg\{            \frac{1}{\big| X^{*} \big| }  \cdot  \big\{ X^* \big\}^{\prime}  \cdot \big| Y^* \big|      -    \mathrm{log}   \big| X  \big|  \cdot \big\{ Y^* \big\}^{\prime }                  \bigg\}                                   }  }    \Longleftrightarrow \big\{ \big| {X} \big| <  \big| {Y}^{*} \big| ,  \big| {Y}^{*} \big| <  \big| {Z} \big|   \big\}               , \\ \\      \big|  Y^* \big|   =    +  \sqrt{  \big| X^* \big|^2    \cdot           \frac{   \bigg[   \frac{  \mathrm{log}\big|  {Y^* }  \big|  \cdot \big| Y^*   \big| }{   \big|  {Y}^{*}  \big|^2  \cdot  \big\{ Y^*  \big\}^{\prime}     }    \bigg]^{-1}     +   \bigg[      \frac{1}{ \big| Y^* \big| \cdot \big\{ Y^* \big\}^{\prime} }  \bigg]^{-1}   }{     \mathrm{log} \bigg[     \frac{\mathrm{log} \big| {Y}^{*} \big|  }{ {\big| {X}^{*} \big| }    }         \bigg]        \cdot     \bigg\{            \frac{1}{\big| Y^{*} \big| }  \cdot  \big\{ Y^* \big\}^{\prime}  \cdot \big| X^* \big|      -    \mathrm{log}   \big| Y^{*} \big|  \cdot \big\{ X^* \big\}^{\prime }                  \bigg\}                                   }  }  \Longleftrightarrow \big\{ \big| {X} \big| >   \big| {Y}^{*} \big| ,  \big| {Y}^{*} \big| <  \big| {Z} \big|   \big\}                    , \\  \\        \big|  Y^* \big|   =    +  \sqrt{  \big| X^* \big|^2    \cdot           \frac{   \bigg[   \frac{  \mathrm{log}\big|  {Z}  \big|  \cdot \big| Z \big| }{   \big|  {Y}^{*}  \big|^2  \cdot  Z^{\prime}     }    \bigg]^{-1}     +  \bigg[      \frac{1}{ \big| Y^* \big| \cdot \big\{ Y^* \big\}^{\prime} }  \bigg]^{-1}   }{     \mathrm{log} \bigg[     \frac{\mathrm{log} \big| {Y}^{*} \big|  }{ {\big| {X}^{*} \big| }    }         \bigg]        \cdot     \bigg\{            \frac{1}{\big| Y^{*} \big| }  \cdot  \big\{ Y^* \big\}^{\prime}  \cdot \big| X^* \big|      -    \mathrm{log}   \big| Y^{*} \big|  \cdot \big\{ X^* \big\}^{\prime }                  \bigg\}                                   }  }  \Longleftrightarrow \big\{ \big| {X} \big| <  \big| {Y}^{*} \big| ,  \big| {Y}^{*} \big| > \big| {Z} \big|   \big\}            . 
\end{array}\right.  \\ \\ \tag{   $r^{\prime} = 0 $          } 
\]  }

\noindent To quantify the points of discontinuity at which the first, or second, derivatives of the converse bit transmission rate defined above could diverge introduce,

{\small       
\[   \mathscr{D} {\tiny \big( r^{\prime} \big)}  \equiv    \left\{\!\begin{array}{ll@{}>{{}}l} 
    \big\{  X, X^* , Y , Y^* , Z , Z^*     :  r^{\prime}_1 \longrightarrow + \infty   \big\}      , \\ \\   \big\{   X, X^* , Y , Y^* , Z , Z^*   : r^{\prime}_2 \longrightarrow + \infty   \big\}        ,  
\end{array}\right.  , \\ 
\]  }

{\small       
\[   \mathscr{D} {\tiny \big( r^{\prime} \big)}  \equiv    \left\{\!\begin{array}{ll@{}>{{}}l} 
    \big\{   X, X^* , Y , Y^* , Z , Z^*   : r^{\prime}_3 \longrightarrow + \infty   \big\}      ,    \\ \\    \big\{   X, X^* , Y , Y^* , Z , Z^*   : r^{\prime}_4 \longrightarrow + \infty           \big\}   ,          
\end{array}\right.  , \\ 
\]  }

{\small  
\[    \mathscr{D} {\tiny \big( r^{\prime\prime } \big)} \equiv  \left\{\!\begin{array}{ll@{}>{{}}l} 
       \big\{  X, X^* , Y , Y^* , Z , Z^*     :  r^{\prime\prime}_1 \longrightarrow + \infty   \big\}      , \\ \\    \big\{   X, X^* , Y , Y^* , Z , Z^*   : r^{\prime\prime}_2 \longrightarrow + \infty   \big\}        , \\ \\    \big\{   X, X^* , Y , Y^* , Z , Z^*   : r^{\prime\prime}_3 \longrightarrow + \infty   \big\}      ,    \\ \\    \big\{   X, X^* , Y , Y^* , Z , Z^*   : r^{\prime\prime}_4 \longrightarrow + \infty           \big\}   , 
\end{array}\right. ,  \\ 
\]  }

\noindent respectively. Analytically, to determine such possible points of discontinuity of the converse bit transmission rate it suffices to write out the first ($r^{\prime}_1 , \cdots , r^{\prime}_4$) and second derivatives ($r^{\prime\prime}_1 , \cdots , r^{\prime\prime}_4$) for each piecewise function used to define $r$ above. As such, to ensure that,

{\small 

\begin{align*}
 \bigg\{ r^{\prime} \bigg( \bigg|  \frac{X}{X^*}  \bigg| ,  \bigg|  \frac{Y}{Y^*}  \bigg|  ,   \bigg|  \frac{Z}{Z^*}  \bigg|   \bigg) \text{ } \textit{is well-defined} \bigg\} \Longleftrightarrow    \bigg\{ r^{\prime} \bigg( \bigg|  \frac{X}{X^*}  \bigg| ,  \bigg|  \frac{Y}{Y^*}  \bigg|  ,   \bigg|  \frac{Z}{Z^*}  \bigg|   \bigg) < + \infty \bigg\} ,
\end{align*}

}

\noindent the piecewise defined functions corresponding to each derivative take the form,

{\small 
\[ r^{\prime} \equiv r^{\prime} \bigg( \bigg|  \frac{X}{X^*}  \bigg| ,  \bigg|  \frac{Y}{Y^*}  \bigg|  ,   \bigg|  \frac{Z}{Z^*}  \bigg|   \bigg)  =   \left\{\!\begin{array}{ll@{}>{{}}l} 
  \boxed{r^{\prime}_1} \equiv 1 \big/  \mathrm{log} \bigg[   {  \mathrm{log} \big| {Y}^{*} \big|  } \big/ {\big| {X}^{*} \big| }          \bigg] \cdot \bigg\{            \big| Y^{*} \big|^{-1} \cdot  \big\{ Y^* \big\}^{\prime} \cdot \big| X^* \big|   -    \mathrm{log}   \big| Y^{*} \big|\cdot \big\{ X^* \big\}^{\prime }                  \bigg\} \big/ \big| X^* \big|^2   \\  \\   + 1 \big/  \big\{   {   \mathrm{log}\big|  {Z}  \big|  } \big/ {  \big|  {Y}^{*}  \big|   }  \big\} \cdot       \bigg\{       \big| Z \big|^{-1}   \cdot  Z^{\prime}  \cdot   \big|  {Y}^{*}  \big|  -   \mathrm{log}\big|  {Z}  \big|  \cdot  \big\{ Y^* \big\}^{\prime}  \bigg\}      \big/    \big|  {Y}^{*}  \big|^2     , \\ \Bigg\Updownarrow \\ \\  \big\{ \big| X^* \big| ,  \big| Y^* \big|  ,   \big|  Z \big|  \text{ }      \textit{are not made arbitrarily small}  \big\} ,  \\         \\  \boxed{r^{\prime}_2} \equiv   1 \big/ \mathrm{log} \bigg[   {  \mathrm{log} \big| {X} \big|  } \big/ { \big| {Y}^{*} \big| }          \bigg] \cdot \bigg\{   \big| X \big|^{-1} \cdot X^{\prime}  \cdot  \big| Y^* \big| -    \mathrm{log} \big| {X} \big|  \cdot    \big\{ Y^* \big\}^{\prime}           \bigg\}  \big/ \big| Y^* \big|^2  \\ \\     + 1 \big/  \big\{   {   \mathrm{log}\big|  {Y}^*  \big|  } \big/ {  \big|  {X}  \big|   }  \big\} \cdot       \bigg\{       \big|  {Y}^*  \big|^{-1}   \cdot  \big\{  {Y}^*  \big\}^{\prime}  \cdot   \big|  {X} \big|  -   \mathrm{log}\big|  {Y}^*   \big|  \cdot  Y^{\prime}  \bigg\}      \big/    \big|  X   \big|^2          ,  \\ \Bigg\Updownarrow \\  \\   \big\{ \big| X^* \big| ,  \big| Y^* \big|  ,   \big|  Z \big|  \text{ }      \textit{are not made arbitrarily small}  \big\}  ,  \\  \\ \boxed{r^{\prime}_3} \equiv 1 \big/   \mathrm{log} \bigg[   {  \mathrm{log} \big| {Y}^{*} \big|  } \big/ {\big| {X}^{*} \big| }          \bigg] \cdot \bigg\{            \big| Y^{*} \big|^{-1} \cdot  \big\{ Y^* \big\}^{\prime} \cdot \big| X^* \big|   -    \mathrm{log}   \big| Y^{*} \big|\cdot \big\{ X^* \big\}^{\prime }                  \bigg\} \big/ \big| X^* \big|^2  \\ \\ + 1 \big/  \big\{   {   \mathrm{log}\big|  {Y}^*  \big|  } \big/ {  \big|  {X}  \big|   }  \big\} \cdot       \bigg\{       \big|  {Y}^*  \big|^{-1}   \cdot  \big\{  {Y}^*  \big\}^{\prime}  \cdot   \big|  {X} \big|  -   \mathrm{log}\big|  {Y}^*   \big|  \cdot  Y^{\prime}  \bigg\}      \big/    \big|  X   \big|^2       ,  
\end{array}\right.  \\ \\ \\ \tag{\textit{First derivative of $r$}}  
\]  }

{\small 
\[ r^{\prime} \equiv r^{\prime} \bigg( \bigg|  \frac{X}{X^*}  \bigg| ,  \bigg|  \frac{Y}{Y^*}  \bigg|  ,   \bigg|  \frac{Z}{Z^*}  \bigg|   \bigg)  =   \left\{\!\begin{array}{ll@{}>{{}}l} 
    \Bigg\Updownarrow \\ \\  \big\{ \big| X^* \big| ,  \big| Y^* \big|  ,   \big|  Z \big|  \text{ }      \textit{are not made arbitrarily small}  \big\}  ,    \\ \\  \boxed{r^{\prime}_4}  \equiv   1 \big/  \mathrm{log} \bigg[   {  \mathrm{log} \big| {Y}^{*} \big|  } \big/ {\big| {X}^{*} \big| }          \bigg] \cdot \bigg\{            \big| Y^{*} \big|^{-1} \cdot  \big\{ Y^* \big\}^{\prime} \cdot \big| X^* \big|  \\ \\  -    \mathrm{log}   \big| Y^{*} \big|\cdot \big\{ X^* \big\}^{\prime }                  \bigg\} \big/ \big| X^* \big|^2   \\ \\   + 1 \big/  \big\{   {   \mathrm{log}\big|  {Z}  \big|  } \big/ {  \big|  {Y}^{*}  \big|   }  \big\} \cdot       \bigg\{       \big| Z \big|^{-1}   \cdot  Z^{\prime}  \cdot   \big|  {Y}^{*}  \big|  -   \mathrm{log}\big|  {Z}  \big|  \cdot  \big\{ Y^* \big\}^{\prime}  \bigg\}      \big/    \big|  {Y}^{*}  \big|^2  , \\ \Bigg\Updownarrow \\ \\  \big\{ \big| X^* \big| ,  \big| Y^* \big|  ,   \big|  Z \big|  \text{ }      \textit{are not made arbitrarily small}  \big\}  . 
\end{array}\right.  \\ \\ \\ \tag{\textit{First derivative of $r$}}  
\]  }  

%{\small 
%\[ r^{\prime} \equiv r^{\prime} \bigg( \bigg|  \frac{X}{X^*}  \bigg| ,  \bigg|  \frac{Y}{Y^*}  \bigg|  ,   \bigg|  \frac{Z}{Z^*}  \bigg|   \bigg) \text{ } \textit{continued}  =   \left\{\!\begin{array}{ll@{}>{{}}l} 
 %  \\ \\ 
%\end{array}\right.  \\ \\ \\ \tag{\textit{First derivative of $r$}}  
%\]  }  

\noindent Similarly, to ensure that,

{\small 

\begin{align*}
  \bigg\{ r^{\prime\prime} \bigg( \bigg|  \frac{X}{X^*}  \bigg| ,  \bigg|  \frac{Y}{Y^*}  \bigg|  ,   \bigg|  \frac{Z}{Z^*}  \bigg|   \bigg) \text{ } \textit{is well-defined} \bigg\} \Longleftrightarrow    \bigg\{ r^{\prime\prime} \bigg( \bigg|  \frac{X}{X^*}  \bigg| ,  \bigg|  \frac{Y}{Y^*}  \bigg|  ,   \bigg|  \frac{Z}{Z^*}  \bigg|   \bigg) < + \infty \bigg\} ,
\end{align*}

}

\noindent one has that,

{\small 
\[  r^{\prime\prime} \equiv r^{\prime\prime} \bigg( \bigg|  \frac{X}{X^*}  \bigg| ,  \bigg|  \frac{Y}{Y^*}  \bigg|  ,   \bigg|  \frac{Z}{Z^*}  \bigg|  \bigg) =  \left\{\!\begin{array}{ll@{}>{{}}l} 
  \boxed{r^{\prime\prime}_1}  \equiv    -  \bigg\{    \mathrm{log} \bigg[   {  \mathrm{log} \big| {Y}^{*} \big|  } \big/ {\big| {X}^{*} \big| }          \bigg] \cdot  \big| X^* \big|^2       \bigg\}^{-2}   \cdot \bigg\{   \big\{   {  \mathrm{log} \big| {Y}^{*} \big|  } \big/ {\big| {X}^{*} \big| } \big\}^{-1} \\ \\ \cdot \bigg[             \frac{\big| Y^* \big|^{-1} \cdot \big\{ Y^* \big\}^{\prime} \cdot \big| X^* \big|   -    \mathrm{log}  \big| Y^* \big|  \cdot \big\{ X^* \big\}^{\prime } }{\big| X^* \big|^2 }             \bigg]  \cdot \big| X^* \big|^2 +    \mathrm{log} \bigg[   {  \mathrm{log} \big| {Y}^{*} \big|  } \big/ {\big| {X}^{*} \big| }          \bigg] \\ \\  \cdot 2 \big| X^* \big| \cdot \big\{ X^* \big\}^{\prime}       \bigg\}          \cdot    \bigg\{            \big| Y^{*} \big|^{-1} \cdot  \big\{ Y^* \big\}^{\prime} \cdot \big| X^* \big|   -    \mathrm{log}   \big| Y^{*} \big|\cdot \big\{ X^* \big\}^{\prime }                  \bigg\}   \\ \\ +    \bigg\{    \mathrm{log} \bigg[   {  \mathrm{log} \big| {Y}^{*} \big|  } \big/ {\big| {X}^{*} \big| }          \bigg] \cdot  \big| X^* \big|^2       \bigg\}^{-1} \cdot     \bigg\{  \big\{   - \big| Y^* \big|^{-2} \cdot \big\{ Y^* \big\}^{\prime }    \cdot  \big\{ Y^* \big\}^{\prime} \cdot \big| X^* \big|  \\ \\   +     \big| Y^{*} \big|^{-1} \cdot  \big\{ Y^* \big\}^{\prime\prime} \cdot \big| X^* \big|    +   \big| Y^{*} \big|^{-1} \cdot  \big\{ Y^* \big\}^{\prime} \cdot \big\{  X^* \big\}^{\prime}       \big\}     -    \big\{   \big| Y^* \big|^{-1} \cdot \big\{ Y^* \big\}^{\prime} \\ \\ \cdot \big\{ X^* \big\}^{\prime}  +   \mathrm{log} \big| Y^* \big| \cdot \big\{ X^* \big\}^{\prime\prime}                  \big\}                     \bigg\}    ,   \\ \Bigg\Updownarrow 
\end{array}\right. \\ \\ \tag{\textit{Second derivative of $r$}}  
\]  }

{\small 
\[  r^{\prime\prime} \equiv r^{\prime\prime} \bigg( \bigg|  \frac{X}{X^*}  \bigg| ,  \bigg|  \frac{Y}{Y^*}  \bigg|  ,   \bigg|  \frac{Z}{Z^*}  \bigg|  \bigg) \text{ } \textit{continued} =  \left\{\!\begin{array}{ll@{}>{{}}l} 
   \big\{         \big| X^* \big| , \big|  Y^* \big|            \text{ }      \textit{are not made arbitrarily small}      \big\}  , \\   \\  \boxed{r^{\prime\prime}_2}   \equiv    -   \bigg\{    \mathrm{log} \bigg[   {  \mathrm{log} \big| {X} \big|  } \big/ {\big| {Y}^{*} \big| }          \bigg] \cdot  \big| Y^* \big|^2            \bigg\}^{-2}         \cdot     \bigg\{   \big\{   {  \mathrm{log} \big| X \big|  } \big/ {\big| {Y}^{*} \big| } \big\}^{-1}  \\ \cdot \bigg[             \frac{\big| X \big|^{-1} \cdot \big\{ X \big\}^{\prime} \cdot \big| Y^* \big|   -    \mathrm{log}  \big| X \big|  \cdot \big\{ Y^* \big\}^{\prime } }{\big| Y^* \big|^2 }             \bigg]  \cdot \big| Y^* \big|^2 +    \mathrm{log} \bigg[   {  \mathrm{log} \big| X \big|  } \big/ {\big| {Y}^{*} \big| }          \bigg] \\ \\  \cdot 2 \big| Y^* \big| \cdot \big\{ Y^* \big\}^{\prime}       \bigg\}          \cdot    \bigg\{            \big| X \big|^{-1} \cdot  \big\{ X \big\}^{\prime} \cdot \big| Y^* \big|   -    \mathrm{log}   \big| X \big|\cdot \big\{ Y^* \big\}^{\prime }                  \bigg\}   \\ \\ +    \bigg\{    \mathrm{log} \bigg[   {  \mathrm{log} \big| X \big|  } \big/ {\big| {Y}^{*} \big| }          \bigg] \cdot  \big| Y^* \big|^2       \bigg\}^{-1} \cdot     \bigg\{  \big\{   - \big| X \big|^{-2} \cdot \big\{ X \big\}^{\prime }    \cdot  \big\{ X  \big\}^{\prime} \cdot \big| Y^* \big|  \\ \\   +     \big| X  \big|^{-1} \cdot  \big\{ X \big\}^{\prime\prime} \cdot \big| Y^* \big|    +   \big| X \big|^{-1} \cdot  \big\{ X \big\}^{\prime} \cdot \big\{  Y^* \big\}^{\prime}       \big\}     -    \big\{   \big| X \big|^{-1} \cdot \big\{ X \big\}^{\prime} \\ \\ \cdot \big\{ Y^* \big\}^{\prime}  +   \mathrm{log} \big| X \big| \cdot \big\{ Y^* \big\}^{\prime\prime}                  \big\}                     \bigg\}   ,  \\ \Bigg\Updownarrow  \\ \\   \big\{          \big| X^* \big| , \big|  Y^* \big|              \text{ }      \textit{are not made arbitrarily small}       \big\}  ,  \\ \\ \boxed{r^{\prime\prime}_3}   \equiv           -   \bigg\{    \mathrm{log} \bigg[   {  \mathrm{log} \big| {Y}^*  \big|  } \big/ {\big| {X}^{*} \big| }          \bigg] \cdot  \big| Y^* \big|^2            \bigg\}^{-2}         \cdot     \bigg\{   \big\{   {  \mathrm{log} \big| Y^*  \big|  } \big/ {\big| {X}^{*} \big| } \big\}^{-1}     \\ \\ \cdot \bigg[             \frac{\big| Y^*  \big|^{-1} \cdot \big\{ Y^*  \big\}^{\prime} \cdot \big| X^* \big|   -    \mathrm{log}  \big| Y^*  \big|  \cdot \big\{ X^* \big\}^{\prime } }{\big| X^* \big|^2 }             \bigg]  \cdot \big| X^* \big|^2 +    \mathrm{log} \bigg[   {  \mathrm{log} \big| X \big|  } \big/ {\big| {X}^{*} \big| }          \bigg] \\ \\  \cdot 2 \big| X^* \big| \cdot \big\{ X^* \big\}^{\prime}       \bigg\}          \cdot    \bigg\{            \big| Y^*  \big|^{-1} \cdot  \big\{ Y^*  \big\}^{\prime} \cdot \big| X^* \big|   -    \mathrm{log}   \big| X \big|\cdot \big\{ X^* \big\}^{\prime }                  \bigg\}   \\ \\ +    \bigg\{    \mathrm{log} \bigg[   {  \mathrm{log} \big| Y^*  \big|  } \big/ {\big| {X} \big| }          \bigg] \cdot  \big| X \big|^2       \bigg\}^{-1} \cdot     \bigg\{  \big\{   - \big| Y^*  \big|^{-2} \cdot \big\{ Y^*  \big\}^{\prime }    \cdot  \big\{ Y^*   \big\}^{\prime} \cdot \big| X \big|  \\ \\   +     \big| Y^*   \big|^{-1} \cdot  \big\{ Y^*  \big\}^{\prime\prime} \cdot \big| X \big|    +   \big| Y^*  \big|^{-1} \cdot  \big\{ Y^*  \big\}^{\prime} \cdot \big\{  X \big\}^{\prime}       \big\}     -    \big\{   \big| Y^*  \big|^{-1} \cdot \big\{ Y^*  \big\}^{\prime} \\ \\ \cdot \big\{ X \big\}^{\prime}  +   \mathrm{log} \big| Y^*  \big| \cdot \big\{ X \big\}^{\prime\prime}                  \big\}                     \bigg\}    ,   \\    \Bigg\Updownarrow  \\ \\ \big\{              \big| X^* \big| , \big|  Y^* \big|     \text{ }      \textit{are not made arbitrarily small}            \big\}  ,  \\ \\ \boxed{r^{\prime\prime}_4}   \equiv               -   \bigg\{    \mathrm{log} \bigg[   {  \mathrm{log} \big| {Y}^*  \big|  } \big/ {\big| {X}^{*} \big| }          \bigg] \cdot  \big| Y^* \big|^2            \bigg\}^{-2}         \cdot     \bigg\{   \big\{   {  \mathrm{log} \big| Y^*  \big|  } \big/ {\big| {X}^{*} \big| } \big\}^{-1}     
\end{array}\right. \\ \\ \tag{\textit{Second derivative of $r$}}  
\]  }

{\small 
\[  r^{\prime\prime} \equiv r^{\prime\prime} \bigg( \bigg|  \frac{X}{X^*}  \bigg| ,  \bigg|  \frac{Y}{Y^*}  \bigg|  ,   \bigg|  \frac{Z}{Z^*}  \bigg|  \bigg) \text{ } \textit{continued} =  \left\{\!\begin{array}{ll@{}>{{}}l} 
 \cdot \bigg[             \frac{\big| Y^*  \big|^{-1} \cdot \big\{ Y^*  \big\}^{\prime} \cdot \big| X^* \big|   -    \mathrm{log}  \big| Y^*  \big|  \cdot \big\{ X^* \big\}^{\prime } }{\big| X^* \big|^2 }             \bigg]  \cdot \big| X^* \big|^2 +    \mathrm{log} \bigg[   {  \mathrm{log} \big| X \big|  } \big/ {\big| {X}^{*} \big| }         \bigg]      \\ \cdot 2 \big| X^* \big| \cdot \big\{ X^* \big\}^{\prime}       \bigg\}          \cdot    \bigg\{            \big| Y^*  \big|^{-1} \cdot  \big\{ Y^*  \big\}^{\prime} \cdot \big| X^* \big|   -    \mathrm{log}   \big| X \big|\cdot \big\{ X^* \big\}^{\prime }                  \bigg\}   \\ \\ +    \bigg\{    \mathrm{log} \bigg[   {  \mathrm{log} \big| Z  \big|  } \big/ {\big| {Y}^{*} \big| }          \bigg] \cdot  \big| Y^* \big|^2       \bigg\}^{-1} \cdot     \bigg\{  \big\{   - \big| Z  \big|^{-2} \cdot \big\{ Z  \big\}^{\prime }    \cdot  \big\{ Z   \big\}^{\prime} \cdot \big| Y^* \big|  \\ \\   +     \big| Z   \big|^{-1} \cdot  \big\{ Z  \big\}^{\prime\prime} \cdot \big| Y^* \big|    +   \big| Y^*  \big|^{-1} \cdot  \big\{ Z^*  \big\}^{\prime} \cdot \big\{  Y^* \big\}^{\prime}       \big\}     -    \big\{   \big| Z   \big|^{-1} \cdot \big\{ Z  \big\}^{\prime} \\ \\ \cdot \big\{ Y^* \big\}^{\prime}  +   \mathrm{log} \big| Z  \big| \cdot \big\{ X^* \big\}^{\prime\prime}                  \big\}                     \bigg\}          \\ \Bigg\Updownarrow \\ \\       \big\{     \big| X^* \big| , \big|  Y^* \big|        \text{ }      \textit{are not made arbitrarily small}                   \big\}  .         
\end{array}\right. \\ \\ \tag{\textit{Second derivative of $r$}}  
\]  }

\bigskip

\noindent To maintain prospective Quantum advantage for Bob and Alice in the presence of more noise over their shared Quantum channel, with the converse bit transmission rate,

\begin{align*}
  (*^{\prime}) \equiv \bigg\{ (*) \text{ } \textit{subject to the constraints that  $r^{\prime} \bigg( \bigg|  \frac{X}{X^*}  \bigg| ,  \bigg|  \frac{Y}{Y^*}  \bigg|  ,   \bigg|  \frac{Z}{Z^*}  \bigg|   \bigg)$  \textit{is well-defined}, }  \textit{$r^{\prime\prime} \bigg( \bigg|  \frac{X}{X^*}  \bigg| ,  \bigg|  \frac{Y}{Y^*}  \bigg|  ,   \bigg|  \frac{Z}{Z^*}  \bigg|   \bigg)$} \text{ } \\ \textit{is well-defined}  , \bigg\{  \big\{  \mathrm{log}  \big| Z \big|^{-1} > 1   \big\}    ,   \big\{  \big| Z \big|^{-1} > 1   \big\}       ,   \big\{      \big\{ \big| Y^* \big|^2 \big\}^{-1} > 1            \big\}  , \big\{ \big\{  Z^{\prime} \big\}^{-1}  > 1   \big\}    ,    \big\{       \big| Y^* \big|^{-1} > 1            \big\}  \\ , \big\{ \big\{ \big\{ Y^* \big\}^{\prime} \big\}^{-1} > 1  \big\}   , \bigg\{   \sqrt{                   \mathrm{log} \bigg\{   {\mathrm{log} \big| {Y}^{*} \big|  }    -  \mathrm{log} \big| {X}^{*} \big|    \bigg\}       } > 0  \bigg\}      ,   \big\{  {\big| Y^* \big|^{-1}} > 0  \big\}                , \big\{ \big| \big\{ Y^* \big\}^{\prime } \big| > 1     \big\}     \\      ,  \big\{ \big| X^*  \big| > 1     \big\}   ,      \big\{ \big| Y^*  \big| > 2     \big\}      ,  \big\{ \big| \big\{ X^* \big\}^{\prime } \big| > 1     \big\}        \bigg\}    \bigg\}  , 
\end{align*}

\noindent in comparison to the channel shared by Bob and Eve, we generalize several features of a counterexample provided in {[38]}. The constraints imposed on the cardinality of Alice, Bob and Eve's alphabets from properties of the overlap function,

{\small \begin{align*}
\bigg\{  \big\{  \mathrm{log}  \big| Z \big|^{-1} > 1   \big\}    ,   \big\{  \big| Z \big|^{-1} > 1   \big\}       ,   \big\{      \big\{ \big| Y^* \big|^2 \big\}^{-1} > 1            \big\}  , \big\{ \big\{  Z^{\prime} \big\}^{-1}  > 1   \big\}    ,    \big\{       \big| Y^* \big|^{-1} > 1            \big\}      , \big\{ \big\{ \big\{ Y^* \big\}^{\prime} \big\}^{-1} \\  > 1  \big\}   , \bigg\{   \sqrt{                   \mathrm{log} \bigg\{   {\mathrm{log} \big| {Y}^{*} \big|  }    -  \mathrm{log} \big| {X}^{*} \big|    \bigg\}       } > 0  \bigg\}      ,   \big\{  {\big| Y^* \big|^{-1}} > 0  \big\}                , \big\{ \big| \big\{ Y^* \big\}^{\prime } \big| > 1     \big\}          ,  \big\{ \big| X^*  \big| > 1     \big\}   ,      \big\{ \big| Y^*  \big| > 2     \big\}   \end{align*}

\begin{align*} ,  \big\{ \big| \big\{ X^* \big\}^{\prime } \big| > 1     \big\}        \bigg\}    \bigg\}  , 
\end{align*}  }

\noindent are obtained in \textbf{Lemma} \textit{2} to argue that the first main result, \textbf{Theorem} \textit{1} stated in the next section, holds. In comparison to the points at which discontinuities emerge for the second derivative of the converse bit transmission rate,

{\small  
\[    \mathscr{D} {\tiny \big( r^{\prime\prime } \big)} \equiv  \left\{\!\begin{array}{ll@{}>{{}}l} 
       \big\{  X, X^* , Y , Y^* , Z , Z^*     :  r^{\prime\prime}_1 \longrightarrow + \infty   \big\}      , \\ \\    \big\{   X, X^* , Y , Y^* , Z , Z^*   : r^{\prime\prime}_2 \longrightarrow + \infty   \big\}        , \\ \\    \big\{   X, X^* , Y , Y^* , Z , Z^*   : r^{\prime\prime}_3 \longrightarrow + \infty   \big\}      ,    \\ \\    \big\{   X, X^* , Y , Y^* , Z , Z^*   : r^{\prime\prime}_4 \longrightarrow + \infty           \big\}   , 
\end{array}\right.  \\ =   \left\{\!\begin{array}{ll@{}>{{}}l} 
    \bigg\{ \mathrm{log}              \sqrt{ \mathrm{log}  \big| Z \big|^{-1} }    \text{ } \textit{diverges}       \bigg\}    \Longleftrightarrow \big\{  \mathrm{log}  \big| Z \big|^{-1} <  1   \big\}                                 , \\ \\  \bigg\{ \mathrm{log}              \sqrt{   \big| Z \big|^{-1} }    \text{ } \textit{diverges}       \bigg\}     \Longleftrightarrow \big\{  \big| Z \big|^{-1} < 1   \big\}                  , \\  \\   \bigg\{    \mathrm{log}              \sqrt{  \big\{ \big| Y^* \big|^2 \big\}^{-1}  }     \text{ } \textit{diverges}          \bigg\}    \Longleftrightarrow \big\{      \big\{ \big| Y^* \big|^2 \big\}^{-1} < 1            \big\}   ,        \\ \\  \bigg\{ \mathrm{log}              \sqrt{  \big\{  Z^{\prime} \big\}^{-1} }    \text{ } \textit{diverges}       \bigg\}     \Longleftrightarrow \big\{ \big\{  Z^{\prime} \big\}^{-1}  < 1   \big\}                   ,  \\ \\    \bigg\{    \mathrm{log}              \sqrt{  \big| Y^* \big|^{-1} }     \text{ } \textit{diverges}          \bigg\}    \Longleftrightarrow \big\{       \big| Y^* \big|^{-1} < 1            \big\}                  ,  \\ \\ 
        \bigg\{  \mathrm{log}   \sqrt{ \big\{ \big\{ Y^* \big\}^{\prime}   \big\}^{-1} }            \text{ } \textit{diverges}    \bigg\}  \Longleftrightarrow \big\{ \big\{ \big\{ Y^* \big\}^{\prime} \big\}^{-1} < 1  \big\}                   ,   \\ \\  \bigg\{ \mathrm{log}           \sqrt{                   \mathrm{log} \bigg\{   {\mathrm{log} \big| {Y}^{*} \big|  }    -  \mathrm{log} \big| {X}^{*} \big|    \bigg\}       }  \text{ } \textit{diverges}    \bigg\}       ,   \\ \Longleftrightarrow \bigg\{   \sqrt{                   \mathrm{log} \bigg\{   {\mathrm{log} \big| {Y}^{*} \big|  }    -  \mathrm{log} \big| {X}^{*} \big|    \bigg\}       } \approx  0  \bigg\}                   , \\   \\     \bigg\{   \mathrm{log}   \sqrt{            \frac{1}{\big| Y^{*} \big| }} \text{ } \textit{diverges}  \bigg\}  \Longleftrightarrow \big\{  {\big| Y^* \big|^{-1}} \approx  0  \big\}                   , \\ \\          \bigg\{  \mathrm{log}  \sqrt{\big\{ Y^* \big\}^{\prime }}  \text{ } \textit{diverges}  \bigg\}  \Longleftrightarrow \big\{ \big| \big\{ Y^* \big\}^{\prime } \big| < 1      \big\}       , \\ \\       \bigg\{     \mathrm{log}  \sqrt{\big| X^* \big|  } \text{ } \textit{diverges}  \bigg\}  \Longleftrightarrow \big\{ \big| X^*  \big| <  1     \big\}    ,  \\ \\ \bigg\{   \mathrm{log}  \sqrt{\mathrm{log}   \big| Y^{*} \big|}  \text{ } \textit{diverges}  \bigg\}  \Longleftrightarrow \big\{ \big| Y^*  \big| <  2     \big\}  , \\ \\    \bigg\{  \mathrm{log}  \sqrt{\big\{ X^* \big\}^{\prime }}  \text{ } \textit{diverges}  \bigg\} \Longleftrightarrow \big\{ \big| \big\{ X^* \big\}^{\prime } \big| <  1     \big\}  .               
\end{array}\right.  
\]  }

%{\small  
%\[    \mathscr{D} {\tiny \big( r^{\prime\prime } \big)} \text{ } \textit{continued} \equiv  \left\{\!\begin{array}{ll@{}>{{}}l} 
 %      \big\{  X, X^* , Y , Y^* , Z , Z^*     :  r^{\prime\prime}_1 \longrightarrow + \infty   \big\}      , \\ \\    \big\{   X, X^* , Y , Y^* , Z , Z^*   : r^{\prime\prime}_2 \longrightarrow + \infty   \big\}        , \\ \\    \big\{   X, X^* , Y , Y^* , Z , Z^*   : r^{\prime\prime}_3 \longrightarrow + \infty   \big\}      ,    \\ \\    \big\{   X, X^* , Y , Y^* , Z , Z^*   : r^{\prime\prime}_4 \longrightarrow + \infty           \big\}   , 
%\end{array}\right.  \\ =   \left\{\!\begin{array}{ll@{}>{{}}l} 
  
%\end{array}\right.  
%\]  }  

\noindent The counterexample provided in [38], given alphabets with only a few letters, was originally provided to demonstrate that the bit transmission rates satisfying,

{\small \begin{align*}
r < \underset{P_{\textbf{X}}}{\mathrm{sup}}  \big\{ \mathrm{min} \big\{  I {\tiny \big(} \textbf{X} , \textbf{Y} {\tiny \big)}  , \underset{z}{\mathrm{min}}  \big\{ H_Q {\tiny \big(} \textbf{Y} \big| \textbf{Z} = z {\tiny \big)}   -  H_P {\tiny \big(}  \textbf{Y} \big| \textbf{X} {\tiny \big)}  \big\}  \big\}    \big\} \text{, }
\end{align*} }

\noindent are dependent upon artifacts of the proof technique, rather than being reflective upon physical limitations, and constraints, of the total number of bits that can be transmitted in various Quantum information processing tasks.

\bigskip

\noindent In the presence of less noise over the channel between Alice and Bob in comparison to that between Bob and Eve, the ability of players to error correction, and accept messages that Eve has not tampered with, is related to the following computations for upper bounding various probabilities,

{\small \begin{align*}
  \underset{i \in \textbf{N}}{\sum}     \textbf{P} \big[     \text{Alice, or Bob, perform a decoding error on input } C_i     \big]      \text{, }
\end{align*} } 

\noindent and also of the upper bound,

{\small \begin{align*}
        \underset{i\in \textbf{N}}{\sum}     \textbf{P} \big[     \text{Alice, or Bob, perform a false acceptance error on input } C_i     \big]             \text{, }
\end{align*} }

\noindent both of which are taken with respect to the measure $\textbf{P} {\tiny \big[ \cdot \big]}$ over messages shared between players over the quantum channel. For infinitely many bits, one can straightforwardly formulate the computation of the behavior of the above probabilities to,

{\small \begin{align*}
       \underset{i \longrightarrow + \infty}{\mathrm{lim}} \bigg\{  \underset{C_i}{\sum}     \textbf{P} \big[     \text{Alice, or Bob, perform a decoding error on input } C_i     \big]  \bigg\}       \text{, }   \end{align*}

       \begin{align*} \underset{i \longrightarrow + \infty}{\mathrm{lim}}  \bigg\{ \underset{C_i}{\sum}     \textbf{P} \big[     \text{Alice, or Bob, perform a false acceptance error on input } C_i     \big]     \bigg\}         \text{, }
\end{align*} }

\noindent asymptotically. As $i \longrightarrow + \infty$, the set of codewords is given by,

{\small \begin{align*}
  C_{+\infty} \equiv \underset{C}{\bigcup} \big\{ \text{infinitely many bit codewords } C \big\}   \text{. }
\end{align*} }

\noindent Paradoxically, to argue that quantum advantage for communication continues to hold if the presence of noise is greater over the quantum channel between Alice and Bob, in comparison to the channel between Eve and Bob, one must characterize various upper bounds to the bit transmission rate, as stated through the converse result. In the original result for the lower bound of the bit transmission rate $r$ before stating the converse, an infimum over the conditional Shannon entropy, and Mutual Information entropy, is performed.

For the converse result on the upper bound of $r$, one must convincingly demonstrate that the probabilities of error correction, and false acceptance, can still be made to occur with smaller probability in comparison to those over the quantum channel between Bob and Eve.

\subsection{Statement of main results}

\noindent We provide a statement of the two main results below, the first of which concerns the upper bound on the bit transmission rate from the converse result, and the second of which concerns the stochastic domination of error correction, and false acceptance, probabilities between the Quantum channel between Alice and Bob, in addition to the Quantum channel between Bob and Eve.

\bigskip

\noindent \textbf{Theorem} \textit{1} (\textit{converse result on the bit transmission rate r for communication between Alice and Bob over a noisy Quantum channel}). One has that a converse result for $r$, $(*^{\prime})$, holds. Such a transmission rate, $r$ is characterized by the two following possible behaviors:

\begin{itemize}
    \item[$\bullet$] \underline{Subcritical bit transmission rates}. Below rates $r$ provided in the converse upper bound in (*), $r < (*^{\prime})$, the probability that Alice and Bob mistakenly accept a bit that is tampered with by Eve, in addition to Bob performing a decoding error on a bit that is initially encoded by Alice, $p_{\mathrm{FA}}$ and $p_{\mathrm{EC}}$, can be made arbitrarily small and large, respectively. 

      \item[$\bullet$] \underline{Supercritical bit transmission rates}. Above rates $r$ provided in the converse upper bound in (*), $r > (*^{\prime})$, the probability that Alice and Bob mistakenly accept a bit that is tampered with by Eve, in addition to Bob performing a decoding error on a bit that is initially encoded by Alice, $p_{\mathrm{FA}}$ and $p_{\mathrm{EC}}$, are strictly positive and small, respectively.
\end{itemize}

\noindent Under this strict upper bound ( for the bit transmission rate, Alice and Bob cannot guarantee, with arbitrarily small probability, that instances of decoding errors, and false acceptance, occur.

\bigskip

\noindent \textbf{Theorem} \textit{2} (\textit{stochastic domination of error correction, and false acceptance, probabilities for communication over Quantum channels between Alice and Bob, and between Bob and Eve}). Fix the following two parameters,

{\small \begin{align*}
N_{A \longleftrightarrow B} \equiv    \text{Noise over the Quantum channel between Alice and Bob}        \text{,} \\ \\  N_{B \longleftrightarrow E} \equiv       \text{Noise over the Quantum channel between Bob and Eve} \text{, }
\end{align*} }

\noindent for the noise over each Quantum channel, with $N_{A \longleftrightarrow B} > N_{B \longleftrightarrow E}$. Denote the probabilities,

{\small \begin{align*}
 P_{\mathrm{EC}, A \longleftrightarrow B} \equiv         \underset{A \longleftrightarrow B}{\underset{\mathrm{ec} \in \mathrm{EC}}{\mathrm{sup}}}    p_{\mathrm{ec}} \equiv          \mathrm{sup} \big\{ \text{success probability of a player using an error correcting} \\ \text{ code over Alice and Bob's Quantum channel} \big\}               \text{, }   \end{align*}

  \begin{align*} P_{\mathrm{FA}, A \longleftrightarrow B} \equiv           \underset{A \longleftrightarrow B}{\underset{\mathrm{fa} \in \mathrm{FA}}{\mathrm{inf}}}    p_{\mathrm{fa}} \equiv           \mathrm{inf} \big\{ \text{failure probability of a player accepting a message} \\    \text{ that should have not been accepted over Alice and Bob's Quantum channel} \big\}                    \text{, }  \end{align*}

  \begin{align*} P_{\mathrm{EC}, B \longleftrightarrow E} \equiv         \underset{B \longleftrightarrow E}{\underset{\mathrm{ec} \in \mathrm{EC}}{\mathrm{sup}}}    p_{\mathrm{ec}} \equiv          \mathrm{sup} \big\{ \text{success probability of a player using an error correcting code over Bob} \\   \text{and Eve's Quantum channel} \big\}               \text{, }   \end{align*}

  \begin{align*} P_{\mathrm{FA}, B \longleftrightarrow E} \equiv           \underset{B \longleftrightarrow E}{\underset{\mathrm{fa} \in \mathrm{FA}}{\mathrm{inf}}}    p_{\mathrm{fa}} \equiv           \mathrm{inf} \big\{ \text{failure probability of a player accepting a message that should have } \\ \text{not been accepted over Bob and Eve's Quantum channel} \big\}                    \text{. }
\end{align*} } 

\noindent corresponding to the probabilities for error correction, and false acceptance, probabilities over the two Quantum channels, in addition to the probabilities,

{\small \begin{align*}
   p_{\mathrm{FA}, A \longleftrightarrow B} \propto  p_{\mathrm{DE}, A \longleftrightarrow B} \equiv \underset{A \longleftrightarrow B}{\underset{\mathrm{de} \in \mathrm{DE}}{\mathrm{sup}}}  \big\{ \text{decoding error probability of a message transmitted over Alice } \\ \text{ and Bob's Quantum channel}      \big\} \text{, }  \end{align*}

   \begin{align*} p_{\mathrm{FA}, B \longleftrightarrow E} \propto  p_{\mathrm{DE}, B \longleftrightarrow E} \equiv \underset{B \longleftrightarrow E}{\underset{\mathrm{de} \in \mathrm{DE}}{\mathrm{sup}}}  \big\{ \text{decoding error probability of a message transmitted over Bob} \\ \text{ and Eve's Quantum channel}      \big\}    \text{. }    
\end{align*} }

\noindent One has the following stochastic domination of error correction, and false acceptance, probabilities,

{\small \begin{align*}
 p_{\mathrm{EC}, A \longleftrightarrow B} >   p_{\mathrm{EC}, B \longleftrightarrow E}  \text{, } \\ \\ \big\{   \exists C^{\prime} \in \textbf{R} :  { p_{\mathrm{EC}, A \longleftrightarrow B}} \big/ {p_{\mathrm{FA}, A \longleftrightarrow B}} \geq C^{\prime} \big\}   \text{, } 
\end{align*} } 

\noindent between Alice and Bob's, and Bob and Eve's, quantum channels.

\bigskip

\noindent The first two results reflect upon potential sources of Quantum advantage in the presence of more noise over the channel between Alice and Bob, in comparison to that over the channel between Bob and Eve. The final main result below, which directly adapts \textbf{Theorem} \textit{3} in [38] for the range of possible transmission rates in the original lower bound result, asserts the existence of suitable policies so that Alice and Bob can map sequences into the Authenticated space.

\bigskip

\noindent \textbf{Theorem} \textit{3} (\textit{the existence of suitable protocols for Alice and Bob so that bit codewords transmitted over the Quantum channel can be mapped into the set of authenticated messages with high probability}). Fix $0 \leq p < q \leq \frac{1}{2}$. For any $r > h \big( q \big) - h \big( p \big)$, and some $\epsilon >0$, for all sufficiently large $n$ there exists a protocol,

{\small \begin{align*}
 {\tiny \pi^n \equiv \big( E^n , D^n \big)}  \equiv {\tiny \big( }  \text{Alice's policy for decoding } n\text{-bit codewords}, \text{Bob's policy for decoding }  n-\text{bit} \\  \text{ codewords} {\tiny \big) }   \text{, }
\end{align*} } 

\noindent so that $\mathcal{N}^n_{p,q} \overset{\{ \pi^n,\epsilon \} }{\longrightarrow} \mathcal{A}^{rn}$. As $n\longrightarrow + \infty$, the policy of Alice and Bob takes the form $\mathcal{N}^{+\infty}_{p,q} \overset{\{ \pi^{+\infty},\epsilon \} }{\longrightarrow} \mathcal{A}^{+\infty}$.

\bigskip

\noindent \textbf{Corollary} \textit{1} (\textit{the probability of false acceptance below the threshold provided in the statement of the converse bit transmission rate is related to a high probability of Alice and Bob being able to perform error correction}). For bit transmission rates $r$ lower than the form provided in (*) one has the correspondence,

{\small \begin{align*}
 \big\{ p_{\mathrm{EC}} \approx 1 \big\} \Longleftrightarrow \big\{ p_{\mathrm{FA}} \approx 0  \big\} \text{. }
\end{align*} }

\bigskip

\noindent 
\noindent \textbf{Corollary} \textit{2} (\textit{stability of the inverse monotonicity of the radius of Hamming balls with respect to channel noise for transmitted codewords with infinitely many bits}). Denote $\mathcal{B}_1$ as the Hamming ball. Under infinitely many bits one has that,

\begin{align*}
   \mathcal{N}^n_{p,q, A \longleftrightarrow B}        \overset{ \{ n\longrightarrow + \infty\} }{\longrightarrow} \mathcal{N}^{+\infty}_{p,q, A \longleftrightarrow B}         \text{, }  \\      \mathcal{N}^n_{p,q, B \longleftrightarrow E} \overset{\{ n\longrightarrow + \infty\} }{\longrightarrow} \mathcal{N}^{+\infty}_{p,q, B \longleftrightarrow E}           \text{, }
\end{align*}

\noindent corresponding to the channels across which codewords with infinitely many bits are transmitted. One has, 

{\small \begin{align*}
  \frac{\textbf{P} \bigg[ \big\{  \forall r_1 > 0, \exists n_1 > 0 :  \big|  \mathscr{C} \cap \mathcal{B}_1 \big( r_1, \mathcal{C}_1 \big)  \big| \equiv n_1       \big\}   \bigg| \big\{  \mathcal{C}_1 \text{ is transmitted through } \mathcal{N}^{+\infty}_{p,q, A \longleftrightarrow B} \big\}   \bigg] }{\textbf{P} \bigg[ \big\{  \forall r_2 > r_1 > 0, \exists n_2 > n_1  > 0 :       \big| \mathscr{C} \cap \mathcal{B}_1 \big( r_2, \mathcal{C}_2 \big) \big| \equiv n_2  \big\}   \bigg| \big\{  \mathcal{C}_2 \text{ is transmitted through } \mathcal{N}^{+\infty}_{p,q, B \longleftrightarrow E} \big\}  \bigg]} \end{align*}

  \begin{align*} \gtrsim   \frac{N^{+\infty}_{A \longleftrightarrow B}}{N^{+\infty}_{B \longleftrightarrow E}}   \text{, }
\end{align*} } 

\noindent for the constants,

\begin{align*}
  N_{A \longleftrightarrow B}    \overset{\{ \text{Infinitely many bits}\} }{\longrightarrow} N^{+\infty}_{A \longleftrightarrow B}   \text{,} \\ N_{B \longleftrightarrow E} \overset{\{ \text{Infinitely many bits}\} }{\longrightarrow} N^{+\infty}_{B \longleftrightarrow E} \text{.}
\end{align*}

\bigskip 

The three results above, followed by \textbf{Corollary} \textit{1} and \textit{2}, demonstrate how the converse result on the bit transmission rate is expected to relate not only to the probability of decoding, and false acceptance, as $n \longrightarrow + \infty$, but also how Eve can tamper with messages sent from Alice and Bob before the set of all authenticated messages is constructed.

In forthcoming arguments on error correcting codes that Alice, or Bob, can make use of for error correction of bit codewords, it is important to study the asymptotic behavior of the probability,

{\small \begin{align*}
 \textbf{P} \bigg[ \bigg\{ \exists c_n > 0 :  \frac{\big| \text{Authenticated } n-\text{bit messages from Alice and Bob, over } A \longleftrightarrow B \big| }{\big| \text{Authenticated } n- \text{bit messages by Eve, over } B \longleftrightarrow E \big| } \approx c_n  \bigg\}    \bigg]   \text{, }
\end{align*} }

\noindent for codewords with $n$ bits. For messages with infinitely many bits as $n \longrightarrow + \infty$, and bit transmission rates satisfying,

{\small \begin{align*}
    r <  \underset{P_{\textbf{X}}}{\mathrm{sup}}  \big\{ \mathrm{min} \big\{  I {\tiny \big(} \textbf{X} , \textbf{Y} {\tiny \big)}  , \underset{z}{\mathrm{min}}  \big\{ H_Q {\tiny \big(} \textbf{Y} \big| \textbf{Z} = z {\tiny \big)}   -  H_P {\tiny \big(}  \textbf{Y} \big| \textbf{X} {\tiny \big)}  \big\}  \big\}    \big\}\text{. }
\end{align*} }

\noindent For upper bounds to the bit transmission rate, as provided in the converse, the probability of determining some strictly positive $c_n$ for which,

{\small \begin{align*}
    \bigg\{  \frac{\big| \text{Authenticated } n-\text{bit messages from Alice and Bob, over } A \longleftrightarrow B \big| }{\big| \text{Authenticated } n- \text{bit messages by Eve, over } B \longleftrightarrow E \big| } \approx c_n \bigg\}  \text{, }
\end{align*}
 } 

\noindent monotonically decreases with respect to channel noise. That is, the overlap between the bit codewords which Alice and Bob authenticate, in comparison to the bit codewords which Eve tampers with, and subsequently attempts to embed within Alice and Bob's authenticated space, satisfies,

{\small \begin{align*}
     \big\{ c_m > c_n > 0  \big\} \Longleftrightarrow           \bigg\{  \frac{\big| \text{Authenticated } m-\text{bit messages from Alice and Bob, over } A \longleftrightarrow B \big| }{\big| \text{Authenticated } m- \text{bit messages by Eve, over } B \longleftrightarrow E \big| } \end{align*}

     \begin{align*} \approx c_m  >    \frac{\big| \text{Authenticated } n-\text{bit messages from Alice and Bob, over } A \longleftrightarrow B \big| }{\big| \text{Authenticated } n- \text{bit messages by Eve, over } B \longleftrightarrow E \big| }   \approx  c_n        \bigg\}            \text{, }
\end{align*} }

\noindent for bit transmission rates in the converse result, namely,

{\small \begin{align*}
   r >  \underset{P_{\textbf{X}}}{\mathrm{sup}}  \big\{ \mathrm{min} \big\{  I {\tiny \big(} \textbf{X} , \textbf{Y} {\tiny \big)}  , \underset{z}{\mathrm{min}}  \big\{ H_Q {\tiny \big(} \textbf{Y} \big| \textbf{Z} = z {\tiny \big)}   -  H_P {\tiny \big(}  \textbf{Y} \big| \textbf{X} {\tiny \big)}  \big\}  \big\}    \big\}   \text{. }
\end{align*} }

\noindent Necessarily, given the expectation that the ratio between the cardinality of the set of authenticated messages by Alice or Bob, in comparison to the cardinality of the set of authenticated messages by Eve, will never vanish, there should exist some real valued limit,

{\small \begin{align*}
 c_{+\infty} \equiv  \underset{n \longrightarrow + \infty}{\mathrm{lim}} c_n \in \textbf{R}   \text{,}
\end{align*} } 

\noindent for which,

{\small \begin{align*}
  c_{+\infty} \equiv \underset{j}{\mathrm{inf}} \bigg\{ \forall j > i > 0 , \nexists \text{ } c_j > c_i :   \frac{\big| \text{Authenticated } m-\text{bit messages from Alice and Bob,} \cdots   }{\big| \text{Authenticated } m- \text{bit messages by Eve,} \cdots }  \\ \frac{\cdots \text{ over } A \longleftrightarrow B  \big| }{\cdots   \text{ over } B \longleftrightarrow E \big| } \\ \approx c_j  >    \frac{\big| \text{Authenticated } n-\text{bit messages from Alice and Bob, over } A \longleftrightarrow B \big| }{\big| \text{Authenticated } n- \text{bit messages by Eve, over } B \longleftrightarrow E \big| }  \approx  c_i            \bigg\}  \text{.}
\end{align*}
 } 

\noindent For infinitely many bits, there could exist a sequence of constants for which,

{\small \begin{align*}
 c_{+\infty} \equiv \underset{m^{\prime}}{\mathrm{inf}} \bigg\{ \forall m^{\prime} > n^{\prime}, \exists c_{m^{\prime}} \equiv c_{n^{\prime}}  :  \frac{\big| \text{Authenticated } m^{\prime}-\text{bit messages from Alice and Bob,} \cdots  }{\big| \text{Authenticated } m^{\prime}- \text{bit messages by Eve,} \cdots  } \end{align*}

 \begin{align*} \frac{\cdots \text{over } A \longleftrightarrow B  \big| }{\cdots \text{over } B \longleftrightarrow E  \big| } \end{align*}

 \begin{align*}  =   \frac{\big| \text{Authenticated } n^{\prime}-\text{bit messages from Alice and Bob, over } A \longleftrightarrow B \big| }{\big| \text{Authenticated } n^{\prime}- \text{bit messages by Eve, over } B \longleftrightarrow E \big| }  = c_{n^{\prime}} = c_{m^{\prime}} \bigg\}   \text{. }
\end{align*} }

\noindent Additionally, properties of error correcting codes for Quantum channels for the converse result on the bit transmission rate,

{\small \begin{align*}
   r >  \underset{P_{\textbf{X}}}{\mathrm{sup}}  \big\{ \mathrm{min} \big\{  I {\tiny \big(} \textbf{X} , \textbf{Y} {\tiny \big)}  , \underset{z}{\mathrm{min}}  \big\{ H_Q {\tiny \big(} \textbf{Y} \big| \textbf{Z} = z {\tiny \big)}   -  H_P {\tiny \big(}  \textbf{Y} \big| \textbf{X} {\tiny \big)}  \big\}  \big\}    \big\}   \text{, }
\end{align*} } 

\noindent are shown to impart Quantum advantage through computation of the derivatives,

{\small \begin{align*}
 \mathscr{P}_1 \equiv   \frac{\mathrm{d} \textbf{P} \big[ \cdot \big| \big\{  \mathcal{C}_1 \text{ transmitted over } A \longleftrightarrow B, \text{ with noise level }  N_{A \longleftrightarrow B} \big\} \big]}{\mathrm{d} \big(  N_{A \longleftrightarrow B} \big) }            \text{, } \\ \mathscr{P}_2 \equiv  \frac{\mathrm{d} \textbf{P} \big[ \cdot \big|  \big\{ \mathcal{C}_2 \text{ transmitted over } B \longleftrightarrow E, \text{ with noise level }  N_{B \longleftrightarrow E} \big\}  \big]}{\mathrm{d} \big(  N_{B \longleftrightarrow E} \big) }         \text{, }
\end{align*} }

\noindent of conditional probabilities, and bit codewords,

{\small \begin{align*}
 \mathcal{C}_1 \equiv 
 \underset{\text{bits}}{\bigcup} \big\{ \text{Codewords transmitted over } A \longleftrightarrow B \text{ with infinitely many bits} \big\}   \text{,} \\ \mathcal{C}_2 \equiv 
 \underset{\text{bits}}{\bigcup} \big\{ \text{Codewords transmitted over } B \longleftrightarrow E \text{ with infinitely many bits} \big\}        \text{, }
\end{align*} }

\noindent with respect to the level of noise over the Quantum channels between Alice and Bob, and between Bob and Eve, respectively. For $N^{\prime}_{A \longleftrightarrow B} > N_{A\longleftrightarrow B}$, and $N^{\prime}_{B \longleftrightarrow E} > N_{B \longleftrightarrow E}$, the fidelity of bit codewords transmitted over each Quantum channel monotonically decrease. Namely, one has that,

{\small \begin{align*}
\frac{1}{N^{\prime}_{A \longleftrightarrow B} -  N_{A \longleftrightarrow B}} \bigg\{   \textbf{P} \bigg[ \cdot \bigg|  \big\{ \mathcal{C}_1 \text{ transmitted over } A \longleftrightarrow B, \text{ with noise level }  N^{\prime}_{A \longleftrightarrow B} \big\} \bigg] \\  -   \textbf{P} \bigg[ \cdot \bigg| \big\{ \mathcal{C}_1 \text{ transmitted over } A \longleftrightarrow B, \text{ with noise level }  N_{A \longleftrightarrow B} \big\}  \bigg] \bigg\}  < 0 \end{align*}

\begin{align*}  \Bigg\Updownarrow \\  \mathscr{P}_1 < 0 \text{, } \end{align*}

\begin{align*} \frac{1}{N^{\prime}_{B \longleftrightarrow E} -  N_{B \longleftrightarrow E}} \bigg\{   \textbf{P} \bigg[ \cdot \bigg| \big\{  \mathcal{C}_2 \text{ transmitted over } B \longleftrightarrow E, \text{ with noise level }  N^{\prime}_{B\longleftrightarrow E} \big\}  \bigg] \\  -   \textbf{P} \bigg[ \cdot \bigg| \big\{ \mathcal{C}_2 \text{ transmitted over } B \longleftrightarrow E, \text{ with noise level }  N_{B \longleftrightarrow E} \big\} \bigg] \bigg\}  < 0 \end{align*}

\begin{align*}  \Bigg\Updownarrow \\   \mathscr{P}_2 < 0 \text{. }
\end{align*} }

\noindent In arguments for characterizing properties, and resiliency, of error correcting codes in the presence of channel noise, the condition above, denoted (MON) for monotonicity, is used to quantify the inversely proportional relationship between the radii of Hamming balls that are centered about bit codewords transmitted over $A \longleftrightarrow B$, and over $B \longleftrightarrow E$, and each respective channel noise threshold.

\subsection{Paper organization}

\noindent Given the prerequisites in previous subsections on transmission rates, the strict upper bound from the converse result, and several related objects, in the next section, \textit{2.1}, we introduce several objects related to cryptography. With such objects, we seek to formalize the setting in which the security of communication across Quantum channels can be characterized; moreover, we comment upon favorable properties of error correcting codes over the channel between Alice and Bob, particularly focusing upon: (1) false acceptance, decoding, and authentication, probabilities for Alice and Bob; (2) the probability that each player can construct a codebook which has an arbitrarily small chance of being decoded incorrectly; (3) properties of good encodings, and decodings, in the presence of noise over the Quantum channel, in \textit{2.4}; as well complications raised by the generalized noise model, ranging from false acceptance, and decoding error probabilities, along with the existence of suitable protocols for Alice and Bob through error correcting codes, in \textit{2.5}. Following the arguments for properties of error correcting codes in the presence of noise, we combine all such properties to argue that \textbf{Theorem} \textit{3} holds, before which arguments for \textbf{Theorem} \textit{1}, and \textit{2}, are provided. All arguments for the main results stated in \textit{1.4} are provided in the final section, \textit{3}. We conclude by demonstrating how \textbf{Corollary} \textit{1} and \textbf{Corollary} \textit{2} hold.

\section{Game-theoretic objects for cryptography}

\subsection{Resources, authentication, and simulators}

\noindent Besides several game-theoretic objects that have been introduced for communication of entangled Quantum information between Alice and Bob, it is also of importance to discuss game-theoretic objects which share connections with cryptography, and hence with error correction, decoding, and false acceptance, probabilities. To this end, we will provide a brief overview of some of the objects introduced in [38], as it was originally a conjecture of Ostrev raised in this work for converse results on $r$.

In particular, the bit transmission rate over the Quantum channel shared by Alice and Bob, along with its inherent noise level, is used to construct more complicated information-theoretic objects, namely codebooks. By a codebook $\mathcal{C}$ of each agent, denoted respectively with $\mathcal{C}_A$ and $\mathcal{C}_B$ for Alice's, and Bob's, codebooks, we mean a collection of finitely many bits, transmitted at the fundamental rate constrained by upper bounds on $r$, that players send, and receive, over the Quantum channel. Adopting the notation originally provided in [38] denote,

{\small  \begin{align*}
      \mathcal{R} \neq \mathcal{S} \equiv \text{Two resources which take inputs from Alice, Bob and Eve}    \text{, } \end{align*}

      \begin{align*} \mathcal{R} \big| \big| \mathcal{S} \equiv \text{A resource of } \mathcal{R}  \text{ and } \mathcal{S}  \text{ simultaneously}           \text{, } \\ \\   d {\tiny \big( \cdot , \cdot \big) } \equiv \text{A metric between two resources}     \text{, } \end{align*}

      \begin{align*}  \mathcal{N}^n_{p,q} \equiv \text{Alice and Bob's resource for sharing } n\text{-bit authenticated messages over}   \\ \text{the Quantum channel with Bernoulli}-p\text{ random variables}    \text{, }  \\ \\ \mathcal{A}^{rn} \equiv    \text{Alice and Bob's resource for sharing } n\text{-bit authenticated message over the} \\ \text{Quantum channel}       \text{. }
\end{align*} }

\noindent Each of the two resources, $\mathcal{R}$ and $\mathcal{S}$, introduced above denote interfaces over the Quantum channel that Alice, Bob, or Eve, can transfer bits into, from which responses can be generated. The metric between resources of the channel, $d {\tiny \big( \cdot, \cdot \big)}$, satisfies, [38],

\begin{itemize}
\item[$\bullet$] \textit{Identity}: The distance, with respect to the metric, of a resource with itself, vanishes,

{\small \begin{align*}
  d \big( \mathcal{R}, \mathcal{R} \big) \equiv 0  \text{. }
\end{align*}} 
\item[$\bullet$] \textit{Symmetry}: The distance, with respect to the metric, between $\mathcal{R}$ and $\mathcal{S}$, and between $\mathcal{S}$ and $\mathcal{R}$, are equal,

{\small \begin{align*}
  d \big( \mathcal{R}, \mathcal{S} \big) \equiv   d \big( \mathcal{S}, \mathcal{R} \big)  \text{. }
\end{align*} }

\item[$\bullet$] \textit{The triangle inequality}. Denote three resources over a Quantum channel with $\mathcal{R}$, $\mathcal{S}$, and $\mathcal{T}$. One has that,

{\small \begin{align*}
   d \big( \mathcal{R}, \mathcal{S} \big)    + d \big( \mathcal{S}, \mathcal{T} \big)         \geq  d \big( \mathcal{R}, \mathcal{T} \big)   \text{. }
\end{align*} } 
\end{itemize}

\noindent Under the assumption that the noise over the Quantum channel between Alice and Bob is strictly less than that over the Quantum channel between Bob and Eve, a version of the stochastic domination, in \textbf{Theorem} \textit{2}, which is related to the converse result, in \textbf{Theorem} \textit{1}, can be anticipated more easily. That is, given the two protocols,

{\small \begin{align*}
  D^n  \overset{\{ n \longrightarrow + \infty \} }{\longrightarrow} D^{+\infty} \equiv \text{Alice's protocol on codewords with infinitely many bits}  \text{, } \\   E^n   \overset{\{ n \longrightarrow + \infty \} }{\longrightarrow} E^{+\infty} \equiv \text{Bob's protocol on codewords with infinitely many bits}  \text{, }
\end{align*} }

\noindent which are implemented by Alice, and Bob, respectively, for communication over the Quantum channel, denote the resource,

{\small \begin{align*}
  E^n D^n \mathcal{N}^n_{p,q}  \text{, }
\end{align*}
 }

\noindent as that which is obtained by applying Bob's, and Alice's, protocols to the bits that are transmitted over the Quantum channel, in addition to the resource,

{\small \begin{align*}
   \sigma_E \mathcal{A}^{rn} \text{, }
\end{align*} }

\noindent obtained by an application of Eve's simulator, $\sigma_E$, to the set of \textit{authenticated} $rn$ bit messages transmitted over the Quantum channel. The distance between resources,

{\small \begin{align*}
    d \big( E^n D^n \mathcal{N}^n_{p,q}, \sigma_E \mathcal{A}^{rn} \big)       \text{, }
\end{align*}
 } 

\noindent is identically,

{\small \begin{align*}
  \mathrm{max} \big( p_{\mathrm{de}} , p_{\mathrm{fa}} \big)  \text{, }
\end{align*} }

\noindent namely, the maximum of the probabilities of a decoding error, and false acceptance, occurring, respectively.

\subsection{Error correction probabilities over $A \longleftrightarrow B$ and $B \longleftrightarrow E$, codebooks}

Counterintuitively, the fact that Alice and Bob can continue to achieve an advantage with regards to the error correction, and false acceptance, probabilities for transmitting authenticated bits, and hence, codewords, over the Quantum channel will be demonstrated by appealing to the fact that the presence of noise does not impede the stochastic domination between either the error correction, or false acceptance probabilities, that are provided in \textbf{Theorem} \textit{2},

{\small \begin{align*}
 p_{\mathrm{EC}, A \longleftrightarrow B} >   p_{\mathrm{EC}, B \longleftrightarrow E}  \text{, } \\ \\   p_{\mathrm{FA}, B \longleftrightarrow E}  >   p_{\mathrm{FA}, A \longleftrightarrow B}  \text{, }
\end{align*} }

\noindent over the Quantum channels between Alice and Bob, and between Bob and Eve, respectively.

Besides objects that have been introduced regarding resources, authentication, and simulators from the adversarial party Eve, it is also important to demonstrate how much complicated cryptographic objects are constructed from bits that are transmitted between two parties over the Quantum channel. Namely, given bits that are transmitted over a noisy Quantum channel, a codeword that an agent receives is given by, a codeword is denoted with.

{\small \begin{align*}
    \mathscr{C} \equiv \underset{\text{Bits } b, \text{ } b \in \{ 0 , 1 \}^n
    }{\bigcup} \big\{ b \text{ transmitted over the Quantum channel at transmission rate } r \big\} \text{. }
\end{align*} }

\noindent Given the simplified model of communication over binary, symmetric Quantum channels, bits that are sent from Alice to Bob, or Bob to Alice, which can be intercepted by Eve can be randomly flipped to the opposing bit (ie, $0 \mapsto 1$, or $1 \mapsto 0$ with a Bernoulli $p$ random-variable). This simplified model of communication across noisy Quantum channels, albeit initially being useful for describing fundamental physical constraints from Classical information theory, can be generalized by incorporating conditional probability distributions for inputs sent to either Alice, or Bob. Given potential interference from Eve, with $z \equiv \big( z_1, \cdots, z_n \big)$, introduce the conditional probability distributions,

{\small   \begin{align*}
P {\tiny \big[ \cdot \big| \cdot \big] }   \text{, } \\ Q {\tiny \big[ \cdot \big| \cdot \big]  } \text{. }
\end{align*}
 } 

\noindent In the first conditional probability distribution above, ${\tiny P \big[ \cdot \big| \cdot \big]}$, it is convenient to take the conditioning as the input from Alice, while for the second conditional probability distribution above, ${\tiny Q \big[ \cdot \big| \cdot \big]}$, it is convenient to take the conditioning as the input from Eve. Under more generalized assumptions on the role that Eve can assume in tampering information sent by Alice, or Bob, over the Quantum channel, previous arguments in [38] proved that the following lower bound,

{\small \begin{align*}
    r <  \underset{P_{\textbf{X}}}{\mathrm{sup}}  \big\{ \mathrm{min} \big\{  I {\tiny \big(} \textbf{X} , \textbf{Y} {\tiny \big)}  , \underset{z}{\mathrm{min}}  \big\{ H_Q {\tiny \big(} \textbf{Y} \big| \textbf{Z} = z {\tiny \big)}   -  H_P {\tiny \big(}  \textbf{Y} \big| \textbf{X} {\tiny \big)}  \big\}  \big\}    \big\}  \text{, }
\end{align*} }

\noindent for the bit transmission rate holds. As described previously in \textit{1.3}, as a constrained optimization problem raised over the set of admissible probability measures, $P_X$, the bit transmission rate not only determines the probability with which error correction, authentication, and false acceptance, can be simultaneously executed by Alice and Bob, but also aspects of Quantum information which can detect when bits transmitted from one player are tampered with.

The possibility of a converse result for the bit transmission rate, over the collection of probability measures,

{\small \begin{align*}
  P_{\textbf{X}} \equiv \underset{X \in \textbf{X}}{\bigcup}  P_X \text{, }
\end{align*} }

\noindent would imply that the transmission rate $r$ satisfies,

{\small \begin{align*}
   r >   \underset{P_{\textbf{X}}}{\mathrm{sup}}  \big\{ \mathrm{min} \big\{  I {\tiny \big(} \textbf{X} , \textbf{Y} {\tiny \big)}  , \underset{z}{\mathrm{min}}  \big\{ H_Q {\tiny \big(} \textbf{Y} \big| \textbf{Z} = z {\tiny \big)}   -  H_P {\tiny \big(}  \textbf{Y} \big| \textbf{X} {\tiny \big)}  \big\}  \big\}    \big\}  \text{, }
\end{align*} } 

\noindent would imply that the strict upper bound for the transmission rate provided in \textbf{Theorem} \textit{1} holds.

\bigskip

\noindent We combine all of the observations provided in the steps above when proving the desired upper bound for the bit transmission rate in the converse result. With respect to the noise threshold over each channel, transmission rates provided in the converse result not only ensure that Quantum advantage for error correction and authentication protocols can be achieve, but also that this advantage is resilient in the presence of noise.

\bigskip

\noindent In the next section, we make use of the observations presented in the list above for establishing that each one of the results, provided in \textit{1.4}, holds.

\section{Arguments for the Main Results}

\noindent We make use of the discussion provided in the previous section in the following arguments of the main results.

\subsection{Theorem $1$}

\subsubsection{General description of the proof}

\noindent In arguments for the first main result, we argue that,

{\small \begin{align*}
 \underset{P_{\textbf{X}}}{\mathrm{sup}}  \big\{ \mathrm{min} \big\{  I {\tiny \big(} \textbf{X} , \textbf{Y} {\tiny \big)}  , \underset{z}{\mathrm{min}}  \big\{ H_Q {\tiny \big(} \textbf{Y} \big| \textbf{Z} = z {\tiny \big)}   -  H_P {\tiny \big(}  \textbf{Y} \big| \textbf{X} {\tiny \big)}  \big\}  \big\}    \big\} 
\end{align*}
\[ <  \left\{\!\begin{array}{ll@{}>{{}}l} 
     \mathrm{log} \text{ }  \mathrm{log} \bigg[   {  \mathrm{log} \big| \textbf{Y}^{*} \big|   } \big/ { \big| \textbf{X}^{*} \big|  }          \bigg]   +   \mathrm{log}  \bigg[ {   \mathrm{log}\big|  \textbf{Z}  \big|  } \big/ {  \big|  \textbf{Y}^{*}  \big|   } \bigg]  \Longleftrightarrow \big| \textbf{X} \big| > \big| \textbf{Y}^{*} \big| ,  \big| \textbf{Y}^{*} \big| > \big| \textbf{Z} \big|    ,  \\    \\ \mathrm{log} \text{ }    \mathrm{log} \bigg[   {  \mathrm{log} \big| \textbf{X} \big|  } \big/ { \big| \textbf{Y}^{*} \big| }          \bigg]   +   \mathrm{log}  \bigg[  { \mathrm{log} \big|  \textbf{Y}^{*} \big|   } \big/ {  \big| \textbf{X} \big|   } \bigg]   \Longleftrightarrow \big| \textbf{X} \big| <  \big| \textbf{Y}^{*} \big| ,  \big|  \textbf{Y}^{*} \big| < \big| \textbf{Z} \big|        ,  \\ \\ \mathrm{log} \text{ }    \mathrm{log} \bigg[    {  \mathrm{log} \big| \textbf{Y}^{*} \big|  } \big/ { \big| \textbf{X}^{*} \big| }          \bigg]   +   \mathrm{log}  \bigg[  {  \mathrm{log}\big|  \textbf{Y}^{*} \big|   } \big/ { \big| \textbf{X} \big|   } \bigg]   \Longleftrightarrow \big| \textbf{X} \big| >   \big| \textbf{Y}^{*} \big| ,  \big|  \textbf{Y}^{*} \big| < \big| \textbf{Z} \big|        ,   \\   \\ \mathrm{log} \text{ }    \mathrm{log} \bigg[   {  \mathrm{log} \big| \textbf{Y}^{*} \big|  } \big/ {\big| \textbf{X}^{*} \big| }          \bigg]   +   \mathrm{log}  \bigg[  {  \mathrm{log}\big|  \textbf{Z}  \big|   } \big/ { \big| \textbf{Y}^{*} \big|   } \bigg]   \Longleftrightarrow \big| \textbf{X} \big| <    \big| \textbf{Y}^{*} \big| ,  \big|  \textbf{Y}^{*} \big| >   \big| \textbf{Z} \big|        .                          
\end{array}\right. \equiv  r \\ \\  \tag{*} 
\]  }

\noindent holds, as a specific instance of the converse result,

{\small \begin{align*}
   r >  \underset{P_{\textbf{X}}}{\mathrm{sup}}  \big\{ \mathrm{min} \big\{  I {\tiny \big(} \textbf{X} , \textbf{Y} {\tiny \big)}  , \underset{z}{\mathrm{min}}  \big\{ H_Q {\tiny \big(} \textbf{Y} \big| \textbf{Z} = z {\tiny \big)}   -  H_P {\tiny \big(}  \textbf{Y} \big| \textbf{X} {\tiny \big)}  \big\}  \big\}    \big\}   \text{. }
\end{align*} }

\noindent The above converse result on $r$ can itself be strictly upper bounded with,

{\small \begin{align*}
  \underset{P_{\textbf{X}}}{\mathrm{sup}}   \big\{ \mathrm{min} \big\{ \mathrm{log} \big(  I \big( \textbf{X} , \textbf{Y} \big)  \big)  , \underset{z}{\mathrm{min}}  \big\{ \mathrm{log} \big( H_Q \big( \textbf{Y} \big| \textbf{Z} = z \big)   -  H_P \big( \textbf{Y} \big| \textbf{X}  \big)  \big) \big\}   \big\}   \big\} \\ \\ \ <     \underset{P_{\textbf{X}}}{\mathrm{sup}}   \big\{ \mathrm{min} \big\{ \mathrm{log} \big(  I \big( \textbf{X} , \textbf{Y} \big)  \big)  , \underset{z}{\mathrm{min}}  \big\{ \mathrm{log} \big( H_Q \big( \textbf{Y} \big| \textbf{Z} = z \big) \big)   -  \mathrm{log} \big(  H_P \big( \textbf{Y} \big| \textbf{X}  \big)  \big) \big\}   \big\}   \big\}  \end{align*} 
  
  \begin{align*}  \equiv    \underset{P_{\textbf{X}}}{\mathrm{sup}}  \big\{  \mathrm{min} \big\{\mathrm{log} \big(   I \big( \textbf{X} , \textbf{Y} \big)  \big)  ,   \underset{z}{\mathrm{min}}  \big\{  \mathrm{log} \big( H_Q \big( \textbf{Y} \big| \textbf{Z} = z \big) \big) \big\}       \big\}  \big\}  -  \underset{P_{\textbf{X}}}{\mathrm{sup}}  \big\{ \mathrm{min} \big\{  \mathrm{log} \big(   I \big( \textbf{X} , \textbf{Y} \big)   \big)   ,    \mathrm{log} \big(  H_P \big( \textbf{Y} \big| \textbf{X}  \big)   \big)     \big\}   \big\}    \text{. }
\end{align*} }

\noindent In \textit{2.5.2}, the alphabets $\textbf{X}^{*}$, $\textbf{Y}^{*}$, and $\textbf{Z}^{*}$, which respectively correspond to the pruned alphabets of Alice, Bob and Eve, were introduced. Given the specification of transition probabilities from the conditional probability distributions $P {\tiny \big[ \cdot \big| \cdot \big]}$, used by Alice and Bob, and $Q {\tiny \big[ \cdot \big| \cdot \big]}$, used by Eve in the generalized noise model over the Quantum channels, instead of a cancellation occurring from,

{\small \begin{align*}
 \mathrm{log} \big| \textbf{Y} \big|    \text{, }
\end{align*} }

\noindent one can introduce can upper bound, with,

{\small \begin{align*}
  \mathrm{log} \bigg[  {\mathrm{log} \big| \textbf{Y} \big|} \big/ {\big| \textbf{Y}^{*} \big|} \bigg]  \equiv  \mathrm{log} \text{ } \mathrm{log} \big| \textbf{Y} \big| - \mathrm{log} \big| \textbf{Y}^{*} \big| >  \mathrm{log} \big| \textbf{Y} \big| - \mathrm{log} \big| \textbf{Y}^{*} \big|     \text{. }
\end{align*}  }

\noindent Moreover, for the remaining contribution in the converse result, from the conditional, and ordinary, Shannon entropy,

{\small \begin{align*}
 \mathrm{log} \bigg[  \frac{H_Q \big( \textbf{Y} \big| \textbf{Z} = z \big) }{H_P  \big( \textbf{Y}^{*} \big| \textbf{X} = x \big) }   \bigg]  >   \frac{H_Q \big( \textbf{Y} \big| \textbf{Z} = z \big) }{H_P  \big( \textbf{Y}^{*} \big| \textbf{X} = x \big) }    >   \frac{H_Q \big( \textbf{Y} \big| \textbf{Z} = z \big) }{H_P  \big( \textbf{Y}\big| \textbf{X} = x \big) }    \end{align*}

 \begin{align*} \Bigg\Updownarrow \\   \bigg[  \mathrm{log} \big[  H_Q \big( \textbf{Y} \big| \textbf{Z} = z \big) \big] - \mathrm{log} \big[ H_P  \big( \textbf{Y}^{*} \big| \textbf{X} = x \big)    \big]   \bigg] \cdot H_P  \big( \textbf{Y}^{*} \big| \textbf{X} = x \big) \\ \\ <  \bigg[  \mathrm{log} \big[  H_Q \big( \textbf{Y} \big| \textbf{Z} = z \big) \big]  - \mathrm{log} \big[ H_P  \big( \textbf{Y}^{*} \big| \textbf{X} = x \big)    \big]   \bigg]   \cdot    H_P  \big( \textbf{Y} \big| \textbf{X} = x \big) \end{align*}

 \begin{align*} <  H_Q  \big( \textbf{Y} \big| \textbf{Z} = z \big)  \text{, }
\end{align*} }

\noindent per the montonicity of the conditional, and ordinary, Shannon entropy, with respect to the cardinality of each participant's alphabet.

\subsubsection{Argument}

\noindent \textit{Proof of Theorem 1}. For bit transmission rates in the converse result,

{\small \begin{align*}
r >  \underset{P_{\textbf{X}}}{\mathrm{sup}}  \big\{ \mathrm{min} \big\{  I {\tiny \big(} \textbf{X} , \textbf{Y} {\tiny \big)}  , \underset{z}{\mathrm{min}}  \big\{ H_Q {\tiny \big(} \textbf{Y} \big| \textbf{Z} = z {\tiny \big)}   -  H_P {\tiny \big(}  \textbf{Y} \big| \textbf{X} {\tiny \big)}  \big\}  \big\}    \big\}     \text{, }
\end{align*} }

\noindent it suffices to argue that upper bounds of the following hold,

{\small \begin{align*}
       \underset{P_{\textbf{X}}}{\mathrm{sup}}      \bigg\{  \mathrm{min} \bigg\{  \mathrm{log} \bigg[  \frac{\mathrm{log} \big[ I \big( \textbf{X}, \textbf{Y} \big) \big]}{I \big( \textbf{X}^{*} , \textbf{Y}^{*} \big) }\bigg]     ,   \underset{z}{\mathrm{min}} \bigg\{     \mathrm{log}  \big[ H_Q \big( \textbf{Y}^{*} \big| \textbf{Z} = z \big)  \big]   \bigg\} \bigg\} \bigg\}   -  \underset{P_{\textbf{X}}}{\mathrm{sup}}      \bigg\{  \mathrm{min} \bigg\{  \mathrm{log} \bigg[  \frac{\mathrm{log} \big[ I \big( \textbf{X}, \textbf{Y} \big) \big]}{I \big( \textbf{X}, \textbf{Y}^{*} \big) }\bigg]   ,   \mathrm{log} \big[ H_P \big( \textbf{Y}^{*} \big| \textbf{X} \big) \big]    \bigg\}  \bigg\}     \text{,}
\end{align*} }

\noindent which can be used to obtain the desired converse result for $r$ from the following arguments. Before performing manipulations with $r$ to show that the desired form of the converse bit transmission rate holds, we present the following sequence of results:

\begin{itemize}
    \item[$\bullet$] \textbf{Lemma} \textit{1}: With respect to channel noise the radius of Hamming balls monotonically decreases,

     \item[$\bullet$] \textbf{Lemma} \textit{2}: The computation of the critical points for the first derivative of the converse bit transmission rate provides an implicit relation between the cardinality of the pruned alphabets for Alice (resp. Bob), with the pruned alphabets of Bob and Eve (resp. Alice and Eve), 

      \item[$\bullet$] \textbf{Lemma} \textit{3}: The second derivative of the converse bit transmission rate given the implicit relations obtained in \textbf{Lemma} \textit{2} is concave, 

      \item[$\bullet$] \textbf{Corollary}: The set of degeneracies for the second derivative of the converse bit transmission rate can be obtained from its discontinuities.

\end{itemize}

\bigskip

\noindent \textbf{Lemma} \textit{1} (\textit{the first property of error correcting codes over noisy Quantum channels: inverse monotonicity of the radius of Hamming balls for codewords with respect to channel noise}). Denote, 

{\small \begin{align*}
   \mathcal{B}_1 \big( r, \mathcal{C}_1  \big)   \equiv  \big\{ \forall r, \exists \epsilon_1 > 0 : \big| \mathcal{C}_1 - r \big| < \epsilon_1       \big\}    \text{, }  \\   \\    \mathcal{B}_2 \big( r, \mathcal{C}_2  \big)  \equiv \big\{  \forall r, \exists \epsilon_2 > 0: \big| \mathcal{C}_2 - r \big| < \epsilon_2  \big\}     \text{, }
\end{align*} }

\noindent as two Hamming balls, $   \mathcal{B}_1$ and $\mathcal{B}_2$,  centered about codewords $\mathcal{C}_1$, and $\mathcal{C}_2$, which are respectively transmitted over the two Quantum channels,

{\small \begin{align*}
   \mathcal{N}^n_{p,q, A \longleftrightarrow B}  \text{, }  \\   \\    \mathcal{N}^n_{p,q, B \longleftrightarrow E}     \text{, }
\end{align*} }

\noindent between Alice and Bob, and between Bob and Eve, respectively. For $N_{A \longleftrightarrow B} > N_{B \longleftrightarrow E}$, the radius of the Hamming balls, $\mathcal{B}_1$ and $\mathcal{B}_2$ is inversely proportional to the presence of noise over a Quantum channel, as,

{\small \begin{align*}
  \frac{\textbf{P} \bigg[ \big\{  \forall r_1 > 0, \exists n_1 > 0 :  \big|  \mathscr{C} \cap \mathcal{B}_1 \big( r_1, \mathcal{C}_1 \big)  \big| \equiv n_1      \big\}    \bigg| \big\{  \mathcal{C}_1 \text{ is transmitted through }  \mathcal{N}^n_{p,q, A \longleftrightarrow B} \big\}   \bigg]  }{\textbf{P} \bigg[ \big\{ \forall r_2 > r_1 > 0, \exists n_2 > n_1  > 0 :       \big| \mathscr{C} \cap \mathcal{B}_1 \big( r_2, \mathcal{C}_2 \big) \big| \equiv n_2 \big\}    \bigg| \big\{ \mathcal{C}_2 \text{ is transmitted through }    \mathcal{N}^n_{p,q, B \longleftrightarrow E}  \big\} \bigg]  }      \equiv \frac{\mathscr{P}_1}{\mathscr{P}_2}  \end{align*} 
  
  \begin{align*}   \gtrsim   \frac{N_{A \longleftrightarrow B}}{N_{B \longleftrightarrow E}}   \text{, }
\end{align*} }

\noindent for the codeword superset,

{\small \begin{align*}
  \mathscr{C} \equiv \underset{\mathcal{C}}{\bigcup} \big\{ n\text{-bit } \mathcal{C} \text{ transmitted over }  \mathcal{N}^n_{p,q, A \longleftrightarrow B}, \text{ and over }   \mathcal{N}^n_{p,q, B \longleftrightarrow E} \big\} \end{align*}
  
  \begin{align*} \equiv \underset{\mathcal{C}, \text{ bits}}{\bigcup} \big\{ \text{infinitely many bit } \mathcal{C} \text{ transmitted over }  \mathcal{N}^{\text{bits}}_{p,q, A \longleftrightarrow B}, \text{ and over }   \mathcal{N}^{\text{bits}}_{p,q, B \longleftrightarrow E} \big\}  \text{. }
\end{align*} }

\noindent \textit{Proof of Lemma 1}. To demonstrate that the desired inequality holds, up to constants, between the ratio of two probabilities corresponding to the conditional probabilities of,

{\small \begin{align*}
   \big\{  \forall r_1 > 0, \exists n_1 > 0 :  \big|  \mathscr{C} \cap \mathcal{B}_1 \big( r_1, \mathcal{C}_1 \big)  \big| \equiv n_1         \big| \mathcal{C}_1 \text{ is transmitted through }   \mathcal{N}^n_{p,q, A \longleftrightarrow B} \big\} , 
\end{align*} }

\noindent and of,

{\small \begin{align*}
   \big\{  \forall r_2 > r_1 > 0, \exists n_2 > n_1  > 0 :       \big| \mathscr{C} \cap \mathcal{B}_1 \big( r_2, \mathcal{C}_2 \big) \big| \equiv n_2    \big| \mathcal{C}_2 \text{ is transmitted through }   \mathcal{N}^n_{p,q, B \longleftrightarrow E} \big\} \text{, }
\end{align*} } 

\noindent observe that the probability corresponding to the number of codewords, belonging to,

{\small \begin{align*}
    \big|  \mathscr{C} \cap \mathcal{B}_1 \big( r_1, \mathcal{C}_1 \big)  \big|   \text{, }
\end{align*} }

\noindent as a subset of $\mathscr{C}$, is monotonoically decreasing with respect to the radius, $r_1$, of the Hamming ball $\mathcal{B}_1$. Explicitly, as the radius of the Hamming ball corresponding to the first codeword for the bit message transmitted over $\mathcal{N}^n_{p,q, A \longleftrightarrow B}$, the sequence of probabilities,

{\small \begin{align*}
  \bigg\{   \frac{\textbf{P} \bigg[ \big\{  \forall r_2 > 0, \exists n_1 > 0 :  \big|  \mathscr{C} \cap \mathcal{B}_1 \big( r_2, \mathcal{C}_1 \big)  \big| \equiv n_1     \big\}     \bigg| \big\{  \mathcal{C}_1 \text{ is transmitted through }   \mathcal{N}^n_{p,q, A \longleftrightarrow B} \big\}  \bigg] }{\textbf{P} \bigg[ \big\{  \forall r_1 > 0, \exists n_1 > 0 :  \big|  \mathscr{C} \cap \mathcal{B}_1 \big( r_1, \mathcal{C}_1 \big)   \big| \equiv n_1      \big\}    \bigg| \big\{  \mathcal{C}_1 \text{ is transmitted through }   \mathcal{N}^n_{p,q, A \longleftrightarrow B} \big\}  \bigg] }    \bigg\}_{r_2 > r_1 > 0}   \text{, }
\end{align*} }

\noindent is monotonically decreasing in the radius of $\mathcal{B}_1$. Denoting,

{\small \begin{align*}
    r_{\mathrm{min}} \equiv \underset{r_c > 0 }{\mathrm{inf}} \bigg\{  r_c : \textbf{P} \bigg[ \big\{  \forall r_c > 0, \exists n_1 > 0 :  \big|  \mathscr{C} \cap \mathcal{B}_1 \big( r_c, \mathcal{C}_1 \big)  \big| \equiv n_1      \big\}    \bigg| \big\{ \mathcal{C}_1 \text{ is transmitted through }     \mathcal{N}^n_{p,q, A \longleftrightarrow B} \big\}  \bigg] \equiv 0       \bigg\} \text{, }
\end{align*}  }

\noindent as the minimum radius of $\mathcal{B}_1$ for which there are no codewords obtained from the intersection of the code superset with $\mathcal{B}_1$, one can construct a sequence of random-variables, where each variable is given by the ratio of probabilities, as indicated above in $r_1$ and $r_2$, a sequence of radius of Hamming balls, $\big\{r_1, r_2, \cdots, r_{\mathrm{min}}-1 \big\}$, such that,

{\small \begin{align*}
  \bigg\{   \frac{\textbf{P} \bigg[ \big\{  \forall r_j > 0, \exists n_1 > 0 :  \big|  \mathscr{C} \cap \mathcal{B}_1 \big( r_j, \mathcal{C}_1 \big)  \big| \equiv n_1       \big\}   \big| \big\{ \mathcal{C}_1 \text{ is transmitted through } \mathcal{N}^n_{p,q, A \longleftrightarrow B} \big\}  \bigg] }{\textbf{P} \bigg[ \big\{ \forall r_i > 0, \exists n_1 > 0 :  \big|  \mathscr{C} \cap \mathcal{B}_1 \big( r_i, \mathcal{C}_1 \big)  \big| \equiv n_1     \big\}     \big| \big\{  \mathcal{C}_1 \text{ is transmitted through }   \mathcal{N}^n_{p,q, A \longleftrightarrow B}  \big\} \bigg]  }    \bigg\}_{r_{\mathrm{min}} - 1 \geq r_j > r_i \geq   0}  \end{align*} 
  
  \begin{align*}  \longrightarrow 0  \text{. }
\end{align*} } 

\noindent Similarly, one can identify the threshold of the radius of the second Hamming ball, $\mathcal{B}_2$, for which,

{\small \begin{align*}
    r^{\prime}_{\mathrm{min}} \equiv \underset{r_c > 0 }{\mathrm{inf}} \bigg\{  r_c : \textbf{P} \bigg[ \big\{ \forall r_c > 0, \exists n_2 > 0 :  \big|  \mathscr{C} \cap \mathcal{B}_1 \big( r_c, \mathcal{C}_2 \big)  \big| \equiv n_2   \big\}      \bigg| \big\{ \mathcal{C}_2 \text{ is transmitted through }   \mathcal{N}^n_{p,q, A \longleftrightarrow B} \big\}   \bigg] \equiv 0       \bigg\} \text{, }
\end{align*}        }

\noindent corresponding to the codeword $\mathcal{C}_2$, with finitely many bits, transmitted over $B \longleftrightarrow E$, with the sequence,

{\small \begin{align*}
  \bigg\{   \frac{\textbf{P} \bigg[ \big\{ \forall r_2 > 0, \exists n_2 > 0 :  \big|  \mathscr{C} \cap \mathcal{B}_2 \big( r_2, \mathcal{C}_2 \big)  \big| \equiv n_2     \big\}    \bigg| \big\{  \mathcal{C}_2 \text{ is transmitted through }   \mathcal{N}^n_{p,q, B \longleftrightarrow E} \big\}   \bigg] }{\textbf{P} \bigg[ \big\{ \forall r_1 > 0, \exists n_2 > 0 :  \big|  \mathscr{C} \cap \mathcal{B}_2 \big( r_2 , \mathcal{C}_2 \big)  \big| \equiv n_2       \big\}   \bigg| \big\{ \mathcal{C}_2 \text{ is transmitted through }   \mathcal{N}^n_{p,q, B \longleftrightarrow E} \big\}  \bigg] }    \bigg\}_{r_2 > r_1 > 0}   \text{. }
\end{align*} } 

\noindent Hence, to upper bound the ratio of conditional probabilities,

{\small \begin{align*}
  \frac{\textbf{P} \bigg[ \big\{ \forall r_1 > 0, \exists n_1 > 0 :  \big|  \mathscr{C} \cap \mathcal{B}_1 \big( r_1, \mathcal{C}_1 \big)  \big| \equiv n_1     \big\}     \bigg| \big\{ \mathcal{C}_1 \text{ is transmitted through }   \mathcal{N}^n_{p,q, A \longleftrightarrow B} \big\}  \bigg] }{\textbf{P} \bigg[ \big\{ \forall r_2 > r_1 > 0, \exists n_2 > n_1  > 0 :       \big| \mathscr{C} \cap \mathcal{B}_1 \big( r_2, \mathcal{C}_2 \big) \big| \equiv n_2  \big\}   \bigg| \big\{ \mathcal{C}_2 \text{ is transmitted through }   \mathcal{N}^n_{p,q, B \longleftrightarrow E}  \big\} \bigg] } \text{, }
\end{align*} } 

\noindent up to constants, one would like to study the rate of decay of the sequence,

{\small \begin{align*}
 \bigg\{  \frac{\textbf{P} \bigg[ \big\{ \forall r_1 > 0, \exists n_1 > 0 :  \big|  \mathscr{C} \cap \mathcal{B}_1 \big( r_1, \mathcal{C}_1 \big)  \big| \equiv n_1        \big\}  \bigg| \big\{  \mathcal{C}_1 \text{ is transmitted through } \mathcal{N}^n_{p,q, A \longleftrightarrow B} \big\}  \bigg]  }{\textbf{P} \bigg[ \big\{  \forall r_2 > r_1 > 0, \exists n_2 > n_1  > 0 :       \big| \mathscr{C} \cap \mathcal{B}_1 \big( r_2, \mathcal{C}_2 \big) \big| \equiv n_2 \big\}    \bigg| \big\{  \mathcal{C}_2 \text{ is transmitted through }  \mathcal{N}^n_{p,q, B \longleftrightarrow E} \big\}   \bigg]   } \bigg\}_{\underset{r^{\prime}_{\mathrm{min}} > r_2}{r_{\mathrm{min}}> r_1}}  \text{, } \\ \\ \tag{A}
\end{align*} }

\noindent of conditional probabilities, given the choice of $r_{\mathrm{min}}$, and of $r^{\prime}_{\mathrm{min}}$. The sequence of ratios of conditional probabilities above, depending upon the occurrence of the conditionings,

{\small \begin{align*}
   \bigg\{  \big\{ \mathcal{C}_1 \text{ is transmitted through }   \mathcal{N}^n_{p,q, A \longleftrightarrow B}  \big\} \text{, } 
  \big\{  \mathcal{C}_2 \text{ is transmitted through }   \mathcal{N}^n_{p,q, B \longleftrightarrow E} \big\} \bigg\}  \text{, }
\end{align*} }

\noindent provided in (A), can be equivalently characterized through the decompositions, 

{\small \begin{align*}
     \textbf{P} \bigg[ \big\{ \forall r_1 > 0, \exists n_1 > 0 :  \big|  \mathscr{C} \cap \mathcal{B}_1 \big( r_1, \mathcal{C}_1 \big)  \big| \equiv n_1    \big\}      \bigg| \big\{  \mathcal{C}_1 \text{ is transmitted through }    \mathcal{N}^n_{p,q, A \longleftrightarrow B} \big\}  \bigg] \end{align*}
     
     \begin{align*} \equiv     \underset{\text{bits in } \mathcal{C}_2}{\sum}  \bigg\{     \textbf{P} \bigg[ \big\{  \forall r_1 > 0, \exists n_1 > 0 :  \big|  \mathscr{C} \cap \mathcal{B}_1 \big( r_1, \mathcal{C}_1 \big)  \big| \equiv n_1      \big\}    \bigg| \big\{  \mathcal{C}_1 \text{ is transmitted through } \mathcal{N}^n_{p,q, A \longleftrightarrow B}  \big\} \bigg]  \bigg\}      \text{, } \end{align*}
     
     \begin{align*}  \textbf{P} \bigg[  \big\{ \forall r_2 > r_1 > 0, \exists n_2 > n_1  > 0 :       \big| \mathscr{C} \cap \mathcal{B}_1 \big( r_2, \mathcal{C}_2 \big) \big| \equiv n_2 \big\}    \bigg| \big\{  \mathcal{C}_2 \text{ is transmitted through }    \mathcal{N}^n_{p,q, B \longleftrightarrow E} \big\}   \bigg] \end{align*}
     
     \begin{align*}  \equiv     \underset{\text{bits in } \mathcal{C}_1}{\sum} \bigg\{   \textbf{P} \bigg[ \big\{ \forall r_2 > r_1 > 0, \exists n_2 > n_1  > 0 :       \big| \mathscr{C} \cap \mathcal{B}_1 \big( r_2, \mathcal{C}_2 \big) \big| \equiv n_2  \big\}   \bigg| \big\{ \mathcal{C}_2 \text{ is transmitted  through }   \mathcal{N}^n_{p,q, B \longleftrightarrow E} \big\}  \bigg] \bigg\}     \text{, }
\end{align*} } 

\noindent for the conditional probabilities of transmitting $\mathcal{C}_1$ and $\mathcal{C}_2$. Fixing $r_1$, and $r_2$, sufficiently far away from $r_{\mathrm{min}}$, and from $r^{\prime}_{\mathrm{min}}$, straightforwardly there exists two strictly positive constants, $C_1$ and $C_2$, for which,

{\small \begin{align*} 
(\mathrm{A}) \equiv    \bigg\{   \underset{\mathcal{C}_2 \neq \emptyset}{\underset{\text{bits in } \mathcal{C}_1, \mathcal{C}_2}{\sum}}  { \frac{\textbf{P} \bigg[ \big\{ \forall r_1 > 0, \exists n_1 > 0 :  \big|  \mathscr{C} \cap \mathcal{B}_1 \big( r_1, \mathcal{C}_1 \big)  \big| \equiv n_1    \big\}      \bigg| \big\{ \mathcal{C}_1 \text{ is transmitted through }     }{\textbf{P} \bigg[ \big\{ \forall r_2 > r_1 > 0, \exists n_2 > n_1  > 0 :       \big| \mathscr{C} \cap \mathcal{B}_1 \big( r_2, \mathcal{C}_2 \big) \big| \equiv n_2  \big\}   \bigg| \big\{ \mathcal{C}_2 \text{ is transmitted through }   \big\}  } }  \end{align*}

\begin{align*} \frac{\mathcal{N}^n_{p,q, A \longleftrightarrow B} \big\}  \bigg]}{ \mathcal{N}^n_{p,q, B \longleftrightarrow E} \big\}  \bigg]}  \bigg\}_{\underset{r^{\prime}_{\mathrm{min}} > r_2}{r_{\mathrm{min}}> r_1}}   \end{align*}

\begin{align*} \overset{(\mathrm{MON})}{>}                      \bigg\{   \underset{\mathcal{C}_2 \neq \emptyset}{\underset{\text{bits in } \mathcal{C}_1, \mathcal{C}_2}{\sum}}   \frac{\textbf{P} \bigg[ \big\{ \forall r_1 > 0, \exists n_1 > 0 :  \big|  \mathscr{C} \cap \mathcal{B}_1 \big( r_1, \mathcal{C}_1 \big)  \big| \equiv n_1    \big\}      \bigg|\big\{  \mathcal{C}_1 \text{ is transmitted through}  }{\textbf{P} \bigg[\big\{  \forall r_2 > r_1 > 0, \exists n_2 > n_1  > 0 :       \big| \mathscr{C} \cap \mathcal{B}_1 \big( r_2, \mathcal{C}_2 \big) \big| \equiv n_2  \big\}   \bigg| \big\{ \mathcal{C}_2 \text{ is transmitted through }  }  \end{align*}

\begin{align*}  \frac{ \mathcal{N}^n_{p^{\prime},q, A \longleftrightarrow B} \big\}  \bigg]}{  \mathcal{N}^n_{p^{\prime},q, B \longleftrightarrow E} \big\}  \bigg]}  \bigg\}_{\underset{r^{\prime}_{\mathrm{min}} > r_2}{r_{\mathrm{min}}> r_1}}  \end{align*}
 } 

\begin{align*}   \overset{  \{ n^{\prime}_2 > n_2 \} }{\overset{ \{ n^{\prime}_1 > n_1 \} }{>}}                       \bigg\{   \underset{\mathcal{C}_2 \neq \emptyset}{\underset{\text{bits in } \mathcal{C}_1, \mathcal{C}_2}{\sum}}   \frac{\textbf{P} \bigg[\big\{  \forall r_1 > 0, \exists n^{\prime}_1 > 0 :  \big|  \mathscr{C} \cap \mathcal{B}_1 \big( r_1, \mathcal{C}_1 \big)  \big| \equiv  n^{\prime}_1    \big\}      \bigg| \big\{ \mathcal{C}_1 \text{ is transmitted through }  }{\textbf{P} \bigg[ \big\{  \forall r_2 > r_1 > 0, \exists n^{\prime}_2 > n^{\prime}_1  > 0 :       \big| \mathscr{C} \cap \mathcal{B}_1 \big( r_2, \mathcal{C}_2 \big) \bigg| \equiv n^{\prime}_2  \big\}  \big| \big\{ \mathcal{C}_2 \text{ is transmitted through }  }    \end{align*}

\begin{align*}   \frac{  \mathcal{N}^n_{p^{\prime},q, A \longleftrightarrow B} \big\}  \bigg]}{ \mathcal{N}^n_{p^{\prime},q, B \longleftrightarrow E} \big\}  \bigg]}  \bigg\}_{\underset{r^{\prime}_{\mathrm{min}} > r_2}{r_{\mathrm{min}}> r_1}}    \end{align*}

\begin{align*} \overset{ \{ r_1, r_2 \longrightarrow + \infty \} }{\longrightarrow}     \underset{\mathcal{C}_2 \neq \emptyset}{\underset{\text{bits in } \mathcal{C}_1, \mathcal{C}_2}{\sum}}                      \frac{C_1 \big( N_{A \longleftrightarrow B}, \mathcal{C}_1  \big)}{C_2 \big( N_{B \longleftrightarrow E}, \mathcal{C}_2 \big)} \equiv                     \frac{ {\underset{\text{bits in } \mathcal{C}_1 }{\sum}}    C_1 \big( N_{A \longleftrightarrow B}, \mathcal{C}_1  \big)}{ \underset{\mathcal{C}_2 \neq \emptyset}{\underset{\text{bits in } \mathcal{C}_2}{\sum}}    C_2 \big( N_{B \longleftrightarrow E}, \mathcal{C}_2 \big)}  \equiv \frac{C_{1, A \longleftrightarrow B}}{C_{2, B \longleftrightarrow E}}                                     \end{align*}

\begin{align*} \gtrsim   \frac{N_{A \longleftrightarrow B}}{N_{B \longleftrightarrow E}}                \text{. } \\ \tag{Noise Lower Bound}
\end{align*}

\noindent In the second line above, (MON), the monotonicity condition, states that the error correcting code associated with communication of Classical bits over the Quantum channels, either between Alice or Bob, or between Bob and Eve, in the converse result on $r$ satisfies,

{\small \begin{align*}
  \frac{\mathrm{d}}{\mathrm{d} \big(  N_{B \longleftrightarrow E} \big) } \bigg\{          \textbf{P} \bigg[ \big\{  \forall r_2 > r_1 > 0, \exists n_2 > n_1  > 0 :       \big| \mathscr{C} \cap \mathcal{B}_2 \big( r_2, \mathcal{C}_2 \big) \big| \equiv n_2  \big\}   \bigg| \big\{ \mathcal{C}_2 \text{ is transmitted through }    \mathcal{N}^n_{p,q, B \longleftrightarrow E} \big\}  \bigg]   \bigg\}  \end{align*} 
  
  \begin{align*} < 0 \text{, } \end{align*} } 
  
  \noindent holds iff, given two probabilities $p$ and $p^{\prime}$ such that $1-p < 1-p^{\prime}$, the stochastic domination,
  
  {\small \begin{align*} \textbf{P} \bigg[ \big\{ \forall r_2 > r_1 > 0, \exists n_2 > n_1  > 0 :       \big| \mathscr{C} \cap \mathcal{B}_1 \big( r_2, \mathcal{C}_2 \big) \big| \equiv n_2   \big\}  \bigg| \big\{  \mathcal{C}_2 \text{ is transmitted through }   \mathcal{N}^n_{p,q, B \longleftrightarrow E} \big\}  \bigg]  \\ \\   >   \textbf{P} \bigg[ \big\{ \forall r_2  > r_1 > 0, \exists n_2 > n_1  > 0 :       \big| \mathscr{C} \cap \mathcal{B}_1 \big( r_2, \mathcal{C}_2 \big) \big| \equiv n_2    \big\} \bigg| \big\{  \mathcal{C}_2 \text{ is transmitted through }   \mathcal{N}^n_{p^{\prime},q, B \longleftrightarrow E} \big\}  \bigg]   \text{, }    \end{align*} } 
  
  \noindent holds. Similarly, for the remaining Quantum channel corresponding to communication of bit codewords between Alice and Bob,

  {\small \begin{align*} \frac{\mathrm{d}}{\mathrm{d} \big(  N_{A \longleftrightarrow B} \big) } \bigg\{         \textbf{P} \bigg[ \big\{ \forall r_1 > 0, \exists n_1 > 0 :  \big|  \mathscr{C} \cap \mathcal{B}_1 \big( r_1, \mathcal{C}_1 \big)  \big| \equiv n_1       \big\}   \bigg| \big\{  \mathcal{C}_1 \text{ is transmitted through }    \mathcal{N}^n_{p,q, A \longleftrightarrow B} \big\}   \bigg]         \bigg\}  \end{align*} 
  
  \begin{align*}  < 0   \text{, }      
\end{align*} }

\noindent holds iff, under choice of two probabilities $p$ and $p^{\prime}$ such that $1-p < 1-p^{\prime}$, the stochastic domination,

  {\small \begin{align*} \textbf{P} \bigg[ \big\{  \forall r_1 > 0, \exists  n_1  > 0 :       \big| \mathscr{C} \cap \mathcal{B}_1 \big( r_1, \mathcal{C}_1 \big) \big| \equiv n_1  \big\}   \bigg| \big\{ \mathcal{C}_1 \text{ is transmitted through }   \mathcal{N}^n_{p,q, A \longleftrightarrow B} \big\}   \bigg]   \\  \\ >   \textbf{P} \bigg[ \big\{  \forall r_1 > 0, \exists n_1   > 0 :       \big| \mathscr{C} \cap \mathcal{B}_1 \big( r_1, \mathcal{C}_1 \big) \big| \equiv n_2 \big\}    \bigg| \big\{  \mathcal{C}_2 \text{ is transmitted through }   \mathcal{N}^n_{p^{\prime},q, A \longleftrightarrow B}\big\}   \bigg] \text{. }    \end{align*} } 

\noindent holds. The constants in the lower bound are proportional to,

{\small \begin{align*}
 \underline{C_{1,n}}  =  \textbf{P} \bigg[ \big\{  \forall r_1 > 0, \exists n_1 > 0 :  \big|  \mathscr{C} \cap \mathcal{B}_1 \big( r_1, \mathcal{C}_1 \big)  \big| \equiv n_1      \big\}    \big| \big\{ \mathcal{C}_1 \text{ is transmitted through }   \mathcal{N}^n_{p,q, A \longleftrightarrow B} \big\}  \bigg]  \\   -  \textbf{P} \bigg[ \big\{ \forall r^{\prime}_1 \  > r_1 > 0, \exists   n_1 > n^{\prime}_1 > 0 :  \big|  \mathscr{C} \cap \mathcal{B}_1 \big( r^{\prime}_1, \mathcal{C}_1 \big)  \big| \equiv n^{\prime}_1  \big\}        \big| \big\{ \mathcal{C}_1 \text{ is transmitted through }  \\   \mathcal{N}^n_{p,q, A \longleftrightarrow B} \big\}  \bigg]    \text{, } \end{align*}
 
 \begin{align*} \underline{C_{2,n} } =   \textbf{P} \bigg[ \big\{ \forall r_2 > r_1 > 0, \exists n_2 > n_1  > 0 :       \big| \mathscr{C} \cap \mathcal{B}_1 \big( r_2, \mathcal{C}_2 \big) \big| \equiv n_2    \big\} \big| \big\{  \mathcal{C}_2 \text{ is transmitted through }    \mathcal{N}^n_{p,q, B \longleftrightarrow E} \big\} \bigg] \\  - \textbf{P} \bigg[ \big\{ \forall r^{\prime}_2   > r_2 > r^{\prime}_1  >  r_1 > 0, \exists n^{\prime}_2 > n_2 > n^{\prime}_1 > n_1  > 0 :       \big|  \mathscr{C} \cap \mathcal{B}_1 \big( r_2, \mathcal{C}_2 \big) \big\}  \bigg| \equiv n^{\prime}_2   \big\}   \big|   \big\{  \mathcal{C}_2 \text{ is transmitted} \\ \text{through }    \mathcal{N}^n_{p,q, B \longleftrightarrow E} \big\}   \bigg]  \text{, }
\end{align*} }

\noindent for the collection of constants,

{\small \begin{align*}
      \bigg\{  \underline{C_{1,n}}  \equiv    \underset{\text{bits}}{\bigcup}  \big\{ C_{1, \text{ infinitely many bits }n}  \big\}     \text{, }   \underline{C_{2,n}}  \equiv   \underset{\text{bits}}{\bigcup}  \big\{ C_{2, \text{ infinitely many bits }n} \big\} \bigg\}  \text{. } 
\end{align*} }

\noindent We conclude the argument, as it has been demonstrated that,

\begin{align*}
  \bigg\{    (\mathrm{A}) \gtrsim   \frac{N_{A \longleftrightarrow B}}{N_{B \longleftrightarrow E}}  \bigg\} \Longleftrightarrow \bigg\{   \frac{\mathscr{P}_1}{\mathscr{P}_2}  \gtrsim   \frac{N_{A \longleftrightarrow B}}{N_{B \longleftrightarrow E}}  \bigg\}     \text{. } \boxed{}
\end{align*}

\noindent \textbf{Lemma} \textit{2} (\textit{critical points of the converse bit transmission rate}). The first derivative of $r$ implies that the set of relations provided in $(r^{\prime} = 0)$ holds.

\bigskip

 \noindent \textit{Proof of Lemma 2}. To argue,

{\small 
\[   \left\{\!\begin{array}{ll@{}>{{}}l} 
  \big|  Y^* \big|   =    +  \sqrt{  \big| X^* \big|^2    \cdot           \frac{   \bigg[   \frac{  \mathrm{log}\big|  {Z}  \big|  \cdot \big| Z \big| }{   \big|  {Y}^{*}  \big|^2  \cdot  Z^{\prime}     }    \bigg]^{-1}     +   \bigg[      \frac{1}{ \big| Y^* \big| \cdot \big\{ Y^* \big\}^{\prime} }  \bigg]^{-1}   }{     \mathrm{log}  \text{ }     \frac{\mathrm{log} \big| {Y}^{*} \big|  }{ {\big| {X}^{*} \big| }    }                \cdot     \bigg\{            \frac{1}{\big| Y^{*} \big| }  \cdot  \big\{ Y^* \big\}^{\prime}  \cdot \big| X^* \big|      -    \mathrm{log}   \big| Y^{*} \big|  \cdot \big\{ X^* \big\}^{\prime }                  \bigg\}                                   }  }  \Longleftrightarrow \big\{ \big| {X} \big| > \big| {Y}^{*} \big| ,  \big| {Y}^{*} \big| > \big| {Z} \big|   \big\}   , \\    \\      \big| X \big| = +           \sqrt{  \big| Y^* \big|^2    \cdot           \frac{  \bigg[   \frac{  \mathrm{log}\big|  {Y^*}  \big|  \cdot \big| Y^*  \big| }{   \big|  {Y}^{*}  \big|^2  \cdot  Z^{\prime}     }    \bigg]^{-1}     +   \bigg[      \frac{1}{ \big| X \big| \cdot \big\{ X  \big\}^{\prime} }  \bigg]^{-1}   }{     \mathrm{log}  \text{ }      \frac{\mathrm{log} \big| {X} \big|  }{ {\big| {Y}^{*} \big| }    }                 \cdot     \bigg\{            \frac{1}{\big| X^{*} \big| }  \cdot  \big\{ X^* \big\}^{\prime}  \cdot \big| Y^* \big|      -    \mathrm{log}   \big| X  \big|  \cdot \big\{ Y^* \big\}^{\prime }                  \bigg\}                                   }  }    \Longleftrightarrow \big\{ \big| {X} \big| <  \big| {Y}^{*} \big| ,  \big| {Y}^{*} \big| <  \big| {Z} \big|   \big\}               , \\ \\      \big|  Y^* \big|   =    +  \sqrt{  \big| X^* \big|^2    \cdot           \frac{   \bigg[   \frac{  \mathrm{log}\big|  {Y^* }  \big|  \cdot \big| Y^*   \big| }{   \big|  {Y}^{*}  \big|^2  \cdot  \big\{ Y^*  \big\}^{\prime}     }    \bigg]^{-1}    +  \bigg[      \frac{1}{ \big| Y^* \big| \cdot \big\{ Y^* \big\}^{\prime} }  \bigg]^{-1}   }{     \mathrm{log}  \text{ }       \frac{\mathrm{log} \big| {Y}^{*} \big|  }{ {\big| {X}^{*} \big| }    }               \cdot     \bigg\{            \frac{1}{\big| Y^{*} \big| }  \cdot  \big\{ Y^* \big\}^{\prime}  \cdot \big| X^* \big|      -    \mathrm{log}   \big| Y^{*} \big|  \cdot \big\{ X^* \big\}^{\prime }                  \bigg\}                                   }  }  \Longleftrightarrow \big\{ \big| {X} \big| >   \big| {Y}^{*} \big| ,  \big| {Y}^{*} \big| <  \big| {Z} \big|   \big\}                    , 
\\ \\ 
       \big|  Y^* \big|   =    +  \sqrt{  \big| X^* \big|^2    \cdot           \frac{   \bigg[   \frac{  \mathrm{log}\big|  {Z}  \big|  \cdot \big| Z \big| }{   \big|  {Y}^{*}  \big|^2  \cdot  Z^{\prime}     }    \bigg]^{-1}     +   \bigg[      \frac{1}{ \big| Y^* \big| \cdot \big\{ Y^* \big\}^{\prime} }  \bigg]^{-1}   }{     \mathrm{log}  \text{ }      \frac{\mathrm{log} \big| {Y}^{*} \big|  }{ {\big| {X}^{*} \big| }    }               \cdot     \bigg\{            \frac{1}{\big| Y^{*} \big| }  \cdot  \big\{ Y^* \big\}^{\prime}  \cdot \big| X^* \big|      -    \mathrm{log}   \big| Y^{*} \big|  \cdot \big\{ X^* \big\}^{\prime }                  \bigg\}                                   }  }  \Longleftrightarrow \big\{ \big| {X} \big| <  \big| {Y}^{*} \big| ,  \big| {Y}^{*} \big| > \big| {Z} \big|   \big\}            , 
\end{array}\right.  \\ 
\]  }

 \noindent recall,

{\small 
\[ r^{\prime} \equiv r^{\prime} \bigg( \bigg|  \frac{X}{X^*}  \bigg| ,  \bigg|  \frac{Y}{Y^*}  \bigg|  ,   \bigg|  \frac{Z}{Z^*}  \bigg|   \bigg)  =   \left\{\!\begin{array}{ll@{}>{{}}l} 
  \boxed{r^{\prime}_1} \equiv 1 \big/  \mathrm{log} \bigg[   {  \mathrm{log} \big| {Y}^{*} \big|  } \big/ {\big| {X}^{*} \big| }          \bigg] \cdot \bigg\{            \big| Y^{*} \big|^{-1} \cdot  \big\{ Y^* \big\}^{\prime} \cdot \big| X^* \big|   -    \mathrm{log}   \big| Y^{*} \big|\cdot \big\{ X^* \big\}^{\prime }                  \bigg\} \big/ \big| X^* \big|^2   \\ + 1 \big/  \big\{   {   \mathrm{log}\big|  {Z}  \big|  } \big/ {  \big|  {Y}^{*}  \big|   }  \big\} \cdot       \bigg\{       \big| Z \big|^{-1}   \cdot  Z^{\prime}  \cdot   \big|  {Y}^{*}  \big|  -   \mathrm{log}\big|  {Z}  \big|  \cdot  \big\{ Y^* \big\}^{\prime}  \bigg\}      \big/    \big|  {Y}^{*}  \big|^2  \\   \Bigg\Updownarrow  
\end{array}\right.  \\ \\ \\ \tag{\textit{First derivative of $r$}}  
\]  }  

{\small 
\[ r^{\prime} \equiv r^{\prime} \bigg( \bigg|  \frac{X}{X^*}  \bigg| ,  \bigg|  \frac{Y}{Y^*}  \bigg|  ,   \bigg|  \frac{Z}{Z^*}  \bigg|   \bigg) \text{ } \textit{continued}  =   \left\{\!\begin{array}{ll@{}>{{}}l} 
 \big\{ \big| X^* \big| ,  \big| Y^* \big|  ,   \big|  Z \big|  \text{ }      \textit{are not made arbitrarily small}  \big\} ,  \\     \boxed{r^{\prime}_2} \equiv   1 \big/ \mathrm{log} \bigg[   {  \mathrm{log} \big| {X} \big|  } \big/ { \big| {Y}^{*} \big| }          \bigg] \cdot \bigg\{   \big| X \big|^{-1} \cdot X^{\prime}  \cdot  \big| Y^* \big| \\ -    \mathrm{log} \big| {X} \big|  \cdot    \big\{ Y^* \big\}^{\prime}           \bigg\}  \big/ \big| Y^* \big|^2  \\ \\     + 1 \big/  \big\{   {   \mathrm{log}\big|  {Y}^*  \big|  } \big/ {  \big|  {X}  \big|   }  \big\} \cdot       \bigg\{       \big|  {Y}^*  \big|^{-1}   \cdot  \big\{  {Y}^*  \big\}^{\prime}  \cdot   \big|  {X} \big|  -   \mathrm{log}\big|  {Y}^*   \big|  \cdot  Y^{\prime}  \bigg\}      \big/    \big|  X   \big|^2          ,   \\ \Bigg\Updownarrow \\  \\   \big\{ \big| X^* \big| ,  \big| Y^* \big|  ,   \big|  Z \big|  \text{ }      \textit{are not made arbitrarily small}  \big\}  ,  \\  \\ \boxed{r^{\prime}_3} \equiv 1 \big/   \mathrm{log} \bigg[   {  \mathrm{log} \big| {Y}^{*} \big|  } \big/ {\big| {X}^{*} \big| }          \bigg] \cdot \bigg\{            \big| Y^{*} \big|^{-1} \cdot  \big\{ Y^* \big\}^{\prime} \cdot \big| X^* \big|  \\  -    \mathrm{log}   \big| Y^{*} \big|\cdot \big\{ X^* \big\}^{\prime }                  \bigg\} \big/ \big| X^* \big|^2         \\ \\  + 1 \big/  \big\{   {   \mathrm{log}\big|  {Y}^*  \big|  } \big/ {  \big|  {X}  \big|   }  \big\} \cdot       \bigg\{       \big|  {Y}^*  \big|^{-1}   \cdot  \big\{  {Y}^*  \big\}^{\prime}  \cdot   \big|  {X} \big|  -   \mathrm{log}\big|  {Y}^*   \big|  \cdot  Y^{\prime}  \bigg\}      \big/    \big|  X   \big|^2       ,  \\ \Bigg\Updownarrow \\ \\  \big\{ \big| X^* \big| ,  \big| Y^* \big|  ,   \big|  Z \big|  \text{ }      \textit{are not made arbitrarily small}  \big\}  ,    \\ \\  \boxed{r^{\prime}_4}  \equiv   1 \big/  \mathrm{log} \bigg[   {  \mathrm{log} \big| {Y}^{*} \big|  } \big/ {\big| {X}^{*} \big| }          \bigg] \cdot \bigg\{            \big| Y^{*} \big|^{-1} \cdot  \big\{ Y^* \big\}^{\prime} \cdot \big| X^* \big|     \\  -    \mathrm{log}   \big| Y^{*} \big|\cdot \big\{ X^* \big\}^{\prime }                  \bigg\} \big/ \big| X^* \big|^2         \\   + 1 \big/  \big\{   {   \mathrm{log}\big|  {Z}  \big|  } \big/ {  \big|  {Y}^{*}  \big|   }  \big\} \cdot       \bigg\{       \big| Z \big|^{-1}   \cdot  Z^{\prime}  \cdot   \big|  {Y}^{*}  \big|  -   \mathrm{log}\big|  {Z}  \big|  \cdot  \big\{ Y^* \big\}^{\prime}  \bigg\}      \big/    \big|  {Y}^{*}  \big|^2  ,    \\ \Bigg\Updownarrow \\ \\  \big\{ \big| X^* \big| ,  \big| Y^* \big|  ,   \big|  Z \big|  \text{ }      \textit{are not made arbitrarily small}  \big\}  .  \\ \\ 
\end{array}\right.  \\ \\ \\ \tag{\textit{First derivative of $r$}}  
\]  }

\noindent Hence it suffices to determine the alphabets for which,

{\small 
\[   \left\{\!\begin{array}{ll@{}>{{}}l} 
 \boxed{r^{\prime}_1} \equiv 0 ,  \\ \\  \boxed{r^{\prime}_2} \equiv 0 ,     \\  \\  \boxed{r^{\prime}_3} \equiv 0 ,    \\  \\  \boxed{r^{\prime}_4} \equiv 0 .
\end{array}\right. 
\]  }

\noindent By direct computation,

{\small

\begin{align*}
     \bigg(  \boxed{r^{\prime}_1} \equiv 0 \bigg)  \Longrightarrow  \bigg(             1 \big/  \mathrm{log} \bigg[   {  \mathrm{log} \big| {Y}^{*} \big|  } \big/ {\big| {X}^{*} \big| }          \bigg] \cdot \bigg\{            \big| Y^{*} \big|^{-1} \cdot  \big\{ Y^* \big\}^{\prime} \cdot \big| X^* \big|   -    \mathrm{log}   \big| Y^{*} \big|\cdot \big\{ X^* \big\}^{\prime }                  \bigg\} \big/ \big| X^* \big|^2  \\  + 1 \big/  \big\{   {   \mathrm{log}\big|  {Z}  \big|  } \big/ {  \big|  {Y}^{*}  \big|   }  \big\} \cdot       \bigg\{       \big| Z \big|^{-1}   \cdot  Z^{\prime}  \cdot   \big|  {Y}^{*}  \big|  -   \mathrm{log}\big|  {Z}  \big|  \cdot  \big\{ Y^* \big\}^{\prime}  \bigg\}      \big/    \big|  {Y}^{*}  \big|^2     = 0            \bigg)  \\ \Bigg\Updownarrow \end{align*}
     
     \begin{align*} \bigg(     \bigg\{            \big| Y^{*} \big|^{-1} \cdot  \big\{ Y^* \big\}^{\prime} \cdot \big| X^* \big|   -    \mathrm{log}   \big| Y^{*} \big|\cdot \big\{ X^* \big\}^{\prime }                  \bigg\} \big/         \bigg\{  \mathrm{log} \bigg[   {  \mathrm{log} \big| {Y}^{*} \big|  } \big/ {\big| {X}^{*} \big| }          \bigg]    \cdot  \big| X^* \big|^2     \bigg\}           = -       \bigg\{       \big| Z \big|^{-1}   \cdot  Z^{\prime}  \\ \cdot   \big|  {Y}^{*}  \big|   -   \mathrm{log}\big|  {Z}  \big|  \cdot  \big\{ Y^* \big\}^{\prime}  \bigg\}     \big/ \bigg\{  \big\{   \mathrm{log}\big|  {Z}  \big|   \big/ {  \big|  {Y}^{*}  \big|   }  \big\}       \cdot    \big| Y^* \big|^2       \bigg\}       \bigg)  \\ \Bigg\Updownarrow \end{align*}
     
     \begin{align*}      \bigg(    \bigg\{  \big\{   \mathrm{log}\big|  {Z}  \big|   \big/ {  \big|  {Y}^{*}  \big|   }  \big\}       \cdot    \big| Y^* \big|^2       \bigg\}   \bigg/    \bigg\{  \mathrm{log} \bigg[   {  \mathrm{log} \big| {Y}^{*} \big|  } \big/ {\big| {X}^{*} \big| }          \bigg]    \cdot  \big| X^* \big|^2     \bigg\}            = -     \bigg\{       \big| Z \big|^{-1}   \cdot  Z^{\prime}  \cdot   \big|  {Y}^{*}  \big|      -   \mathrm{log}\big|  {Z}  \big| \\  \cdot  \big\{ Y^* \big\}^{\prime}  \bigg\}     \bigg/     \bigg\{            \big| Y^{*} \big|^{-1}  \cdot  \big\{ Y^* \big\}^{\prime} \cdot \big| X^* \big|    -    \mathrm{log}   \big| Y^{*} \big|\cdot \big\{ X^* \big\}^{\prime }                  \bigg\}               \bigg)  \\ \Bigg\Updownarrow \end{align*}
     
     \begin{align*}    \bigg(       \bigg\{  \big\{   \mathrm{log}\big|  {Z}  \big|   \big/ {  \big|  {Y}^{*}  \big|   }  \big\}       \cdot    \big| Y^* \big|^2       \bigg\}   \bigg/    \bigg\{  \mathrm{log} \bigg[   {  \mathrm{log} \big| {Y}^{*} \big|  } \big/ {\big| {X}^{*} \big| }          \bigg]    \cdot  \big| X^* \big|^2     \bigg\}            = -     \bigg\{       \big| Z \big|^{-1}   \cdot  Z^{\prime}  \cdot   \big|  {Y}^{*}  \big|    \bigg\}  \bigg/     \bigg\{            \big| Y^{*} \big|^{-1} \\  \cdot  \big\{ Y^* \big\}^{\prime} \cdot \big| X^* \big|     -    \mathrm{log}   \big| Y^{*} \big|\cdot \big\{ X^* \big\}^{\prime }                  \bigg\}  -  \bigg\{  \mathrm{log}\big|  {Z}  \big|  \cdot  \big\{ Y^* \big\}^{\prime}  \bigg\}     \bigg/     \bigg\{            \big| Y^{*} \big|^{-1}  \\ \cdot  \big\{ Y^* \big\}^{\prime} \cdot \big| X^* \big|     -    \mathrm{log}   \big| Y^{*} \big|  \cdot \big\{ X^* \big\}^{\prime }                  \bigg\}                    \bigg)    \\ \Bigg\Updownarrow  
\end{align*}

\begin{align*}
\bigg(  \big| Y^* \big/ X^* \big|^2 = -   \bigg[    \mathrm{log}\big|  {Z}  \big|   \big/ {  \big|  {Y}^{*}  \big|   }       \bigg/    \mathrm{log} \bigg[   {  \mathrm{log} \big| {Y}^{*} \big|  } \big/ {\big| {X}^{*} \big| }          \bigg]          \bigg]^{-1 }     \cdot        \bigg[  \bigg\{       \big| Z \big|^{-1}   \cdot  Z^{\prime}  \cdot   \big|  {Y}^{*}  \big|    \bigg\}  \bigg/     \bigg\{            \big| Y^{*} \big|^{-1}  \\ \cdot  \big\{ Y^* \big\}^{\prime} \cdot \big| X^* \big|     -    \mathrm{log}   \big| Y^{*} \big|\cdot \big\{ X^* \big\}^{\prime }                  \bigg\}  -  \bigg\{  \mathrm{log}\big|  {Z}  \big|  \cdot  \big\{ Y^* \big\}^{\prime}  \bigg\}     \bigg/     \bigg\{            \big| Y^{*} \big|^{-1}  \cdot  \big\{ Y^* \big\}^{\prime} \\ \cdot \big| X^* \big|     -    \mathrm{log}   \big| Y^{*} \big|  \cdot \big\{ X^* \big\}^{\prime }                  \bigg\}  \bigg]       \bigg)  \\  \Bigg\Updownarrow   
\end{align*}

\begin{align*}
   \bigg(   \big| Y^* \big/ X^* \big|^2 = -              \bigg[   \big\{  \mathrm{log}\big|  {Z}  \big| \big\} \cdot \big| Z \big|   \big/ \big\{  \big|  {Y}^{*}  \big|^2  \cdot  Z^{\prime}   \big\}       \bigg/    \mathrm{log} \bigg[   {  \mathrm{log} \big| {Y}^{*} \big|  } \big/ {\big| {X}^{*} \big| }          \bigg]          \bigg]^{-1 }        \cdot      \bigg\{           \big| Y^{*} \big|^{-1}  \\  \cdot  \big\{ Y^* \big\}^{\prime} \cdot \big| X^* \big|      -    \mathrm{log}   \big| Y^{*} \big|\cdot \big\{ X^* \big\}^{\prime }             \bigg\}^{-1}                -  \bigg[      1 \big/ \big\{ \big| Y^* \big| \cdot \big\{ Y^* \big\}^{\prime} \big\} \bigg/     \mathrm{log} \bigg[   {  \mathrm{log} \big| {Y}^{*} \big|  } \big/ {\big| {X}^{*} \big| }          \bigg]         \bigg]^{-1}  \\  \cdot     \bigg\{            \big| Y^{*} \big|^{-1}   \cdot  \big\{ Y^* \big\}^{\prime}  \cdot \big| X^* \big|      -    \mathrm{log}   \big| Y^{*} \big|  \cdot \big\{ X^* \big\}^{\prime }                  \bigg\}^{-1}                                                  \bigg)      \\  \Bigg\Updownarrow    
\end{align*}

\begin{align*}
  \bigg(    \big| Y^* \big/ X^* \big|^2 = -                    \bigg[   \big\{  \mathrm{log}\big|  {Z}  \big| \big\} \cdot \big| Z \big|   \big/ \big\{  \big|  {Y}^{*}  \big|^2  \cdot  Z^{\prime}   \big\}       \bigg/    \mathrm{log} \bigg[   {  \mathrm{log} \big| {Y}^{*} \big|  } \big/ {\big| {X}^{*} \big| }  \bigg]                         \cdot      \bigg\{           \big| Y^{*} \big|^{-1}  \\  \cdot  \big\{ Y^* \big\}^{\prime} \cdot \big| X^* \big|      -    \mathrm{log}   \big| Y^{*} \big|\cdot \big\{ X^* \big\}^{\prime }             \bigg\}    \bigg]^{-1 }             -  \bigg[      1 \big/ \big\{ \big| Y^* \big| \cdot \big\{ Y^* \big\}^{\prime} \big\} \bigg/     \mathrm{log} \bigg[   {  \mathrm{log} \big| {Y}^{*} \big|  } \big/ {\big| {X}^{*} \big| }            \bigg]      \\  \cdot     \bigg\{            \big| Y^{*} \big|^{-1}   \cdot  \big\{ Y^* \big\}^{\prime}  \cdot \big| X^* \big|      -    \mathrm{log}   \big| Y^{*} \big|  \cdot \big\{ X^* \big\}^{\prime }                  \bigg\}      \bigg]^{-1}                                                           \bigg)     \\ \Bigg\Updownarrow  
\end{align*}

\begin{align*}
         \bigg(  \big| Y^* \big/ X^* \big|^2 =                   \bigg\{ -  \bigg[   \big\{  \mathrm{log}\big|  {Z}  \big| \big\} \cdot \big| Z \big|   \big/ \big\{  \big|  {Y}^{*}  \big|^2  \cdot  Z^{\prime}   \big\}      \bigg]^{-1}     -  \bigg[      1 \big/ \big\{ \big| Y^* \big| \cdot \big\{ Y^* \big\}^{\prime} \big\}  \bigg]^{-1} \bigg\}  \\  \bigg/  \bigg[    \mathrm{log} \bigg[   {  \mathrm{log} \big| {Y}^{*} \big|  } \big/ {\big| {X}^{*} \big| }            \bigg]        \cdot     \bigg\{            \big| Y^{*} \big|^{-1}   \cdot  \big\{ Y^* \big\}^{\prime}  \cdot \big| X^* \big|      -    \mathrm{log}   \big| Y^{*} \big|  \cdot \big\{ X^* \big\}^{\prime }                  \bigg\}    \bigg]                              \bigg)                                      \\ \Bigg\Updownarrow  
\end{align*}

\begin{align*}
         \bigg(  \big|  Y^* \big|   =    +  \sqrt{  \big| X^* \big|^2    \cdot           \frac{   -\bigg[   \frac{  \mathrm{log}\big|  {Z}  \big|  \cdot \big| Z \big| }{   \big|  {Y}^{*}  \big|^2  \cdot  Z^{\prime}     }    \bigg]^{-1}     -   \bigg[      \frac{1}{ \big| Y^* \big| \cdot \big\{ Y^* \big\}^{\prime} }  \bigg]^{-1}   }{     \mathrm{log}  \text{ }       \frac{\mathrm{log} \big| {Y}^{*} \big|  }{ {\big| {X}^{*} \big| }    }            \cdot     \bigg\{            \frac{1}{\big| Y^{*} \big| }  \cdot  \big\{ Y^* \big\}^{\prime}  \cdot \big| X^* \big|      -    \mathrm{log}   \big| Y^{*} \big|  \cdot \big\{ X^* \big\}^{\prime }                  \bigg\}                                   }  } \end{align*}

\begin{align*}   \Longleftrightarrow \big\{ \big| {X} \big| > \big| {Y}^{*} \big| ,  \big| {Y}^{*} \big| > \big| {Z} \big|   \big\}    \bigg)      \\ \Bigg\Updownarrow    
\end{align*}

\begin{align*}
         \bigg(  \big|  Y^* \big|   =    +  \sqrt{  \big| X^* \big|^2    \cdot           \frac{   \bigg[   \frac{  \mathrm{log}\big|  {Z}  \big|  \cdot \big| Z \big| }{   \big|  {Y}^{*}  \big|^2  \cdot  Z^{\prime}     }    \bigg]^{-1}     +   \bigg[      \frac{1}{ \big| Y^* \big| \cdot \big\{ Y^* \big\}^{\prime} }  \bigg]^{-1}   }{     \mathrm{log}  \text{ }       \frac{\mathrm{log} \big| {Y}^{*} \big|  }{ {\big| {X}^{*} \big| }    }            \cdot     \bigg\{            \frac{1}{\big| Y^{*} \big| }  \cdot  \big\{ Y^* \big\}^{\prime}  \cdot \big| X^* \big|      -    \mathrm{log}   \big| Y^{*} \big|  \cdot \big\{ X^* \big\}^{\prime }                  \bigg\}                                   }  }  \end{align*}

         \begin{align*} \Longleftrightarrow \big\{ \big| {X} \big| > \big| {Y}^{*} \big| ,  \big| {Y}^{*} \big| > \big| {Z} \big|   \big\}    \bigg)   . 
\end{align*}

}

\noindent Hence the roots to the numerator of the above square root,

{\small

\begin{align*}
     \bigg(     \bigg[   \frac{  \mathrm{log}\big|  {Z}  \big|  \cdot \big| Z \big| }{   \big|  {Y}^{*}  \big|^2  \cdot  Z^{\prime}     }    \bigg]^{-1}     +  \bigg[      \frac{1}{ \big| Y^* \big| \cdot \big\{ Y^* \big\}^{\prime} }  \bigg]^{-1}  = 0       \bigg) \\   \Bigg\Updownarrow  \\      \bigg(     \frac{    \big|  {Y}^{*}  \big|^2  \cdot  Z^{\prime}  }{  \mathrm{log}\big|  {Z}  \big|  \cdot \big| Z \big|    }       = -   { \big| Y^* \big| \cdot \big\{ Y^* \big\}^{\prime} }       \bigg) \\   \Bigg\Updownarrow      \\             \bigg(     \frac{    \big|  {Y}^{*}  \big|  \cdot  Z^{\prime}  }{  \mathrm{log}\big|  {Z}  \big|  \cdot \big| Z \big|    }       = -   { \big\{ Y^* \big\}^{\prime} }       \bigg) . 
\end{align*}

}

\noindent correspond to the critical points of $r_1$ for which $\big| Y^* \big| \equiv 0$. Along similar lines, repeating the above rearrangements for $r^{\prime}_2 \equiv 0$, instead of $r^{\prime}_1 \equiv 0$, yields,

{\small

\begin{align*}
     \bigg( \boxed{r^{\prime}_2} \equiv 0 \bigg) \Longrightarrow  \bigg(       1 \big/ \mathrm{log} \bigg[   {  \mathrm{log} \big| {X} \big|  } \big/ { \big| {Y}^{*} \big| }          \bigg] \cdot \bigg\{   \big| X \big|^{-1} \cdot X^{\prime}  \cdot  \big| Y^* \big| -    \mathrm{log} \big| {X} \big|  \cdot    \big\{ Y^* \big\}^{\prime}           \bigg\}  \big/ \big| Y^* \big|^2  \\      + 1 \big/  \big\{   {   \mathrm{log}\big|  {Y}^*  \big|  } \big/ {  \big|  {X}  \big|   }  \big\} \cdot       \bigg\{       \big|  {Y}^*  \big|^{-1}   \cdot  \big\{  {Y}^*  \big\}^{\prime}  \cdot   \big|  {X} \big|  -   \mathrm{log}\big|  {Y}^*   \big|  \cdot  Y^{\prime}  \bigg\}      \big/    \big|  X   \big|^2         = 0            \bigg) \\ \Bigg\Updownarrow \end{align*}

     \begin{align*} \vdots \\  \bigg(          \big| X \big| = +           \sqrt{  \big| Y^* \big|^2    \cdot           \frac{ -  \bigg[   \frac{  \mathrm{log}\big|  {Y^*}  \big|  \cdot \big| Y^*  \big| }{   \big|  {Y}^{*}  \big|^2  \cdot  Z^{\prime}     }    \bigg]^{-1}     -  \bigg[      \frac{1}{ \big| X \big| \cdot \big\{ X  \big\}^{\prime} }  \bigg]^{-1}   }{     \mathrm{log} \text{ }        \frac{\mathrm{log} \big| {X} \big|  }{ {\big| {Y}^{*} \big| }    }                 \cdot     \bigg\{            \frac{1}{\big| X^{*} \big| }  \cdot  \big\{ X^* \big\}^{\prime}  \cdot \big| Y^* \big|      -    \mathrm{log}   \big| X  \big|  \cdot \big\{ Y^* \big\}^{\prime }                  \bigg\}                                   }  }      \\  \Longleftrightarrow \big\{ \big| {X} \big| <  \big| {Y}^{*} \big| ,  \big| {Y}^{*} \big| <  \big| {Z} \big|   \big\}              \bigg) 
\end{align*}

  \begin{align*} \Bigg\Updownarrow  \\  \bigg(          \big| X \big| = +           \sqrt{  \big| Y^* \big|^2    \cdot           \frac{   \bigg[   \frac{  \mathrm{log}\big|  {Y^*}  \big|  \cdot \big| Y^*  \big| }{   \big|  {Y}^{*}  \big|^2  \cdot  Z^{\prime}     }    \bigg]^{-1}   +   \bigg[      \frac{1}{ \big| X \big| \cdot \big\{ X  \big\}^{\prime} }  \bigg]^{-1}   }{     \mathrm{log} \text{ }        \frac{\mathrm{log} \big| {X} \big|  }{ {\big| {Y}^{*} \big| }    }                 \cdot     \bigg\{            \frac{1}{\big| X^{*} \big| }  \cdot  \big\{ X^* \big\}^{\prime}  \cdot \big| Y^* \big|      -    \mathrm{log}   \big| X  \big|  \cdot \big\{ Y^* \big\}^{\prime }                  \bigg\}                                   }  } \end{align*}

         \begin{align*}   \Longleftrightarrow \big\{ \big| {X} \big| <  \big| {Y}^{*} \big| ,  \big| {Y}^{*} \big| <  \big| {Z} \big|   \big\}              \bigg) . 
\end{align*} 

}

\noindent Similarly,

{\small

\begin{align*}
     \bigg( \boxed{r^{\prime}_3} \equiv 0 \bigg) \Longrightarrow  \bigg(       1 \big/   \mathrm{log} \bigg[     {  \mathrm{log} \big| {Y}^{*} \big|  } \big/ {\big| {X}^{*} \big| }     \bigg]     \cdot \bigg\{            \big| Y^{*} \big|^{-1} \cdot  \big\{ Y^* \big\}^{\prime} \cdot \big| X^* \big|    -    \mathrm{log}   \big| Y^{*} \big|\cdot \big\{ X^* \big\}^{\prime }                  \bigg\} \big/ \big| X^* \big|^2        \\ + 1 \big/  \big\{   {   \mathrm{log}\big|  {Y}^*  \big|  } \big/ {  \big|  {X}  \big|   }  \big\} \cdot       \bigg\{       \big|  {Y}^*  \big|^{-1}   \cdot  \big\{  {Y}^*  \big\}^{\prime}  \cdot   \big|  {X} \big|  -   \mathrm{log}\big|  {Y}^*   \big|  \cdot  Y^{\prime}  \bigg\}      \big/    \big|  X   \big|^2     = 0            \bigg) \\ \Bigg\Updownarrow  \end{align*}

     \begin{align*} \vdots \\  \bigg(                    \big|  Y^* \big|   =    +  \sqrt{  \big| X^* \big|^2    \cdot           \frac{ -  \bigg[   \frac{  \mathrm{log}\big|  {Y^* }  \big|  \cdot \big| Y^*   \big| }{   \big|  {Y}^{*}  \big|^2  \cdot  \big\{ Y^*  \big\}^{\prime}     }    \bigg]^{-1}     -  \bigg[      \frac{1}{ \big| Y^* \big| \cdot \big\{ Y^* \big\}^{\prime} }  \bigg]^{-1}   }{     \mathrm{log} \text{ }       \frac{\mathrm{log} \big| {Y}^{*} \big|  }{ {\big| {X}^{*} \big| }    }            \cdot     \bigg\{            \frac{1}{\big| Y^{*} \big| }  \cdot  \big\{ Y^* \big\}^{\prime}  \cdot \big| X^* \big|      -    \mathrm{log}   \big| Y^{*} \big|  \cdot \big\{ X^* \big\}^{\prime }                  \bigg\}                                   }  } \\  \Longleftrightarrow \big\{ \big| {X} \big| >   \big| {Y}^{*} \big| ,  \big| {Y}^{*} \big| <  \big| {Z} \big|   \big\}                     \bigg) . 
\end{align*}

}

{\small

\begin{align*}
     \bigg( \boxed{r^{\prime}_4} \equiv 0 \bigg) \Longrightarrow  \bigg(     1 \big/  \mathrm{log} \bigg[   {  \mathrm{log} \big| {Y}^{*} \big|  } \big/ {\big| {X}^{*} \big| }          \bigg] \cdot \bigg\{            \big| Y^{*} \big|^{-1} \cdot  \big\{ Y^* \big\}^{\prime} \cdot \big| X^* \big|    -    \mathrm{log}   \big| Y^{*} \big|\cdot \big\{ X^* \big\}^{\prime }                  \bigg\} \big/ \big| X^* \big|^2   \\  + 1 \big/  \big\{   {   \mathrm{log}\big|  {Z}  \big|  } \big/ {  \big|  {Y}^{*}  \big|   }  \big\} \cdot       \bigg\{       \big| Z \big|^{-1}   \cdot  Z^{\prime}  \cdot   \big|  {Y}^{*}  \big|  -   \mathrm{log}\big|  {Z}  \big|  \cdot  \big\{ Y^* \big\}^{\prime}  \bigg\}      \big/    \big|  {Y}^{*}  \big|^2        = 0            \bigg) \\ \Bigg\Updownarrow \\ \vdots \end{align*}

     \begin{align*} \bigg(       \big|  Y^* \big|   =    +  \sqrt{  \big| X^* \big| ^2    \cdot           \frac{  - \bigg[   \frac{  \mathrm{log}\big|  {Z}  \big|  \cdot \big| Z \big| }{   \big|  {Y}^{*}  \big|^2  \cdot  Z^{\prime}     }    \bigg]^{-1}    -   \bigg[      \frac{1}{ \big| Y^* \big| \cdot \big\{ Y^* \big\}^{\prime} }  \bigg]^{-1}   }{     \mathrm{log} \text{ }        \frac{\mathrm{log} \big| {Y}^{*} \big|  }{ {\big| {X}^{*} \big| }    }                \cdot     \bigg\{            \frac{1}{\big| Y^{*} \big| }  \cdot  \big\{ Y^* \big\}^{\prime}  \cdot \big| X^* \big|      -    \mathrm{log}   \big| Y^{*} \big|  \cdot \big\{ X^* \big\}^{\prime }                  \bigg\}                                   }  }  \\ \Longleftrightarrow \big\{ \big| {X} \big| <  \big| {Y}^{*} \big| ,  \big| {Y}^{*} \big| > \big| {Z} \big|   \big\}                            \bigg)  
\end{align*}

     \begin{align*} \Bigg\Updownarrow  \\ \bigg(       \big|  Y^* \big|   =    +  \sqrt{  \big| X^* \big| ^2    \cdot           \frac{   \bigg[   \frac{  \mathrm{log}\big|  {Z}  \big|  \cdot \big| Z \big| }{   \big|  {Y}^{*}  \big|^2  \cdot  Z^{\prime}     }    \bigg]^{-1}    +  \bigg[      \frac{1}{ \big| Y^* \big| \cdot \big\{ Y^* \big\}^{\prime} }  \bigg]^{-1}   }{     \mathrm{log} \text{ }        \frac{\mathrm{log} \big| {Y}^{*} \big|  }{ {\big| {X}^{*} \big| }    }                \cdot     \bigg\{            \frac{1}{\big| Y^{*} \big| }  \cdot  \big\{ Y^* \big\}^{\prime}  \cdot \big| X^* \big|      -    \mathrm{log}   \big| Y^{*} \big|  \cdot \big\{ X^* \big\}^{\prime }                  \bigg\}                                   }  }  \end{align*}

         \begin{align*}   \Longleftrightarrow \big\{ \big| {X} \big| <  \big| {Y}^{*} \big| ,  \big| {Y}^{*} \big| > \big| {Z} \big|   \big\}                            \bigg)  
\end{align*}

}

\noindent implying that the collection of four expressions provided in $(r^{\prime} = 0)$ holds, from which we conclude the argument. \boxed{}

\bigskip

\noindent To make use of the above expressions obtained from the first derivative of the converse bit transmission rate being equal to $0$, below we compute the value of $r_1, \cdots, r_4$ at the respective expressions obtained above for $\big| Y^* \big|$, $\big| X \big|$, $\big| Y^* \big|$, $\big| Y^* \big|$.

\bigskip

 \noindent \textbf{Lemma} \textit{3} (\textit{the second derivative of the proposed expression for the converse bit transmission rate over Alice and Bob's shared Quantum channel is strictly positive given the expressions obtained in $\textbf{Lemma}$ \textit{2} when the first derivative vanishes}). $r^{\prime\prime}$ is strictly positive given the expressions obtained from \textbf{Lemma} \textit{2}.

\bigskip

\noindent \textit{Proof of Lemma 3}. It suffices to evaluate the second derivatives of the converse bit transmission rate $r^{\prime\prime}$. From the fact that,

{\small 
\[  r^{\prime\prime} \equiv r^{\prime\prime} \bigg( \bigg|  \frac{X}{X^*}  \bigg| ,  \bigg|  \frac{Y}{Y^*}  \bigg|  ,   \bigg|  \frac{Z}{Z^*}  \bigg|  \bigg) =  \left\{\!\begin{array}{ll@{}>{{}}l} 
  \boxed{r^{\prime\prime}_1}  \equiv  r^{\prime\prime}_{1,1} +  r^{\prime\prime}_{1,2}  , \\   \\    \boxed{r^{\prime\prime}_2}  \equiv  r^{\prime\prime}_{2,1} +  r^{\prime\prime}_{2,2}   , \\ \\   \boxed{r^{\prime\prime}_3}  \equiv  r^{\prime\prime}_{3,1} +  r^{\prime\prime}_{3,2}  , \\ \\   \boxed{r^{\prime\prime}_4}  \equiv r^{\prime\prime}_{4,1} +  r^{\prime\prime}_{4,2}   ,
\end{array}\right. \\ \\ \tag{\textit{Second derivative of $r$}}  
\]  }

\noindent write,

\begin{align*}
      r^{\prime\prime}_1 =   -  \bigg\{ \underbrace{\mathrm{log}  \text{ }  \mathrm{log}    \bigg(   {\sqrt{  \big| X^* \big|^2    \cdot           \frac{   \bigg[   \frac{  \mathrm{log}\big|  {Z}  \big|  \cdot \big| Z \big| }{   \big|  {Y}^{*}  \big|^2  \cdot  Z^{\prime}     }    \bigg]^{-1}     +  \bigg[      \frac{1}{ \big| Y^* \big| \cdot \big\{ Y^* \big\}^{\prime} }  \bigg]^{-1}   }{     \mathrm{log} \text{ }   {\mathrm{log} \big| {Y}^{*} \big|  } \big/ { {\big| {X}^{*} \big| }    }              \cdot     \bigg\{            \frac{1}{\big| Y^{*} \big| }  \cdot  \big\{ Y^* \big\}^{\prime}  \cdot \big| X^* \big|      -    \mathrm{log}   \big| Y^{*} \big|  \cdot \big\{ X^* \big\}^{\prime }                  \bigg\}                       }  }} \times {\big| X^* \big|^{-1}}\bigg)}_{\{ ( r^{\prime\prime}_1 )-1     \}   }  \end{align*}

        \begin{align*} \cdot  \big| X^* \big|^2  \bigg\}^{-2} \end{align*}

        \begin{align*} \times       \bigg\{ \bigg\{              \underbrace{\mathrm{log}   \bigg(   {\sqrt{  \big| X^* \big|^2    \cdot           \frac{   \bigg[   \frac{  \mathrm{log}\big|  {Z}  \big|  \cdot \big| Z \big| }{   \big|  {Y}^{*}  \big|^2  \cdot  Z^{\prime}     }    \bigg]^{-1}     +  \bigg[      \frac{1}{ \big| Y^* \big| \cdot \big\{ Y^* \big\}^{\prime} }  \bigg]^{-1}   }{     \mathrm{log} \text{ }   {\mathrm{log} \big| {Y}^{*} \big|  } \big/ { {\big| {X}^{*} \big| }    }              \cdot     \bigg\{            \frac{1}{\big| Y^{*} \big| }  \cdot  \big\{ Y^* \big\}^{\prime}  \cdot \big| X^* \big|      -    \mathrm{log}   \big| Y^{*} \big|  \cdot \big\{ X^* \big\}^{\prime }                  \bigg\}                                   }  }} \times {\big| X^* \big|^{-1}}\bigg)                \bigg\}^{-1}}_{\{ (   r^{\prime\prime}_1) -2 \} }      \end{align*}

        \begin{align*}
         \times    \bigg[ \mathcal{T}_1  \mathcal{T}_2 \mathcal{T}_3                 - \mathrm{log} \big[ \mathcal{T}_1 \big]   \big\{ X^* \big\}^{\prime}                  \bigg]    
        \end{align*}

        \begin{align*} \cdot  \big| X^* \big|^2  \end{align*}

\begin{align*}
  +     \underbrace{\mathrm{log}  \text{ }  \mathrm{log}    \bigg(   {\sqrt{  \big| X^* \big|^2    \cdot           \frac{   \bigg[   \frac{  \mathrm{log}\big|  {Z}  \big|  \cdot \big| Z \big| }{   \big|  {Y}^{*}  \big|^2  \cdot  Z^{\prime}     }    \bigg]^{-1}     +  \bigg[      \frac{1}{ \big| Y^* \big| \cdot \big\{ Y^* \big\}^{\prime} }  \bigg]^{-1}   }{     \mathrm{log} \text{ }   {\mathrm{log} \big| {Y}^{*} \big|  } \big/ { {\big| {X}^{*} \big| }    }               \cdot     \bigg\{            \frac{1}{\big| Y^{*} \big| }  \cdot  \big\{ Y^* \big\}^{\prime}  \cdot \big| X^* \big|      -    \mathrm{log}   \big| Y^{*} \big|  \cdot \big\{ X^* \big\}^{\prime }                  \bigg\}                                   }  }} \times {\big| X^* \big|^{-1}}\bigg)}_{ \{ (  r^{\prime\prime}_1 )-3 \} } 
\end{align*}

\begin{align*}
\cdot 2 \big| X^* \big| \cdot \big\{ X^* \big\}^{\prime} \bigg\} 
\end{align*} 

\begin{align*}
    \times  \bigg\{   \mathcal{T}_1 \mathcal{T}_2 \mathcal{T}_3      -               \mathrm{log} \big[ \mathcal{T}_1 \big]                \big\{ X^* \big\}^{\prime}       \bigg\} 
\end{align*}

\begin{align*}
+  \bigg\{ \underbrace{\mathrm{log}  \text{ }  \mathrm{log}    \bigg(   {\sqrt{  \big| X^* \big|^2    \cdot           \frac{   \bigg[   \frac{  \mathrm{log}\big|  {Z}  \big|  \cdot \big| Z \big| }{   \big|  {Y}^{*}  \big|^2  \cdot  Z^{\prime}     }    \bigg]^{-1}     +  \bigg[      \frac{1}{ \big| Y^* \big| \cdot \big\{ Y^* \big\}^{\prime} }  \bigg]^{-1}   }{     \mathrm{log} \text{ }   {\mathrm{log} \big| {Y}^{*} \big|  } \big/ { {\big| {X}^{*} \big| }    }              \cdot     \bigg\{            \frac{1}{\big| Y^{*} \big| }  \cdot  \big\{ Y^* \big\}^{\prime}  \cdot \big| X^* \big|      -    \mathrm{log}   \big| Y^{*} \big|  \cdot \big\{ X^* \big\}^{\prime }                  \bigg\}                       }  }} \times {\big| X^* \big|^{-1}}\bigg)}_{\{ (  r^{\prime\prime}_1 )-4 \} } \end{align*}

        \begin{align*} \cdot  \big| X^* \big|^2  \bigg\}^{-1} \end{align*}

        \begin{align*} \times              \bigg\{        \big\{  - \mathcal{T}^{-1}_1  \mathcal{T}^2_2 \mathcal{T}_3   +     \mathcal{T}_1    \mathcal{T}^{\prime}_2 \mathcal{T}_3 + \mathcal{T}_1 \mathcal{T}_2         \big\{ X^* \big\}^{\prime}       \big\}\end{align*}
        
        \begin{align*} -     \big\{        \mathcal{T}_1 \mathcal{T}_2 \big\{ X^* \big\}^{\prime } + \mathrm{log} \big[ \mathcal{T}_1 \big] \big\{ X^* \big\}^{\prime\prime}     \big\}                                          \bigg\}      \end{align*}

\noindent corresponding to the first piecewise defined function $r_1$ for the converse transmission rate $r$, where,

{\small 
\[   \left\{\!\begin{array}{ll@{}>{{}}l} 
    \mathcal{T}_1 \mathcal{T}_2 \mathcal{T}_3 =    \sqrt{  \big| X^* \big|^2    \cdot           \frac{   \bigg[   \frac{  \mathrm{log}\big|  {Z}  \big|  \cdot \big| Z \big| }{   \big|  {Y}^{*}  \big|^2  \cdot  Z^{\prime}     }    \bigg]^{-1}     +  \bigg[      \frac{1}{ \big| Y^* \big| \cdot \big\{ Y^* \big\}^{\prime} }  \bigg]^{-1}   }{     \mathrm{log} \text{ }    {\mathrm{log} \big| {Y}^{*} \big|  } \big/ { {\big| {X}^{*} \big| }    }              \cdot     \bigg\{            \frac{1}{\big| Y^{*} \big| }  \cdot  \big\{ Y^* \big\}^{\prime}  \cdot \big| X^* \big|      -    \mathrm{log}   \big| Y^{*} \big|  \cdot \big\{ X^* \big\}^{\prime }                  \bigg\}                                   }  }^{-1}  \\ \\   \times      \bigg\{  -     \sqrt{  \big| X^* \big|^2    \cdot           \frac{   \bigg[   \frac{  \mathrm{log}\big|  {Z}  \big|  \cdot \big| Z \big| }{   \big|  {Y}^{*}  \big|^2  \cdot  Z^{\prime}     }    \bigg]^{-1}     +  \bigg[      \frac{1}{ \big| Y^* \big| \cdot \big\{ Y^* \big\}^{\prime} }  \bigg]^{-1}   }{     \mathrm{log} \text{ }    {\mathrm{log} \big| {Y}^{*} \big|  } \big/ { {\big| {X}^{*} \big| }    }              \cdot     \bigg\{            \frac{1}{\big| Y^{*} \big| }  \cdot  \big\{ Y^* \big\}^{\prime}  \cdot \big| X^* \big|      -    \mathrm{log}   \big| Y^{*} \big|  \cdot \big\{ X^* \big\}^{\prime }                  \bigg\}                                   }  }^{-2} \bigg\} \\ \\ \cdot   \big| X^* \big|^2  \bigg[    \frac{  \bigg\{           2 \frac{\big| Y^* \big| \cdot \big\{ Y^* \big\}^{\prime} \cdot Z^{\prime}}{ \mathrm{log} \big| Z \big| \cdot \big| Z \big|        }  +   \big| Y^* \big| \cdot  \big[ \big\{ Y^* \big\}^{\prime }  \big]^2  + \big| Y^* \big| \cdot  \big\{ Y^* \big\}^{\prime\prime} \bigg\}                                               \cdot \bigg[  \mathrm{log} \text{ }    {\mathrm{log} \big| {Y}^{*} \big|  } \big/ { {\big| {X}^{*} \big| }    }              \cdot     \bigg\{            \frac{1}{\big| Y^{*} \big| }  \cdot  \big\{ Y^* \big\}^{\prime}  \cdot \big| X^* \big|      -    \mathrm{log}   \big| Y^{*} \big|  \cdot \big\{ X^* \big\}^{\prime }                  \bigg\}                            \bigg]}{\bigg[  \mathrm{log} \text{ }    {\mathrm{log} \big| {Y}^{*} \big|  } \big/ { {\big| {X}^{*} \big| }    }              \cdot     \bigg\{            \frac{1}{\big| Y^{*} \big| }  \cdot  \big\{ Y^* \big\}^{\prime}  \cdot \big| X^* \big|      -    \mathrm{log}   \big| Y^{*} \big|  \cdot \big\{ X^* \big\}^{\prime }                  \bigg\}                            \bigg]^2 }       \\ 
             -     \frac{   \bigg\{    \bigg[   \frac{  \mathrm{log}\big|  {Z}  \big|  \cdot \big| Z \big| }{   \big|  {Y}^{*}  \big|^2  \cdot  Z^{\prime}     }    \bigg]^{-1}    +   \bigg[      \frac{1}{ \big| Y^* \big| \cdot \big\{ Y^* \big\}^{\prime} }  \bigg]^{-1}            \bigg\}  \cdot            \bigg[     \bigg\{   \frac{ \big\{ Y^* \big\}^{\prime} \big/ \big| X^* \big|} {\mathrm{log \big| Y^*  \big| \big/ \big|  X^* \big| }} \cdot \frac{1}{ \big| Y^* \big| } \cdot \big\{ Y^* \big\}^{\prime} \cdot \big| X^* \big|      -   \mathrm{log} \text{ }    {\mathrm{log} \big| {Y}^{*} \big|  } \big/ { {\big| {X}^{*} \big| }    }    \cdot  \frac{1}{ \big| Y^* \big|^2 } \cdot \big[ \big\{      Y^*     \big\}^{\prime}  \big]^2     \cdot \big| X^* \big|       \bigg\}                    \bigg]       }{{\bigg[  \mathrm{log} \text{ }    {\mathrm{log} \big| {Y}^{*} \big|  } \big/ { {\big| {X}^{*} \big| }    }              \cdot     \bigg\{            \frac{1}{\big| Y^{*} \big| }  \cdot  \big\{ Y^* \big\}^{\prime}  \cdot \big| X^* \big|      -    \mathrm{log}   \big| Y^{*} \big|  \cdot \big\{ X^* \big\}^{\prime }                  \bigg\}                            \bigg]^2 } }  \\ \\       -     \frac{   \bigg\{    \bigg[   \frac{  \mathrm{log}\big|  {Z}  \big|  \cdot \big| Z \big| }{   \big|  {Y}^{*}  \big|^2  \cdot  Z^{\prime}     }    \bigg]^{-1}    +   \bigg[      \frac{1}{ \big| Y^* \big| \cdot \big\{ Y^* \big\}^{\prime} }  \bigg]^{-1}            \bigg\}  \cdot            \bigg[         \big| Y^* \big|^{-1} \cdot \big\{ Y^* \big\}^{\prime} \cdot \big\{ X^* \big\}^{\prime }                              \bigg]       }{{\bigg[  \mathrm{log} \text{ }    {\mathrm{log} \big| {Y}^{*} \big|  } \big/ { {\big| {X}^{*} \big| }    }              \cdot     \bigg\{            \frac{1}{\big| Y^{*} \big| }  \cdot  \big\{ Y^* \big\}^{\prime}  \cdot \big| X^* \big|      -    \mathrm{log}   \big| Y^{*} \big|  \cdot \big\{ X^* \big\}^{\prime }                  \bigg\}                            \bigg]^2 } }          \bigg]                     \\   \\  \times     \big| X^* \big|     , \\ \\      \mathcal{T}^{\prime}_2 \equiv   \frac{\partial \mathcal{T}_2}{\partial Y^* }     .      
\end{array}\right. 
\]  }

%   
% \end{array}\right. 
% \]  } 

% {\small 
% \[   \left\{\!\begin{array}{ll@{}>{{}}l} 

\noindent Denoting,

{\small \begin{align*}
  \underline{\mathcal{T}_{2,1}} \equiv  -     \sqrt{  \big| X^* \big|^2    \cdot           \frac{   \bigg[   \frac{  \mathrm{log}\big|  {Z}  \big|  \cdot \big| Z \big| }{   \big|  {Y}^{*}  \big|^2  \cdot  Z^{\prime}     }    \bigg]^{-1}     +  \bigg[      \frac{1}{ \big| Y^* \big| \cdot \big\{ Y^* \big\}^{\prime} }  \bigg]^{-1}   }{     \mathrm{log} \text{ }    {\mathrm{log} \big| {Y}^{*} \big|  } \big/ { {\big| {X}^{*} \big| }    }              \cdot     \bigg\{            \frac{1}{\big| Y^{*} \big| }  \cdot  \big\{ Y^* \big\}^{\prime}  \cdot \big| X^* \big|      -    \mathrm{log}   \big| Y^{*} \big|  \cdot \big\{ X^* \big\}^{\prime }                  \bigg\}                                   }  }^{-2}  , \end{align*}

  \begin{align*} {\tiny \underline{\mathcal{T}_{2,2}}  \equiv          \big| X^* \big|^2  \cdot    \frac{  \bigg\{           2 \frac{\big| Y^* \big| \cdot \big\{ Y^* \big\}^{\prime} \cdot Z^{\prime}}{ \mathrm{log} \big| Z \big| \cdot \big| Z \big|        }  +   \big| Y^* \big| \cdot  \big[ \big\{ Y^* \big\}^{\prime }  \big]^2  + \big| Y^* \big| \cdot  \big\{ Y^* \big\}^{\prime\prime} \bigg\}                                               \cdot \bigg[  \mathrm{log} \text{ }    {\mathrm{log} \big| {Y}^{*} \big|  } \big/ { {\big| {X}^{*} \big| }    }              \cdot     \bigg\{            \frac{1}{\big| Y^{*} \big| }  \cdot  \big\{ Y^* \big\}^{\prime}  \cdot \big| X^* \big|      -    \mathrm{log}   \big| Y^{*} \big|  \cdot \big\{ X^* \big\}^{\prime }                  \bigg\}                            \bigg]}{\bigg[  \mathrm{log} \text{ }    {\mathrm{log} \big| {Y}^{*} \big|  } \big/ { {\big| {X}^{*} \big| }    }              \cdot     \bigg\{            \frac{1}{\big| Y^{*} \big| }  \cdot  \big\{ Y^* \big\}^{\prime}  \cdot \big| X^* \big|      -    \mathrm{log}   \big| Y^{*} \big|  \cdot \big\{ X^* \big\}^{\prime }                  \bigg\}                            \bigg]^2 }      }    ,  \end{align*}

  \begin{align*}   \underline{\mathcal{T}_{2,3}}  \equiv     {\tiny \big| X^* \big|^2  \cdot       \frac{   \bigg\{    \bigg[   \frac{  \mathrm{log}\big|  {Z}  \big|  \cdot \big| Z \big| }{   \big|  {Y}^{*}  \big|^2  \cdot  Z^{\prime}     }    \bigg]^{-1}    +   \bigg[      \frac{1}{ \big| Y^* \big| \cdot \big\{ Y^* \big\}^{\prime} }  \bigg]^{-1}            \bigg\}  \cdot               \bigg\{   \frac{ \big\{ Y^* \big\}^{\prime} \big/ \big| X^* \big|} {\mathrm{log \big| Y^*  \big| \big/ \big|  X^* \big| }} \cdot \frac{1}{ \big| Y^* \big| } \cdot \big\{ Y^* \big\}^{\prime} \cdot \big| X^* \big|  \bigg\}    }{{\bigg[  \mathrm{log} \text{ }    {\mathrm{log} \big| {Y}^{*} \big|  } \big/ { {\big| {X}^{*} \big| }    }              \cdot     \bigg\{            \frac{1}{\big| Y^{*} \big| }  \cdot  \big\{ Y^* \big\}^{\prime}  \cdot \big| X^* \big|      -    \mathrm{log}   \big| Y^{*} \big|  \cdot \big\{ X^* \big\}^{\prime }                  \bigg\}                            \bigg]^2 } }       } \end{align*}

  \begin{align*} {\tiny - \big| X^* \big|^2 \cdot       \frac{  \bigg\{    \bigg[   \frac{  \mathrm{log}\big|  {Z}  \big|  \cdot \big| Z \big| }{   \big|  {Y}^{*}  \big|^2  \cdot  Z^{\prime}     }    \bigg]^{-1}    +   \bigg[      \frac{1}{ \big| Y^* \big| \cdot \big\{ Y^* \big\}^{\prime} }  \bigg]^{-1}            \bigg\}  \cdot \bigg\{ \mathrm{log} \text{ }    {\mathrm{log} \big| {Y}^{*} \big|  } \big/ { {\big| {X}^{*} \big| }    }    \cdot  \frac{1}{ \big| Y^* \big|^2 } \cdot \big[ \big\{      Y^*     \big\}^{\prime}  \big]^2     \cdot \big| X^* \big|       \bigg\}  }{{\bigg[  \mathrm{log} \text{ }    {\mathrm{log} \big| {Y}^{*} \big|  } \big/ { {\big| {X}^{*} \big| }    }              \cdot     \bigg\{            \frac{1}{\big| Y^{*} \big| }  \cdot  \big\{ Y^* \big\}^{\prime}  \cdot \big| X^* \big|      -    \mathrm{log}   \big| Y^{*} \big|  \cdot \big\{ X^* \big\}^{\prime }                  \bigg\}                            \bigg]^2 } }                     }          , \end{align*}

  \begin{align*}    \underline{\mathcal{T}_{2,4} }  \equiv {\tiny           - \big| X^* \big|^2  \cdot          \frac{   \bigg\{    \bigg[   \frac{  \mathrm{log}\big|  {Z}  \big|  \cdot \big| Z \big| }{   \big|  {Y}^{*}  \big|^2  \cdot  Z^{\prime}     }    \bigg]^{-1}    +   \bigg[      \frac{1}{ \big| Y^* \big| \cdot \big\{ Y^* \big\}^{\prime} }  \bigg]^{-1}            \bigg\}  \cdot            \bigg[         \big| Y^* \big|^{-1} \cdot \big\{ Y^* \big\}^{\prime} \cdot \big\{ X^* \big\}^{\prime }                              \bigg]       }{{\bigg[  \mathrm{log} \text{ }    {\mathrm{log} \big| {Y}^{*} \big|  } \big/ { {\big| {X}^{*} \big| }    }              \cdot     \bigg\{            \frac{1}{\big| Y^{*} \big| }  \cdot  \big\{ Y^* \big\}^{\prime}  \cdot \big| X^* \big|      -    \mathrm{log}   \big| Y^{*} \big|  \cdot \big\{ X^* \big\}^{\prime }                  \bigg\}                            \bigg]^2 } }            }   , 
\end{align*} }

\noindent further rearrangements to demonstrate that conditions such as,

{\small

\begin{align*}
\bigg\{ \mathcal{T}_1  \mathcal{T}_2 \mathcal{T}_3                 - \mathrm{log} \big[ \mathcal{T}_1 \big]   \big\{ X^* \big\}^{\prime} < + \infty \bigg\} \Longrightarrow  \end{align*}

\begin{align*} \bigg\{  \bigg(   \bigg\{              \mathrm{log}   \bigg(   {\sqrt{  \big| X^* \big|^2    \cdot           \frac{   \bigg[   \frac{  \mathrm{log}\big|  {Z}  \big|  \cdot \big| Z \big| }{   \big|  {Y}^{*}  \big|^2  \cdot  Z^{\prime}     }    \bigg]^{-1}     +  \bigg[      \frac{1}{ \big| Y^* \big| \cdot \big\{ Y^* \big\}^{\prime} }  \bigg]^{-1}   }{     \mathrm{log} \text{ }   {\mathrm{log} \big| {Y}^{*} \big|  } \big/ { {\big| {X}^{*} \big| }    }              \cdot     \bigg\{            \frac{1}{\big| Y^{*} \big| }  \cdot  \big\{ Y^* \big\}^{\prime}  \cdot \big| X^* \big|      -    \mathrm{log}   \big| Y^{*} \big|  \cdot \big\{ X^* \big\}^{\prime }                  \bigg\}                                   }  }} \times {\big| X^* \big|^{-1}}\bigg)                \bigg\}^{-1} \end{align*}

\begin{align*}  \times    \bigg[ \mathcal{T}_1  \mathcal{T}_2 \mathcal{T}_3                 - \mathrm{log} \big[ \mathcal{T}_1 \big]   \big\{ X^* \big\}^{\prime}                  \bigg]      \bigg)   < + \infty \bigg\} ,
\end{align*}

}

\noindent imply $r^{\prime\prime}_1 < + \infty$ observe, from $(1)$-$(11)$ below,

{\small

\begin{align*}
\big|    \big\{  \big( r^{\prime\prime}_1 \big) -1 \big\}     \big| = \mathrm{log}  \text{ }  \mathrm{log}      {\sqrt{              \frac{     \bigg|    \bigg[   \frac{  \mathrm{log}\big|  {Z}  \big|  \cdot \big| Z \big| }{   \big|  {Y}^{*}  \big|^2  \cdot  Z^{\prime}     }    \bigg]^{-1}     +  \bigg[      \frac{1}{ \big| Y^* \big| \cdot \big\{ Y^* \big\}^{\prime} }  \bigg]^{-1}   \bigg|   }{    \bigg|    \mathrm{log} \text{ }   {\mathrm{log} \big| {Y}^{*} \big|  } \big/ { {\big| {X}^{*} \big| }    }              \cdot     \bigg\{            \frac{1}{\big| Y^{*} \big| }  \cdot  \big\{ Y^* \big\}^{\prime}  \cdot \big| X^* \big|      -    \mathrm{log}   \big| Y^{*} \big|  \cdot \big\{ X^* \big\}^{\prime }                  \bigg\}                   \bigg|       }  }}  
\end{align*}

\begin{align*}
\\ \overset{(1)}{=}  \mathrm{log}  \text{ }  \mathrm{log}      {\sqrt{              \frac{     \bigg|    \bigg[   \frac{  \mathrm{log}\big|  {Z}  \big|  \cdot \big| Z \big| }{   \big|  {Y}^{*}  \big|^2  \cdot  Z^{\prime}     }    \bigg]^{-1}     +  \bigg[      \frac{1}{ \big| Y^* \big| \cdot \big\{ Y^* \big\}^{\prime} }  \bigg]^{-1}   \bigg|   }{     \mathrm{log} \text{ }   {\mathrm{log} \big| {Y}^{*} \big|  } \big/ { {\big| {X}^{*} \big| }    }              \cdot     \bigg\{            \frac{1}{\big| Y^{*} \big| }  \cdot  \big\{ Y^* \big\}^{\prime}  \cdot \big| X^* \big|      -    \mathrm{log}   \big| Y^{*} \big|  \cdot \big\{ X^* \big\}^{\prime }                  \bigg\}                       }  }}  
\end{align*}

\begin{align*} \\ \overset{(2)}{\equiv}   \mathrm{log}  \text{ }  \mathrm{log}      {\sqrt{              \frac{   \bigg[   \frac{  \mathrm{log}\big|  {Z}  \big|  \cdot \big| Z \big| }{   \big|  {Y}^{*}  \big|^2  \cdot  Z^{\prime}     }    \bigg]^{-1}   +     \bigg[      \frac{1}{ \big| Y^* \big| \cdot \big\{ Y^* \big\}^{\prime} }  \bigg]^{-1}   }{     \mathrm{log} \text{ }   {\mathrm{log} \big| {Y}^{*} \big|  } \big/ { {\big| {X}^{*} \big| }    }              \cdot     \bigg\{            \frac{1}{\big| Y^{*} \big| }  \cdot  \big\{ Y^* \big\}^{\prime}  \cdot \big| X^* \big|      -    \mathrm{log}   \big| Y^{*} \big|  \cdot \big\{ X^* \big\}^{\prime }                  \bigg\}                       }  }}  \end{align*}

\begin{align*}
   \\  \overset{(3)}{=}    \mathrm{log}  \text{ }  \mathrm{log}      {              \frac{ \sqrt{  \bigg[   \frac{  \mathrm{log}\big|  {Z}  \big|  \cdot \big| Z \big| }{   \big|  {Y}^{*}  \big|^2  \cdot  Z^{\prime}     }    \bigg]^{-1}   +   \bigg[      \frac{1}{ \big| Y^* \big| \cdot \big\{ Y^* \big\}^{\prime} }  \bigg]^{-1}  } }{ \sqrt{    \mathrm{log} \text{ }   {\mathrm{log} \big| {Y}^{*} \big|  } \big/ { {\big| {X}^{*} \big| }    }              \cdot     \bigg\{            \frac{1}{\big| Y^{*} \big| }  \cdot  \big\{ Y^* \big\}^{\prime}  \cdot \big| X^* \big|      -    \mathrm{log}   \big| Y^{*} \big|  \cdot \big\{ X^* \big\}^{\prime }                  \bigg\}                       }  }}   
\end{align*}

  \begin{align*} \\ \overset{(4)}{=}       \mathrm{log} \bigg\{    \mathrm{log}       {\sqrt{              {   \bigg[   \frac{  \mathrm{log}\big|  {Z}  \big|  \cdot \big| Z \big| }{   \big|  {Y}^{*}  \big|^2  \cdot  Z^{\prime}     }    \bigg]^{-1}     +   \bigg[      \frac{1}{ \big| Y^* \big| \cdot \big\{ Y^* \big\}^{\prime} }  \bigg]^{-1}   } }}      \end{align*}
  
  \begin{align*}  -      \mathrm{log}     {\sqrt{              {     \mathrm{log} \text{ }   {\mathrm{log} \big| {Y}^{*} \big|  } \big/ { {\big| {X}^{*} \big| }    }              \cdot     \bigg\{            \frac{1}{\big| Y^{*} \big| }  \cdot  \big\{ Y^* \big\}^{\prime}  \cdot \big| X^* \big|      -    \mathrm{log}   \big| Y^{*} \big|  \cdot \big\{ X^* \big\}^{\prime }                  \bigg\}                       }  }}   \bigg\}    \end{align*}
  
  \begin{align*}            \\  \overset{(5)}{\geq }              \mathrm{log} \bigg\{             C_1 \cdot   \bigg[  \mathrm{log}            \sqrt{                 \bigg[   \frac{  \mathrm{log}\big|  {Z}  \big|  \cdot \big| Z \big| }{   \big|  {Y}^{*}  \big|^2  \cdot  Z^{\prime}     }    \bigg]^{-1}  }         +     \mathrm{log}         \sqrt{\bigg[      \frac{1}{ \big| Y^* \big| \cdot \big\{ Y^* \big\}^{\prime} }  \bigg]^{-1}  }               \bigg]          -  C_2 \cdot          \bigg[     \mathrm{log} \bigg(            \sqrt{                   \mathrm{log} \text{ }   {\mathrm{log} \big| {Y}^{*} \big|  } \big/ { {\big| {X}^{*} \big| }    }            } \end{align*}

     \begin{align*}    \cdot   \sqrt{  \bigg\{            \frac{1}{\big| Y^{*} \big| }  \cdot  \big\{ Y^* \big\}^{\prime}  \cdot \big| X^* \big|      -    \mathrm{log}   \big| Y^{*} \big|  \cdot \big\{ X^* \big\}^{\prime }                  \bigg\}                         } \bigg)                                   \bigg]         \bigg\}            
\end{align*}

\begin{align*}
  \\  \overset{(6)}{=}       \mathrm{log} \bigg\{             C_1 \cdot   \bigg[  \mathrm{log}                              \frac{ \sqrt{  \big\{  \mathrm{log}\big|  {Z}  \big|  \cdot \big| Z \big| \big\}^{-1 } } }{  \sqrt{ \big\{ \big|  {Y}^{*}  \big|^2  \cdot  Z^{\prime} \big\}^{-1}   }    }               +     \mathrm{log}              \frac{1}{ \sqrt{ \big\{ \big| Y^* \big| \cdot \big\{ Y^* \big\}^{\prime} }  \big\}^{-1} }                \bigg]          -  C_2 \cdot          \bigg[     \mathrm{log}           \sqrt{                   \mathrm{log} \text{ }   {\mathrm{log} \big| {Y}^{*} \big|  } \big/ { {\big| {X}^{*} \big| }    }            }   \end{align*}

     \begin{align*}   \cdot \mathrm{log}  \sqrt{  \bigg\{            \frac{1}{\big| Y^{*} \big| }  \cdot  \big\{ Y^* \big\}^{\prime}  \cdot \big| X^* \big|      -    \mathrm{log}   \big| Y^{*} \big|  \cdot \big\{ X^* \big\}^{\prime }                  \bigg\}                         }                                   \bigg]         \bigg\}  
\end{align*}

\begin{align*}
  \\  \overset{(7)}{=}       \mathrm{log} \bigg\{             C_1 \cdot   \bigg[  \mathrm{log}                              \sqrt{  \big\{  \mathrm{log}\big|  {Z}  \big|  \cdot \big| Z \big| \big\}^{-1 } }  - \mathrm{log}   \sqrt{ \big\{ \big|  {Y}^{*}  \big|^2  \cdot  Z^{\prime} \big\}^{-1}   }                -      \mathrm{log}              \sqrt{ \big\{ \big| Y^* \big| \cdot \big\{ Y^* \big\}^{\prime}   \big\}^{-1} }                       \bigg]          -  C_2 \end{align*}

     \begin{align*}  \cdot          \bigg[     \mathrm{log}           \sqrt{                   \mathrm{log} \text{ }   {\mathrm{log} \big| {Y}^{*} \big|  } \big/ { {\big| {X}^{*} \big| }    }            }      \cdot \mathrm{log}  \sqrt{  \bigg\{            \frac{1}{\big| Y^{*} \big| }  \cdot  \big\{ Y^* \big\}^{\prime}  \cdot \big| X^* \big|      -    \mathrm{log}   \big| Y^{*} \big|  \cdot \big\{ X^* \big\}^{\prime }                  \bigg\}                         }                                   \bigg]         \bigg\}  
\end{align*}

 \begin{align*}
    \\  \overset{(8)}{=}      \mathrm{log} \bigg\{             C_1 C^{\prime}_1 \cdot   \bigg[  \mathrm{log}      \bigg\{                         \sqrt{  \big\{  \mathrm{log}\big|  {Z}  \big| \big\}^{-1} } \cdot \sqrt{ \big\{    \big| Z \big| \big\}^{-1 } } \bigg\}   - \mathrm{log}   \bigg\{ \sqrt{ \big\{ \big|  {Y}^{*}  \big|^2 \big\}^{-1} }  \cdot  \sqrt{ \big\{ Z^{\prime} \big\}^{-1}   }   \bigg\}              -      \mathrm{log}            \bigg\{   \sqrt{ \big\{ \big| Y^* \big|^{-1} } \end{align*}

     \begin{align*} \cdot \sqrt{\big\{ Y^* \big\}^{\prime}   \big\}^{-1} }  \bigg\}                                                 \bigg]         -  C_2  \cdot          \bigg[     \mathrm{log}           \sqrt{                   \mathrm{log} \text{ }   {\mathrm{log} \big| {Y}^{*} \big|  } \big/ { {\big| {X}^{*} \big| }    }            }   \end{align*}

     \begin{align*}    \cdot \mathrm{log}  \sqrt{  \bigg\{            \frac{1}{\big| Y^{*} \big| }  \cdot  \big\{ Y^* \big\}^{\prime}  \cdot \big| X^* \big|      -    \mathrm{log}   \big| Y^{*} \big|  \cdot \big\{ X^* \big\}^{\prime }                  \bigg\}                         }                                   \bigg]         \bigg\}   
 \end{align*}

}

 \begin{align*} 
 \\   \overset{(9)}{=}       \mathrm{log} \bigg\{             C_1 C^{\prime}_1 \cdot   \bigg[  \mathrm{log}                            \sqrt{  \big\{  \mathrm{log}\big|  {Z}  \big| \big\}^{-1} } + \mathrm{log}  \sqrt{ \big\{    \big| Z \big| \big\}^{-1 } }    - \mathrm{log}   \sqrt{ \big\{ \big|  {Y}^{*}  \big|^2 \big\}^{-1} }   - \mathrm{log}    \sqrt{ \big\{ Z^{\prime} \big\}^{-1}   }                 -      \mathrm{log}              \sqrt{ \big\{ \big| Y^* \big|^{-1} } \end{align*}

     \begin{align*}  -      \mathrm{log}   \sqrt{\big\{ Y^* \big\}^{\prime}   \big\}^{-1} }            \bigg]         -  C_2  \cdot          \bigg[     \mathrm{log}           \sqrt{                   \mathrm{log} \text{ }   {\mathrm{log} \big| {Y}^{*} \big|  } \big/ { {\big| {X}^{*} \big| }    }            }   \end{align*}

     \begin{align*}    \cdot \mathrm{log}  \sqrt{  \bigg\{            \frac{1}{\big| Y^{*} \big| }  \cdot  \big\{ Y^* \big\}^{\prime}  \cdot \big| X^* \big|      -    \mathrm{log}   \big| Y^{*} \big|  \cdot \big\{ X^* \big\}^{\prime }                  \bigg\}                         }                                   \bigg]         \bigg\}        
  \end{align*}

 \begin{align*} 
 \\  \overset{(10)}{\geq }     \mathrm{log} \bigg\{             C_1 C^{\prime}_1 \cdot   \bigg[  \mathrm{log}                            \sqrt{  \big\{  \mathrm{log}\big|  {Z}  \big| \big\}^{-1} } + \mathrm{log}  \sqrt{ \big\{    \big| Z \big| \big\}^{-1 } }    - \mathrm{log}   \sqrt{ \big\{ \big|  {Y}^{*}  \big|^2 \big\}^{-1} }   - \mathrm{log}    \sqrt{ \big\{ Z^{\prime} \big\}^{-1}   }                 -      \mathrm{log}              \sqrt{ \big\{ \big| Y^* \big|^{-1} } \end{align*}

     \begin{align*}  -      \mathrm{log}   \sqrt{\big\{ Y^* \big\}^{\prime}   \big\}^{-1} }            \bigg]         -  C_2  \cdot          \bigg[     \mathrm{log}           \sqrt{                   \mathrm{log} \bigg\{   {\mathrm{log} \big| {Y}^{*} \big|  }    -  \mathrm{log} \big| {X}^{*} \big|    \bigg\}       }   \end{align*}

     \begin{align*} \times C^{\prime}_2  \cdot \mathrm{log} \bigg\{  \sqrt{            \frac{1}{\big| Y^{*} \big| }}   \cdot  \sqrt{\big\{ Y^* \big\}^{\prime}}  \cdot \sqrt{\big| X^* \big| }     -    \sqrt{\mathrm{log}   \big| Y^{*} \big|}  \cdot \sqrt{\big\{ X^* \big\}^{\prime }}            \bigg\}                                                                   \bigg]         \bigg\}        
        \end{align*}

 \begin{align*} 
 \\   \overset{(11)}{=}       \mathrm{log} \bigg\{             C_1 C^{\prime}_1 \cdot   \bigg[  \mathrm{log}                            \sqrt{  \big\{  \mathrm{log}\big|  {Z}  \big| \big\}^{-1} } + \mathrm{log}  \sqrt{ \big\{    \big| Z \big| \big\}^{-1 } }    - \mathrm{log}   \sqrt{ \big\{ \big|  {Y}^{*}  \big|^2 \big\}^{-1} }   - \mathrm{log}    \sqrt{ \big\{ Z^{\prime} \big\}^{-1}   }                 -      \mathrm{log}              \sqrt{  \big| Y^* \big|^{-1} } \end{align*}

     \begin{align*}  -      \mathrm{log}   \sqrt{ \big\{ \big\{ Y^* \big\}^{\prime}   \big\}^{-1} }            \bigg]         -  C_2  \cdot          \bigg[     \mathrm{log}           \sqrt{                   \mathrm{log} \bigg\{   {\mathrm{log} \big| {Y}^{*} \big|  }    -  \mathrm{log} \big| {X}^{*} \big|    \bigg\}       }   \end{align*}

     \begin{align*} \times C^{\prime}_2  \cdot \bigg[  \mathrm{log}   \sqrt{            \frac{1}{\big| Y^{*} \big| }}  + \mathrm{log}  \sqrt{\big\{ Y^* \big\}^{\prime}}  + \mathrm{log}  \sqrt{\big| X^* \big| }     -    \mathrm{log}  \sqrt{\mathrm{log}   \big| Y^{*} \big|}  -  \mathrm{log}  \sqrt{\big\{ X^* \big\}^{\prime }}        \bigg]                                                                   \bigg]         \bigg\}  .       
        \end{align*}

\noindent The above expression for $ \big\{  \big( r^{\prime\prime}_1 \big) -1 \big\} $, from the fact that,

\begin{align*}
   \big\{  \big( r^{\prime\prime}_1 \big) -1 \big\}  =  \big\{  \big( r^{\prime\prime}_1 \big) -2 \big\} =  \big\{  \big( r^{\prime\prime}_1 \big) -3 \big\} =  \big\{  \big( r^{\prime\prime}_1 \big) -4 \big\}    , 
\end{align*}

\noindent implies,

\begin{align*}
\bigg(   \bigg\{  \big\{  \big( r^{\prime\prime}_1 \big) -1 \big\}  \cdot \big| X^* \big|^2 \bigg\}^{-2}     \times  \bigg\{     \big\{  \big( r^{\prime\prime}_1 \big) -1 \big\}^{-1} \cdot     \bigg[ \mathcal{T}_1  \mathcal{T}_2 \mathcal{T}_3                 - \mathrm{log} \big[ \mathcal{T}_1 \big]   \big\{ X^* \big\}^{\prime}                  \bigg]   \cdot \big| X^* \big|^2 +      \big\{  \big( r^{\prime\prime}_1 \big) -1 \big\}  \end{align*}
  
  \begin{align*} \cdot 2 \big| X^* \big| \cdot \big\{ X^* \big\}^{\prime} \bigg\}        \cdot \bigg\{    \mathcal{T}_1  \mathcal{T}_2 \mathcal{T}_3                 - \mathrm{log} \big[ \mathcal{T}_1 \big]   \big\{ X^* \big\}^{\prime}                                                                  \bigg\}    +    \bigg\{  \big\{  \big( r^{\prime\prime}_1 \big) -1 \big\}  \cdot \big| X^* \big|^2 \bigg\}^{-1}           
\end{align*}

  \begin{align*} 
  \times  \bigg\{    \frac{- \mathcal{T}^2_2 \mathcal{T}_3 +   \mathcal{T}^2_1 \mathcal{T}^{\prime}_2 \mathcal{T}_3 + \mathcal{T}^2_1 \mathcal{T}_2 \big\{ X^* \big\}^{\prime} -  \mathcal{T}^2_1 \mathcal{T}_2 \big\{ X^* \big\}^{\prime} +  \mathcal{T}_1   \mathrm{log} \big[ \mathcal{T}_1 \big] \big\{ X^* \big\}^{\prime\prime}}{\mathcal{T}_1 }                        \bigg\}         \bigg) \longrightarrow \text{Constant} < + \infty ,   
\end{align*}

\noindent given the choice of alphabets for which,

{\small 
\[   \left\{\!\begin{array}{ll@{}>{{}}l} 
    \bigg\{ \mathrm{log}              \sqrt{ \mathrm{log}  \big| Z \big|^{-1} }    \text{ } \textit{does not diverge}       \bigg\}   \Longleftrightarrow \bigg\{  \mathrm{log}  \big| Z \big|^{-1}  \text{ }    \textit{is not taken sufficiently small}    \bigg\}   \Longleftrightarrow \big\{  \mathrm{log}  \big| Z \big|^{-1} > 1   \big\}                                 , \\ \\  \bigg\{ \mathrm{log}              \sqrt{   \big| Z \big|^{-1} }    \text{ } \textit{does not diverge}       \bigg\}   \Longleftrightarrow \bigg\{  \big| Z \big|^{-1}  \text{ }    \textit{is not taken sufficiently small}    \bigg\}   \Longleftrightarrow \big\{  \big| Z \big|^{-1} > 1   \big\}                  ,  \\ \\ \bigg\{    \mathrm{log}              \sqrt{  \big\{ \big| Y^* \big|^2 \big\}^{-1}  }     \text{ } \textit{does not diverge}          \bigg\}   \Longleftrightarrow \bigg\{ \big\{ \big| Y^* \big|^2 \big\}^{-1} \text{ }    \text{ }    \textit{is not taken sufficiently small}      \bigg\}   \Longleftrightarrow \big\{      \big\{ \big| Y^* \big|^2 \big\}^{-1} > 1            \big\}   ,  
\end{array}\right. 
\]  }    

{\small 
\[   \left\{\!\begin{array}{ll@{}>{{}}l}     \bigg\{ \mathrm{log}              \sqrt{  \big\{  Z^{\prime} \big\}^{-1} }    \text{ } \textit{does not diverge}       \bigg\}   \Longleftrightarrow \bigg\{ \big\{  Z^{\prime} \big\}^{-1}  \text{ }    \textit{is not taken sufficiently small}    \bigg\}   \Longleftrightarrow \big\{ \big\{  Z^{\prime} \big\}^{-1}  > 1   \big\}                   ,  \\ \\    \bigg\{    \mathrm{log}              \sqrt{  \big| Y^* \big|^{-1} }     \text{ } \textit{does not diverge}          \bigg\}   \Longleftrightarrow \bigg\{  \big| Y^* \big|^{-1} \text{ }    \text{ }    \textit{is not taken sufficiently small}      \bigg\}   \Longleftrightarrow \big\{       \big| Y^* \big|^{-1} > 1            \big\}                  ,  \\ \\ 
        \bigg\{  \mathrm{log}   \sqrt{ \big\{ \big\{ Y^* \big\}^{\prime}   \big\}^{-1} }            \text{ } \textit{does not diverge}    \bigg\}    \Longleftrightarrow \bigg\{ \big\{ Y^* \big\}^{\prime}        \text{ } \textit{is not taken sufficiently small} \bigg\}  \Longleftrightarrow \big\{ \big\{ \big\{ Y^* \big\}^{\prime} \big\}^{-1} > 1  \big\}                   ,    \\ \\ \bigg\{ \mathrm{log}           \sqrt{                   \mathrm{log} \bigg\{   {\mathrm{log} \big| {Y}^{*} \big|  }    -  \mathrm{log} \big| {X}^{*} \big|    \bigg\}       }  \text{ } \textit{does not diverge}    \bigg\} \Longleftrightarrow \bigg\{   \big| Y^* \big| , \big| X^* \big| \text{ } \textit{are not taken sufficiently small}          \bigg\}   \\ \Longleftrightarrow \bigg\{   \sqrt{                   \mathrm{log} \bigg\{   {\mathrm{log} \big| {Y}^{*} \big|  }    -  \mathrm{log} \big| {X}^{*} \big|    \bigg\}       } > 0  \bigg\}                   , \\   \\     \bigg\{   \mathrm{log}   \sqrt{            \frac{1}{\big| Y^{*} \big| }} \text{ } \textit{does not diverge}  \bigg\} \Longleftrightarrow \bigg\{ \big| Y^* \big| \text{ } \textit{is not taken sufficiently large}  \bigg\}   \Longleftrightarrow \big\{  {\big| Y^* \big|^{-1}} > 0  \big\}                   , \\ \\     \bigg\{  \mathrm{log}  \sqrt{\big\{ Y^* \big\}^{\prime }}  \text{ } \textit{does not diverge}  \bigg\} \Longleftrightarrow \bigg\{ \big\{ Y^* \big\}^{\prime } \text{ } \textit{is not taken sufficiently small}     \bigg\} \Longleftrightarrow \big\{ \big| \big\{ Y^* \big\}^{\prime } \big| > 1     \big\}       , \\ \\       \bigg\{     \mathrm{log}  \sqrt{\big| X^* \big|  } \text{ } \textit{does not diverge}  \bigg\}    \Longleftrightarrow \bigg\{ \big|  X^* \big| \text{ } \textit{is not taken sufficiently small}     \bigg\} \Longleftrightarrow \big\{ \big| X^*  \big| > 1     \big\}    , \\ \\         \bigg\{   \mathrm{log}  \sqrt{\mathrm{log}   \big| Y^{*} \big|}  \text{ } \textit{does not diverge}  \bigg\} \Longleftrightarrow \bigg\{ \big|  Y^* \big| \text{ } \textit{is not taken sufficiently small}     \bigg\} \Longleftrightarrow \big\{ \big| Y^*  \big| > 2     \big\}  , \\ \\    \bigg\{  \mathrm{log}  \sqrt{\big\{ X^* \big\}^{\prime }}  \text{ } \textit{does not diverge}  \bigg\} \Longleftrightarrow \bigg\{ \big\{ X^* \big\}^{\prime } \text{ } \textit{is not taken sufficiently small}     \bigg\} \Longleftrightarrow \big\{ \big| \big\{ X^* \big\}^{\prime } \big| > 1     \big\}  .   
\end{array}\right. 
\]  }

\noindent Making use of the above computations yields a similar set of expressions for which $r^{\prime\prime}_2 , r^{\prime\prime}_3, r^{\prime\prime}_4$ are convergent, from which we conclude the argument. \boxed{}

\bigskip

\noindent \textbf{Corollary} (\textit{the set of degeneracies for the second derivative of the converse bit transmission rate}). One has that,

{\small  
\[    \mathscr{D} {\tiny \big( r^{\prime\prime } \big)} \equiv  \left\{\!\begin{array}{ll@{}>{{}}l} 
       \big\{  X, X^* , Y , Y^* , Z , Z^*     :  r^{\prime\prime}_1 \longrightarrow + \infty   \big\}      , \\ \\    \big\{   X, X^* , Y , Y^* , Z , Z^*   : r^{\prime\prime}_2 \longrightarrow + \infty   \big\}        , \\ \\    \big\{   X, X^* , Y , Y^* , Z , Z^*   : r^{\prime\prime}_3 \longrightarrow + \infty   \big\}      ,    \\ \\    \big\{   X, X^* , Y , Y^* , Z , Z^*   : r^{\prime\prime}_4 \longrightarrow + \infty           \big\}   , 
\end{array}\right.  \\ =   \left\{\!\begin{array}{ll@{}>{{}}l} 
    \bigg\{ \mathrm{log}              \sqrt{ \mathrm{log}  \big| Z \big|^{-1} }    \text{ } \textit{diverges}       \bigg\}    \Longleftrightarrow \big\{  \mathrm{log}  \big| Z \big|^{-1} <  1   \big\}                                 ,   \\ \\  \bigg\{ \mathrm{log}              \sqrt{   \big| Z \big|^{-1} }    \text{ } \textit{diverges}       \bigg\}     \Longleftrightarrow \big\{  \big| Z \big|^{-1} < 1   \big\}                  , \\  \\   \bigg\{    \mathrm{log}              \sqrt{  \big\{ \big| Y^* \big|^2 \big\}^{-1}  }     \text{ } \textit{diverges}          \bigg\}    \Longleftrightarrow \big\{      \big\{ \big| Y^* \big|^2 \big\}^{-1} < 1            \big\}   ,   \\ \\    \bigg\{ \mathrm{log}              \sqrt{  \big\{  Z^{\prime} \big\}^{-1} }    \text{ } \textit{diverges}       \bigg\}     \Longleftrightarrow \big\{ \big\{  Z^{\prime} \big\}^{-1}  < 1   \big\}                   ,    
\end{array}\right.  
\]  }

{\small  
\[    \mathscr{D} {\tiny \big( r^{\prime\prime } \big)} \text{ } \textit{cont.} \equiv  \left\{\!\begin{array}{ll@{}>{{}}l} 
       \big\{  X, X^* , Y , Y^* , Z , Z^*     :  r^{\prime\prime}_1 \longrightarrow + \infty   \big\}      , \\ \\    \big\{   X, X^* , Y , Y^* , Z , Z^*   : r^{\prime\prime}_2 \longrightarrow + \infty   \big\}        , \\ \\    \big\{   X, X^* , Y , Y^* , Z , Z^*   : r^{\prime\prime}_3 \longrightarrow + \infty   \big\}      ,    \\ \\    \big\{   X, X^* , Y , Y^* , Z , Z^*   : r^{\prime\prime}_4 \longrightarrow + \infty           \big\}   , 
\end{array}\right.  \\ =   \left\{\!\begin{array}{ll@{}>{{}}l}  \bigg\{    \mathrm{log}              \sqrt{  \big| Y^* \big|^{-1} }     \text{ } \textit{diverges}          \bigg\}    \Longleftrightarrow \big\{       \big| Y^* \big|^{-1} < 1            \big\}                  ,  \\ \\ \bigg\{  \mathrm{log}   \sqrt{ \big\{ \big\{ Y^* \big\}^{\prime}   \big\}^{-1} }            \text{ } \textit{diverges}    \bigg\}  \Longleftrightarrow \big\{ \big\{ \big\{ Y^* \big\}^{\prime} \big\}^{-1} < 1  \big\}                   ,  \\ \\  \bigg\{ \mathrm{log}           \sqrt{                   \mathrm{log} \bigg\{   {\mathrm{log} \big| {Y}^{*} \big|  }    -  \mathrm{log} \big| {X}^{*} \big|    \bigg\}       }  \text{ } \textit{diverges}    \bigg\}       ,           
 \\  \Longleftrightarrow \bigg\{   \sqrt{                   \mathrm{log} \bigg\{   {\mathrm{log} \big| {Y}^{*} \big|  }    -  \mathrm{log} \big| {X}^{*} \big|    \bigg\}       } \approx  0  \bigg\}                   , \\   \\     \bigg\{   \mathrm{log}   \sqrt{            \frac{1}{\big| Y^{*} \big| }} \text{ } \textit{diverges}  \bigg\}  \Longleftrightarrow \big\{  {\big| Y^* \big|^{-1}} \approx  0  \big\}                   , \\ \\          \bigg\{  \mathrm{log}  \sqrt{\big\{ Y^* \big\}^{\prime }}  \text{ } \textit{diverges}  \bigg\}  \Longleftrightarrow \big\{ \big| \big\{ Y^* \big\}^{\prime } \big| < 1      \big\}       ,  \\ \\      \bigg\{     \mathrm{log}  \sqrt{\big| X^* \big|  } \text{ } \textit{diverges}  \bigg\}  \Longleftrightarrow \big\{ \big| X^*  \big| <  1     \big\}    ,  \\ \\ \bigg\{   \mathrm{log}  \sqrt{\mathrm{log}   \big| Y^{*} \big|}  \text{ } \textit{diverges}  \bigg\}  \Longleftrightarrow \big\{ \big| Y^*  \big| <  2     \big\}  , \\ \\    \bigg\{  \mathrm{log}  \sqrt{\big\{ X^* \big\}^{\prime }}  \text{ } \textit{diverges}  \bigg\} \Longleftrightarrow \big\{ \big| \big\{ X^* \big\}^{\prime } \big| <  1     \big\}  .            
\end{array}\right.  
\]  }

\noindent \textit{Proof of Corollary}. The desired set above follows directly from inspection at which the second derivative of the converse bit transmission rate diverges (given the computations provided in \textbf{Lemma} \textit{3}) from which we conclude the argument. \boxed{}

\bigskip 

\noindent In light of the previous results provided for characterizing the converse bit transmission rate, observe,

 {\small \begin{align*}
 \underset{P_{\textbf{X}}}{\mathrm{sup}}  \big\{ \mathrm{min} \big\{  I \big( \textbf{X} , \textbf{Y} \big)  , \underset{z}{\mathrm{min}}  \big\{ H_Q \big( \textbf{Y} \big| \textbf{Z} = z \big)   -  H_P \big( \textbf{Y} \big| \textbf{X} \big)  \big\}  \big\}  \end{align*}
 
 \begin{align*}    < \mathrm{log} \bigg[ \underset{P_{\textbf{X}}}{\mathrm{sup}}  \big\{ \mathrm{min} \big\{  I \big( \textbf{X} , \textbf{Y} \big)  , \underset{z}{\mathrm{min}}  \big\{ H_Q \big( \textbf{Y} \big| \textbf{Z} = z \big)    -  H_P \big( \textbf{Y} \big| \textbf{X} \big)  \big\}  \big\}    \big\} \bigg] \end{align*}
 
 \begin{align*} \equiv   \underset{P_{\textbf{X}}}{\mathrm{sup}}  \big\{ \mathrm{min} \big\{  \mathrm{log} \big[  I \big( \textbf{X} , \textbf{Y} \big) \big]   ,  \mathrm{log} \big[ \underset{z}{\mathrm{min}}  \big\{ H_Q \big( \textbf{Y} \big| \textbf{Z} = z \big)    -  H_P \big( \textbf{Y} \big| \textbf{X} \big)  \big\}  \big] \big\}   \big\}     \end{align*}
 
 \begin{align*}   \equiv   \underset{P_{\textbf{X}}}{\mathrm{sup}}  \big\{ \mathrm{min} \big\{  \mathrm{log} \big[  I \big( \textbf{X} , \textbf{Y} \big) \big]   ,  \underset{z}{\mathrm{min}}  \big\{  \mathrm{log}  \big[ H_Q \big( \textbf{Y} \big| \textbf{Z} = z \big)    -  H_P \big( \textbf{Y} \big| \textbf{X} \big) \big]  \big\}  \big\}       \big\}  \end{align*}
 
 \begin{align*}      <    \underset{P_{\textbf{X}}}{\mathrm{sup}}  \big\{ \mathrm{min} \big\{  \mathrm{log} \big[  I \big( \textbf{X} , \textbf{Y} \big) \big]   ,  \underset{z}{\mathrm{min}}  \big\{  \mathrm{log}  \big[ H_Q \big( \textbf{Y} \big| \textbf{Z} = z \big)  \big]   -  \mathrm{log} \big[ H_P \big( \textbf{Y} \big| \textbf{X} \big) \big]  \big\}  \big\} \big\}  \end{align*}
 
 \begin{align*}   \overset{(H)}{<}  \underset{P_{\textbf{X}}}{\mathrm{sup}}  \big\{ \mathrm{min} \big\{  \mathrm{log} \big[  I \big( \textbf{X} , \textbf{Y} \big) \big]   ,  \underset{z}{\mathrm{min}}  \big\{  \mathrm{log}  \big[ H_Q \big( \textbf{Y} \big| \textbf{Z} = z \big)  \big]   -  \mathrm{log} \big[ H_P \big( \textbf{Y}^{*} \big| \textbf{X} \big) \big]  \big\}  \big\} \big\} \end{align*}
 
 \begin{align*}   \equiv \underset{P_{\textbf{X}}}{\mathrm{sup}}  \big\{ \mathrm{min} \big\{  \mathrm{log} \big[  I \big( \textbf{X} , \textbf{Y} \big) \big]   ,  \underset{z}{\mathrm{min}}  \big\{  \mathrm{log}  \big[ H_Q \big( \textbf{Y} \big| \textbf{Z} = z \big)  \big] \big\} \big\}  \big\}  - \underset{P_{\textbf{X}}}{\mathrm{sup}}   \big\{   \mathrm{min} \big\{ \mathrm{log} \big[  I \big( \textbf{X} , \textbf{Y} \big) \big]       , \mathrm{log} \big[ H_P \big( \textbf{Y}^{*} \big| \textbf{X} \big) \big]  \big\} \big\}  \end{align*}
 
 \begin{align*}  <  \underset{P_{\textbf{X}}}{\mathrm{sup}}  \big\{ \mathrm{min} \big\{  \mathrm{log} \big[  I \big( \textbf{X} , \textbf{Y} \big) \big]   ,  \underset{z}{\mathrm{min}}  \big\{  \mathrm{log}  \big[ H_Q \big( \textbf{Y}^{*} \big| \textbf{Z} = z \big)  \big] \big\} \big\} \big\}   - \underset{P_{\textbf{X}}}{\mathrm{sup}}   \big\{   \mathrm{min} \big\{ \mathrm{log} \big[  I \big( \textbf{X} , \textbf{Y} \big) \big]       , \mathrm{log} \big[ H_P \big( \textbf{Y}^{*} \big| \textbf{X} \big) \big]  \big\} \big\}    \end{align*}

 \begin{align*}        <  \underset{P_{\textbf{X}}}{\mathrm{sup}}  \big\{ \mathrm{min} \big\{  \mathrm{log} \mathrm{log} \big[  I \big( \textbf{X} , \textbf{Y} \big) \big]   ,  \underset{z}{\mathrm{min}}  \big\{  \mathrm{log}  \big[ H_Q \big( \textbf{Y}^{*} \big| \textbf{Z} = z \big)  \big] \big\} \big\}   - \underset{P_{\textbf{X}}}{\mathrm{sup}} \big\{ \mathrm{min}  \big\{   \mathrm{log} \big[  I \big( \textbf{X} , \textbf{Y} \big) \big]          , \mathrm{log} \big[ H_P \big( \textbf{Y}^{*} \big| \textbf{X} \big) \big]  \big\} \big\}  \end{align*}
 
 \begin{align*}               \equiv   \underset{P_{\textbf{X}}}{\mathrm{sup}}  \big\{ \mathrm{min} \big\{  \mathrm{log} \mathrm{log} \big[  I \big( \textbf{X} , \textbf{Y} \big) \big]   ,  \underset{z}{\mathrm{min}}  \big\{  \mathrm{log}  \big[ H_Q \big( \textbf{Y}^{*} \big| \textbf{Z} = z \big)  \big] \big\} \big\}   - \underset{P_{\textbf{X}}}{\mathrm{sup}} \big\{ \mathrm{min}  \big\{   \mathrm{log} \big[  I \big( \textbf{X} , \textbf{Y} \big) \big]      , \mathrm{log} \big[ H_P \big( \textbf{Y}^{*} \big| \textbf{X} \big) \big]  \big\}   \big\} \end{align*}
 
 \begin{align*}        <   \underset{P_{\textbf{X}}}{\mathrm{sup}}     \bigg\{  \mathrm{min} \bigg\{  \mathrm{log} \bigg[  \frac{\mathrm{log} \big[ I \big( \textbf{X}, \textbf{Y} \big) \big]}{I \big( \textbf{X}, \textbf{Y} \big) }\bigg]     ,    \underset{z}{\mathrm{min}}  \big\{   \mathrm{log}  \big[ H_Q \big( \textbf{Y}^{*} \big| \textbf{Z} = z \big)  \big]   -    \mathrm{log} \big[ H_P \big( \textbf{Y}^{*} \big| \textbf{X} \big) \big] \big\}    \bigg\}       \bigg\}   \end{align*}
 
 \begin{align*}               <    \underset{P_{\textbf{X}}}{\mathrm{sup}}      \bigg\{ \mathrm{min} \bigg\{   \mathrm{log} \bigg[  \frac{\mathrm{log} \big[ I \big( \textbf{X}, \textbf{Y} \big) \big]}{I \big( \textbf{X}, \textbf{Y}^{*} \big) }\bigg]     ,    \underset{z}{\mathrm{min}}  \big\{   \mathrm{log}  \big[ H_Q \big( \textbf{Y}^{*} \big| \textbf{Z} = z \big)  \big]   -    \mathrm{log} \big[ H_P \big( \textbf{Y}^{*} \big| \textbf{X} \big) \big] \big\}  \bigg\}   \bigg\}           \end{align*}
 
 \begin{align*} 
     \equiv  \underset{P_{\textbf{X}}}{\mathrm{sup}}      \bigg\{   \mathrm{min} \bigg\{ \mathrm{log} \bigg[  \frac{\mathrm{log} \big[ I \big( \textbf{X}, \textbf{Y} \big) \big]}{I \big( \textbf{X}, \textbf{Y}^{*} \big) }\bigg]     ,    \underset{z}{\mathrm{min}}  \big\{   \mathrm{log}  \big[ H_Q \big( \textbf{Y}^{*} \big| \textbf{Z} = z \big)  \big] \big\}   \bigg\} \bigg\}    -  \underset{P_{\textbf{X}}}{\mathrm{sup}}      \bigg\{   \mathrm{min}  \bigg\{ \mathrm{log} \bigg[  \frac{\mathrm{log} \big[ I \big( \textbf{X}, \textbf{Y} \big) \big]}{I \big( \textbf{X}, \textbf{Y}^{*} \big) }\bigg]  \\ ,   \mathrm{log} \big[ H_P \big( \textbf{Y}^{*} \big| \textbf{X} \big) \big]  \bigg\}    \bigg\}   \end{align*}
 
 \begin{align*}   \equiv  \underset{P_{\textbf{X}}}{\mathrm{sup}}      \bigg\{   \mathrm{min} \bigg\{   \mathrm{log} \bigg[  \frac{\mathrm{log} \big[ I \big( \textbf{X}, \textbf{Y} \big) \big]}{I \big( \textbf{X}, \textbf{Y}^{*} \big) }\bigg]     ,   \underset{z}{\mathrm{min}} \big\{      \mathrm{log}  \big[ H_Q \big( \textbf{Y}^{*} \big| \textbf{Z} = z \big)  \big] \big\}   \bigg\} \bigg\}  -  \underset{P_{\textbf{X}}}{\mathrm{sup}}      \bigg\{   \mathrm{min}  \bigg\{ \mathrm{log} \bigg[  \frac{\mathrm{log} \big[ I \big( \textbf{X}, \textbf{Y} \big) \big]}{I \big( \textbf{X}, \textbf{Y}^{*} \big) }\bigg]  \\  ,   \mathrm{log} \big[ H_P \big( \textbf{Y}^{*} \big| \textbf{X} \big) \big]   \bigg\}  \bigg\}         \end{align*}
 
 \begin{align*}              <     \underset{P_{\textbf{X}}}{\mathrm{sup}}      \bigg\{  \mathrm{min} \bigg\{   \mathrm{log} \bigg[  \frac{\mathrm{log} \big[ I \big( \textbf{X}, \textbf{Y} \big) \big]}{I \big( \textbf{X}^{*} , \textbf{Y}^{*} \big) }\bigg]     ,          \underset{z}{\mathrm{min}} \big\{    \mathrm{log}  \big[ H_Q \big( \textbf{Y}^{*} \big| \textbf{Z} = z \big)  \big] \big\}  \bigg\}   \bigg\}    -  \underset{P_{\textbf{X}}}{\mathrm{sup}}      \bigg\{   \mathrm{min}   \bigg\{ \mathrm{log} \bigg[  \frac{\mathrm{log} \big[ I \big( \textbf{X}, \textbf{Y} \big) \big]}{I \big( \textbf{X}, \textbf{Y}^{*} \big) }\bigg]    \\   ,   \mathrm{log} \big[ H_P \big( \textbf{Y}^{*} \big| \textbf{X} \big) \big]    \bigg\}   \bigg\}  \end{align*}

  \begin{align*} \equiv          \underset{P_{\textbf{X}}}{\mathrm{sup}}      \bigg\{  \mathrm{min} \bigg\{  \mathrm{log} \bigg[  \frac{\mathrm{log} \big[ I \big( \textbf{X}, \textbf{Y} \big) \big] I \big( \textbf{X}, \textbf{Y}^{*} \big)}{I \big( \textbf{X}^{*} , \textbf{Y}^{*} \big)        \mathrm{log} \big[ I \big( \textbf{X}, \textbf{Y} \big) \big]     }\bigg]   ,    \underset{z}{\mathrm{min}} \bigg\{    \mathrm{log}  \bigg[ \frac{H_Q \big( \textbf{Y}^{*} \big| \textbf{Z} = z \big)}{  H_P \big( \textbf{Y}^{*} \big| \textbf{X} \big) } \bigg] \bigg\}  \bigg\}    \bigg\}                                  \text{, }
 \end{align*} }

 \noindent where, in (H), we made use of the fact that properties of the conditional Shannon entropy imply,

 {\small \begin{align*}
\big\{  H_P \big( \textbf{Y} \big| \textbf{X} \big) > H_P \big( \textbf{Y}^{*} \big| \textbf{X} \big) \big\}  \Longrightarrow  \bigg\{ \mathrm{log} \big[ H_Q \big( \textbf{Y} \big| \textbf{Z} = z \big) \big] -  \mathrm{log} \big[ H_P \big( \textbf{Y}^{*} \big| \textbf{X} \big) \big]   > \mathrm{log} \big[ H_Q \big( \textbf{Y} \big| \textbf{Z} = z \big) \big]   -  \mathrm{log} \big[ H_P \big( \textbf{Y} \big| \textbf{X} \big) \big]  \bigg\}    \text{. }
 \end{align*} } 

 \noindent The Mutual Information entropy terms,

 {\small \begin{align*}
    I \big( \textbf{X}, \textbf{Y}^{*} \big)    \text{,}
\\ \\ 
   I \big( \textbf{X}^{*}, \textbf{Y}^{*} \big)       \text{,}
 \end{align*}
}

\noindent included in the supremum over $P_{\textbf{X}}$, and $z$, can each respectively be strictly upper bounded with,

 {\small \begin{align*}
     \mathrm{min} \big\{ H  \big( \textbf{X} \big) , H \big( \textbf{Y}^{*} \big) \big\}    \text{,}  \tag{H-1}
 \end{align*} }

 \noindent and with,

 {\small \begin{align*}
       \mathrm{min} \big\{ H  \big( \textbf{X}^{*} \big) , H \big( \textbf{Y}^{*} \big) \big\}     \text{.} \tag{H-2}
 \end{align*}
 }

\noindent Furthermore, each Shannon entropy that is used to bound each Mutual Information entropy can be upper bounded with the logarithm of the cardinality of each alphabet. Namely,

{\small \begin{align*}
  (H-1) <    \mathrm{min} \big\{ \mathrm{log} \big| \textbf{X} \big| , \mathrm{log} \big| \textbf{Y}^{*} \big| \big\}   \text{, } \\  \\ (H-2) <   \mathrm{min} \big\{ \mathrm{log}\big|  \textbf{X}^{*} \big| , \mathrm{log} \big| \textbf{Y}^{*} \big| \big\}   \text{. }
\end{align*} }

\noindent Altogether,

{\small \[
 \mathrm{min} \big\{ \mathrm{log} \big| \textbf{X} \big| , \mathrm{log} \big| \textbf{Y}^{*} \big| \big\}  \equiv \left\{\!\begin{array}{ll@{}>{{}}l} \mathrm{log}  \big| \textbf{X} \big| \Longleftrightarrow  \big|   \textbf{X} \big| < \big| \textbf{Y}^{*} \big|                                  \text{,}    \\ \mathrm{log}  \big| \textbf{Y}^{*} \big| \Longleftrightarrow  \big|   \textbf{X} \big| >   \big| \textbf{Y}^{*} \big|                                  \text{,} \end{array}\right.  \] 

\bigskip

 \[ \mathrm{min} \big\{ \mathrm{log}\big|  \textbf{X}^{*} \big| , \mathrm{log} \big| \textbf{Y}^{*} \big| \big\} \equiv \left\{\!\begin{array}{ll@{}>{{}}l} \mathrm{log}  \big| \textbf{X}^{*} \big| \Longleftrightarrow  \big|   \textbf{X}^{*} \big| < \big| \textbf{Y}^{*} \big|                                  \text{,}    \\ \mathrm{log}  \big| \textbf{Y}^{*} \big| \Longleftrightarrow  \big|   \textbf{X}^{*} \big| >  \big| \textbf{Y}^{*} \big|                                  \text{,} \end{array}\right.      \text{. }
\] 
 } 

\noindent From the previous rearrangements of the strict upper bound of $r$,

{\small \begin{align*}
 \underset{P_{\textbf{X}}}{\mathrm{sup}}  \big\{ \mathrm{min} \big\{  I \big( \textbf{X} , \textbf{Y} \big)  , \underset{z}{\mathrm{min}}  \big\{ H_Q \big( \textbf{Y} \big| \textbf{Z} = z \big)   -  H_P \big( \textbf{Y} \big| \textbf{X} \big)  \big\}  \big\} \text{,}
\end{align*}
 } 

\noindent The desired upper bound for the converse result can be read from the computation,

{\small \begin{align*}
    \underset{P_{\textbf{X}}}{\mathrm{sup}}      \bigg\{  \mathrm{min} \bigg\{  \mathrm{log} \bigg[  \frac{ I \big( \textbf{X}, \textbf{Y}^{*} \big)}{I \big( \textbf{X}^{*} , \textbf{Y}^{*} \big)           }\bigg]   ,    \underset{z}{\mathrm{min}} \bigg\{     \mathrm{log}  \bigg[ \frac{H_Q \big( \textbf{Y}^{*} \big| \textbf{Z} = z \big)}{  H_P \big( \textbf{Y}^{*} \big| \textbf{X} \big) } \bigg]   \bigg\}   \bigg\}          \bigg\}  \end{align*}
    
    \begin{align*} < \underset{P_{\textbf{X}}}{\mathrm{sup}}      \bigg\{    \mathrm{min} \bigg\{  \mathrm{log} \bigg[  \frac{ \mathrm{min} \big\{ H  \big( \textbf{X} \big) , H \big( \textbf{Y}^{*} \big) \big\}  }{ \underset{\textbf{X}^{*} \neq \emptyset, \textbf{Y}^{*} \neq \emptyset }{\mathrm{min}} \big\{ H  \big( \textbf{X}^{*} \big) , H \big( \textbf{Y}^{*} \big) \big\} }         \bigg]    ,   \underset{z}{\mathrm{min}} \bigg\{        \mathrm{log}  \bigg[ \frac{H_Q \big( \textbf{Y}^{*} \big| \textbf{Z} = z \big)}{  H_P \big( \textbf{Y}^{*} \big| \textbf{X} \big) } \bigg]  \bigg\}   \bigg\}  \bigg\}      \end{align*}
    
    \begin{align*}  < \underset{P_{\textbf{X}}}{\mathrm{sup}}      \bigg\{    \mathrm{min} \bigg\{  \mathrm{log} \bigg[  \underset{\textbf{X}^{*} \neq \emptyset, \textbf{Y}^{*} \neq \emptyset }{\mathrm{min}} \bigg\{   \frac{  \big\{ H  \big( \textbf{X} \big) ,  H \big( \textbf{Y}^{*} \big) \big\}  }{  \big\{ H  \big( \textbf{X}^{*} \big) , H \big( \textbf{Y}^{*} \big) \big\} }   \bigg\} \bigg]  ,   \underset{z}{\mathrm{min}} \bigg\{       \mathrm{log}  \bigg[ \frac{H_Q \big( \textbf{Y}^{*} \big| \textbf{Z} = z \big)}{  H_P \big( \textbf{Y}^{*} \big| \textbf{X} \big) } \bigg]     \bigg\}   \bigg\} \bigg\}    \end{align*}
    
    \begin{align*} <               \underset{P_{\textbf{X}}}{\mathrm{sup}}      \bigg\{    \mathrm{min} \bigg\{ \mathrm{log} \mathrm{log} \bigg[  \underset{\textbf{X}^{*} \neq \emptyset, \textbf{Y}^{*} \neq \emptyset }{\mathrm{min}} \bigg\{   \frac{  \big\{ H  \big( \textbf{X} \big) , H \big( \textbf{Y}^{*} \big) \big\}  }{  \big\{ H  \big( \textbf{X}^{*} \big) , H \big( \textbf{Y}^{*} \big) \big\} }       \bigg\}   \bigg]        ,    \underset{z}{\mathrm{min}} \bigg\{       \mathrm{log}  \bigg[ \frac{H_Q \big( \textbf{Y}^{*} \big| \textbf{Z} = z \big)}{  H_P \big( \textbf{Y}^{*} \big| \textbf{X} \big) } \bigg]    \bigg\} \bigg\}  \bigg\}    \end{align*}
    
    \begin{align*}  <   \underset{P_{\textbf{X}}}{\mathrm{sup}}      \bigg\{     \mathrm{min} \bigg\{ \mathrm{log} \mathrm{log} \bigg[  \underset{\textbf{X}^{*} \neq \emptyset, \textbf{Y}^{*} \neq \emptyset }{\mathrm{min}} \bigg\{   \frac{  \big\{ H  \big( \textbf{X} \big) , H \big( \textbf{Y}^{*} \big) \big\}  }{  \big\{ H  \big( \textbf{X}^{*} \big) , H \big( \textbf{Y}^{*} \big) \big\} }       \bigg\}   \bigg]    ,  \underset{z}{\mathrm{min}} \bigg\{       \mathrm{log}  \bigg[ \frac{ \mathrm{min} \big\{ \mathrm{log}\big|  \textbf{Y}^{*} \big| ,  \underset{z}{\mathrm{min}} \big\{  \mathrm{log} \big| \textbf{Z}  \big| \big\}  }{  \underset{\textbf{X} \neq \emptyset, \textbf{Y}^{*} \neq \emptyset}{\mathrm{min}} \big\{ \mathrm{log}\big|  \textbf{X}  \big| , \mathrm{log} \big| \textbf{Y}^{*} \big| \big\}  } \bigg]     \bigg\}  \bigg\} \bigg\} \end{align*}
  
  \begin{align*}        <   \underset{P_{\textbf{X}}}{\mathrm{sup}}      \bigg\{   \mathrm{min} \bigg\{  \mathrm{log} \mathrm{log} \bigg[  \underset{\textbf{X}^{*} \neq \emptyset, \textbf{Y}^{*} \neq \emptyset }{\mathrm{min}} \bigg\{   \frac{  \big\{ H  \big( \textbf{X} \big) , H \big( \textbf{Y}^{*} \big) \big\}  }{  \big\{ H  \big( \textbf{X}^{*} \big) , H \big( \textbf{Y}^{*} \big) \big\} }       \bigg\}   \bigg]     ,    \underset{z}{\mathrm{min}} \bigg\{      \mathrm{log}  \bigg[ \underset{\textbf{X} \neq \emptyset, \textbf{Y}^{*} \neq \emptyset}{\mathrm{min}} \bigg\{  \frac{  \big\{ \mathrm{log}\big|  \textbf{Y}^{*} \big| , \mathrm{log} \big| \textbf{Z}  \big| \big\}  }{  \big\{ \mathrm{log}\big|  \textbf{X}  \big| , \mathrm{log} \big| \textbf{Y}^{*} \big| \big\}  } \bigg\}  \bigg]    \bigg\} \bigg\}               \bigg\} \end{align*}
    
    \begin{align*}        <   \underset{P_{\textbf{X}}}{\mathrm{sup}}      \bigg\{   \mathrm{min} \bigg\{   \mathrm{log} \mathrm{log} \bigg[  \underset{\textbf{X}^{*} \neq \emptyset, \textbf{Y}^{*} \neq \emptyset }{\mathrm{min}} \bigg\{   \frac{  \big\{ \mathrm{log}   \big| \textbf{X} \big| , \mathrm{log} \big| \textbf{Y}^{*} \big| \big\}  }{  \big\{ \mathrm{log}  \big| \textbf{X}^{*} \big| , \mathrm{log} \big|  \textbf{Y}^{*} \big| \big) \big\} }       \bigg\}   \bigg]      ,     \underset{z}{\mathrm{min}}  \bigg\{     \mathrm{log}  \bigg[ \underset{\textbf{X} \neq \emptyset, \textbf{Y}^{*} \neq \emptyset}{\mathrm{min}} \bigg\{  \frac{  \big\{ \mathrm{log}\big|  \textbf{Y}^{*} \big| ,   \mathrm{log} \big| \textbf{Z}  \big| \big\}  }{  \big\{ \mathrm{log}\big|  \textbf{X}  \big| , \mathrm{log} \big| \textbf{Y}^{*} \big| \big\}  }  \bigg\} \bigg]   \bigg\} \bigg\}   \bigg\}    \text{. }    \\ \\   \tag{$**$}     \end{align*} }

    \noindent Hence,

    {\small \[ (**)  < \left\{\!\begin{array}{ll@{}>{{}}l} 
     \mathrm{log} \mathrm{log} \bigg[   {  \mathrm{log} \big| \textbf{Y}^{*} \big|   } \big/ { \mathrm{log} \big| \textbf{X}^{*} \big|  }          \bigg]   +   \mathrm{log}  \bigg[ {   \mathrm{log}\big|  \textbf{Z}  \big|  } \big/ {  \mathrm{log}\big|  \textbf{Y}^{*}  \big|   } \bigg]  \Longleftrightarrow \big| \textbf{X} \big| > \big| \textbf{Y}^{*} \big| ,  \big| \textbf{Y}^{*} \big| > \big| \textbf{Z} \big|    , \\     \\ \mathrm{log} \mathrm{log} \bigg[   {  \mathrm{log} \big| \textbf{X} \big|  } \big/ { \mathrm{log} \big| \textbf{Y}^{*} \big| }          \bigg]   +   \mathrm{log}  \bigg[  {  \mathrm{log}\big|  \textbf{Y}^{*} \big|   } \big/ { \mathrm{log} \big| \textbf{X} \big|   } \bigg]   \Longleftrightarrow \big| \textbf{X} \big| <  \big| \textbf{Y}^{*} \big| ,  \big|  \textbf{Y}^{*} \big| < \big| \textbf{Z} \big|        ,  \\  \\ \mathrm{log} \mathrm{log} \bigg[   {  \mathrm{log} \big| \textbf{Y}^{*} \big|  } \big/ { \mathrm{log} \big| \textbf{X}^{*} \big| }          \bigg]   +   \mathrm{log}  \bigg[  {  \mathrm{log}\big|  \textbf{Y}^{*} \big|   } \big/ { \mathrm{log} \big| \textbf{X} \big|   } \bigg]   \Longleftrightarrow \big| \textbf{X} \big| >   \big| \textbf{Y}^{*} \big| ,  \big|  \textbf{Y}^{*} \big| < \big| \textbf{Z} \big|        ,   \\   \\ \mathrm{log} \mathrm{log} \bigg[     {  \mathrm{log} \big| \textbf{Y}^{*} \big|  } \big/ { \mathrm{log} \big| \textbf{X}^{*} \big| }          \bigg]   +   \mathrm{log}  \bigg[  {  \mathrm{log}\big|  \textbf{Z}  \big|   } \big/ { \mathrm{log} \big| \textbf{Y}^{*} \big|   } \bigg]   \Longleftrightarrow \big| \textbf{X} \big| <    \big| \textbf{Y}^{*} \big| ,  \big|  \textbf{Y}^{*} \big| >   \big| \textbf{Z} \big|        .                          
\end{array}\right. 
\] 

 \[  < \left\{\!\begin{array}{ll@{}>{{}}l} 
     \mathrm{log} \mathrm{log} \bigg[   {  \mathrm{log} \big| \textbf{Y}^{*} \big|   } \big/ { \big| \textbf{X}^{*} \big|  }          \bigg]   +   \mathrm{log}  \bigg[ {   \mathrm{log}\big|  \textbf{Z}  \big|  } \big/ {  \big|  \textbf{Y}^{*}  \big|   } \bigg]  \Longleftrightarrow \big| \textbf{X} \big| > \big| \textbf{Y}^{*} \big| ,  \big| \textbf{Y}^{*} \big| > \big| \textbf{Z} \big|    ,  \\    \\ \mathrm{log} \mathrm{log} \bigg[   {  \mathrm{log} \big| \textbf{X} \big|  } \big/ { \big| \textbf{Y}^{*} \big| }          \bigg]   +   \mathrm{log}  \bigg[  { \mathrm{log} \big|  \textbf{Y}^{*} \big|   } \big/ {  \big| \textbf{X} \big|   } \bigg]   \Longleftrightarrow \big| \textbf{X} \big| <  \big| \textbf{Y}^{*} \big| ,  \big|  \textbf{Y}^{*} \big| < \big| \textbf{Z} \big|        ,  \\ \\ \mathrm{log} \mathrm{log} \bigg[    {  \mathrm{log} \big| \textbf{Y}^{*} \big|  } \big/ { \big| \textbf{X}^{*} \big| }          \bigg]   +   \mathrm{log}  \bigg[  {  \mathrm{log}\big|  \textbf{Y}^{*} \big|   } \big/ { \big| \textbf{X} \big|   } \bigg]   \Longleftrightarrow \big| \textbf{X} \big| >   \big| \textbf{Y}^{*} \big| ,  \big|  \textbf{Y}^{*} \big| < \big| \textbf{Z} \big|        ,   \\   \\ \mathrm{log} \mathrm{log} \bigg[   {  \mathrm{log} \big| \textbf{Y}^{*} \big|  } \big/ {\big| \textbf{X}^{*} \big| }          \bigg]   +   \mathrm{log}  \bigg[  {  \mathrm{log}\big|  \textbf{Z}  \big|   } \big/ { \big| \textbf{Y}^{*} \big|   } \bigg]   \Longleftrightarrow \big| \textbf{X} \big| <    \big| \textbf{Y}^{*} \big| ,  \big|  \textbf{Y}^{*} \big| >   \big| \textbf{Z} \big|        .                          
\end{array}\right. 
  \equiv r  , \\ \\  \tag{*} 
\]  }

\noindent from which we conclude the argument. \boxed{}

\subsection{Theorem $2$}

\subsubsection{General description of the proof}

\noindent We obtain the stochastic domination of false acceptance, and decoding error, probabilities by making use of the previously defined function,

{\small \begin{align*}
   \mathscr{O} {\tiny \big( }  \textbf{X}, \textbf{Y}, \textbf{Z} {\tiny \big)}  \equiv  \underset{ x \in \textbf{X}, y \in \textbf{Y}, z \in  \textbf{Z}}{\bigcup} \mathscr{O} {\tiny \big( x, y, z \big) }  \text{, }
\end{align*}
 } 

\noindent corresponding to the overlap of alphabets $\textbf{X}$, $\textbf{Y}$, and $\textbf{Z}$. In particular, one has,

{\small \begin{align*}
   \mathscr{O} {\tiny \big( }  \textbf{X}, \textbf{Y}, \textbf{Z} {\tiny \big) }  \bigg|_{\textbf{X} \equiv \emptyset} \equiv  \underset{ y \in \textbf{Y}, z \in  \textbf{Z}}{\bigcup} \mathscr{O} {\tiny \big(  y, z \big)  } \text{, } \\   \mathscr{O} {\tiny \big( } \textbf{X}, \textbf{Y}, \textbf{Z} {\tiny \big) } \bigg|_{\textbf{Z} \equiv \emptyset}  \equiv  \underset{ x \in \textbf{X}, y \in \textbf{Y}}{\bigcup} \mathscr{O} {\tiny \big( x, y \big) } \text{, }
\end{align*} }

\noindent which can be leveraged to obtain the first, and second, stochastic domination results. Specifically, the argument makes significant use of the fact that over the more noisy Quantum channel, $A \longleftrightarrow B$, the overlap function is dependent upon Eve's alphabet, which she can use in attempts to introduced corrupted codewords that Alice and Bob mistakenly place into their shared authenticated set. 

Additionally, the second stochastic domination result in the proof establishes that the probabilities of error correction, and false acceptance, simultaneously occur whp. This is achieved by making use of the computations introduced for the first stochastic domination result, which indicates that the probability of Alice and Bob executing error correcting codes over their shared channel is greater than the probability of Bob and Eve executing error correcting codes over their shared channel. That is, from the fact that,

{\small \begin{align*}
   \mathscr{O} {\tiny \big( } \textbf{X}, \textbf{Y}, \textbf{Z} {\tiny \big) }  \equiv  \underset{ x \in \textbf{X}, y \in \textbf{Y}, z \in  \textbf{Z}}{\bigcup} \mathscr{O} {\tiny \big( x, y, z \big) }  \equiv  \underset{ x \in \textbf{X}, y \in \textbf{Y}}{\bigcup} \mathscr{O} {\tiny \big( x, y \big) }    \cup   \underset{z \in  \textbf{Z}}{\bigcup} \mathscr{O} {\tiny \big( z \big) }  \equiv  \underset{ y \in \textbf{Y}, z \in \textbf{Z}}{\bigcup} \mathscr{O} {\tiny \big( y,z \big) }     \cup   \underset{x \in  \textbf{X}} {\bigcup} \mathscr{O} {\tiny \big( x \big)}  \end{align*}
   
   \begin{align*} \equiv \underset{ x \in \textbf{X}, z \in \textbf{Z}}{\bigcup} \mathscr{O} {\tiny \big( y,z \big)  }    \cup   \underset{y \in  \textbf{Y}}{\bigcup} \mathscr{O} {\tiny \big( y \big) }   \text{, }
\end{align*} }

\noindent one can examine the overlap functions,

{\small \begin{align*}
\bigg\{  \underset{x \in \textbf{X}}{\bigcup} \mathscr{O} {\tiny \big( x \big)  }   \text{, }  \underset{y \in \textbf{Y}}{\bigcup} \mathscr{O} {\tiny \big( y \big)  }  \text{, }   \underset{z \in \textbf{Z}}{\bigcup} \mathscr{O} {\tiny \big( z \big) } \bigg\} \text{, }
\end{align*}
 } 

\noindent of letters within a single alphabet given several responses from Alice, Bob, and Eve. Explicitly, the three above functions determine,

{\small \[
 \underset{x \in \textbf{X}}{\bigcup} \mathscr{O} {\tiny \big( x \big) }  \equiv \left\{\!\begin{array}{ll@{}>{{}}l}  c \Longleftrightarrow \bigg[ \underset{\text{Questions } \mathcal{Q}}{\sum} \bigg\{ \textbf{1}_{\{\text{Alice uses a letter } x \text{  in response to some } \mathcal{Q} \sim \pi \}} \bigg\}  \equiv c \bigg] \\  \\ 0  \text{ otherwise,}\end{array}\right.        \text{,} \]

 \bigskip \[   \underset{y \in \textbf{Y}}{\bigcup} \mathscr{O} {\tiny \big( y \big) }  \equiv \left\{\!\begin{array}{ll@{}>{{}}l}  c \Longleftrightarrow \bigg[ \underset{\text{Questions } \mathcal{Q}}{\sum} \bigg\{  \textbf{1}_{\{\text{Bob uses a letter } y \text{ in response to some } \mathcal{Q}\sim \pi\}} \bigg\} \equiv c  \bigg] \\  \\ 0  \text{ otherwise,}   \end{array}\right.    \text{,} \]
 
 \bigskip \[ \underset{z \in \textbf{Z}}{\bigcup} \mathscr{O} {\tiny \big( z \big) }  \equiv \left\{\!\begin{array}{ll@{}>{{}}l} c \Longleftrightarrow \bigg[ \underset{\text{Questions } \mathcal{Q}}{\sum} \bigg\{  \textbf{1}_{\{\text{Eve uses a letter } z \text{ in response to some } \mathcal{Q}\sim \pi\}} \bigg\} \equiv c  \bigg] \\   \\ 0  \text{ otherwise,} \end{array}\right.    \text{, }
\]  }

\noindent from the referee's probability distribution of questions, $\pi$. In the following arguments, we express the error correction probability over the Quantum channel between Alice and Bob, $p_{\mathrm{EC}, A \longleftrightarrow B}$, in terms of conditions on the overlap function between the alphabets of the three players, as well as the existence of a suitable class of error correction codes, $\mathrm{EC}$. Explicitly, the supremum over all admissible error correcting codes that is further manipulated in the forthcoming arguments takes the form,

{\small \begin{align*}
 p_{\mathrm{EC}, A\longleftrightarrow B} \approx  \underset{\mathrm{EC}, A\longleftrightarrow B}{\mathrm{sup}} \bigg\{  \textbf{P} \big[ \big\{   \forall \mathscr{C}^{*} \in \textbf{R}, \exists \text{ } \mathrm{ EC } :   \mathrm{sup} \big\{ \mathscr{O} {\tiny \big(} \textbf{X}^{*}, \textbf{Y}, \textbf{Z} {\tiny \big),}  \mathscr{O} {\tiny \big( } \textbf{X}, \textbf{Y}^{*}, \textbf{Z} {\tiny \big)}  \big\}   > \mathscr{C}^{*}  , p_{\mathrm{EC}} > 0   \big\}   \big| \big\{ \text{alpha-} \\ \text{bets } \textbf{X}^{*}   , \textbf{Y},  \textbf{Z},  \text{ or alphabets } \textbf{X}, \textbf{Y}^{*}, \textbf{Z} \big\} \big] \bigg\}   \text{. }
\end{align*} }

\noindent Beginning from the conditional probability above, we repeatedly make use of Bayes' Rule, three times in fact, in addition to taking a supremum over the class $\mathrm{EC}$ of error correcting codes. By making use of the fact that the overlap function, $\mathscr{O}$, over $A \longleftrightarrow B$ is dependent upon the alphabets of all three players, namely $\textbf{X}$, $\textbf{Y}$, and $\textbf{Z}$, we obtain the desired lower bound, $p_{\mathrm{EC},B \longleftrightarrow E}$, by taking the limit of, 

{\small \begin{align*}
CC^{-1} \cdot     \underset{x \in \textbf{X}}{\bigcup}    \textbf{P} \big[  \big\{  \forall \mathscr{C}^{**} \in \textbf{R}, \exists \text{ } \mathrm{ ec } : \mathscr{O} {\tiny \big( x  \big)  } <<   \big| \textbf{X} \big| , p_{\mathrm{ec}} > 0   \big\}    \big| \big\{ \text{alphabet } x \big\} \big]    \text{,}
 \end{align*}  } 

\noindent as $\big| \textbf{X} \big| \longrightarrow + \infty$, and some strictly positive constant, $CC$. For the remaining contributions over the alphabets $\textbf{Y}$ and $\textbf{Z}$ which Alice and Eve use for responses that are introduced over $A \longleftrightarrow B$, we lower bound $p_{\mathrm{EC},A \longleftrightarrow B }$ with,

{\small \begin{align*}
\underset{\mathrm{EC}, B \longleftrightarrow E}{\mathrm{sup}} \bigg\{  \underset{y \in \textbf{Y}, z \in \textbf{Z}}{\bigcup}    \bigg\{ \textbf{P} \big[  \big\{  \forall \mathscr{C}^{*} \in \textbf{R}, \exists \text{ } \mathrm{ ec } : \mathscr{O} {\tiny \big( y,z  \big) }  > \mathscr{C}^{*}, p_{\mathrm{ec}} > 0   \big\}    \big| \big\{ \text{alphabets } y, z  \big\} \big]     \bigg\} \bigg\} \text{,} \end{align*} } 

\noindent which is approximately $p_{\mathrm{EC}, B \longleftrightarrow E}$. With these computations that are used for relating $p_{\mathrm{EC},A \longleftrightarrow B}$, and $p_{\mathrm{EC}, B \longleftrightarrow E}$, to obtain the second stochastic domination, we lower bound,

{\small \begin{align*}
  \frac{p_{\mathrm{EC}, A \longleftrightarrow B}}{p_{\mathrm{FA}, A \longleftrightarrow B}}  \text{. }
\end{align*}
 }

\noindent We make use of computations for obtaining the first  stochastic domination result, between the probability of error correction over $A \longleftrightarrow B$ and $B \longleftrightarrow E$, to obtain the second stochastic domination result, between the probability of false acceptance, $p_{\mathrm{FA}}$. The computations associated withh the second stochastic domination are more technical, as they are related to manipulating,

 {\small \begin{align*} \frac{ \underset{\mathrm{EC}, B \longleftrightarrow E}{\mathrm{sup}} \bigg\{  \underset{y \in \textbf{Y}, z \in \textbf{Z}}{\bigcup}    \bigg\{ \textbf{P} \big[  \big\{  \forall \mathscr{C}^{*} \in \textbf{R}, \exists \text{ } \mathrm{ ec } : \mathscr{O} {\tiny \big( y,z  \big) }  > \mathscr{C}^{*}, p_{\mathrm{ec}} > 0   \big\}    \big| \big\{ \text{alphabets } y, z  \big\} \big]     \bigg\} \bigg\}   +     CC^{-1}   }{p_{\mathrm{FA}, A \longleftrightarrow B }}     \text{. }
\end{align*}
 } 

\noindent As $\big| \textbf{Y} \big|$, $\big| \textbf{Z} \big| \longrightarrow + \infty$, the second stochastic domination result can be obtained by making use of Bayes' rule, first through the equality,

{\small \begin{align*}
 \textbf{P} \big[  \big\{  \forall \mathscr{C}^{*} \in \textbf{R}, \exists \text{ } \mathrm{ ec } : \mathscr{O} {\tiny \big( y,z  \big) }   > \mathscr{C}^{*}, p_{\mathrm{ec}} > 0   \big\}    \big| \big\{ \text{alphabets } y, z  \big\} \big]    \\  \\  \equiv   \textbf{P} \big[ \big\{ \text{alphabets } y, z  \big\} \big|  \big\{  \forall \mathscr{C}^{*} \in \textbf{R}  , \exists \text{ } \mathrm{ ec } : \mathscr{O} {\tiny \big( y,z  \big) }  > \mathscr{C}^{*}  , p_{\mathrm{ec}} > 0   \big\}   \big] \\ \times  \big\{  \textbf{P} \big[ \big\{ \text{alphabets } y, z  \big\} \big] \big\}^{-1}  \textbf{P} \big[  \big\{  \forall \mathscr{C}^{*} \in \textbf{R}, \exists \text{ } \mathrm{ ec } : \mathscr{O} {\tiny \big( y,z  \big) }  > \mathscr{C}^{*}  \\ , p_{\mathrm{ec}} > 0   \big\}  \big]      \text{, }
\end{align*} }

\noindent in addition to the fact that,

{\small \begin{align*}
  \textbf{P} \big[ \text{alphabets } y,z \big]  \text{, }
\end{align*}
 } 

\noindent occurs with positive probability (namely, that Bob, and Eve, can form an alphabet with letters $y,z$). The above probability is equivalent to,

{\small \begin{align*}
 \textbf{P} \big[ \text{Bob and Eve use alphabets } y,z \big] \equiv  \textbf{P} \big[ \text{alphabets } y,z \big]  \text{, }
\end{align*} } 

\noindent which is related to the conditioning introduced in the expression that is approximately $p_{\mathrm{EC}, A \longleftrightarrow B}$,

{\small \begin{align*}
  \textbf{P} \big[ \big\{ \text{alphabets } \textbf{X}^{*}  , \textbf{Y},  \textbf{Z},  \text{ or alphabets } \textbf{X}, \textbf{Y}^{*}, \textbf{Z} \big\} \big] \\ \\ \equiv \textbf{P} \big[ \big\{ \text{Alice, Bob, and Eve, respectively use alphabets } \\  \textbf{X}^{*}  , \textbf{Y},  \textbf{Z},  \text{or, Alice, Bob, and Eve, respectively} \\ \text{use  alphabets }  \textbf{X}, \textbf{Y}^{*}, \textbf{Z} \big\} \big]   \text{. }
\end{align*} }

\noindent For the second stochastic domination result between the probabilities of error correction, and false acceptance, the lower bound crucially depends upon the fact that,

{\small \begin{align*}
 \underset{\mathscr{C}^{***} \longrightarrow 0^{+}}{\underset{| \textbf{Z} | \longrightarrow + \infty}{\underset{| \textbf{Y} | \longrightarrow + \infty}{\mathrm{lim}}}}     \bigg\{     \underset{\mathrm{fa} \in \mathrm{FA}}{\underset{\mathrm{EC}, B \longleftrightarrow E}{\mathrm{sup}}}  \bigg\{        \textbf{P} \big[           \big\{  \forall \mathscr{C}^{***} > \mathscr{C}^{**} > \mathscr{C}^{*} \in \textbf{R} , \exists \text{ } \mathrm{ ec } : \mathscr{O} {\tiny \big( } \textbf{Y},\textbf{Z}  {\tiny \big) }  > \mathscr{C}^{***}  , p_{\mathrm{ec}} > 0   \big\}  \big|   \big\{ \text{alphabets } \textbf{Y}, \textbf{Z}  \big\} \big]             \big\} \bigg\}  \bigg\} \\   \\ >  \underset{\mathscr{C}^{**} \longrightarrow 0^{+}}{\underset{| \textbf{Z} | \longrightarrow + \infty}{\underset{| \textbf{Y} | \longrightarrow + \infty}{\mathrm{lim}}}}     \bigg\{     \underset{\mathrm{fa} \in \mathrm{FA}}{\underset{\mathrm{EC}, B \longleftrightarrow E}{\mathrm{sup}}}  \bigg\{        \textbf{P} \big[           \big\{  \forall \mathscr{C}^{**} > \mathscr{C}^{*} \in \textbf{R} , \exists \text{ } \mathrm{ ec } : \mathscr{O} {\tiny \big( } \textbf{Y},\textbf{Z}  {\tiny \big) }  > \mathscr{C}^{**}  , p_{\mathrm{ec}} > 0   \big\}  \big|  \big\{ \text{alphabets } \textbf{Y}, \textbf{Z}  \big\} \big]             \big\} \bigg\}  \bigg\}   \text{, }
\end{align*}
 } 

\noindent namely that, as the size of the alphabets that Bob and Eve use approach $+ \infty$, the conditional probability that the overlap function of their alphabets approximately vanishes occurs with arbitrarily small probability.

\subsubsection{Argument}

\noindent \textit{Proof of Theorem 2}. To show that each desired probability can be made arbitrarily small at the same time, it suffices to argue,

{\small \begin{align*}
\bigg\{  \textbf{P} \big[ p_{\mathrm{FA}} > 0 \big| p_{\mathrm{DE}} > 0  \big] > 0 \bigg\}  \Longleftrightarrow  \bigg\{ \textbf{P} \big[ p_{\mathrm{DE}} > 0 \big| p_{\mathrm{FA}} > 0  \big] >  0  \bigg\}    \text{. }
\end{align*} } 

\noindent That is, conditionally upon $p_{\mathrm{DE}}$, or $p_{\mathrm{FA}}$ occurring, $p_{\mathrm{FA}}$ and $p_{\mathrm{DE}}$ can occur. If the probabilities in the conditioning above are set to be very close to $0$,

{\small \begin{align*}
\bigg\{  \textbf{P} \big[ p_{\mathrm{FA}} > 0 \big| p_{\mathrm{DE}} \approx 0  \big] \approx 0 \bigg\}  \Longleftrightarrow  \bigg\{ \textbf{P} \big[ p_{\mathrm{DE}} > 0 \big| p_{\mathrm{FA}} \approx 0   \big] \approx 0  \bigg\}   \text{. }
\end{align*}
 } 

\noindent Fix some constant $CC>0$. Under the assumption that $N_{A \longleftrightarrow B} > N_{B \longleftrightarrow E}$, to prove the desired result,

{\small \begin{align*}
 p_{\mathrm{EC}, A \longleftrightarrow B} >   p_{\mathrm{EC}, B \longleftrightarrow E}  \text{, } \\ \\   p_{\mathrm{FA}, B \longleftrightarrow E}  >   p_{\mathrm{FA}, A \longleftrightarrow B}  \text{, }
\end{align*}

 } 

\noindent for,

{\small \begin{align*}
 \underset{\mathrm{EC}, A \longleftrightarrow B}{\mathrm{sup}}  \big\{ \cdot \big\} \equiv   \underset{\mathrm{ec} \in \mathrm{EC}, A \longleftrightarrow B}{\mathrm{sup}}  \big\{ \cdot \big\} \text{, }
\end{align*} }

\noindent write,

{\small \begin{align*}
    \underline{p_{\mathrm{EC}, A\longleftrightarrow B}} \approx  \underset{\mathrm{EC}, A\longleftrightarrow B}{\mathrm{sup}} \bigg\{  \textbf{P} \big[ \big\{   \forall \mathscr{C}^{*} \in \textbf{R}, \exists \text{ } \mathrm{ EC } :   \mathrm{sup} \big\{ \mathscr{O} {\tiny  \big( }  \textbf{X}^{*}, \textbf{Y}, \textbf{Z} {\tiny \big)} , \mathscr{O} {\tiny \big(}  \textbf{X}, \textbf{Y}^{*}, \textbf{Z} {\tiny \big) }  \big\}  > \mathscr{C}^{*} , p_{\mathrm{EC}} > 0   \big\}     \big| \big\{ \text{alpha-} \\ \text{bets } \textbf{X}^{*}  , \textbf{Y},  \textbf{Z},  \text{ or alphabets } \textbf{X}, \textbf{Y}^{*}, \textbf{Z} \big\} \big] \bigg\} \tag{$*-1$} \end{align*}

    \begin{align*} > \underset{\mathrm{EC}, A\longleftrightarrow B}{\mathrm{sup}} \bigg\{  \textbf{P} \big[  \big\{  \forall \mathscr{C}^{*} \in \textbf{R}, \exists \text{ } \mathrm{ EC } : \mathscr{O} {\tiny \big( } \textbf{X}, \textbf{Y}, \textbf{Z} {\tiny \big)}   > \mathscr{C}^{*}, p_{\mathrm{EC}} > 0   \big\}    \big| \big\{ \text{alphabets } \textbf{X}, \textbf{Y}, \textbf{Z} \big\} \big] \bigg\} \tag{$*-2$}   \end{align*}

    \begin{align*}  \equiv \underset{\mathrm{EC}, A\longleftrightarrow B}{\mathrm{sup}} \bigg\{ \underset{x \in \textbf{X}, y \in \textbf{Y}, z \in \textbf{Z}}{\bigcup}  \bigg\{   \textbf{P} \big[  \big\{  \forall \mathscr{C}^{*} \in \textbf{R}, \exists \text{ } \mathrm{ ec } : \mathscr{O} {\tiny \big( x, y, z \big)  } > \mathscr{C}^{*}, p_{\mathrm{ec}} > 0   \big\}    \big| \big\{ \text{alphabets } x, y, z \big\} \big]     \bigg\}    \bigg\}                  \tag{$*-3$}        \end{align*}

    \begin{align*}         >  \underset{\mathrm{EC}, A\longleftrightarrow B}{\mathrm{sup}} \bigg\{  \underset{y \in \textbf{Y}, z \in \textbf{Z}}{\bigcup}    \bigg\{ \textbf{P} \big[  \big\{  \forall \mathscr{C}^{*} \in \textbf{R}, \exists \text{ } \mathrm{ ec } : \mathscr{O} {\tiny \big( y,z \big) }  > \mathscr{C}^{*}, p_{\mathrm{ec}} > 0   \big\}    \big| \big\{ \text{alphabets } x,y  \big\} \big]     \bigg\}   \end{align*}

    \begin{align*}    +    C^{-1}   \cdot  \underset{x \in \textbf{X}}{\bigcup}    \textbf{P} \big[  \big\{  \forall \mathscr{C}^{**} \in \textbf{R}, \exists \text{ } \mathrm{ ec } : \mathscr{O} {\tiny \big( x  \big)  } > \mathscr{C}^{**}, p_{\mathrm{ec}} > 0   \big\}    \big| \big\{ \text{alphabet } x \big\} \big]      \bigg\}  \tag{$*-4$}   \end{align*}

    \begin{align*}      >   \underset{\textbf{X} |\longrightarrow + \infty}{\mathrm{lim}}        \bigg[  \underset{\mathrm{EC}, A\longleftrightarrow B}{\mathrm{sup}} \bigg\{  \underset{y \in \textbf{Y}, z \in \textbf{Z}}{\bigcup}    \bigg\{ \textbf{P} \big[  \big\{  \forall \mathscr{C}^{*} \in \textbf{R}, \exists \text{ } \mathrm{ ec } : \mathscr{O} \big( y,z \big)  > \mathscr{C}^{*}, p_{\mathrm{ec}} > 0   \big\}    \big|   \big\{ \text{alphabets } x,y  \big\} \big]     \bigg\}  \end{align*}

    \begin{align*} +   {C}^{-1}  \cdot  \underset{x \in \textbf{X}}{\bigcup}    \textbf{P} \big[  \big\{  \forall \mathscr{C}^{**} \in \textbf{R}, \exists \text{ } \mathrm{ ec } : \mathscr{O} \big( x  \big) <<   \big| \textbf{X} \big| , p_{\mathrm{ec}} > 0   \big\}    \big| \big\{ \text{alphabet } x \big\} \big]   \bigg\}     \bigg] \tag{$*-5$} \end{align*}

    \begin{align*}   \equiv    \underset{\mathrm{EC}, A\longleftrightarrow B}{\mathrm{sup}} \bigg\{  \underset{y \in \textbf{Y}, z \in \textbf{Z}}{\bigcup}    \bigg\{ \textbf{P} \big[  \big\{  \forall \mathscr{C}^{*} \in \textbf{R}, \exists \text{ } \mathrm{ ec } : \mathscr{O} {\tiny \big( y,z \big) }  > \mathscr{C}^{*}, p_{\mathrm{ec}} > 0   \big\}    \big| \big\{ \text{alphabets } x,y  \big\} \big]     \bigg\}  \end{align*}

       \begin{align*}    +    CC^{-1}  \cdot  \underset {| \textbf{X} |\longrightarrow \infty}{\mathrm{lim}}      \underset{x \in \textbf{X}}{\bigcup}    \textbf{P} \big[  \big\{  \forall \mathscr{C}^{**} \in \textbf{R}, \exists \text{ } \mathrm{ ec } : \mathscr{O} {\tiny \big( x  \big)  } <<   \big| \textbf{X} \big| , p_{\mathrm{ec}} > 0   \big\}    \big| \big\{ \text{alphabet } x \big\} \big]   \bigg\}    \tag{$*-6$} \end{align*}

    \begin{align*}     {>}     \underset{\mathrm{EC}, B \longleftrightarrow E}{\mathrm{sup}} \bigg\{  \underset{y \in \textbf{Y}, z \in \textbf{Z}}{\bigcup}    \bigg\{ \textbf{P} \big[  \big\{  \forall \mathscr{C}^{*} \in \textbf{R}, \exists \text{ } \mathrm{ ec } : \mathscr{O} {\tiny \big( y,z  \big) }  > \mathscr{C}^{*}, p_{\mathrm{ec}} > 0   \big\}    \big| \big\{ \text{alphabets } y, z  \big\} \big]     \bigg\}        +   CC^{-1}   \bigg\}  \tag{$*-7$} \end{align*}

    \begin{align*}      \equiv  \underset{\mathrm{EC}, B \longleftrightarrow E}{\mathrm{sup}} \bigg\{  \underset{y \in \textbf{Y}, z \in \textbf{Z}}{\bigcup}    \bigg\{ \textbf{P} \big[  \big\{  \forall \mathscr{C}^{*} \in \textbf{R}, \exists \text{ } \mathrm{ ec } : \mathscr{O} {\tiny \big( y,z  \big) }  > \mathscr{C}^{*}, p_{\mathrm{ec}} > 0   \big\}    \big| \big\{ \text{alphabets } y, z  \big\} \big]     \bigg\} \bigg\}        +     CC^{-1}     \\ \tag{$*-8$} \end{align*}

    \begin{align*}     \approx    p_{\mathrm{EC}, B \longleftrightarrow E} + CC^{-1}    \overset{\{ CC \longrightarrow + \infty \} }{>} \underline{p_{\mathrm{EC}, B \longleftrightarrow E}}        \text{, }
\end{align*} } 

\noindent corresponding to the first stochastic domination. For the remaining stochastic domination, namely for, 

{\small \begin{align*}
  p_{\mathrm{FA}, B \longleftrightarrow E}  \equiv \underset{\mathrm{fa} \in \mathrm{FA}}{\mathrm{sup}}   \big\{ p_{\mathrm{fa}, B \longleftrightarrow E} \big\}                \text{, }       
\end{align*} }

\noindent write,

{\small \begin{align*}
    \frac{ \underline{p_{\mathrm{EC}, A \longleftrightarrow B}}}{\underline{p_{\mathrm{FA}, A \longleftrightarrow B}}} \equiv \frac{(\mathrm{*-1})}{p_{\mathrm{FA}, A \longleftrightarrow B }}       > \frac{\mathrm{(*-2)}}{p_{\mathrm{FA}, A \longleftrightarrow B }} >   \frac{\mathrm{(*-3)}}{p_{\mathrm{FA}, A \longleftrightarrow B }}  > \frac{\mathrm{(*-4)}}{p_{\mathrm{FA}, A \longleftrightarrow B }} > \frac{\mathrm{(*-5)}}{p_{\mathrm{FA}, A \longleftrightarrow B }}  > \frac{\mathrm{(*-6)}}{p_{\mathrm{FA}, A \longleftrightarrow B }} \\ > \frac{\mathrm{(*-7)}}{p_{\mathrm{FA}, A \longleftrightarrow B }} \equiv  \frac{\mathrm{(*-8)}}{p_{\mathrm{FA}, A \longleftrightarrow B }} \text{. }  \end{align*} }

\noindent The final expression,
    
    {\small \begin{align*} \frac{ \underset{\mathrm{EC}, B \longleftrightarrow E}{\mathrm{sup}} \bigg\{  \underset{y \in \textbf{Y}, z \in \textbf{Z}}{\bigcup}    \bigg\{ \textbf{P} \big[  \big\{  \forall \mathscr{C}^{*} \in \textbf{R}, \exists \text{ } \mathrm{ ec } : \mathscr{O} {\tiny \big( y,z  \big) }  > \mathscr{C}^{*}, p_{\mathrm{ec}} > 0   \big\}    \big| \big\{ \text{alphabets } y, z  \big\} \big]     \bigg\} \bigg\}   +     CC^{-1}  }{p_{\mathrm{FA}, A \longleftrightarrow B }}    \tag{$*-9$}   \text{, }
\end{align*} }

\noindent can be lower bounded from the following observations,

{\small \begin{align*}
  (\mathrm{*-9}) \\ \equiv  \frac{ \underset{\mathrm{EC}, B \longleftrightarrow E}{\mathrm{sup}} \bigg\{  \underset{y \in \textbf{Y}, z \in \textbf{Z}}{\bigcup}    \bigg\{ \textbf{P} \big[  \big\{  \forall \mathscr{C}^{*} \in \textbf{R}, \exists \text{ } \mathrm{ ec } : \mathscr{O} {\tiny \big( y,z  \big) }  > \mathscr{C}^{*}, p_{\mathrm{ec}} > 0   \big\}    \big| \big\{ \text{alphabets } y, z  \big\} \big]     \bigg\} \bigg\}   +   CC^{-1} }{\underset{\mathrm{fa} \in \mathrm{FA}}{\mathrm{sup}}   \big\{ p_{\mathrm{fa}, B \longleftrightarrow E} \big\}        }  \end{align*}

    \begin{align*}       \equiv   \frac{ \underset{\mathrm{EC}, B \longleftrightarrow E}{\mathrm{sup}} \bigg\{  \underset{y \in \textbf{Y}, z \in \textbf{Z}}{\bigcup}    \bigg\{ \textbf{P} \big[  \big\{  \forall \mathscr{C}^{*} \in \textbf{R}, \exists \text{ } \mathrm{ ec } : \mathscr{O} {\tiny \big( y,z  \big) }  > \mathscr{C}^{*}, p_{\mathrm{ec}} > 0   \big\}    \big| \big\{ \text{alphabets } y, z  \big\} \big]     \bigg\} \bigg\}   +     CC^{-1} }{\underset{\mathrm{fa} \in \mathrm{FA}}{\mathrm{sup}}   \bigg\{ \underset{y \in \textbf{Y}, z \in \textbf{Z}}{\bigcup} \big\{ p_{\mathrm{fa}, B \longleftrightarrow E} \big\}  \bigg\}        }   \end{align*}

    \begin{align*}       \equiv     \frac{ \underset{\mathrm{EC}, B \longleftrightarrow E}{\mathrm{sup}} \bigg\{  \underset{y \in \textbf{Y}, z \in \textbf{Z}}{\bigcup}    \bigg\{ \textbf{P} \big[  \big\{  \forall \mathscr{C}^{*} \in \textbf{R}, \exists \text{ } \mathrm{ ec } : \mathscr{O} {\tiny \big( y,z  \big) }   > \mathscr{C}^{*}, p_{\mathrm{ec}} > 0   \big\}    \big| \big\{ \text{alphabets } y, z  \big\} \big]     \bigg\} \bigg\}   +     CC^{-1} }{\underset{\mathrm{fa} \in \mathrm{FA}}{\mathrm{sup}}   \bigg\{ \underset{y \in \textbf{Y}, z \in \textbf{Z}}{\bigcup} \big\{ p_{\mathrm{fa}, B \longleftrightarrow E} \big\}  \bigg\}        }   \end{align*}

    \begin{align*}      >                    \underset{\mathrm{fa} \in \mathrm{FA}}{\underset{\mathrm{EC}, B \longleftrightarrow E}{\mathrm{sup}}} \bigg\{  \frac{\underset{y \in \textbf{Y}, z \in \textbf{Z}}{\bigcup}    \big\{ \textbf{P} \big[  \big\{  \forall \mathscr{C}^{*} \in \textbf{R}, \exists \text{ } \mathrm{ ec } : \mathscr{O} {\tiny \big( y,z  \big) }  > \mathscr{C}^{*}, p_{\mathrm{ec}} > 0   \big\}    \big| \big\{ \text{alphabets } y, z  \big\} \big]     \big\}}{\underset{y \in \textbf{Y}, z \in \textbf{Z}}{\bigcup} \big\{ p_{\mathrm{fa}, B \longleftrightarrow E} \big\}  } \bigg\}       \\ +  \bigg[  CC\bigg[ \underset{\mathrm{fa} \in \mathrm{FA}}{\mathrm{sup}}   \bigg\{ \underset{y \in \textbf{Y}, z \in \textbf{Z}}{\bigcup} \big\{ p_{\mathrm{fa}, B \longleftrightarrow E} \big\}  \bigg\}      \bigg] \bigg]^{-1}  \end{align*}

    \begin{align*}     \equiv  \underset{\mathrm{fa} \in \mathrm{FA}}{\underset{\mathrm{EC}, B \longleftrightarrow E}{\mathrm{sup}}} \bigg\{ \underset{y \in \textbf{Y}, z \in \textbf{Z}}{\bigcup}  \frac{    \big\{ \textbf{P} \big[  \big\{  \forall \mathscr{C}^{*} \in \textbf{R}, \exists \text{ } \mathrm{ ec } : \mathscr{O} {\tiny \big( y,z  \big) }  > \mathscr{C}^{*}, p_{\mathrm{ec}} > 0   \big\}    \big| \big\{ \text{alphabets } y, z  \big\} \big]     \big\}}{ \big\{ p_{\mathrm{fa}, B \longleftrightarrow E} \big\}  } \bigg\} \\      +  \bigg[  CC     \bigg[ \underset{\mathrm{fa} \in \mathrm{FA}}{\mathrm{sup}}   \bigg\{ \underset{y \in \textbf{Y}, z \in \textbf{Z}}{\bigcup} \big\{ p_{\mathrm{fa}, B \longleftrightarrow E} \big\}  \bigg\}      \bigg] \bigg]^{-1}  \end{align*}

    \begin{align*}    \overset{(\text{Bayes' Rule})}{\equiv}    \underset{\mathrm{fa} \in \mathrm{FA}}{\underset{\mathrm{EC}, B \longleftrightarrow E}{\mathrm{sup}}} \bigg\{ \underset{y \in \textbf{Y}, z \in \textbf{Z}}{\bigcup}  \frac{    \big\{ \textbf{P} \big[ \big\{ \text{alphabets } y, z  \big\}  \big|   \big\{  \forall \mathscr{C}^{*} \in \textbf{R} , \exists \text{ } \mathrm{ ec } : \mathscr{O} {\tiny \big( y,z  \big) }  > \mathscr{C}^{*}  , p_{\mathrm{ec}} > 0   \big\}       \big]         \big\}  }{ \big\{ p_{\mathrm{fa}, B \longleftrightarrow E} \big\} \big\{ \textbf{P} \big[ \text{alphabets }y,z \big] \big\}   }  \\      \times    \big\{  \textbf{P} \big[    \big\{  \forall \mathscr{C}^{*} \in \textbf{R} , \exists \text{ } \mathrm{ ec } : \mathscr{O} {\tiny \big( y,z  \big) }  > \mathscr{C}^{*}  , p_{\mathrm{ec}} > 0   \big\}               \big]          \big\} \bigg\}      +  \bigg[  CC  \bigg[ \underset{\mathrm{fa} \in \mathrm{FA}}{\mathrm{sup}}   \bigg\{ \underset{y \in \textbf{Y}, z \in \textbf{Z}}{\bigcup} \big\{ p_{\mathrm{fa}, B \longleftrightarrow E} \big\}  \bigg\}      \bigg] \bigg]^{-1}  \end{align*}

    \begin{align*}        >           \underset{| \textbf{Z} | \longrightarrow + \infty}{\underset{| \textbf{Y} | \longrightarrow + \infty}{\mathrm{lim}}}    \bigg\{        \underset{\mathrm{fa} \in \mathrm{FA}}{\underset{\mathrm{EC}, B \longleftrightarrow E}{\mathrm{sup}}} \bigg\{ \underset{y \in \textbf{Y}, z \in \textbf{Z}}{\bigcup}  \frac{    \big\{ \textbf{P} \big[ \big\{ \text{alphabets } y, z  \big\}  \big|   \big\{  \forall \mathscr{C}^{*} \in \textbf{R} , \exists \text{ } \mathrm{ ec } : \mathscr{O} {\tiny \big( y,z  \big) }  > \mathscr{C}^{*}  , p_{\mathrm{ec}} > 0   \big\}       \big]         \big\}  }{ \big\{ p_{\mathrm{fa}, B \longleftrightarrow E} \big\} \big\{ \textbf{P} \big[ \text{alphabets }y,z \big] \big\}   }  \\      \times   \big\{  \textbf{P} \big[    \big\{  \forall \mathscr{C}^{*} \in \textbf{R} , \exists \text{ } \mathrm{ ec } : \mathscr{O} {\tiny \big( y,z  \big) }  > \mathscr{C}^{*}  , p_{\mathrm{ec}} > 0   \big\}               \big]          \big\} \bigg\}           +  \bigg[  CC   \bigg[ \underset{\mathrm{fa} \in \mathrm{FA}}{\mathrm{sup}}   \bigg\{ \underset{y \in \textbf{Y}, z \in \textbf{Z}}{\bigcup} \big\{ p_{\mathrm{fa}, B \longleftrightarrow E} \big\}  \bigg\}      \bigg] \bigg]^{-1}    \bigg\}  \end{align*}

    \begin{align*}         \equiv   \underset{| \textbf{Z} | \longrightarrow + \infty}{\underset{| \textbf{Y} | \longrightarrow + \infty}{\mathrm{lim}}}     \bigg\{     \underset{\mathrm{fa} \in \mathrm{FA}}{\underset{\mathrm{EC}, B \longleftrightarrow E}{\mathrm{sup}}} \bigg\{ \underset{y \in \textbf{Y}, z \in \textbf{Z}}{\bigcup}  \frac{   \big\{ \textbf{P} \big[ \big\{ \text{alphabets } y, z  \big\}  \big|   \big\{  \forall \mathscr{C}^{*} \in \textbf{R} , \exists \text{ } \mathrm{ ec } : \mathscr{O} {\tiny \big( y,z  \big) }  > \mathscr{C}^{*}  , p_{\mathrm{ec}} > 0   \big\}       \big]         \big\} }{ \big\{ p_{\mathrm{fa}, B \longleftrightarrow E} \big\} \big\{ \textbf{P} \big[ \text{alphabets }y,z \big] \big\}   }   \\      \times   \big\{  \textbf{P} \big[    \big\{  \forall \mathscr{C}^{*} \in \textbf{R} , \exists \text{ } \mathrm{ ec } : \mathscr{O} {\tiny \big( y,z  \big) }  > \mathscr{C}^{*}  , p_{\mathrm{ec}} > 0   \big\}               \big]          \big\} \bigg\}     +  \underset{| \textbf{Z} | \longrightarrow + \infty}{\underset{| \textbf{Y} | \longrightarrow + \infty}{\mathrm{lim}}}    \bigg[  CC      \bigg[ \underset{\mathrm{fa} \in \mathrm{FA}}{\mathrm{sup}}   \bigg\{ \underset{y \in \textbf{Y}, z \in \textbf{Z}}{\bigcup} \big\{ p_{\mathrm{fa}, B \longleftrightarrow E} \big\}  \bigg\}      \bigg] \bigg]^{-1}    \end{align*}

    \begin{align*}       \overset{\{  CC \longrightarrow +\infty \}}{>}   \underset{| \textbf{Z} | \longrightarrow + \infty}{\underset{| \textbf{Y} | \longrightarrow + \infty}{\mathrm{lim}}}     \bigg\{     \underset{\mathrm{fa} \in \mathrm{FA}}{\underset{\mathrm{EC}, B \longleftrightarrow E}{\mathrm{sup}}} \bigg\{ \underset{y \in \textbf{Y}, z \in \textbf{Z}}{\bigcup}  \frac{   \big\{   \textbf{P} \big[ \big\{ \text{alphabets } y, z  \big\}  \big|   \big\{  \forall \mathscr{C}^{*} \in \textbf{R} , \exists \text{ } \mathrm{ ec } : \mathscr{O} {\tiny \big( y,z  \big) }   > \mathscr{C}^{*}  , p_{\mathrm{ec}} > 0   \big\}       \big]             \big\} }{ \big\{ p_{\mathrm{fa}, B \longleftrightarrow E} \big\} \big\{ \textbf{P} \big[ \text{alphabets }y,z \big] \big\}   }  \\ \times   \big\{  \textbf{P} \big[    \big\{  \forall \mathscr{C}^{*} \in \textbf{R} , \exists \text{ } \mathrm{ ec } : \mathscr{O} {\tiny \big( y,z  \big) }  > \mathscr{C}^{*}  , p_{\mathrm{ec}} > 0   \big\}               \big]          \big\} \bigg\}   \bigg\}     \tag{$*-10$}         \text{. }
\end{align*} }

\noindent To obtain the final desired lower bound, observe,

{\small \begin{align*}
 (*-10)  \end{align*}

    \begin{align*}     \equiv     \underset{| \textbf{Z} | \longrightarrow + \infty}{\underset{| \textbf{Y} | \longrightarrow + \infty}{\mathrm{lim}}}     \bigg\{     \underset{\mathrm{fa} \in \mathrm{FA}}{\underset{\mathrm{EC}, B \longleftrightarrow E}{\mathrm{sup}}} \bigg\{ \underset{y \in \textbf{Y}, z \in \textbf{Z}}{\bigcup}  \frac{   \big\{   \textbf{P} \big[ \big\{ \text{alphabets } y, z  \big\}  \big|   \big\{  \forall \mathscr{C}^{*} \in \textbf{R} , \exists \text{ } \mathrm{ ec } : \mathscr{O} {\tiny \big( y,z  \big) }  > \mathscr{C}^{*}  , p_{\mathrm{ec}} > 0   \big\}       \big]             \big\} }{ \big\{ p_{\mathrm{fa}, B \longleftrightarrow E} \big\} \big\{ \textbf{P} \big[ \text{alphabets }y,z \big] \big\}   } \bigg\} \bigg\}    \\ \times   \underset{| \textbf{Z} | \longrightarrow + \infty}{\underset{| \textbf{Y} | \longrightarrow + \infty}{\mathrm{lim}}} \bigg\{  \underset{\mathrm{fa} \in \mathrm{FA}}{\underset{\mathrm{EC}, B \longleftrightarrow E}{\mathrm{sup}}}  \bigg\{    \big\{  \textbf{P} \big[    \big\{  \forall \mathscr{C}^{*} \in \textbf{R} , \exists \text{ } \mathrm{ ec } : \mathscr{O} {\tiny \big( y,z  \big) }  > \mathscr{C}^{*}  , p_{\mathrm{ec}} > 0   \big\}               \big]          \big\} \bigg\}   \bigg\}        \text{, } \tag{$*-11$}
\end{align*}
 } 

\noindent from the fact that,

{\small \begin{align*}
  \underset{| \textbf{Z} | \longrightarrow + \infty}{\underset{| \textbf{Y} | \longrightarrow + \infty}{\mathrm{lim}}}     \bigg\{     \underset{\mathrm{fa} \in \mathrm{FA}}{\underset{\mathrm{EC}, B \longleftrightarrow E}{\mathrm{sup}}} \bigg\{ \underset{y \in \textbf{Y}, z \in \textbf{Z}}{\bigcup}  \frac{   \big\{   \textbf{P} \big[ \big\{ \text{alphabets } y, z  \big\}  \big|   \big\{  \forall \mathscr{C}^{*} \in \textbf{R} , \exists \text{ } \mathrm{ ec } : \mathscr{O} {\tiny \big( y,z  \big) }  > \mathscr{C}^{*}  , p_{\mathrm{ec}} > 0   \big\}       \big]             \big\} }{ \big\{ p_{\mathrm{fa}, B \longleftrightarrow E} \big\} \big\{ \textbf{P} \big[ \text{alphabets }y,z \big] \big\}   } \bigg\} \bigg\}    \end{align*}

 \begin{align*}  \longrightarrow     \mathcal{C}^{*} < + \infty      \text{, }  \end{align*}

 \begin{align*}        \underset{| \textbf{Z} | \longrightarrow + \infty}{\underset{| \textbf{Y} | \longrightarrow + \infty}{\mathrm{lim}}} \bigg\{  \underset{\mathrm{fa} \in \mathrm{FA}}{\underset{\mathrm{EC}, B \longleftrightarrow E}{\mathrm{sup}}} \bigg\{  \underset{y \in \textbf{Y}, z \in \textbf{Z}}{\bigcup}  \bigg\{    \big\{  \textbf{P} \big[    \big\{  \forall \mathscr{C}^{*} \in \textbf{R} , \exists \text{ } \mathrm{ ec } : \mathscr{O} {\tiny \big( y,z  \big) }  > \mathscr{C}^{*}  , p_{\mathrm{ec}} > 0   \big\}               \big]          \big\} \bigg\}   \bigg\} \bigg\}     \longrightarrow    \mathcal{C}^{**} < + \infty       \text{. }
\end{align*} }

\noindent Hence, as a result of the two above limits being finite, one has that,

{\small \begin{align*}
 (*-11) \longrightarrow \mathcal{C}^{**}  \end{align*}

 \begin{align*} \times  \underset{| \textbf{Z} | \longrightarrow + \infty}{\underset{| \textbf{Y} | \longrightarrow + \infty}{\mathrm{lim}}}     \bigg\{     \underset{\mathrm{fa} \in \mathrm{FA}}{\underset{\mathrm{EC}, B \longleftrightarrow E}{\mathrm{sup}}}  \bigg\{ \underset{y \in \textbf{Y}, z \in \textbf{Z}}{\bigcup}  \frac{   \big\{   \textbf{P} \big[ \big\{ \text{alphabets } y, z  \big\}  \big|   \big\{  \forall \mathscr{C}^{*} \in \textbf{R} , \exists \text{ } \mathrm{ ec } : \mathscr{O} {\tiny \big( y,z  \big) }  > \mathscr{C}^{*}  , p_{\mathrm{ec}} > 0   \big\}       \big]             \big\} }{ \big\{ p_{\mathrm{fa}, B \longleftrightarrow E} \big\} \big\{ \textbf{P} \big[ \text{alphabets }y,z \big] \big\}   } \bigg\} \bigg\}   \end{align*}

 \begin{align*}   \overset{(\text{Bayes' Rule})}{= }                   \mathcal{C}^{**}   \end{align*}

    \begin{align*}  \times                           \underset{| \textbf{Z} | \longrightarrow + \infty}{\underset{| \textbf{Y} | \longrightarrow + \infty}{\mathrm{lim}}}     \bigg\{     \underset{\mathrm{fa} \in \mathrm{FA}}{\underset{\mathrm{EC}, B \longleftrightarrow E}{\mathrm{sup}}}  \bigg\{ \underset{y \in \textbf{Y}, z \in \textbf{Z}}{\bigcup}           \frac{ \big\{   \textbf{P} \big[   \big\{  \forall \mathscr{C}^{*} \in \textbf{R} , \exists \text{ } \mathrm{ ec } : \mathscr{O} {\tiny \big( y,z  \big) }  > \mathscr{C}^{*}  , p_{\mathrm{ec}} > 0   \big\}     \big|     \big\{ \text{alphabets } y, z  \big\}   \big]             \big\}   }{ \textbf{P} \big[ \big\{ \text{alphabets } y, z  \big\}  \big]  }          \end{align*}

    \begin{align*}   \times \textbf{P} \big[  \big\{  \forall \mathscr{C}^{*} \in \textbf{R} , \exists \text{ } \mathrm{ ec } : \mathscr{O} {\tiny \big( y,z  \big)  } > \mathscr{C}^{*}  , p_{\mathrm{ec}} > 0  \big\} \big]        \big\{ p_{\mathrm{fa}, B \longleftrightarrow E} \big\}^{-1}  \\ \times \big\{ \textbf{P} \big[ \text{alphabets }y,z \big] \big\}^{-1}      \bigg\} \bigg\}   \\    \overset{(**)}{=}              \underset{| \textbf{Z} | \longrightarrow + \infty}{\underset{| \textbf{Y} | \longrightarrow + \infty}{\mathrm{lim}}}     \bigg\{     \underset{\mathrm{fa} \in \mathrm{FA}}{\underset{\mathrm{EC}, B \longleftrightarrow E}{\mathrm{sup}}}  \bigg\{ \underset{y \in \textbf{Y}, z \in \textbf{Z}}{\bigcup}           \frac{ \big\{   \textbf{P} \big[   \big\{  \forall \mathscr{C}^{*} \in \textbf{R} , \exists \text{ } \mathrm{ ec } : \mathscr{O} {\tiny \big( y,z  \big) }  > \mathscr{C}^{*}  , p_{\mathrm{ec}} > 0   \big\}     \big|     \big\{ \text{alphabets } y, z  \big\}   \big]             \big\}   }{ \textbf{P} \big[ \big\{ \text{alphabets } y, z  \big\}  \big]  }   \bigg\} \bigg\}           \end{align*}
 
 \begin{align*}   \times   \underset{| \textbf{Z} | \longrightarrow + \infty}{\underset{| \textbf{Y} | \longrightarrow + \infty}{\mathrm{lim}}} \bigg\{ {\underset{\mathrm{EC}, B \longleftrightarrow E}{\mathrm{sup}}}    \bigg\{   \textbf{P} \big[  \big\{  \forall \mathscr{C}^{*} \in \textbf{R} , \exists \text{ } \mathrm{ ec } : \mathscr{O} {\tiny \big( y,z  \big) }  > \mathscr{C}^{*}  , p_{\mathrm{ec}} > 0  \big\} \big]        \big\{ p_{\mathrm{fa}, B \longleftrightarrow E} \big\}^{-1}  \big\{ \textbf{P} \big[ \text{alphabets }y,z \big] \big\}^{-1}    \bigg\} \bigg\}  \end{align*}

 \begin{align*}       \overset{(***)}{=}   \underset{| \textbf{Z} | \longrightarrow + \infty}{\underset{| \textbf{Y} | \longrightarrow + \infty}{\mathrm{lim}}}     \bigg\{     \underset{\mathrm{fa} \in \mathrm{FA}}{\underset{\mathrm{EC}, B \longleftrightarrow E}{\mathrm{sup}}}  \bigg\{ \underset{y \in \textbf{Y}, z \in \textbf{Z}}{\bigcup}           \frac{ \big\{   \textbf{P} \big[   \big\{  \forall \mathscr{C}^{*} \in \textbf{R} , \exists \text{ } \mathrm{ ec } : \mathscr{O} {\tiny \big( y,z  \big) }   > \mathscr{C}^{*}  , p_{\mathrm{ec}} > 0   \big\}     \big|     \big\{ \text{alphabets } y, z  \big\}   \big]             \big\}   }{ \textbf{P} \big[ \big\{ \text{alphabets } y, z  \big\}  \big]  }   \bigg\} \bigg\}       \end{align*}

 \begin{align*}       \times   \underset{| \textbf{Z} | \longrightarrow + \infty}{\underset{| \textbf{Y} | \longrightarrow + \infty}{\mathrm{lim}}}   \bigg\{  {\underset{\mathrm{EC}, B \longleftrightarrow E}{\mathrm{sup}}}  \bigg\{   \textbf{P} \big[  \big\{  \forall \mathscr{C}^{*} \in \textbf{R} , \exists \text{ } \mathrm{ ec } : \mathscr{O} {\tiny \big( y,z  \big) }  > \mathscr{C}^{*}  , p_{\mathrm{ec}} > 0  \big\} \big]    \bigg\}     \bigg\}   \underset{| \textbf{Z} | \longrightarrow + \infty}{\underset{| \textbf{Y} | \longrightarrow + \infty}{\mathrm{lim}}}  \big\{ p_{\mathrm{fa}, B \longleftrightarrow E}  \textbf{P} \big[ \text{alphabets }y,z \big] \big\}^{-1}     \text{, }  \\                \tag{$*-12$}
\end{align*} }

\noindent where, in $(**)$, we made use of the observation that,

{\small \begin{align*}
   \big\{   \mathcal{C}^{*} > 1 \big\}  \Longleftrightarrow    \bigg\{    \mathcal{C}^{*}  \equiv   \underset{y \in \textbf{Y}, z \in \textbf{Z}}{\bigcup} \mathcal{C}^{*}_{y,z} \equiv     \underset{y \in \textbf{Y}, z \in \textbf{Z}}{\sum} \mathcal{C}^{*}_{y,z} > \big( c^{*}_{y, z} \big)^{| \textbf{Y} | + | \textbf{Z} |}    > 1  \bigg\} \Longleftrightarrow \bigg\{ c^{*}_{y, z} < 1,   \forall y \in \textbf{Y}, z \in \textbf{Z} \bigg\}  \text{, } \end{align*} }

\noindent to remove the prefactor of $\mathcal{C}^{**}$ which is $>1$, in addition to the fact that,

 {\small \begin{align*}
          \underset{| \textbf{Z} | \longrightarrow + \infty}{\underset{| \textbf{Y} | \longrightarrow + \infty}{\mathrm{lim}}}     \bigg\{     \underset{\mathrm{fa} \in \mathrm{FA}}{\underset{\mathrm{EC}, B \longleftrightarrow E}{\mathrm{sup}}}  \bigg\{ \underset{y \in \textbf{Y}, z \in \textbf{Z}}{\bigcup}           \frac{ \big\{   \textbf{P} \big[   \big\{  \forall \mathscr{C}^{*} \in \textbf{R} , \exists \text{ } \mathrm{ ec } : \mathscr{O} {\tiny \big( y,z  \big) }  > \mathscr{C}^{*}  , p_{\mathrm{ec}} > 0   \big\}     \big|     \big\{ \text{alphabets } y, z  \big\}   \big]             \big\}   }{ \textbf{P} \big[ \big\{ \text{alphabets } y, z  \big\}  \big]  }   \bigg\} \bigg\}     \end{align*}

 \begin{align*}   \longrightarrow \mathcal{C}^{***} < + \infty  \text{, } \end{align*}
 
 \begin{align*}  \underset{| \textbf{Z} | \longrightarrow + \infty}{\underset{| \textbf{Y} | \longrightarrow + \infty}{\mathrm{lim}}} \bigg\{ {\underset{\mathrm{EC}, B \longleftrightarrow E}{\mathrm{sup}}}   \bigg\{    \underset{y \in \textbf{Y}, z \in \textbf{Z}}{\bigcup}  \bigg\{ \textbf{P} \big[  \big\{  \forall \mathscr{C}^{*} \in \textbf{R} , \exists \text{ } \mathrm{ ec } : \mathscr{O} {\tiny \big( y,z  \big) }  > \mathscr{C}^{*}  , p_{\mathrm{ec}} > 0  \big\} \big]        \big\{ p_{\mathrm{fa}, B \longleftrightarrow E} \big\}^{-1} \big\{ \textbf{P} \big[ \text{alphabets }y,z \big] \big\}^{-1}    \bigg\}  \bigg\} \bigg\}   \end{align*}

 \begin{align*}  \longrightarrow \mathcal{C}^{****} < + \infty \text{, }
\end{align*} }

\noindent while, in (***), we made use of the observation that,

{\small \begin{align*}
     \underset{| \textbf{Z} | \longrightarrow + \infty}{\underset{| \textbf{Y} | \longrightarrow + \infty}{\mathrm{lim}}}   \bigg\{  {\underset{\mathrm{EC}, B \longleftrightarrow E}{\mathrm{sup}}} \bigg\{  \underset{y \in \textbf{Y}, z \in \textbf{Z}}{\bigcup}  \bigg\{   \textbf{P} \big[  \big\{  \forall \mathscr{C}^{*} \in \textbf{R} , \exists \text{ } \mathrm{ ec } : \mathscr{O} {\tiny \big( y,z  \big) }  > \mathscr{C}^{*}  , p_{\mathrm{ec}} > 0  \big\} \big]         \bigg\} \bigg\} \bigg\}        \longrightarrow \mathcal{C}^{*****} < + \infty   \text{, } \end{align*}

     \begin{align*} \underset{| \textbf{Z} | \longrightarrow + \infty}{\underset{| \textbf{Y} | \longrightarrow + \infty}{\mathrm{lim}}}   \bigg\{  \underset{y \in \textbf{Y}, z \in \textbf{Z}}{\bigcup}  \big\{ p_{\mathrm{fa}, B \longleftrightarrow E} \cdot \textbf{P} \big[ \text{alphabets }y,z \big] \big\}^{-1} \bigg\}     \longrightarrow \mathcal{C}^{******} < + \infty         \text{. }
\end{align*}
 }

\noindent To conclude the argument corresponding to the second stochastic domination, observe that it suffices to argue that,

{\small \begin{align*}
        \underset{| \textbf{Z} | \longrightarrow + \infty}{\underset{| \textbf{Y} | \longrightarrow + \infty}{\mathrm{lim}}}     \bigg\{     \underset{\mathrm{fa} \in \mathrm{FA}}{\underset{\mathrm{EC}, B \longleftrightarrow E}{\mathrm{sup}}}  \bigg\{ \underset{y \in \textbf{Y}, z \in \textbf{Z}}{\bigcup}           \frac{ \big\{   \textbf{P} \big[   \big\{  \forall \mathscr{C}^{*} \in \textbf{R} , \exists \text{ } \mathrm{ ec } : \mathscr{O} {\tiny \big( y,z  \big)  }  > \mathscr{C}^{*}  , p_{\mathrm{ec}} > 0   \big\}     \big|     \big\{ \text{alphabets } y, z  \big\}   \big]             \big\}   }{ \textbf{P} \big[ \big\{ \text{alphabets } y, z  \big\}  \big]  }   \bigg\} \bigg\}      > 0        \text{, } \end{align*}

     \begin{align*}    \underset{| \textbf{Z} | \longrightarrow + \infty}{\underset{| \textbf{Y} | \longrightarrow + \infty}{\mathrm{lim}}}   \bigg\{  {\underset{\mathrm{EC}, B \longleftrightarrow E}{\mathrm{sup}}}  \bigg\{  \underset{y \in \textbf{Y}, z \in \textbf{Z}}{\bigcup} \bigg\{  \textbf{P} \big[  \big\{  \forall \mathscr{C}^{*} \in \textbf{R} , \exists \text{ } \mathrm{ ec } : \mathscr{O} {\tiny \big( y,z  \big) }  > \mathscr{C}^{*}  , p_{\mathrm{ec}} > 0  \big\} \big]    \bigg\}   \bigg\}   \bigg\}  > 0   \text{, } \end{align*}

     \begin{align*} \underset{| \textbf{Z} | \longrightarrow + \infty}{\underset{| \textbf{Y} | \longrightarrow + \infty}{\mathrm{lim}}}  \bigg\{  \underset{y \in \textbf{Y}, z \in \textbf{Z}}{\bigcup}   \big\{ p_{\mathrm{fa}, B \longleftrightarrow E}  \textbf{P} \big[ \text{alphabets }y,z \big] \big\}^{-1}  \bigg\} \equiv   \underset{| \textbf{Z} | \longrightarrow + \infty}{\underset{| \textbf{Y} | \longrightarrow + \infty}{\mathrm{lim}}}  \bigg\{  \underset{y \in \textbf{Y}, z \in \textbf{Z}}{\bigcup}      p_{\mathrm{fa}, B \longleftrightarrow E}^{-1}  \bigg\}   \end{align*}

 \begin{align*}  \times     \underset{| \textbf{Z} | \longrightarrow + \infty}{\underset{| \textbf{Y} | \longrightarrow + \infty}{\mathrm{lim}}}   \bigg\{  \underset{y \in \textbf{Y}, z \in \textbf{Z}}{\bigcup} \textbf{P} \big[ \text{alphabets }y,z \big]^{-1} \bigg\} > 0     \text{. }
\end{align*} }

\noindent The final inequality provided holds, namely that the product of the two above limits, as the cardinality of the alphabets for Bob and Eve approach $+ \infty$, as,

{\small \begin{align*}
 \underset{| \textbf{Z} | \longrightarrow + \infty}{\underset{| \textbf{Y} | \longrightarrow + \infty}{\mathrm{lim}}}  \bigg\{  \underset{y \in \textbf{Y}, z \in \textbf{Z}}{\bigcup} \big\{     p_{\mathrm{fa}, B \longleftrightarrow E}^{-1}   \big\} \bigg\}  \equiv      \underset{y \in \textbf{Y}, z \in \textbf{Z}}{\bigcup}  \bigg\{  {{ \underset{| \textbf{Z} | \longrightarrow + \infty}{\underset{| \textbf{Y} | \longrightarrow + \infty}{\mathrm{lim}}} 1}} \bigg/ { \underset{| \textbf{Z} | \longrightarrow + \infty}{\underset{| \textbf{Y} | \longrightarrow + \infty}{\mathrm{lim}}} \big\{  p_{\mathrm{fa}, B \longleftrightarrow E} \big\} } \bigg\} \end{align*}
        
        \begin{align*}   \equiv   {{ \underset{| \textbf{Z} | \longrightarrow + \infty}{\underset{| \textbf{Y} | \longrightarrow + \infty}{\mathrm{lim}}} 1}} \bigg/ { \underset{| \textbf{Z} | \longrightarrow + \infty}{\underset{| \textbf{Y} | \longrightarrow + \infty}{\mathrm{lim}}} \bigg\{  \underset{y \in \textbf{Y}, z \in \textbf{Z}}{\bigcup} p_{\mathrm{fa}, B \longleftrightarrow E} \bigg\} }   \equiv   {1} \bigg/ { \underset{| \textbf{Z} | \longrightarrow + \infty}{\underset{| \textbf{Y} | \longrightarrow + \infty}{\mathrm{lim}}} \bigg\{  \underset{y \in \textbf{Y}, z \in \textbf{Z}}{\bigcup} p_{\mathrm{fa}, B \longleftrightarrow E} \bigg\} }      \end{align*}

    \begin{align*}   \longrightarrow \mathscr{C}^{*} < + \infty \Longleftrightarrow   p_{\mathrm{fa}, B \longleftrightarrow E} > 0  \text{, }   \end{align*}

 \begin{align*}      \underset{| \textbf{Z} | \longrightarrow + \infty}{\underset{| \textbf{Y} | \longrightarrow + \infty}{\mathrm{lim}}} \bigg\{ \underset{y \in \textbf{Y}, z \in \textbf{Z}}{\bigcup}  \textbf{P} \big[ \text{alphabets }y,z \big]^{-1} \bigg\}  \equiv   \underset{y \in \textbf{Y}, z \in \textbf{Z}}{\bigcup}  \bigg\{   { \underset{| \textbf{Z} | \longrightarrow + \infty}{\underset{| \textbf{Y} | \longrightarrow + \infty}{\mathrm{lim}}} 1} \bigg/ { \underset{| \textbf{Z} | \longrightarrow + \infty}{\underset{| \textbf{Y} | \longrightarrow + \infty}{\mathrm{lim}}}  \textbf{P} \big[ \text{alphabets }y,z \big]} \bigg\}  \end{align*}

 \begin{align*}   \equiv {\underset{| \textbf{Z} | \longrightarrow + \infty}{\underset{| \textbf{Y} | \longrightarrow + \infty}{\mathrm{lim}}}   1} \bigg/ { \underset{| \textbf{Z} | \longrightarrow + \infty}{\underset{| \textbf{Y} | \longrightarrow + \infty}{\mathrm{lim}}} \bigg\{ \underset{y \in \textbf{Y}, z \in \textbf{Z}}{\bigcup}  \textbf{P} \big[ \text{alphabets }y,z \big] \bigg\} }  \equiv  1 \bigg/ { \underset{| \textbf{Z} | \longrightarrow + \infty}{\underset{| \textbf{Y} | \longrightarrow + \infty}{\mathrm{lim}}} \bigg\{ \underset{y \in \textbf{Y}, z \in \textbf{Z}}{\bigcup}  \textbf{P} \big[ \text{alphabets }y,z \big] \bigg\} } \end{align*}

    \begin{align*} \longrightarrow \bigg\{  \mathscr{C}^{**} < + \infty \bigg\}  \Longleftrightarrow  \bigg\{ \textbf{P} \big[ \text{alphabets }y,z \big] \neq 0  \bigg\} \text{, }
\end{align*} }

\noindent for  constants $\mathscr{C}^{*} \neq \mathscr{C}^{**}$, each of which are strictly positive, satisfying,

{\small \begin{align*}
\mathscr{C}^* \mathscr{C}^{**} > 0 \text{.}
\end{align*} }

\noindent The second inequality provided above,

{\small \begin{align*}
   \underset{| \textbf{Z} | \longrightarrow + \infty}{\underset{| \textbf{Y} | \longrightarrow + \infty}{\mathrm{lim}}}   \bigg\{  {\underset{\mathrm{EC}, B \longleftrightarrow E}{\mathrm{sup}}}  \bigg\{  \underset{y \in \textbf{Y}, z \in \textbf{Z}}{\bigcup} \bigg\{  \textbf{P} \big[  \big\{  \forall \mathscr{C}^{*} \in \textbf{R} , \exists \text{ } \mathrm{ ec } : \mathscr{O} {\tiny \big( y,z  \big) }   > \mathscr{C}^{*}  , p_{\mathrm{ec}} > 0  \big\} \big]    \bigg\}   \bigg\}   \bigg\} > 0   \text{, } \tag{$\mathscr{P}-2$}
\end{align*} }

\noindent independently of the limit as $\big| \textbf{Y} \big|, \big| \textbf{Z} \big| \longrightarrow + \infty$,

{\small \begin{align*}
 {\underset{\mathrm{EC}, B \longleftrightarrow E}{\mathrm{sup}}}  \bigg\{  \underset{y \in \textbf{Y}, z \in \textbf{Z}}{\bigcup} \bigg\{  \textbf{P} \big[  \big\{  \forall \mathscr{C}^{*} \in \textbf{R} , \exists \text{ } \mathrm{ ec } : \mathscr{O} {\tiny \big( y,z  \big) }  > \mathscr{C}^{*}  , p_{\mathrm{ec}} > 0  \big\} \big]    \bigg\}   \bigg\}     \text{, }
\end{align*} }

\noindent can be manipulated to show the the second desired inequality above holds. More specifically, the fact that,

{\small \begin{align*}
 \underset{\mathscr{C}^{***} \longrightarrow 0^{+}}{\underset{| \textbf{Z} | \longrightarrow + \infty}{\underset{| \textbf{Y} | \longrightarrow + \infty}{\mathrm{lim}}}}     \bigg\{     \underset{\mathrm{fa} \in \mathrm{FA}}{\underset{\mathrm{EC}, B \longleftrightarrow E}{\mathrm{sup}}}  \bigg\{        \textbf{P} \big[           \big\{  \forall \mathscr{C}^{***} > \mathscr{C}^{**} > \mathscr{C}^{*} \in \textbf{R} , \exists \text{ } \mathrm{ ec } : \mathscr{O} {\tiny \big( }\textbf{Y},\textbf{Z}  {\tiny \big) }  > \mathscr{C}^{***}  , p_{\mathrm{ec}} > 0   \big\}  \big|   \big\{ \text{alphabets } \textbf{Y}, \textbf{Z}  \big\} \big]             \big\} \bigg\}  \bigg\}   \end{align*}

 \begin{align*} >  \underset{\mathscr{C}^{**} \longrightarrow 0^{+}}{\underset{| \textbf{Z} | \longrightarrow + \infty}{\underset{| \textbf{Y} | \longrightarrow + \infty}{\mathrm{lim}}}}     \bigg\{     \underset{\mathrm{fa} \in \mathrm{FA}}{\underset{\mathrm{EC}, B \longleftrightarrow E}{\mathrm{sup}}}  \bigg\{        \textbf{P} \big[           \big\{  \forall \mathscr{C}^{**} > \mathscr{C}^{*} \in \textbf{R} , \exists \text{ } \mathrm{ ec } : \mathscr{O} {\tiny \big(} \textbf{Y},\textbf{Z}  {\tiny \big) }   > \mathscr{C}^{**}  , p_{\mathrm{ec}} > 0   \big\}  \big|   \big\{ \text{alphabets } \textbf{Y}, \textbf{Z}  \big\} \big]             \big\} \bigg\}  \bigg\}   \text{, }
\end{align*}
 }

\noindent implies the existence of a suitable threshold for the overlap function, $\mathscr{C}^{***}$, for which,

{\small \begin{align*}
\underset{\mathscr{C}^{***} \longrightarrow 0^{+}}{\underset{| \textbf{Z} | \longrightarrow + \infty}{\underset{| \textbf{Y} | \longrightarrow + \infty}{\mathrm{lim}}}}     \bigg\{     \underset{\mathrm{fa} \in \mathrm{FA}}{\underset{\mathrm{EC}, B \longleftrightarrow E}{\mathrm{sup}}}  \bigg\{        \textbf{P} \big[           \big\{  \forall \mathscr{C}^{***} > \mathscr{C}^{**} > \mathscr{C}^{*} \in \textbf{R} , \exists \text{ } \mathrm{ ec } : \mathscr{O} {\tiny \big(}  \textbf{Y},\textbf{Z}  {\tiny \big) }   > \mathscr{C}^{***}  , p_{\mathrm{ec}} > 0   \big\}  \big|   \big\{ \text{alphabets } \textbf{Y}, \textbf{Z}  \big\} \big]             \big\} \bigg\}  \bigg\}  \end{align*}

\begin{align*} \neq 0 \text{,}
\end{align*} } 

\noindent and hence for which,

{\small \begin{align*}
C_2 \equiv \underset{\mathscr{C}^{***} \longrightarrow 0^{+}}{\underset{| \textbf{Z} | \longrightarrow + \infty}{\underset{| \textbf{Y} | \longrightarrow + \infty}{\mathrm{lim}}}}     \bigg\{     \underset{\mathrm{fa} \in \mathrm{FA}}{\underset{\mathrm{EC}, B \longleftrightarrow E}{\mathrm{sup}}}  \bigg\{        \textbf{P} \big[           \big\{  \forall \mathscr{C}^{***} > \mathscr{C}^{**} > \mathscr{C}^{*} \in \textbf{R} , \exists \text{ } \mathrm{ ec } : \mathscr{O} {\tiny \big(} \textbf{Y},\textbf{Z}  {\tiny \big) }   > \mathscr{C}^{***}   , p_{\mathrm{ec}} > 0   \big\}  \big|   \big\{ \text{alphabets } \textbf{Y}, \textbf{Z}  \big\} \big]             \big\} \bigg\}  \bigg\}  \end{align*}

\begin{align*}     > 0 \text{,}
\end{align*}
 }

\noindent which is the desired strictly positive constant. Hence $(\mathscr{P}-2)$ holds.

\bigskip

\noindent Finally, the first inequality provided above can be written as,

{\small \begin{align*}
       \underset{| \textbf{Z} | \longrightarrow + \infty}{\underset{| \textbf{Y} | \longrightarrow + \infty}{\mathrm{lim}}}     \bigg\{     \underset{\mathrm{fa} \in \mathrm{FA}}{\underset{\mathrm{EC}, B \longleftrightarrow E}{\mathrm{sup}}}  \bigg\{ \underset{y \in \textbf{Y}, z \in \textbf{Z}}{\bigcup}           \frac{ \big\{   \textbf{P} \big[   \big\{  \forall \mathscr{C}^{*} \in \textbf{R} , \exists \text{ } \mathrm{ ec } : \mathscr{O} {\tiny \big( y,z  \big) }   > \mathscr{C}^{*}  , p_{\mathrm{ec}} > 0   \big\}     \big|     \big\{ \text{alphabets } y, z  \big\}   \big]             \big\}   }{ \textbf{P} \big[ \big\{ \text{alphabets } y, z  \big\}  \big]  }   \bigg\} \bigg\} \end{align*}

       \begin{align*} \equiv           \frac{ \underset{| \textbf{Z} | \longrightarrow + \infty}{\underset{| \textbf{Y} | \longrightarrow + \infty}{\mathrm{lim}}}     \bigg\{     \underset{\mathrm{fa} \in \mathrm{FA}}{\underset{\mathrm{EC}, B \longleftrightarrow E}{\mathrm{sup}}}  \bigg\{ \underset{y \in \textbf{Y}, z \in \textbf{Z}}{\bigcup}          \big\{   \textbf{P} \big[   \big\{  \forall \mathscr{C}^{*} \in \textbf{R} , \exists \text{ } \mathrm{ ec } : \mathscr{O} {\tiny \big( y,z  \big)}   > \mathscr{C}^{*}  , p_{\mathrm{ec}} > 0   \big\}     \big|     \big\{ \text{alphabets } y, z  \big\}   \big]             \big\}          \bigg\} \bigg\}    }{ \underset{| \textbf{Z} | \longrightarrow + \infty}{\underset{| \textbf{Y} | \longrightarrow + \infty}{\mathrm{lim}}}     \bigg\{      \underset{y \in \textbf{Y}, z \in \textbf{Z}}{\bigcup}      \textbf{P} \big[ \big\{ \text{alphabets } y, z  \big\}  \big]   \bigg\}  }       \text{. }
\end{align*}
 }

\noindent The probability in the denominator above is strictly positive. To argue that the probability in the numerator above is also strictly positive, observe,

{\small \begin{align*}
        \underset{| \textbf{Z} | \longrightarrow + \infty}{\underset{| \textbf{Y} | \longrightarrow + \infty}{\mathrm{lim}}}     \bigg\{     \underset{\mathrm{fa} \in \mathrm{FA}}{\underset{\mathrm{EC}, B \longleftrightarrow E}{\mathrm{sup}}}  \bigg\{ \underset{y \in \textbf{Y}, z \in \textbf{Z}}{\bigcup}          \big\{   \textbf{P} \big[   \big\{  \forall \mathscr{C}^{*} \in \textbf{R} , \exists \text{ } \mathrm{ ec } : \mathscr{O} {\tiny \big( y,z  \big) }   > \mathscr{C}^{*}  , p_{\mathrm{ec}} > 0   \big\}      \big|     \big\{ \text{alphabets } y, z  \big\}   \big]             \big\}          \bigg\} \bigg\}   \end{align*}
        
        \begin{align*} \overset{(\mathrm{Bayes'})}{\equiv}          \underset{| \textbf{Z} | \longrightarrow + \infty}{\underset{| \textbf{Y} | \longrightarrow + \infty}{\mathrm{lim}}}     \bigg\{     \underset{\mathrm{fa} \in \mathrm{FA}}{\underset{\mathrm{EC}, B \longleftrightarrow E}{\mathrm{sup}}}  \bigg\{ \underset{y \in \textbf{Y}, z \in \textbf{Z}}{\bigcup}      \frac{\textbf{P} \big[   \big\{  \forall \mathscr{C}^{*} \in \textbf{R} , \exists \text{ } \mathrm{ ec } : \mathscr{O} {\tiny \big( y,z  \big) }  > \mathscr{C}^{*}  , p_{\mathrm{ec}} > 0   \big\} \big] }{\textbf{P} \big[   \big\{ \text{alphabets } y, z  \big\}    \big]}          \end{align*}

     \begin{align*}   \times       \big\{   \textbf{P} \big[         \big\{ \text{alphabets } y, z  \big\}  \big|   \big\{  \forall \mathscr{C}^{*} \in \textbf{R} , \exists \text{ } \mathrm{ ec } : \mathscr{O} {\tiny \big( y,z  \big) }   > \mathscr{C}^{*}  , p_{\mathrm{ec}} > 0   \big\}  \big]             \big\}                               \bigg\} \bigg\}       \end{align*}
        
        \begin{align*}                    \overset{(****)}{=}    \underset{| \textbf{Z} | \longrightarrow + \infty}{\underset{| \textbf{Y} | \longrightarrow + \infty}{\mathrm{lim}}}     \bigg\{     \underset{\mathrm{fa} \in \mathrm{FA}}{\underset{\mathrm{EC}, B \longleftrightarrow E}{\mathrm{sup}}}  \bigg\{ \underset{y \in \textbf{Y}, z \in \textbf{Z}}{\bigcup}     \textbf{P} \big[   \big\{  \forall \mathscr{C}^{*} \in \textbf{R} , \exists \text{ } \mathrm{ ec } : \mathscr{O} {\tiny \big( y,z  \big) }  > \mathscr{C}^{*}  , p_{\mathrm{ec}} > 0   \big\} \big] \bigg\} \bigg\}          \end{align*}
        
        \begin{align*}   \times               \underset{| \textbf{Z} | \longrightarrow + \infty}{\underset{| \textbf{Y} | \longrightarrow + \infty}{\mathrm{lim}}}     \bigg\{     \underset{\mathrm{fa} \in \mathrm{FA}}{\underset{\mathrm{EC}, B \longleftrightarrow E}{\mathrm{sup}}}  \bigg\{ \underset{y \in \textbf{Y}, z \in \textbf{Z}}{\bigcup}         \big\{   \textbf{P} \big[         \big\{ \text{alphabets } y, z  \big\}  \big|   \big\{  \forall \mathscr{C}^{*} \in \textbf{R} , \exists \text{ } \mathrm{ ec } : \mathscr{O} {\tiny \big( y,z  \big)  } > \mathscr{C}^{*}   , p_{\mathrm{ec}} > 0   \big\}  \big]             \big\} \bigg\} \bigg\}   \end{align*}

     \begin{align*}   \times    \underset{| \textbf{Z} | \longrightarrow + \infty}{\underset{| \textbf{Y} | \longrightarrow + \infty}{\mathrm{lim}}}       \bigg\{  \underset{y \in \textbf{Y}, z \in \textbf{Z}}{\bigcup}     \big\{  \textbf{P} \big[ \big\{ \text{alphabets } y,z  \big\} \big]   \big\}^{-1}  \bigg\}   \text{,} \end{align*}} 

        \noindent which can be bound below with a suitable, strictly positive constant. The last expression above can be strictly lower bounded by,
        
        {\small \begin{align*}     \mathcal{C}_{*}   \cdot      \underset{| \textbf{Z} | \longrightarrow + \infty}{\underset{| \textbf{Y} | \longrightarrow + \infty}{\mathrm{lim}}}     \bigg\{     \underset{\mathrm{fa} \in \mathrm{FA}}{\underset{\mathrm{EC}, B \longleftrightarrow E}{\mathrm{sup}}}  \bigg\{ \underset{y \in \textbf{Y}, z \in \textbf{Z}}{\bigcup}     \textbf{P} \big[   \big\{  \forall \mathscr{C}^{*} \in \textbf{R} , \exists \text{ } \mathrm{ ec } : \mathscr{O} {\tiny \big( y,z  \big) }   > \mathscr{C}^{*}  , p_{\mathrm{ec}} > 0   \big\} \big] \bigg\} \bigg\}        \end{align*}

     \begin{align*}  \times               \underset{| \textbf{Z} | \longrightarrow + \infty}{\underset{| \textbf{Y} | \longrightarrow + \infty}{\mathrm{lim}}}     \bigg\{     \underset{\mathrm{fa} \in \mathrm{FA}}{\underset{\mathrm{EC}, B \longleftrightarrow E}{\mathrm{sup}}}  \bigg\{ \underset{y \in \textbf{Y}, z \in \textbf{Z}}{\bigcup}         \big\{   \textbf{P} \big[         \big\{ \text{alphabets } y, z  \big\}  \big|   \big\{  \forall \mathscr{C}^{*} \in \textbf{R} , \exists \text{ } \mathrm{ ec } : \mathscr{O} {\tiny \big( y,z  \big) }  > \mathscr{C}^{*}   , p_{\mathrm{ec}} > 0   \big\}  \big]             \big\} \bigg\} \bigg\}   \end{align*}

     \begin{align*} >  \mathcal{C}_{*}  \mathcal{C}_{**}              \text{,}
\end{align*} }

\noindent for,

{\small \begin{align*}
  \bigg\{  + \infty >  \underset{| \textbf{Z} | \longrightarrow + \infty}{\underset{| \textbf{Y} | \longrightarrow + \infty}{\mathrm{lim}}}       \bigg\{  \underset{y \in \textbf{Y}, z \in \textbf{Z}}{\bigcup}     \big\{  \textbf{P} \big[ \big\{ \text{alphabets } y,z  \big\} \big]   \big\}^{-1}  \bigg\}   > \mathcal{C}_{*}   > {100000}^{-1}    \bigg\}       \Longleftrightarrow \big\{ \mathcal{C}_{*} \in \textbf{R} \big\}  \text{, }
\end{align*} }

\noindent and for,

{\small \begin{align*}
            \underset{| \textbf{Z} | \longrightarrow + \infty}{\underset{| \textbf{Y} | \longrightarrow + \infty}{\mathrm{lim}}}     \bigg\{     \underset{\mathrm{fa} \in \mathrm{FA}}{\underset{\mathrm{EC}, B \longleftrightarrow E}{\mathrm{sup}}}  \bigg\{ \underset{y \in \textbf{Y}, z \in \textbf{Z}}{\bigcup}         \big\{   \textbf{P} \big[         \big\{ \text{alphabets } y, z  \big\}  \big|   \big\{  \forall \mathscr{C}^{*} \in \textbf{R} , \exists \text{ } \mathrm{ ec } : \mathscr{O} {\tiny \big( y,z  \big)  }  > \mathscr{C}^{*}    , p_{\mathrm{ec}} > 0   \big\}  \big]             \big\} \bigg\} \bigg\}    \end{align*}
            
            \begin{align*} \overset{(\text{Bayes' Rule})}{\equiv}   \underset{| \textbf{Z} | \longrightarrow + \infty}{\underset{| \textbf{Y} | \longrightarrow + \infty}{\mathrm{lim}}}     \bigg\{     \underset{\mathrm{fa} \in \mathrm{FA}}{\underset{\mathrm{EC}, B \longleftrightarrow E}{\mathrm{sup}}}  \bigg\{ \underset{y \in \textbf{Y}, z \in \textbf{Z}}{\bigcup}         \bigg\{   \textbf{P} \big[           \big\{  \forall \mathscr{C}^{*} \in \textbf{R} , \exists \text{ } \mathrm{ ec } : \mathscr{O} {\tiny \big( y,z  \big) }  > \mathscr{C}^{*}    , p_{\mathrm{ec}} > 0   \big\}  \big|  \big\{ \text{alphabets } y, z  \big\} \big]             \big\} \end{align*}

     \begin{align*}   \times     \big\{ \textbf{P} \big[           \big\{  \forall \mathscr{C}^{*} \in \textbf{R} , \exists \text{ } \mathrm{ ec } : \mathscr{O} \big( y,z  \big)  > \mathscr{C}^{*}  , p_{\mathrm{ec}} > 0   \big\} \big] \big\}^{-1} \textbf{P} \big[ \text{alphabets } y,z \big]                      \bigg\} \bigg\} \bigg\}     \end{align*}
        
        \begin{align*}  >          \mathcal{C}_{*}  \cdot    \underset{| \textbf{Z} | \longrightarrow + \infty}{\underset{| \textbf{Y} | \longrightarrow + \infty}{\mathrm{lim}}}     \bigg\{     \underset{\mathrm{fa} \in \mathrm{FA}}{\underset{\mathrm{EC}, B \longleftrightarrow E}{\mathrm{sup}}}  \bigg\{ \underset{y \in \textbf{Y}, z \in \textbf{Z}}{\bigcup}         \bigg\{   \textbf{P} \big[           \big\{  \forall \mathscr{C}^{*} \in \textbf{R} , \exists \text{ } \mathrm{ ec } : \mathscr{O} \big( y,z  \big)  > \mathscr{C}^{*}    , p_{\mathrm{ec}} > 0   \big\}  \big|  \big\{ \text{alphabets } y, z  \big\} \big]             \big\}       \\ \times     \big\{ \textbf{P} \big[           \big\{  \forall \mathscr{C}^{*} \in \textbf{R} , \exists \text{ } \mathrm{ ec } : \mathscr{O} \big( y,z  \big)  > \mathscr{C}^{*}  , p_{\mathrm{ec}} > 0   \big\} \big] \big\}^{-1} \bigg\} \bigg\} \bigg\}     \end{align*}
        
        \begin{align*}    \overset{(*****)}{\equiv}             \mathcal{C}_{*}  \cdot    \underset{| \textbf{Z} | \longrightarrow + \infty}{\underset{| \textbf{Y} | \longrightarrow + \infty}{\mathrm{lim}}}     \bigg\{     \underset{\mathrm{fa} \in \mathrm{FA}}{\underset{\mathrm{EC}, B \longleftrightarrow E}{\mathrm{sup}}}  \bigg\{ \underset{y \in \textbf{Y}, z \in \textbf{Z}}{\bigcup}         \bigg\{  \textbf{P} \big[           \big\{  \forall \mathscr{C}^{*} \in \textbf{R} , \exists \text{ } \mathrm{ ec } : \mathscr{O} \big( y,z  \big)  > \mathscr{C}^{*}  , p_{\mathrm{ec}} > 0   \big\}   \big|  \big\{ \text{alphabets } y, z  \big\} \big]             \big\}    \bigg\} \bigg\} \bigg\}   \end{align*}

     \begin{align*}   \times    \underset{| \textbf{Z} | \longrightarrow + \infty}{\underset{| \textbf{Y} | \longrightarrow + \infty}{\mathrm{lim}}}     \bigg\{     \underset{\mathrm{fa} \in \mathrm{FA}}{\underset{\mathrm{EC}, B \longleftrightarrow E}{\mathrm{sup}}}  \bigg\{ \underset{y \in \textbf{Y}, z \in \textbf{Z}}{\bigcup}         \bigg\{      \big\{ \textbf{P} \big[           \big\{  \forall \mathscr{C}^{*} \in \textbf{R} , \exists \text{ } \mathrm{ ec } : \mathscr{O} \big( y,z  \big)  > \mathscr{C}^{*}  , p_{\mathrm{ec}} > 0   \big\} \big] \big\}^{-1} \bigg\} \bigg\}    \bigg\}   \text{.} \end{align*} }

      \noindent which can be bound below with a suitable, strictly positive constant. The last expression obtained above can be strictly lower bounded by,   
      
   {\small  \begin{align*}
     \mathcal{C}_{*}^2 \cdot 
              \underset{| \textbf{Z} | \longrightarrow + \infty}{\underset{| \textbf{Y} | \longrightarrow + \infty}{\mathrm{lim}}}     \bigg\{     \underset{\mathrm{fa} \in \mathrm{FA}}{\underset{\mathrm{EC}, B \longleftrightarrow E}{\mathrm{sup}}}  \bigg\{ \underset{y \in \textbf{Y}, z \in \textbf{Z}}{\bigcup}         \bigg\{  \textbf{P} \big[           \big\{  \forall \mathscr{C}^{*} \in \textbf{R} , \exists \text{ } \mathrm{ ec } : \mathscr{O} {\tiny \big( y,z  \big) }   > \mathscr{C}^{*}  , p_{\mathrm{ec}} > 0   \big\}  \big|   \big\{ \text{alphabets } y, z  \big\} \big]             \big\}    \bigg\} \bigg\} \bigg\}    \end{align*}
        
        \begin{align*}   \equiv \mathcal{C}_{*}^2 \cdot 
              \underset{| \textbf{Z} | \longrightarrow + \infty}{\underset{| \textbf{Y} | \longrightarrow + \infty}{\mathrm{lim}}}     \bigg\{     \underset{\mathrm{fa} \in \mathrm{FA}}{\underset{\mathrm{EC}, B \longleftrightarrow E}{\mathrm{sup}}}  \bigg\{        \textbf{P} \big[           \big\{  \forall \mathscr{C}^{*} \in \textbf{R} , \exists \text{ } \mathrm{ ec } : \mathscr{O} {\tiny \big(} \textbf{Y},\textbf{Z}  {\tiny \big) }   > \mathscr{C}^{*}  , p_{\mathrm{ec}} > 0   \big\}  \big|  \big\{ \text{alphabets } \textbf{Y}, \textbf{Z}  \big\} \big]             \big\} \bigg\}  \bigg\}  \end{align*}

              \begin{align*}  >   \mathcal{C}_{*}^2 \cdot 
              \underset{\mathscr{C}^{**} \longrightarrow 0^{+}}{\underset{| \textbf{Z} | \longrightarrow + \infty}{\underset{| \textbf{Y} | \longrightarrow + \infty}{\mathrm{lim}}}}     \bigg\{     \underset{\mathrm{fa} \in \mathrm{FA}}{\underset{\mathrm{EC}, B \longleftrightarrow E}{\mathrm{sup}}}  \bigg\{        \textbf{P} \big[           \big\{  \forall \mathscr{C}^{**} > \mathscr{C}^{*} \in \textbf{R} , \exists \text{ } \mathrm{ ec } : \mathscr{O} {\tiny \big( }  \textbf{Y},\textbf{Z}  {\tiny \big) }  > \mathscr{C}^{**}  , p_{\mathrm{ec}} > 0   \big\}  \big|  \big\{ \text{alphabets } \textbf{Y}, \textbf{Z}  \big\} \big]             \big\} \bigg\}  \bigg\}      \end{align*}
        
        \begin{align*}     > \mathcal{C}^2_* \cdot  C \big( \mathscr{C}^{**} , \big| \textbf{Y} \big| , \big|  \textbf{Z} \big| \big)   \equiv \mathcal{C}_{***}   \in \textbf{R}                     \text{, }
\end{align*} }

\noindent for $C {\tiny \big( \mathscr{C}^{**} , \big| }  \textbf{Y} {\tiny \big| , \big| }   \textbf{Z} {\tiny \big| \big) }  \in \textbf{R}$, where, in (****), we made use of the observation that,

{\small \begin{align*}
       \underset{| \textbf{Z} | \longrightarrow + \infty}{\underset{| \textbf{Y} | \longrightarrow + \infty}{\mathrm{lim}}} \bigg\{  \underset{y \in \textbf{Y}, z \in \textbf{Z}}{\bigcup}  \bigg\{        \big\{   \textbf{P} \big[         \big\{ \text{alphabets } y, z  \big\}  \big|   \big\{  \forall \mathscr{C}^{*} \in \textbf{R} , \exists \text{ } \mathrm{ ec } : \mathscr{O} {\tiny \big( y,z  \big) }  > \mathscr{C}^{*}  , p_{\mathrm{ec}} > 0   \big\}  \big]             \big\}  \bigg\}  \bigg\}  \longrightarrow \mathscr{C}^{***} < + \infty     \text{, }      \end{align*}
        
        \begin{align*}  \underset{| \textbf{Z} | \longrightarrow + \infty}{\underset{| \textbf{Y} | \longrightarrow + \infty}{\mathrm{lim}}} \bigg\{ \underset{y \in \textbf{Y}, z \in \textbf{Z}}{\bigcup}   \textbf{P} \big[ \text{alphabets }y,z \big]^{-1} \bigg\}  \equiv  \underset{y \in \textbf{Y}, z \in \textbf{Z}}{\bigcup}  \bigg\{ { { \underset{| \textbf{Z} | \longrightarrow + \infty}{\underset{| \textbf{Y} | \longrightarrow + \infty}{\mathrm{lim}}} 1}} \bigg/ { \underset{| \textbf{Z} | \longrightarrow + \infty}{\underset{| \textbf{Y} | \longrightarrow + \infty}{\mathrm{lim}}}  \textbf{P} \big[ \text{alphabets }y,z \big]} \bigg\} \end{align*}
        
        \begin{align*}   \equiv   { 1} \bigg/ { \underset{| \textbf{Z} | \longrightarrow + \infty}{\underset{| \textbf{Y} | \longrightarrow + \infty}{\mathrm{lim}}} \bigg\{ \underset{y \in \textbf{Y}, z \in \textbf{Z}}{\bigcup}   \textbf{P} \big[ \text{alphabets }y,z \big] \bigg\} }   \longrightarrow \mathscr{C}^{**} < + \infty  \Longleftrightarrow \textbf{P} \big[ \text{alphabets }y,z \big] \neq 0     \text{, }
\end{align*}

\begin{align*}
  \underset{| \textbf{Z} | \longrightarrow + \infty}{\underset{| \textbf{Y} | \longrightarrow + \infty}{\mathrm{lim}}} \bigg\{   \underset{y \in \textbf{Y}, z \in \textbf{Z}}{\bigcup} \bigg\{  \textbf{P} \big[   \big\{  \forall \mathscr{C}^{*} \in \textbf{R} , \exists \text{ } \mathrm{ ec } : \mathscr{O} {\tiny \big( y,z  \big) }  > \mathscr{C}^{*}  , p_{\mathrm{ec}} > 0   \big\} \big] \bigg\} \bigg\}     \longrightarrow \mathscr{C}^{****} < + \infty   \text{. }  
\end{align*} }

\noindent To conclude, one obtains the desired stochastic domination, corresponding to the second result in $\textbf{Theorem}$ \textit{2}, by taking a lower bound equaling the final constant, $C_{\mathrm{Final}}$, as $\mathcal{C}_{*}  \mathcal{C}_{**}  \mathcal{C}_{***} C_2 \mathscr{C}^{*} \mathscr{C}^{**} \equiv \mathcal{C}_{*} \mathcal{C}_{**} \mathcal{C}^2_{*} C {\tiny \big( \mathscr{C}^{**} , \big|}  \textbf{Y} {\tiny \big| , \big| } \textbf{Z} {\tiny \big| \big) }  C_2 \mathscr{C}^{*} \mathscr{C}^{**} \gtrsim  \mathcal{C}_{*} \mathcal{C}_{**} \mathcal{C}^2_{*} C_2 \mathscr{C}^{*} \big( \mathscr{C}^{**} \big)^2 >  0$, from which we conclude the argument. \boxed{}

\subsection{Theorem $3$}

\subsubsection{General description of the proof}

\noindent For the last remaining main result, we apply previous observations for expressing the decoding, and false acceptance, probabilities. Namely, recall, 

 {\small \begin{align*}  \underset{n \longrightarrow + \infty}{\mathrm{lim}}  \textbf{P} \big[  \text{Alice and Bob can encode messages across } \mathcal{N}^n           \big]   > 0     \text{, }   \end{align*}  } 

    \noindent and that,

    {\small \begin{align*}  \underset{n \longrightarrow + \infty}{\mathrm{lim}}  \textbf{P} \big[   \text{Alice and Bob can decode, and authenticate, codewords sent over } \mathcal{N}^n          \big]    > 0     \text{. } 
    \end{align*} } 

    \noindent Also, one has that,

   {\small  \begin{align*}
\underset{\text{bits} \longrightarrow + \infty }{\mathrm{lim}} \bigg\{  \underset{\text{bits}}{\bigcup} \textbf{P} \big[ \forall \text{ encoding, } \exists \text{ decoding}: p_{\mathrm{encoding}} \cap p_{\mathrm{decoding } \in      \mathcal{A}^{rn}_{A \longleftrightarrow B}   } > 0 \big]  \bigg\} \equiv 0    \text{, }
\end{align*} }

\noindent from the set of possible error correcting codes,

{\small \begin{align*}
  \mathscr{E}\mathscr{C} \equiv \underset{n \longrightarrow + \infty}{\mathrm{lim}} \bigg\{ \underset{\text{codes}}{\bigcup} \big\{         \text{codes for correcting } n-\text{bit codewords}      \big\}  \bigg\}    \text{, }
\end{align*} } 

\noindent and the set of all possible codewords,

{\small \begin{align*}
 \text{Codebook} \equiv \underset{\text{Codewords } \mathscr{C}}{\bigcup} \mathscr{C} \equiv \underset{\mathscr{C}}{\bigcup}  \bigg\{  \underset{\text{Bits } b, \text{ } b \in \{ 0 , 1 \}^n
    }{\bigcup} \big\{ b \text{ transmitted over the Quantum channel at rate } r \big\}   \bigg\}  \text{,}
\end{align*} }

\noindent transmitted at rate $r$. In the following arguments for the last main result, protocols so that Alice and Bob map into the authenticated space over their shared Quantum channel with high probability directly follows from an argument in {[38]}.

\subsubsection{Argument}

\noindent \textit{Proof of Theorem 3}. To argue that good enough protocols $\pi$ exist so that encodings, and decodings, of bit codewords by Alice and Bob can be mapped into the authenticated space whp, one can directly apply the arguments from the following result, which is independent of the upper and lower bounds on $r$:

\bigskip

\noindent \textbf{Proposition} \textit{4} (\textit{mapping into the authenticated set of sequences over the noisy Quantum channel between Alice and Bob, with high probability}, {[38]}). Let $\pi^n = \big( E^n , D^n \big) $ be a protocol encoding $rn$ bit mesages into $n$ bit codewords. Suppose the real system $E^n_A D^n_B \mathcal{N}^n_{p,q}$ has probability of decoding error $p_{\mathrm{de}}$ and probability of false acceptance $p_{\mathrm{fa}}$. Then,

\begin{itemize}
    \item[$\bullet$] \textit{(1)}: There exists a simulator for Eve, $*_E$, for which $d \big( E^n_A D^n_B \mathcal{N}^n_{p,q} , *_E \mathcal{A}^{rn} \big) = p_{\mathrm{de}}$.

    \bigskip

    \item[$\bullet$] \textit{(2)}: There exists a simulator $\sigma$, for Eve, for which $d \big( E^n_A D^n_B \mathcal{N}^n_{p,q} , \sigma_E \mathcal{A}^{rn} \big) = \mathrm{max} \big\{ p_{\mathrm{de}} , p_{\mathrm{fa}} \big\} $
\end{itemize}

\noindent With the result above, the desired result, through the existence of a protocol for each of the two players excluding Eve, implies that the desired result holds for upper bounds to the bit transmission rate,

{\small \begin{align*}
\underset{P_{\textbf{X}}}{\mathrm{sup}}  \big\{ \mathrm{min} \big\{  I \big( \textbf{X} , \textbf{Y} \big)  , \underset{z}{\mathrm{min}}  \big\{ H_Q \big( \textbf{Y} \big| \textbf{Z} = z \big)   -  H_P \big( \textbf{Y} \big| \textbf{X} \big)  \big\}  \big\}    \big\} 
\end{align*}
\[ <  \left\{\!\begin{array}{ll@{}>{{}}l} 
     \mathrm{log} \mathrm{log} \bigg[   \frac{  \mathrm{log} \big| \textbf{Y}^{*} \big|   }{ \big| \textbf{X}^{*} \big|  }          \bigg]   +   \mathrm{log}  \bigg[   \frac{   \mathrm{log}\big|  \textbf{Z}  \big|  }{  \big|  \textbf{Y}^{*}  \big|   } \bigg]  \Longleftrightarrow \big| \textbf{X} \big| > \big| \textbf{Y}^{*} \big| ,  \big| \textbf{Y}^{*} \big| > \big| \textbf{Z} \big|    ,     \\ \mathrm{log} \mathrm{log} \bigg[     \frac{  \mathrm{log} \big| \textbf{X} \big|  }{ \big| \textbf{Y}^{*} \big| }          \bigg]   +   \mathrm{log}  \bigg[  \frac{ \mathrm{log} \big|  \textbf{Y}^{*} \big|   }{  \big| \textbf{X} \big|   } \bigg]   \Longleftrightarrow \big| \textbf{X} \big| <  \big| \textbf{Y}^{*} \big| ,  \big|  \textbf{Y}^{*} \big| < \big| \textbf{Z} \big|        ,  \\ \mathrm{log} \mathrm{log} \bigg[     \frac{  \mathrm{log} \big| \textbf{Y}^{*} \big|  }{ \big| \textbf{X}^{*} \big| }          \bigg]   +   \mathrm{log}  \bigg[  \frac{  \mathrm{log}\big|  \textbf{Y}^{*} \big|   }{ \big| \textbf{X} \big|   } \bigg]   \Longleftrightarrow \big| \textbf{X} \big| >   \big| \textbf{Y}^{*} \big| ,  \big|  \textbf{Y}^{*} \big| < \big| \textbf{Z} \big|        ,     \\ \mathrm{log} \mathrm{log} \bigg[     \frac{  \mathrm{log} \big| \textbf{Y}^{*} \big|  }{\big| \textbf{X}^{*} \big| }          \bigg]   +   \mathrm{log}  \bigg[  \frac{  \mathrm{log}\big|  \textbf{Z}  \big|   }{ \big| \textbf{Y}^{*} \big|   } \bigg]   \Longleftrightarrow \big| \textbf{X} \big| <    \big| \textbf{Y}^{*} \big| ,  \big|  \textbf{Y}^{*} \big| >   \big| \textbf{Z} \big|        .                          
\end{array}\right. \equiv r  \tag{*} 
\] } 
\noindent provided in \textbf{Theorem} \textit{1}, from which we conclude the argument. \boxed{}

\subsection{Corollary 1}

\subsubsection{General description of the proof}

\noindent With a generalized notion of noise over a Quantum channel, the probability of Alice or Bob performing a decoding error, normalized in the power set of all possible codewords, can be made arbitrarily small. The parameter used for a threshold for upper bounding the decoding error probability, as in the model before incorporating generalized aspects of noise, essentially is composed of two contributions: the first of which is dependent upon whether the codeword that Alice or Bob receives belongs to the set,

\begin{align*}
  T \big( n, p, \delta \big) \bigotimes T \big( n, p, \delta \big)   \text{, }
\end{align*}

\noindent of two-dimensional $\delta$ exact sequences introduced in [38]. As the sequence length tends to infinity, with $n \longrightarrow + \infty$, given the probability measure $P_{\textbf{X}}$, for defining a joint distribution on the tensor product of two sets of $\delta$-exact sequences, introduce,

\begin{align*}
 P_{\textbf{Y}} \equiv \underset{Y \in \textbf{Y}}{\bigcup} P_Y   \text{, }
\end{align*}

\noindent for the alphabet $\textbf{Y}$ of the second player. The joint probability measure over the alphabets of the two players hence takes the form,

\begin{align*}
  P_{\textbf{X} \times \textbf{Y}} \equiv      \underset{X \in \textbf{X}, Y \in \textbf{Y}}{\bigcup} P_{X \times Y}         \text{. }
\end{align*}

\noindent As the length of the sequence that is distributed, through bits, over the Quantum channel approaches $+\infty$, the set of possible encodings provided by Alice and Bob,

\begin{align*}
 \mathcal{E}_n \equiv \underset{n>0}{\bigcup}    \big\{ n\text{-bit encodings} \big\}      \overset{n \longrightarrow + \infty}{\longrightarrow} \mathcal{E}_{+\infty} \supsetneq \mathcal{E}_{\text{good }+\infty \text{ encodings}} \text{, }
\end{align*}

\noindent can be tampered with by the adversary Eve through the actions specified by the generalized noise model,

{\small \begin{align*}
  \mathcal{N}_{+\infty} = \underset{n \longrightarrow + \infty}{\mathrm{lim}} \bigg\{ \underset{n\text{-bit codewords}}{\bigcup}  \big\{    \text{Random bit flips along all positions of an } n-\text{bit codeword}             \big\} \bigg\}   \text{. }
\end{align*} }

\noindent As a result, whether the first conditional probability,

{\small \begin{align*}
  \textbf{P} \bigg[  \cdot     \bigg| \big( x^{\prime\prime}_1, \cdots, x^{\prime\prime}_n, \cdots, x^{\prime\prime}_{+\infty} \big) \times    \big( x^{\prime\prime\prime}_1, \cdots, x^{\prime\prime\prime}_n, \cdots, x^{\prime\prime\prime}_{+\infty} \big) \not\in    T \big( n, p, \delta \big) \bigotimes T \big( n, p, \delta \big)   \bigg]  > 0   \text{, }
\end{align*} }

\noindent or the second conditional probability,

{\small \begin{align*}
   \textbf{P} \bigg[  \cdot     \bigg| \big( x_1, \cdots, x_n, \cdots, x_{+\infty} \big) \times    \big( x^{\prime}_1, \cdots, x^{\prime}_n, \cdots, x^{\prime}_{+\infty} \big) \in    T \big( n, p, \delta \big) \bigotimes T \big( n, p, \delta \big)   \bigg]  > 0   \text{, }
\end{align*} }

\noindent occurs determines whether secure communication can be carried out from bits transmitted over a Quantum channel. In the first conditional probability provided above, the sequences in the joint distribution, sampled from $\textbf{X} \times \textbf{Y}$, $\big(x^{\prime\prime}_1, \cdots, x^{\prime\prime}_n, \cdots, x^{\prime\prime}_{+\infty} \big)$ and $\big( x^{\prime\prime\prime}_1, \cdots, x^{\prime\prime\prime}_n, \cdots, x^{\prime\prime\prime}_{+\infty} \big)$, are obtained from the actions of the generalized noise model,

\begin{align*}
   \mathcal{N}_{+\infty}  \curvearrowright       \big( x_1, \cdots, x_n, \cdots, x_{+\infty} \big)  \text{, } \\  \mathcal{N}_{+\infty}    \curvearrowright          \big( x^{\prime}_1, \cdots, x^{\prime}_n, \cdots, x^{\prime}_{+\infty} \big)    \text{. }
\end{align*}

\subsubsection{Argument}

\noindent 
\noindent \textit{Proof of Corollary 1}. For the forward direction of the correspondence, we argue that the desired conclusion holds, namely that the false acceptance probability, approximately vanishes, with the following. We establish a collection of several equivalent conditions, which is expressed as a union over all instances of error correcting correction and false acceptance. By taking a summation over all bits in the codeword that is transmitted and then sending the number of bits to $+ \infty$, we conclude that $p_{\mathrm{FA}} \approx 0$ by . ($\Longrightarrow$) Suppose that $p_{\mathrm{EC}} \approx 1$. It suffices to prove that the stochastic domination,

\begin{align*}
    p_{\mathrm{EC},n} > p_{\mathrm{FA},n}       \text{, }
\end{align*}

\noindent holds between implementing an error correcting scheme, from the class of all suitable codes $\mathrm{EC}$, is \textit{stable} as $n \longrightarrow + \infty$. \textit{Stability} in the limit of infinitely many bits for a codeword requires,

{\small \begin{align*}
  \bigg\{    p_{\mathrm{EC},+ \infty} > p_{\mathrm{FA},+\infty} \bigg\}  \Longleftrightarrow \underset{N \longrightarrow + \infty}{\mathrm{lim}} \bigg\{   \forall N   ,  \big\{    p_{\mathrm{EC}, N} > p_{\mathrm{FA}, N } \big\}          \Longleftrightarrow \big\{     p_{\mathrm{EC}, N-1}  > p_{\mathrm{FA}, N-1}  \big\}  \bigg\}                \text{. }
\end{align*} }

\noindent From an application of Bayes' Rule, one can demonstrate that the first direction of the iff correspondence holds, upon manipulating the conditional probability as the number of bits $\longrightarrow + \infty$, which equals,

{\small \begin{align*}
   (\mathcal{C}\text{-Probab.}) =     \underset{\text{bits}^{\prime} \longrightarrow + \infty}{\mathrm{lim}}         \big\{   \mathscr{P}_1 \mathscr{P}_2 \mathscr{P}_3     \big\}     \text{, }
\end{align*} } 

\noindent for,

{\small \begin{align*}
 \mathscr{P}_1 \equiv \textbf{P} \bigg[   \big\{ p_{\mathrm{EC}, \text{bits}^{\prime}-1} \approx 1   \big\}  \bigg|  \big\{ \forall \epsilon > 0,  \exists \text{ codewords } \mathcal{C} \in \mathscr{C} \text{ with } \text{bits}^{\prime} \text{ bits}  :    \big| \mathcal{B} \big( \mathcal{C}, r_{\mathcal{C}} \big) \big|   \approx {1} \big/ {\epsilon} \cdot \big| \mathscr{C} \big|  \big\}    \bigg]        \text{, } \end{align*}
 
 \begin{align*}  \mathscr{P}_2 \equiv    \frac{1}{\textbf{P} \big[   p_{\mathrm{EC}, \text{bits}^{\prime}-1} \approx 1  \big] }        \text{, } \end{align*}
 
 \begin{align*} \mathscr{P}_3 \equiv     \textbf{P} \bigg[   \forall \epsilon > 0,  \exists \text{ codewords } \mathcal{C} \in \mathscr{C} \text{ with } \text{bits}^{\prime} \text{ bits}   :   \big| \mathcal{B} \big( \mathcal{C}, r_{\mathcal{C}} \big) \big|     \approx {1} \big/ {\epsilon} \cdot \big| \mathscr{C} \big| \bigg]       \text{. }
\end{align*} }

\noindent Observe, for $\mathrm{bits}^{\prime\prime} >> \mathrm{bits}^{\prime}$,

\begin{align*}
   \bigg\{  \mathscr{P}_2 < \frac{1}{\textbf{P} \big[ p_{\mathrm{EC}, \mathrm{bits}^{\prime\prime}-1} \approx 1  \big]} \equiv \mathscr{P}^{\prime}_2  \bigg\}    \Longrightarrow  \bigg\{   \underset{\text{bits}^{\prime} \longrightarrow + \infty}{\mathrm{lim}}         \big\{   \mathscr{P}_1 \mathscr{P}_2 \mathscr{P}_3     \big\}   <  \underset{\text{bits}^{\prime} \longrightarrow + \infty}{\mathrm{lim}}         \big\{   \mathscr{P}_1 \mathscr{P}^{\prime}_2 \mathscr{P}_3     \big\}   \bigg\}  \text{, }
\end{align*}

\noindent from the error correction probability being monotonically decreasing with respect to the number of bits in a transmitted codeword. To argue that,

\begin{align*}
           \underset{\text{bits}^{\prime} \longrightarrow + \infty}{\mathrm{lim}}         \big\{   \mathscr{P}_1 \mathscr{P}^{\prime}_2 \mathscr{P}_3     \big\}       \text{, }
\end{align*}

\noindent can be upper bounded by the error correction probability, observe,

\begin{align*}
     \mathscr{P}_3   \lesssim r_{\mathcal{C}} \equiv C^{\prime} \big( r_{\mathcal{C}} , \mathrm{bits}^{\prime} \big) \equiv C^{\prime}  \in \big[ 0, 1 \big]       \text{, }
\end{align*}

\noindent readily implying, 

{\small \begin{align*}
0 \approx  p_{\mathrm{FA}} <  (\mathcal{C}\text{-Probab.})  <  \underset{\text{bits}^{\prime} \longrightarrow + \infty}{\mathrm{lim}}         \big\{   \mathscr{P}_1 \mathscr{P}^{\prime}_2 \mathscr{P}_3     \big\} <   \underset{\text{bits}^{\prime} \longrightarrow + \infty}{\mathrm{lim}}         \big\{   \mathscr{P}_1 \mathscr{P}^{\prime}_2 C^{\prime}   \big\}   \overset{(\mathscr{P}_1 \mathscr{P}_2 \mathscr{P}_3)}{\equiv} \big\{ \underset{\text{bits}^{\prime} \longrightarrow + \infty}{\mathrm{lim}}        \mathscr{P}_1 \big\}  \\  \cdot  \big\{ \underset{\text{bits}^{\prime} \longrightarrow + \infty}{\mathrm{lim}}      \mathscr{P}^{\prime}_2  \big\} \cdot \big\{ \underset{\text{bits}^{\prime} \longrightarrow + \infty}{\mathrm{lim}}      C^{\prime} \big\}  \lesssim     \big\{ \underset{\text{bits}^{\prime} \longrightarrow + \infty}{\mathrm{lim}}        \mathscr{P}_1 \big\} \cdot  \big\{ \underset{\text{bits}^{\prime} \longrightarrow + \infty}{\mathrm{lim}}      \mathscr{P}^{\prime}_2  \big\}  \\    \lesssim             \underset{\text{bits}^{\prime} \longrightarrow + \infty}{\mathrm{lim}}        \mathscr{P}_1    <   \underset{\text{bits}^{\prime} \longrightarrow + \infty}{\mathrm{lim}}   \textbf{P} \big[   p_{\mathrm{EC}, \text{bits}^{\prime}-1} \approx 1   \big]  \\  <   p_{\mathrm{EC}} \approx 1          \text{, }
\end{align*} }

\noindent where, in $(\mathscr{P}_1 \mathscr{P}_2 \mathscr{P}_3)$, the limit of the product $\mathscr{P}_1 \mathscr{P}^{\prime}_2 C^{\prime}$ equals the product of the individual limits as $\mathrm{bits}^{\prime} \longrightarrow + \infty$ because each limit is finite.

\bigskip

\noindent Altogether,

\begin{align*}
 \big\{ p_{\mathrm{FA}, \mathrm{bits}} \longrightarrow 0  \big\}  \Longrightarrow \big\{ p_{\mathrm{EC}, \mathrm{bits}} \longrightarrow 1 \text{, as bits} \longrightarrow + \infty \big\} \Longrightarrow        \big\{ p_{\mathrm{FA}, + \infty} \approx  0  \big\}  \Longrightarrow \big\{ p_{\mathrm{EC}, + \infty } \approx  1  \big\} \text{, }
\end{align*}

\noindent for codewords with infinitely many bits, from which we conclude the forward direction of the argument. ($\Longleftarrow$) Suppose that $p_{\mathrm{FA}} \approx 0$. To argue that the remaining direction of the correspondence holds, introduce the following expression for the false acceptance probability, as a supremum over all instances over $A \longrightarrow B$,

\begin{align*}
p_{\mathrm{FA}} \equiv \underset{\mathrm{FA}, A \longleftrightarrow B}{\mathrm{sup}}  \textbf{P} \bigg[   \mathrm{FA} :  \textbf{P} \big[      \text{Alice, or Bob, accepts a message corrupted by Eve that is placed} \\ \text{into } \mathcal{A}^{rn}_{A \longleftrightarrow B}     \big] > 0  \bigg]  \text{, }
\end{align*}

\noindent for which Alice and Bob can mistakenly accept a message from Eve that has been corrupted. Moreover,

\begin{align*}
 p_{\mathrm{FA}} \equiv   \underset{n \longrightarrow + \infty}{\mathrm{lim}}  \bigg\{       \underset{\text{FA on } n-\text{bits}, A \longleftrightarrow B}{\mathrm{sup}}    p_{\mathrm{FA}}             \bigg\} \equiv \underset{n \longrightarrow + \infty}{\mathrm{lim}} \bigg\{ \underset{\mathrm{fa} \in \mathrm{FA}}{\bigcup} \bigg\{         \underset{\text{fa on } n-\text{bits}, A \longleftrightarrow B}{\mathrm{sup}}    p_{\mathrm{FA}}             \bigg\} \bigg\}     <   \underset{n \longrightarrow + \infty}{\mathrm{lim}}  \bigg\{ \underset{\mathrm{fa} \in \mathrm{FA}}{\bigcup} p^{*}_{\mathrm{fa},n}   \bigg\}  \end{align*}
 
 \begin{align*} \equiv p^{*}_{\mathrm{FA}}              \text{,}
\end{align*}

\noindent where $p^{*}_{\mathrm{FA}}     $ denotes the instance of false acceptance for which the probability of Alice, or Bob, falsely accepting a corrupted message is maximized. This probability can be lower bounded with the error correction probability, from the fact that,

\begin{align*}
  0 \approx  p_{\mathrm{FA},n} <   p^{*}_{\mathrm{FA},n}  <   \textbf{P} \big[   p_{\mathrm{EC},n} \big|   p^{*}_{\mathrm{FA},n}        \big]             \overset{(\text{Bayes' Rule})}{\equiv}   \textbf{P} \big[ p_{\mathrm{EC},n}     \big] \cdot \bigg[  \frac{\textbf{P} \big[    p^{*}_{\mathrm{FA},n}  \big| p_{\mathrm{EC},n}       \big]}{\textbf{P} \big[  p^{*}_{\mathrm{FA},n}       \big]} \bigg]    \\ <   \frac{\textbf{P} \big[    p^{*}_{\mathrm{FA},n}  \big| p_{\mathrm{EC},n}       \big]}{\textbf{P} \big[  p^{*}_{\mathrm{FA},n}       \big]}           \overset{(1)}{\approx}  \textbf{P} \big[ p_{\mathrm{EC},n} \big] \equiv  p_{\mathrm{EC},n} \overset{(2)}{<} p^{*}_{\mathrm{EC},n}  \overset{(3)}{<} p^{*}_{\mathrm{EC}}      \text{, }
\end{align*}

\noindent where, in (1), we made use of the observation that,

{\small \begin{align*}
  \frac{\textbf{P} \big[    p^{*}_{\mathrm{FA},n}  \big| p_{\mathrm{EC},n}       \big]}{\textbf{P} \big[  p^{*}_{\mathrm{FA},n}       \big]}  \approx  \frac{\textbf{P} \big[    p^{*}_{\mathrm{FA},n}  ,  p_{\mathrm{EC},n}       \big]}{\textbf{P} \big[  p^{*}_{\mathrm{FA},n}       \big]} \equiv  \frac{\textbf{P} \big[    p^{*}_{\mathrm{FA},n}  \cap   p_{\mathrm{EC},n}       \big]}{\textbf{P} \big[  p^{*}_{\mathrm{FA},n}       \big]} \equiv \frac{\textbf{P} \big[    p^{*}_{\mathrm{FA},n} \big] \textbf{P} \big[   p_{\mathrm{EC},n}       \big]}{\textbf{P} \big[  p^{*}_{\mathrm{FA},n}       \big]} =  \textbf{P} \big[   p_{\mathrm{EC},n}       \big] \text{, }
\end{align*} }

\noindent while, in (2), we made use of the observation that,

{\small \begin{align*}
 p^{*}_{\mathrm{EC},n} \equiv \bigg\{         \underset{\mathrm{EC}, A \longleftrightarrow B}{\mathrm{sup}} p_{\mathrm{EC}}            \bigg\}  \equiv \underset{\mathrm{ec} \in \mathrm{EC}}{\bigcup} \bigg\{         \underset{\mathrm{ec}, A \longleftrightarrow B}{\mathrm{sup}} p_{\mathrm{ec}}            \bigg\} <    \underset{\mathrm{ec} \in \mathrm{EC}}{\bigcup}  \big\{ p^{*}_{\mathrm{ec},n} \big\} \equiv p^*_{\mathrm{EC},n} \text{,}
\end{align*} }

\noindent and finally, in (3), that,

{\small \begin{align*}
 \underset{n \longrightarrow + \infty}{\mathrm{lim}}  p^*_{\mathrm{EC},n} \equiv p^*_{\mathrm{EC}} \longrightarrow 1 \text{. }
\end{align*} } 

\noindent Hence,

{\small \begin{align*}
 0 \approx  p_{\mathrm{FA}} < p_{\mathrm{EC}} \approx 1   \text{, }
\end{align*} }

\noindent from which we conclude the reverse direction of the argument. With both directions of the correspondence established, we conclude the argument. \boxed{}

\subsection{Corollary 2}

\subsubsection{General description of the proof}

\noindent To exhibit that the transmission mechanisms for classical bits distributed over a Quantum channel are stable under infinitely many bits, in light of the previously obtained converse result on $r$, WLOG, the following situations can occur with positive probability:

\begin{itemize}
\item[$\bullet$] \textit{There fails to exist good decodings over the noisy Quantum channel, which Eve intercepts for tampering}. The existence of a codeword, encoded in classical bits, that is transmitted with an unfavorable decoding over the Quantum channel that Eve intercepts is given by,

{\small \begin{align*}
  \textbf{P} \bigg[ \big\{ \exists \text{  encoding} : p_{\text{encoding}} \cap p_{\mathrm{FA}} \equiv 0 \big\} , \big\{   \exists \text{ encoding} : p_{\mathrm{encoding}} \cap p_{\text{encoding} \in \mathcal{A}} \equiv 0       \big\}  \bigg]   \text{, }
\end{align*} }

\noindent which corresponds to the occurrence of an encoding, which satisfies,

\begin{align*}
  p_{\text{encoding}}  \equiv \underset{\text{bits}}{\bigcup}p_{\text{encoding infinitely many bits}} \text{, }
\end{align*}

\noindent that cannot have a strictly positive false acceptance probability, as well as not belonging to the set of authenticated messages $\mathcal{A}$ of Alice, or of Bob.

\item[$\bullet$] \textit{There exists good decodings over the more noisy Quantum channel from Alice and Bob, in comparison to the Quantum channel from Bob to Eve, that Eve can tamper with that belong to the set of authenticated messages}. Related to the first possibility, one can also have that the probability,

{\small \begin{align*}
   \textbf{P} \bigg[  \big\{ \exists \text{ encoding} : p_{\mathrm{encoding}} \cap p_{\mathrm{FA}} \neq 0\big\}, \big\{       \exists \text{ encoding} : p_{\mathrm{encoding}}   \cap p_{\mathrm{encoding} \in \mathcal{A}} > 0           \big\}         \bigg] \text{, }
\end{align*} }

\noindent which corresponds to the occurrence of a good encoding over the more noisy Quantum channel between Alice and Bob, for which there is a nonzero probability of false acceptance, due to the fact that the encoding for the bit message over the Quantum channel has an appropriate encoding, in addition to the existence of encodings provided by Eve, which on top of the encoding introduce by Alice or Bob, belong to the set of authenticated messages, $\mathcal{A}$.

\item[$\bullet$] \textit{There exists good decodings over the more noisy Quantum channel from Alice to Bob, such that an encoding provided by Eve does not belong to the set of authenticated messages}. From the events introduced in the previous probability, one has,

{\small \begin{align*}
   \textbf{P} \bigg[  \big\{ \exists \text{ encoding} : p_{\mathrm{encoding}} \cap p_{\mathrm{FA}} \neq 0\big\}, \big\{       \exists \text{ encoding} : p_{\mathrm{encoding}}  \cap p_{\mathrm{encoding} \in \mathcal{A}} \equiv  0           \big\}         \bigg] \end{align*}
   
   \begin{align*} \equiv \underset{\mathrm{fa} \in \mathrm{FA}}{\bigcup} \bigg\{   \textbf{P} \bigg[  \big\{ \exists \text{ encoding} : p_{\mathrm{encoding}} \cap p_{\mathrm{fa}} \neq 0\big\}, \big\{       \exists \text{ encoding} : p_{\mathrm{encoding}}  \cap p_{\mathrm{encoding} \in \mathcal{A}} \equiv  0           \big\}         \bigg]       \bigg\}  \text{, }
\end{align*} } 

\noindent which corresponds to the occurrence of a good encoding over the more noisy Quantum channel between Alice and Bob, for which there is a nonzero probability of false acceptance, in addition to the lack of existence of a decoding, provided by Eve, so that the tampered messages belongs to the set of authenticated messages, $\mathcal{A}$.
 
\item[$\bullet$] \textit{Bob performs a decoding error after receiving a bit codeword sent over the Quantum channel by Alice}. One has,

{\small \begin{align*}
    \textbf{P} \bigg[   \big\{ \exists \text{ encoding} : p_{\mathrm{encoding}} \cap p_{\mathrm{DE}} \neq 0  \big\}, \big\{   \exists \text{ encoding}:     p_{\mathrm{DE}}  \cap p_{\mathrm{FA}} > 0      \big\}       \bigg] \end{align*}
    
    \begin{align*} \equiv \underset{\mathrm{de} \in \mathrm{DE}}{\underset{\mathrm{fa} \in \mathrm{FA}}{\bigcup}}   \bigg\{  \textbf{P} \bigg[   \big\{ \exists \text{ encoding} : p_{\mathrm{encoding}} \cap p_{\mathrm{de}} \neq 0  \big\}, \big\{   \exists \text{ encoding}:     p_{\mathrm{de}}  \cap p_{\mathrm{fa}} > 0      \big\}       \bigg]    \bigg\}     \text{, }
\end{align*} }

\noindent which corresponds to the probability of Alice or Bob performing a decoding error, given some encoding over the Quantum channel,

{\small \begin{align*}
  p_{\mathrm{DE}} \equiv \underset{\mathrm{de} \in \mathrm{DE}}{\bigcup} \textbf{P} \bigg[ \forall \text{ encoding}, \exists \text{ decoding de}: \big\{ p_{\text{encoding is good}}  \big\} \\ \cap  \big\{  p_{\text{the encoding is decoded with some positive error, in de}} \big\}  > 0  \bigg]  \text{, }
\end{align*} }

\noindent and to the probability of preparing a good encoding over the Quantum channel,

\begin{align*}
   p_{\text{encoding is good}} \equiv p_{\mathrm{ec}} \cap p_{\text{ec is transmitted with high enough fidelity } \mathcal{F}} \\ \equiv \underset{f \in \mathcal{F}}{\bigcup} \big\{  p_{\mathrm{ec}} \cap p_{\text{ec is transmitted with high enough fidelity f}}  \big\} \text{. }
\end{align*}

\item[$\bullet$] \textit{There fails to exist good decodings for Bob to receive a message that has a good encoding provided by Alice}. Given the existence of good encodings for bit codewords transmitted over the Quantum channel by Alice, one has,

{\small \begin{align*}
  \textbf{P} \bigg[ \big\{ \exists \text{  encoding} :  p_{\mathrm{encoding}} \cap p_{\mathrm{FA}} \neq 0      \big\} , \big\{  \exists \text{ encoding, decoding }, p_{\mathrm{encoding}} > 0  : \\ p_{\mathrm{encoding}} \cap   p_{\mathrm{decoding}} \equiv 0          \big\}  \bigg] \end{align*}
  
  \begin{align*} \equiv  \underset{\mathrm{fa} \in \mathrm{FA}}{\bigcup} \bigg\{  \textbf{P} \bigg[ \big\{ \exists \text{  encoding} :  p_{\mathrm{encoding}} \cap p_{\mathrm{fa}} \neq 0      \big\} , \big\{  \exists \text{ encoding, decoding }, p_{\mathrm{encoding}} > 0  : \\ p_{\mathrm{encoding}} \cap   p_{\mathrm{decoding}} \equiv 0          \big\}  \bigg] \bigg\}  \text{, } \tag{1}
\end{align*} }

\noindent which corresponds to the failure of the occurrence of a good decoding for Bob, specified by the probability,

\begin{align*}
  p_{\mathrm{decoding}} \equiv \underset{\text{bits}}{\bigcup} p_{\text{decoding infinitely many bits}}  \text{. }
\end{align*}

\item[$\bullet$] \textit{There exists good decodings for Bob}. Related to the items introduced in the previously probability corresponding to the lack of existence of good decodings, one has,
 
{\small \begin{align*}
  \textbf{P} \bigg[ \big\{ \exists \text{  encoding}:  p_{\mathrm{encoding}} \cap p_{\mathrm{FA}} \neq 0      \big\} , \big\{  \exists \text{ encoding, decoding }, p_{\mathrm{encoding}} > 0  : p_{\mathrm{encoding}} \cap   p_{\mathrm{decoding}} > 0          \big\}  \bigg] \end{align*}
  
  \begin{align*} \equiv \underset{\mathrm{fa} \in \mathrm{FA}}{\bigcup} \bigg\{   \textbf{P} \bigg[ \big\{ \exists \text{  encoding}:  p_{\mathrm{encoding}} \cap p_{\mathrm{fa}} \neq 0      \big\} , \big\{  \exists \text{ encoding, decoding }, p_{\mathrm{encoding}} > 0  : p_{\mathrm{encoding}} \\ \cap   p_{\mathrm{decoding}} > 0          \big\}  \bigg]  \bigg\} \text{. } \\ \tag{2}
\end{align*} }

\item[$\bullet$] \textit{The set of messages intercepted by Eve, captured through (1), and (2), belongs to the authenticated set of messages}. Given previously defined quantities, one has,

\begin{align*}
  (1) \cap \textbf{P} \big[ \exists \text{ encoding}:  p_{\text{encoding tampered by Eve} \in \mathcal{A}} \big]  \text{, }
\end{align*}

\noindent which corresponds to the occurrence of Eve's intercepted codeword belonging to the authenticated set of messages, given the occurrence, of (1). Also, one has, 

\begin{align*}
   (2) \cap  \textbf{P} \big[ \exists \text{ encoding} :  p_{\text{encoding tampered by Eve} \in \mathcal{A}} \big]     \text{, }
\end{align*}

which corresponds to the occurrence of Eve's intercepted codeword belonging to the authenticated set of messages, given the occurrence, of (2).

\item[$\bullet$] \textit{The sets of messages intercepted by Eve, captured through (1), and (2), does not belong to the authenticated set of messages}. Straightforwardly, from the probability introduced in the previous item, one has,

\begin{align*}
  (1) \cap \textbf{P} \big[ \exists \text{ encoding}:  p_{\text{encoding tampered by Eve} \not\in \mathcal{A}} \big]  \text{, }
\end{align*}

\noindent and,

\begin{align*}
   (2) \cap  \textbf{P} \big[ \exists \text{ encoding}:  p_{\text{encoding tampered by Eve} \not\in \mathcal{A}} \big]     \text{, }
\end{align*}

\item[$\bullet$] \textit{Miscellaneous probabilities corresponding to communication over Quantum channels}. One can introduce probabilities, such as those provided in (1) and (2), for codewords that are first encoded, before being sent over the Quantum channel, from Bob, instead of from Alice.

\end{itemize}

\subsubsection{Argument}

\noindent \textit{Proof of Corollary 2}. Take the limit of infinitely many bits, as $n \longrightarrow +\infty$, of the sequence of computations performed for obtaining (\text{Noise Lower Bound}) provided in \textbf{Lemma}, from which we conclude the argument. \boxed{}

\section{Conclusion}

\noindent In this work, bit transmission rates for secure communication across Quantum channels, which are resilient to noise, were investigated. In comparison to transmission rates originally provided in {[38]}, bit transmission rates analyzed in the converse result, hence constituting possible strict upper bounds for $\underset{P_{\textbf{X}}}{\mathrm{sup}}  \big\{ \mathrm{min} \big\{  I \big( \textbf{X} , \textbf{Y} \big)  , \underset{z}{\mathrm{min}}  \big\{ H_Q \big( \textbf{Y} \big| \textbf{Z} = z \big)   -  H_P \big( \textbf{Y} \big| \textbf{X} \big)  \big\}  \big\} $, paradoxically hold under the assumption that $N_{A \longleftrightarrow B} > N_{B \longleftrightarrow E}$. Besides the strict upper bound for $r$ provided in the converse result stated in \textbf{Theorem} \textit{1}, \textbf{Theorem} \textit{2}, and \textit{3}, arguments for upper bounding $r$ characterize implications of transmitting Classical bits over noisy Quantum channels that the eavesdropper Eve attempts to intercept, and manipulate. Despite the fact that $N_{A \longleftrightarrow B} > N_{B \longleftrightarrow E}$, Alice and Bob can still map bit codewords into the authenticated space $\mathcal{A}^{rn}_{A \longleftrightarrow B}$ whp.

Fundamentally, this prospective Quantum advantage in transmission, and communication, protocols is realized by generalizing the counterexample originally provided in {[38]}; the counterexample for demonstrating that Alice and Bob need not sacrifice error correction, and false acceptance, probabilities in the presence of greater noise over their channel relies upon the fact that Bob can send messages to Alice, or vice versa, using letters of his or her alphabet that Eve does not use when sending her corrupted messages to $A \longleftrightarrow B$. Hence, simply enough, Bob, or Alice, can determine with probability very close to $1$, that there is neither a false acceptance error, nor a bit codeword mapped into $\mathcal{A}^{rn}_{A \longleftrightarrow B}$ that was actually sent by Eve. This simple counterexample can be generalized with the overlap function, $\mathscr{O}$, which was first introduced in \textit{1.2}. As a function of the alphabets for Alice, Bob and Eve, $\mathscr{O}$ determines whether there is any \textit{overlap}, ie any nonempty subset of common letters, that players use from their respective alphabets when responding to questions drawn from the referee's probability distribution. Albeit the fact that the overlap function could be nonempty for alphabets with an arbitrary number of letters for each participant, procedures for \textit{pruning}, namely eliminating, letters from each player's alphabet were introduced so that Alice and Bob need not sacrifice their respective probabilities of error correction, and false acceptance, hence preserving prospective sources of Quantum advantage. It continues to remain of interest to further investigate closely related variants of the communication protocols discussed in this work, in addition to other possible directions of research interest mentioned at the end of {[38]}.

\section{Declarations}

\subsection{Availability of data and materials}

Not applicable.

\subsection{Competing interests}

Not applicable.

\subsection{Funding}

Not applicable.

\subsection{Authors' contributions}

PR wrote the entire manuscript and performed several rounds of editing.

\subsection{Acknowledgments}

Not applicable.

\section{References}

\noindent [1] Amr, A., Villanueva, I. Quantum one way vs. classical two way communication in XOR games. \textit{Quantum Information Processing} \textbf{20}, 79 (2021).

\bigskip

\noindent [2] Bannik, T. et al. Bounding Quantum-Classical Separations for Classes of Nonlocal Games. \textit{STACS} \textbf{12}: 1-12 (2019). $\mathrm{https://doi.org/10.4230/LIPIcs.STACS.2019.12}$.

\bigskip

\noindent [3] Briet, J. Buhrman, H., Toner, B. A generalized Grothendieck inequality and entanglement in XOR games. \textit{Comm. Math. Phys.} \textbf{305}: 827-843 (2011). $\mathrm{https://doi.org/ 10.1007/s00220-011-1280-3}$.

\bigskip

\noindent [4] Broadbent, A., Methot, A.A. On the power of non-local boxes. \textit{Theoretical Computer Science} \textbf{358}: 3-14 (2006). $\mathrm{
https://doi.org/10.1016/j.tcs.2005.08.035}$.

\bigskip

\noindent [5] Brassard, G., Broadbent, A., Tapp, A. Quantum Pseudo-Telepathy. \textit{Found. Phys.} \textbf{35}: 1877-1907 (2005). $\mathrm{https://philpapers.org/rec/BRAQP}$.

\bigskip

\noindent [6] Benedetti, M. and Coyle, B. and Fiorentini, M. and Lubasch, M. and Rosenkranz, M. Variational Inference with a Quantum Computer. \textit{Phys Rev Applied} 16: 044057 (2021) https://doi.org/10.1103/PhysRevApplied

\noindent .16.044057. 

\bigskip

% Benedetti, M., Coyle, B., Fiorentini, M., Lubasch, M., Rosenkranz, M. Variational Inference with a Quantum Computer. \textit{Phys Rev Applied} \textbf{16}, 044057 (2021).

% https://doi.org/10.1103/PhysRevApplied.16.044057 

\noindent [7] Bittel, L. and Kliesch, M. Training Variational Quantum Algorithms is NP-Hard. \textit{Physical Review Letters} 127: 120502 (2021). https://doi.org/10.1103/PhysRevLett.127.120502

% Bittel, L., and Kliesch, M. Training Variational Quantum Algorithms is NP-Hard. \textit{Physical Review Letters} \textbf{127}, 120502 (2021).

% https://doi.org/10.1103/PhysRevLett.127.120502

\bigskip

\noindent [8] Catani, L. and Faleiro, R. and Emeriau,P.E. and Mansfield,S. and Pappa, A. Connecting XOR and XOR* games. \textit{Phys. Rev. A.} 109: 012427 (2024). https://doi.org/10.1103/PhysRevA.109.012427. 

% Catani, L., Faleiro, R., Emeriau,P.E., Mansfield,S., Pappa, A.. Connecting XOR and XOR* games. Phys. Rev. A 109, 012427

% https://doi.org/10.1103/PhysRevA.109.012427

\bigskip

\noindent [9] Chen, H. and Vives, M. and Metcalf, M. Parametric amplification of an optomechanical quantum interconnect. \textit{Physical Review Research} 4: 043119 (2022). https://doi.org/10.1103/PhysRevResearch.4.043119

% Chen, H., Vives, M., Metcalf, M. Parametric amplification of an optomechanical quantum interconnect. \textit{Physical Review Research} \textbf{4}, 043119 (2022).

% https://doi.org/10.1103/PhysRevResearch.4.043119

\bigskip

\noindent [10] Cong, I. and Duan, L. Quantum discriminant analysis for dimensionality reduction and classification. \textit{New Journal of Physics} 18: 073011 (2016). https://doi.org/10.1088/1367-2630/18/7/073011. 

% Cong, I., Duan, L. Quantum discriminant analysis for dimensionality reduction and classification. \textit{New Journal of Physics} \textbf{18}, 073011 (2016).

\bigskip

\noindent [11] Cleve, R., Hoyer, P., Toner, B., Watrous, J. Consequences and Limits of Nonlocal Strategies. \textit{19th IEEE Annual Conference on Computational Complexity Proceedings}: 236-249 (2004). $\mathrm{https://doi.org/10.1109}$$\mathrm{/CCC.2004.1313847}$.

\bigskip

\noindent [12] Culf, E., Mousavi, H., and Spirig, T. "Approximation Algorithms for Noncommutative CSPs," 2024 \textit{IEEE 65th Annual Symposium on Foundations of Computer Science (FOCS)}, 920-929 (2024). $\mathrm{https://doi.org}$ $\mathrm{/ 10.1109/FOCS61266.2024.00061}$.

\bigskip

\noindent [13] Cui, D., Malavolta, G., Mehta, A., Natarajan, A., Paddock, C., Schmidt, S., Walter, M., Zhang, T. A Computational Tsireslson's Theorem for the Value of Compiled XOR games. \textit{arXiv: 2402.17301} (2024).

\bigskip

\noindent [14] Doherty, A.C., Liang, Y.C., Toner, B., Wehner, S. The Quantum Moment Problem and Bounds on Entangled Multi-Prover Games. \textit{23rd Annual IEEE Conference on Computational Complexity} \textbf{8}: 1093-0159/08 (2018).

\bigskip

\noindent [15] Drmota, P., Main, D., Ainley, E.M., Agrawal, A., Araneda, G., Nadlinger, Srinivas, R., Cabello, A.,  et al. Experimental Quantum Advantage in the Odd-Cycle Game. \textit{Phys. Rev. Lett.} \textbf{134}: 070201 (2025).
$\mathrm{https://doi.org/10.1103/PhysRevLett.134.070201}$.

% doi:10.1088/1367-2630/18/7/073011

\bigskip

\noindent [16] Ewe, W-B. and Koh, D. E. and Goh, S. T. and Chu, H-S and Png, C. E. Variational Quantum-Based Simulation of Waveguide Modes. \textit{IEEE Transactions on Microwave Theory and Techniques} 70 (5): 2517-2525 (2022). https://doi.org/10.1109/TMTT.2022.3151510.

\bigskip

\noindent [17] Pierre-Emmanuel Emeriau, P-E., Howard, M. and Mansfield, S. Quantum Advantage in Information Retrieval. \textit{PRX Quantum} \textbf{3}, 020307 (2022). $\mathrm{ https://doi.org/10.1103/PRXQuantum.3.020307}$.

% Ewe, W-B., Koh, D. E., Goh, S. T., Chu, H-S, Png, C. E. Variational Quantum-Based Simulation of Waveguide Modes. \textit{IEEE Transactions on Microwave Theory and Techniques} \textbf{70}(5): 2517-2525 (2022). 

% doi:10.1109/TMTT.2022.3151510

\bigskip

\noindent [18] Faleiro, R. Quantum strategies for simple 2-player XOR games. \textit{Quantum Inf Process} \textbf{19}, 229 (2020). $\mathrm{doi:10.1007/s11128-020-02717-2}$

\bigskip

\noindent [19] Garg, D. and Ikbal, S. and Srivastava, S.K. and Vishwakarma, H. and Karanam, H. and Subramaniam, L.V. Quantum Embedding of Knowledge for Reasoning. \textit{Advance in Neural Information Processing Systems} 32 (2019). 

\noindent $\mathrm{https://papers.nips.cc/paper_files/paper/2019/hash/cb12d7f933e7d102c52231bf62b8a678-Abstract.html}$.

% Garg, D., Ikbal, S., Srivastava, S.K., Vishwakarma, H., Karanam, H., Subramaniam, L.V. Quantum Embedding of Knowledge for Reasoning. \textit{Advance in Neural Information Processing Systems 32} (2019).

\bigskip

\noindent [20] Genoni, M.G. and Tufarelli, T. Non-orthogonal bases for quantum metrology. \textit{Journal of Physics A: Mathematical and Theoretical} 52: 43 (2019). https://doi.org/10.1088/1751-8121/ab3fe0.

% Genoni, M.G. and Tufarelli, T. Non-orthogonal bases for quantum metrology. \textit{Journal of Physics A: Mathematical and Theoretical} \textbf{52}, 43 (2019).

\bigskip

% https://doi.org/10.1088/1751-8121/ab3fe0

\noindent [21] Gidi, J.A. and Candia, B. and Munoz-Moller, A.D. and Rojas, A. and Pereira, L. and Munoz, M. and Zambrano, L. and Delgado, A. Stochastic optimization algorithms for quantum applications. \textit{Phys.Rev.A} 108: 032409 (2023). https://doi.org/10.1103/PhysRevA.108.032409. 

% Gidi, J.A., Candia, B., Munoz-Moller, A.D., Rojas, A., Pereira, L., Munoz, M., Zambrano, L. and Delgado, A. Stochastic optimization algorithms for quantum applications. Phys.Rev.A 108 (2023) 3, 032409

% https://doi.org/10.1103/PhysRevA.108.032409 

\bigskip

\noindent [22] Givi, P. and Daley, A.J. and Mavriplis, D. and Malik, M. Quantum Speedup for Aeroscience and Engineering. \textit{AIAA} 58:8 (2020). 

https://ntrs.nasa.gov/api/citations/20200003505/downloads/20200003505.pdf.

% Givi, P., Daley, A.J., Mavriplis, D. and Malik, M. Quantum Speedup for Aeroscience and Engineering. \textit{AIAA Journal} \textbf{58}, 8 (2020).

\bigskip

\noindent [23] Helton, J.W., Mousavi, H., Nezhadi, S.S. et al. Synchronous Values of Games. \textit{Ann. Henri Poincaré} \textbf{25}, 4357–4397 (2024). https://doi.org/10.1007/s00023-024-01426-1

\bigskip

\noindent [24] Hadiashar, S.B. and Nayak, A. and Sinha, P. Optimal lower bounds for Quantum Learning via Information Theory. \textit{IEEE Transactions on Information Theory} 70(3): 1876--1896 (2024). https://doi.org/10.1109/

\noindent TIT.2023.3324527. 

% Hadiashar, S.B., Nayak, A., Sinha, P. Optimal lower bounds for Quantum Learning via Information Theory. IEEE Transactions on Information Theory, vol. 70 (3): 1876-1896 (2024).

% https://doi.org/10.1109/TIT.2023.3324527

\bigskip

\noindent [25] Hur, T. and Kim, L. and Park, D.K. Quantum convolutional neural network for classical data classification. \textit{Quantum Machine Intelligence} 4: 3 (2022). https://doi.org/10.1007/s42484-021-00061-x. 

% Hur, T. Kim, L., Park, D.K. Quantum convolutional neural network for classical data classification. \textit{Quantum Machine Intelligence} \textbf{4}, 3 (2022).

% https://doi.org/10.1007/s42484-021-00061-x

\bigskip

\noindent [26] Holmes, Z. and Coble, N.J. and Sornborger, A.T. and Subasi, Y. On nonlinear transformations in quantum computation. \textit{Phys. Rev. Research} 5: 013105 (2023). https://doi.org/10.1103/PhysRevResearch.5.013105. 

% Holmes, Z., Coble, N.J., Sornborger, A.T. and Subasi, Y. On nonlinear transformations in quantum computation. \textit{Phys. Rev. Research} \textbf{5}, 013105 (2023).

% https://doi.org/10.1103/PhysRevResearch.5.013105

\bigskip

\noindent [27] Jing, H. and Wang, Y. and Li, Y. Data-Driven Quantum Approximate Optimization Algorithm for Cyber-Physical Power Systems. \textit{arXiv}: 2204.00738 (2022). https://doi.org/10.48550/arXiv.2204.00738.

% Jing, H., Wang, Y., Li, Y. Data-Driven Quantum Approximate Optimization Algorithm for Cyber-Physical Power Systems. \textit{arXiv: 2204.00738} (2022).

\bigskip

\noindent [28] Junge, M., Palazuelos, C. On the power of quantum entanglement in multipartite quantum XOR games. \textit{Journal of the London Mathematical Society} \textbf{110}(5) (2024).

\bigskip

\noindent [29] Kubo, K. and Nakagawa, Y.O. and Endo, S. and Nagayama, S. Variational quantum simulations of stochastic differential equations. \textit{Physical Review A} 103: 052425 (2021). https://doi.org/10.1103/PhysRevA.

\noindent 103.052425.

% Kubo, K., Nakagawa, Y.O., Endo, S. and Nagayama, S. Variational quantum simulations of stochastic differential equations. \textit{Physical Review A} \textbf{103}, 052425 (2021). 

\bigskip

% https://doi.org/10.1103/PhysRevA.103.052425

\noindent [30] Kribs, D.W. A quantum computing primer for operator theorists. \textit{Linear Algebra and its Applications} 400: 147-167 (2005). https://doi.org/10.48550/arXiv.math/0404553.

% Kribs, D.W. A quantum computing primer for operator theorists. \textit{Linear Algebra and its Applications} \textbf{400}: 147-167 (2005).

\bigskip

\noindent [31] Li, R. Y. and Di Felice, R. and Rohs, R. and Lidar, D.A. Quantum annealing versus classical machine learning applied to a simplied computational biology problem. \textit{NPJ Quantum Information} 4: 14 (2008). https://doi.org/10.1038/s41534-018-0060-8. 

% Li, R. Y., Di Felice, R., Rohs, R. and Lidar, D.A. Quantum annealing versus classical machine learning applied to a simplied computational biology problem. \textit{npj Quantum Information} \textbf{4}: 14 (2008).

% 

\bigskip

\noindent [32] Mahdian, M. and Yeganeh, H.D. Toward a quantum computing algorithm to quantify classical and quantum correlation of system states. \textit{Quantum Information Processing} 20: 393 (2021). https://doi.org/10.1007/

\noindent s11128-021-03331-6.

%  Mahdian, M., Yeganeh, H.D. Toward a quantum computing algorithm to quantify classical and quantum correlation of system states. \textit{Quantum Information Processing} \textbf{20}: 393 (2021).

\bigskip

\noindent [33] Maldonado, T.J. and Flick, J. and Krastanov, S. and Galda, A. Error rate reduction of single-qubit gates via noise-aware decomposition into native gates. \textit{Scientific Reports} 12: 6379 (2022). https://doi.org/10.1038

\noindent /s41598-022-10339-0.

% Maldonado, T.J., Flick, J., Krastanov, S., and Galda, A. Error rate reduction of single-qubit gates via noise-aware decomposition into native gates. \textit{Scientific Reports} \textbf{12}, 6379 (2022).

% doi https://doi.org/10.1038/s41598-022-10339-0

\bigskip

\noindent [34] Manby, F.R. and Stella, M. and Goodpaster, J.D. and Miller, T.F. A Simple, Exact Density-Functional-Theory Embedding Scheme. \textit{Journal of Chemical Theory and Computation} 8 (8): 2564-2568 (2012). https://doi.org/10.1021/ct300544e.

% Manby, F.R., Stella, M., Goodpaster, J.D., and Miller, T.F. A Simple, Exact Density-Functional-Theory Embedding Scheme. \textit{Journal of Chemical Theory and Computation} \textbf{8}(8): 2564-2568 (2012).

\bigskip

% https://doi.org/10.1021/ct300544e

\noindent [35] Mensa, S. and Sahin, E. and Tacchino, F. and Barkoutsos, P.K. and Tavernelli, I. Quantum Machine Learning Framework for Virtual Screening in Drug Discovery: a Prospective Quantum Advantage. \textit{Mach. Learn.: Sci. Technol.} 4: 015023 (2023) https://doi.org/10.1088/2632-2153/acb900.

%  Mensa, S., Sahin, E., Tacchino, F., Barkoutsos, P.K., and Tavernelli, I. Quantum Machine Learning Framework for Virtual Screening in Drug Discovery: a Prospective Quantum Advantage. Mach. Learn.: Sci. Technol. 4 015023.

\bigskip

% https://doi.org/10.1088/2632-2153/acb900

\noindent [36] Nan Sheng, H.M. and Govono, M. and Galli, G. Quantum Embedding Theory for Strongly-Correlated States in Materials. \textit{J. Chem. Theory Comput.} 17 (4): 2116-2125 (2021). https://doi.org/10.1021/acs.jctc.

\noindent 0c01258. 

% Nan Sheng, H.M., Govono, M., and Galli, G. Quantum Embedding Theory for Strongly-Correlated States in Materials. \textit{J. Chem. Theory Comput.} \textbf{17}(4), 2116-2125 (2021). 

\bigskip

\noindent [37] Ostrev, D. The structure of optimal and nearly-optimal quantum strategies for non-local XOR games. \textit{Quantum Information and Computation} 16 (13-14): 1191-1211 (2016). https://doi.org/10.26421/QIC16.13-14-6.

% Ostrev, D. The structure of nearly-optimal quantum strategies for the $\mathrm{CHSH(n)}$ XOR games. \textit{Quantum Information and Computation} \textbf{16}(13-14): 1191-1211 (2016).

\bigskip

\noindent [38] Ostrev, D. Composable, Unconditionally Secure Message Authentication without any Secret Key. \textit{IEEE International Symposium on Information Theory} 10 (1109): 622–626 (2019). https://doi.org/10.1109/ISIT.2019.8849510

\bigskip

% https://doi.org/10.26421/QIC16.13-14-6

\noindent [39] Paine, A.E. and Elfving, V.E. and Kyriienko, O. Quantum Kernel Methods for Solving Differential Equations. \textit{Physical Review A} 107: 032428 (2023). https://doi.org/10.1103/PhysRevA.107.032428. 

% Paine, A.E., Elfving, V.E., and Kyriienko, O. Quantum Kernel Methods for Solving Differential Equations. \textit{Physical Review A} \textbf{107}, 032428 (2023).

\bigskip

% https://doi.org/10.1103/PhysRevA.107.032428

\noindent [40] Paudel, H.P., Syamlal, M., Crawford, S.E., Lee, Y-L, Shugayev, R.A., Lu, P., Ohodnicki, P.R., Mollot, D., Duan, Y. \textit{Quantum Computing and Simulations for Energy Applications: Review and Perspective. ACS Eng. Au}: 3 151-196 (2022). $\mathrm{https://doi.org/10.1021/ac}$$\mathrm{sengineeringau.1c00033}$.

% Paudel, H.P., Syamlal, M., Crawford, S.E., Lee, Y-L, Shugayev, R.A., Lu, P., Ohodnicki, P.R., Mollot, D., Duan, Y. Quantum Computing and Simulations for Energy Applications: Review and Perspective. \textit{ACS Eng. Au} \textbf{3}: 151-196 (2022).

\bigskip

% https://doi.org/10.1021/acsengineeringau.1c00033 

\noindent [41] Przhiyalkovskiy, Y.V. Quantum process in probability representation of quantum mechanics. \textit{Journal of Physics A: Mathematical and Theoretical} 55: 085301 (2022). https://doi.org/10.1088/1751-8121/ac4b15.

% Przhiyalkovskiy, Y.V. Quantum process in probability representation of quantum mechanics. \textit{Journal of Physics A: Mathematical and Theoretical} \textbf{55}, 085301 (2022).

\bigskip

\noindent [42] Perc, M. Statistical physics of human cooperation. \textit{Physics Reports} \textbf{687}: 1-51 (2017). $\mathrm{https://papers.ssrn}$ $\mathrm{.com/sol3/papers.cfm?abstract_id=2972841}$.

\bigskip

\noindent [43] Ravishankar Ramanathan, R., Augusiak, R., and Murta, G. Generalized XOR games with $d$ outcomes and the task of nonlocal computation. \textit{Phys. Rev. A} \textbf{93}, 022333 (2016). $\mathrm{https://doi.org/10.1103/PhysRevA}$ $\mathrm{.93.022333}$.

\bigskip

\noindent [44] Rigas, P. Optimal, and approximately optimal, quantum strategies for $\mathrm{XOR^{*}}$ and $\mathrm{FFL}$ games. \textit{arXiv: 2311.12887} (2023), submitted.

\bigskip

% https://doi.org/10.1088/1751-8121/ac4b15 

\noindent [45] Rigas, P. Variational quantum algorithm for measurement extraction from the Navier-Stokes, Einstein, Maxwell, B-type, Lin-Tsien, Camassa-Holm, DSW, H-S, KdV-B, non-homogeneous KdV, generalized KdV, KdV, translational KdV, sKdV, B-L and Airy equations, shortened version including simulation results from only 3 PDEs submitted (2025).

% Rigas, P. Variational quantum algorithm for measurement extraction from the Navier-Stokes, Einstein, Maxwell, B-type, Lin-Tsien, Camassa-Holm, DSW, H-S, KdV-B, non-homogeneous KdV, generalized KdV, KdV, translational KdV, sKdV, B-L and Airy equations. \textit{arXiv: 2209.07714 v3} (2022).

\bigskip

\noindent [46] Rigas, P. Quantum error bounds, optimality, and duality gaps for multiplayer XOR, XOR*, compiled XOR, XOR*, and strong parallel repetition of XOR, XOR*, and FFL games. arXiv:2505.06322, submitted (2025).

\bigskip

\noindent [47] Roscika, M., Mazurek, P., Grudka, A., Horodecki, M. Generalized XOR non-locality games with graph description on a square lattice. \textit{Journal of Phys A: Math. Theor.} \textbf{53} 265302 (2020). $\mathrm{https://doi.org/10.1088/1751}$ $\mathrm{-8121/ab8f3e}$

\bigskip

\noindent [48] Slofstra, W.Lower bounds on the entanglement needed to play XOR non-local games \textit{Journal of Mathematical Physics} 52 (10): 102202 (2011). https://doi.org/10.1063/1.3652924.

\bigskip

% https://doi.org/10.1063/1.3652924

\noindent [49] van Dam, W. and Sasaki, Y. Quantum Algorithms for Problems in Number Theory, Algebraic Geometry, and Group Theory. \textit{Diversities in Quantum Computation and Quantum Information}: 79-105 (2012). https://doi.org/10.1142/9789814425988$\mathrm{-}$0003.

% van Dam, W. and Sasaki, Y. Quantum Algorithms for Problems in Number Theory, Algebraic Geometry, and Group Theory. \textit{Diversities in Quantum Computation and Quantum Information}, 79-105 (2012).

\bigskip

% https://doi.org/10.1142/9789814425988_0003

\noindent [50] Wang, Y. and Krstic, P.S. Multistate Transition Dynamics by Strong Time-Dependent Perturbation in NISQ era. \textit{J. Phys.} Commun.7: 075004 (2023). https://doi.org/10.1088/2399-6528/ace67a.

% Wang, Y., Krstic, P.S. Multistate Transition Dynamics by Strong Time-Dependent Perturbation in NISQ era. J. Phys. Commun. 7 075004.

\bigskip

% https://doi.org/10.1088/2399-6528/ace67a

\noindent [51] Zhao, L. and Zhao, Z. and Rebentrost, P. and Fitzsimons, J. Compiling basic linear algebra subroutines for quantum computers,
 \textit{Quantum Machine Intelligence} 3: 21 (2021). https://doi.org/10.1007/s42484-021-00048-8"

\end{document}